\documentclass[12pt]{article}
\usepackage{amssymb}
\usepackage{amsfonts}
\usepackage{amsmath}
\usepackage{eurosym}
\usepackage{makeidx}
\usepackage{amssymb}
\usepackage{times}
\usepackage{anysize}
\usepackage{tikz-cd,mathtools}
\usepackage{cite}
\usepackage{tikz-feynman,contour}

\marginsize{2cm}{2cm}{1cm}{1cm}
\newtheorem{theorem}{Theorem}[section]

\newtheorem{corollary}[theorem]{Corollary}

\newtheorem{definition}[theorem]{Definition}

\newtheorem{lemma}[theorem]{Lemma}

\newtheorem{proposition}[theorem]{Proposition}
\newtheorem{remark}[theorem]{Remark}

\newenvironment{proof}[1][Proof]{\noindent\textbf{#1.} }{\ \rule{0.5em}{0.5em}}
\input{tcilatex}

\begin{document}

\title{Scattering and Pairing by Exchange Interactions}
\author{J.-B. Bru \and W. de Siqueira Pedra \and A. Ramer dos Santos}
\date{\today }
\maketitle

\begin{abstract}
Quantum interactions exchanging different types of particles play a pivotal r%
\^{o}le in quantum many-body theory, but they are not sufficiently
investigated from a mathematical perspective. Here, we consider a system
made of two\ fermions and one boson, in order to study the effect of such an
off-diagonal interaction term, having in mind the physics of cuprate
superconductors. Additionally, our model also includes a generalized Hubbard
interaction (i.e., a general local repulsion term for the fermions).
Regarding pairing, exponentially localized dressed bound fermion pairs are
shown to exist and their effective dispersion relation is studied in detail.
Scattering properties of the system are derived for two channels: the
unbound and bound pair channels. We give particular attention to the regime
of very large on-site (Hubbard) \ repulsions, because this situation is
relevant for cuprate superconductors. \bigskip


\end{abstract}

\tableofcontents%

\section{Introduction\label{introduction}}

\subsection{Exchange Interactions and High-Tc Superconductivity}

\noindent \textbf{Exchange interactions in Mathematical Physics. }%
Off-diagonal interaction terms of the form 
\begin{equation}
B^{\ast }A+A^{\ast }B\ ,  \label{off-diagonal}
\end{equation}%
with $A,B$ being two monomials of annihilation operators of two species ($a$%
) and ($b$) of quantum particles, play a pivotal r\^{o}le in the rigorous
understanding of quantum many-body systems at low temperatures. Such terms
are also named \textquotedblleft exchange\textquotedblright\ terms, because
they encode (quantum) processes destroying a set of particles of one specie
to create another kind of particles.

For instance, for the Bogoliubov model, an off-diagonal term of the form 
\begin{equation}
\sum_{k}f_{1}\left( k\right) \left( b_{k}^{\ast }b_{-k}^{\ast }a^{2}+\left(
a^{\ast }\right) ^{2}b_{k}b_{-k}\right) \ ,\qquad f_{1}\left( k\right) \geq
0\ ,  \label{off-diagonal terms bogo}
\end{equation}%
exchanging two bosons ($a=b_{0}$) having zero momentum ($k=0$)\ with a pair
of boson having non-zero momentum of opposite sign ($b_{k\neq 0}$), is shown
in \cite{BruZagrebnov2} to imply a non-conventional Bose condensation. Made
of dressed bound pairs of (zero-momentum) bosons, the non-conventional
condensate is structurally different from the Bose-Einstein condensate of
the ideal Bose gas. In particular, it must be depleted to take advantage of
the effective attraction induced by the exchange interaction (\ref%
{off-diagonal terms bogo}). See \cite{BruZagrebnov3}. This is reminiscent of
liquid helium physics, where 100\% superfluid helium occurs at zero
temperature with only 9\% Bose condensate \cite%
{Dubnagroup1,Dubnagroup2,densite1,BoseGasbook,Griffin}. Off-diagonal
interaction terms (\ref{off-diagonal terms bogo}) are conjectured in \cite%
{bru3} to be relevant to explain the macroscopic behavior of weakly
interacting Bose gases.

Another example from quantum statistical mechanics is given by the
spin-boson model within the so-called \textquotedblleft rotating wave
approximation\textquotedblright . In this approximation the model has terms
of the form%
\begin{equation*}
\sum_{k}f_{2}\left( k\right) \left( b_{k}^{\ast }\sigma _{-}+\sigma
_{+}b_{k}\right) \ ,\qquad f_{2}\left( k\right) \geq 0\ ,
\end{equation*}%
with $\sigma _{\pm }=\sigma _{x}\pm \sigma _{y}$ ($\sigma _{x},\sigma _{y}$
being Pauli matrices) and $b_{k}$ being the annihilation operator of a
boson. Note that $\sigma _{-}$ ($\sigma _{+}$) can be related to the
annihilation (creation) operator $a$ ($a^{\ast }$) of a fermion, via a
so-called Jordan-Wigner transformation. Such off-diagonal terms make
impossible the diagonalization of the quantum Hamiltonian with usual
methods. In particular, the impact of these interaction terms on the
properties of the model is expected to be major. For a general presentation
of spin-boson models, see, e.g., \cite[Introduction and Section 2.3]%
{spinboson1}.

More recently, using the Hubbard model with nearest neighbor interaction
near its Hartree-Fock ground state, Bach and Rauch demonstrate \cite%
{Bach-Rauch} that interaction terms of the form 
\begin{equation}
\sum_{x,y}\sum_{\mathrm{s},\mathrm{t}\in \{\uparrow ,\downarrow
\}}f_{3}\left( x-y\right) \left( b_{x,\mathrm{s}}^{\ast }b_{y,\mathrm{t}%
}^{\ast }a_{y,\mathrm{t}}a_{x,\mathrm{s}}+a_{x,\mathrm{s}}^{\ast }a_{y,%
\mathrm{t}}^{\ast }b_{y,\mathrm{t}}b_{x,\mathrm{s}}\right) \ ,\qquad
f_{2}\left( x-y\right) \geq 0\ ,  \label{off-diagonal terms bach}
\end{equation}%
exchanging fermions inside ($a$) and outside ($b$) of the Fermi surface are
the only ones that can prevent from getting \emph{uniform}\footnote{%
We mean a relative bound that is uniform with respect to the length of the
discrete $d$-dimensional torus where the Hubbard model is defined.} relative
bounds of the effective interaction with respect to the effective kinetic
energy. See \cite[Theorems III.1, III.2 and III.3]{Bach-Rauch} for more
details. In other words, (\ref{off-diagonal terms bach}) should again have a
drastic impact on the corresponding quantum many-body system.\bigskip

\noindent \textbf{Three-body fermion-boson exchange interactions.} In the
present paper, for a fairly general function $\upsilon :\mathbb{Z}%
^{2}\rightarrow \mathbb{R}$ we study the effect of the off-diagonal
interaction term 
\begin{equation}
{\sum\limits_{x,y}}\upsilon \left( x-y\right) \left( c_{y}^{\ast
}\,b_{x}+b_{x}^{\ast }c_{y}\right) \ ,  \label{model off diag}
\end{equation}%
where $b_{x}$ is the annihilation operator of a spinless boson on the site $%
x $ of the two-dimensional (square) lattice $\mathbb{Z}^{2}$, while $c_{y}$
represents the annihilation of a fermion pair of zero total spin, the two
components of which are spread around the lattice position $y\in \mathbb{Z}%
^{2}$. See Fig. \ref{figure0}. 
\begin{figure}[!hbtp]
\begin{center}
\begin{tikzpicture}
\begin{feynman}
\vertex (a) {$f$};
\vertex [below right=of a] (b);
\vertex [below left =of b] (c)  {$f$};
\vertex [right =of b] (d)  {$b$};
\vertex [right =of d] (e)  {$b$};
\vertex [right =of e] (f) ;
\vertex [above right =of f] (g) {$f$};;
\vertex [below right =of f] (h) {$f$};;

\diagram* {
(a) -- [fermion] (b),
(c) -- [fermion] (b),
(b) -- [photon, edge label=$\upsilon$] (d),
(e) -- [photon, edge label=$\upsilon$] (f),
(f) -- [fermion] (g),
(f) -- [fermion] (h)
};
\end{feynman}

\end{tikzpicture}
\end{center}
\caption{Illustration of fermion-boson exchange interactions in the form of two
Feynman diagrams. In theoretical physics, a Feynman diagram visually represents the
mathematical expressions that describe the behavior and interactions of
quantum particles. In the example on the left, the two arrows indicate that
two fermions, named $(f)$, \textquotedblleft collide\textquotedblright\ to
create a new particle, the boson $(b)$. The oscillating line is generally used
to describe an interaction with a mediator, which can be seen by combining
the two diagrams: two fermions $(f)$ interact to produce a boson, which
annihilates again to produce two fermions $(f)$. This can lead to an effective
interaction between fermions. In particular, this process could
produce a pair of fermions ($f-f$) bonded by the exchange of a bosonic field
($b$), according to the coupling function $\upsilon $. This is typically what we are going to show. Note that the opposite combination can also be made: a boson $(b)$ is
destroyed to create two fermions $(f)$, which annihilate to recreate a boson $(b)
$. This does not really create an interaction as such, but a kinetic term,
or seen another way, a self-interaction on the boson $(b)$. The combination of two diagrams
refers to a perturbative approach of second order, but we can also combine
several of the same diagrams (perturbative approach of order $n$). Note, however, that no such perturbative argument is used here.}
\label{figure0}
\end{figure}%

Note that the opposite combination can also be made: a boson $b$ is
destroyed to create two fermions $f$, which annihilate to recreate a boson $%
b $. This does not really create an interaction as such, but a kinetic term,
or seen another way, a self-interaction. The combination of two diagrams
refers to a perturbative approach of second order, but we can also combine
several of the same diagrams (perturbative approach of order $n$).

The purely fermionic part of the considered model corresponds to the \emph{%
extended} Hubbard Hamiltonian, as used in the context of ultracold atoms,
ions, and molecules \cite{extended-hubbard0}, while the purely bosonic
component refers to an ideal gas, i.e., it has only a kinetic part (or
\textquotedblleft hopping term\textquotedblright ), without interbosonic
interactions. Because of the fermionic part, which is not exactly
diagonalizable, the behavior of the full quantum many-body system, outside
perturbative regimes, is almost inaccessible with the mathematical tools at
our disposal.

We thus consider only a three-body problem, by restricting the model to the
sector of one boson and two fermions of opposite spins. In fact, the system
restricted to this particular sector is very interesting, both
mathematically and physically. Note that such sector restrictions in Fock
spaces are also performed for the study of the Pauli-Fierz and Nelson models 
\cite{Marco1,Marco2,Marco3} in non-relativistic Quantum Field Theory (QFT).
\bigskip

\noindent \textbf{Physical context: High-}$T_{c}$\textbf{\ superconductivity
of cuprates.} Physically, the model is related to cuprate superconductors%
\footnote{%
You can take for instance the cuprate $\mathrm{La}_{2}\mathrm{CuO}_{4}$,
which is a Mott insulator with an antiferromagnetic phase at low
temperature. As with semiconductors, it is doped with atoms like $\mathrm{Sr}
$ or $\mathrm{Ba}$, which add a few charge carriers (in this case, holes).
Then, with moderate doping $x$, the material becomes superconducting at low
temperatures. This is the meaning of the chemical formulae $\mathrm{La}_{2-x}%
\mathrm{Sr}_{x}\mathrm{CuO}_{4}$ and $\mathrm{La}_{2-x}\mathrm{Ba}_{x}%
\mathrm{CuO}_{4}$, $x$ being a small number characterizing the cuprate
doping. See also Section \ref{Section physics}.}, like for instance $\mathrm{%
La}_{2-x}\mathrm{Sr}_{x}\mathrm{CuO}_{4}$ (LaSr 214 or LSCO) and $\mathrm{La}%
_{2-x}\mathrm{Ba}_{x}\mathrm{CuO}_{4}$. It is known \cite%
{Saxena,tinpou1,unitcoherence} that in such crystals charge transport occurs
within two-dimensional isotropic layers of copper oxides. This is why we
consider here quantum particles on $2$-dimensional lattices $\mathbb{Z}^{2}$.

A convincing microscopic mechanism behind superconductivity at high critical
temperature is still lacking even after almost four decades of intensive
theoretical and experimental studies. See Section \ref{Section physics} for
more details. Many physicists believe that the celebrated Hubbard model
could be pivotal, one way or another, in order to get a microscopic theory
of high-temperature superconductivity, but many alternative explanations or
research directions have also been considered in theoretical physics. For
some of the more popular models for cuprate superconductors, see, e.g., \cite%
[Chap. 7]{tinpou1}.

In many theoretical approaches to this problem, the existence of polaronic
quasiparticles in relation with the very strong Jahn-Teller (JT) effect
associated with copper ions is neglected, as stressed in \cite[Part VII]%
{Jan-teller}. The role of polarons is however highlighted in \cite{BM86},
since the JT effect actually led to the discovery of superconductivity in
cuprates in 1986. See \cite[p. 2]{M07} or \cite{65,AZencore}.

Our theoretical approach differs from most popular ones, being based on the
existence of JT bipolarons in copper oxides, as is discussed in the
literature \cite{SS90} at least as early as 1990. The physics behind this
approach is explained in detail in \cite{articulo2}, where a simplified
version of the model studied here is considered. In our microscopic model
for cuprate superconductors, as presented in \cite{articulo,articulo2}, the
bosonic operator $b_{x}$ ($b_{x}^{\ast }$) in (\ref{model off diag}) refers
to the annihilation (creation) of a JT bipolaron, whereas the fermionic one $%
c_{y}$ ($c_{y}^{\ast }$) annihilates (creates) a fermion pair, which is
reminiscent of Cooper pairs in conventional superconductivity.\bigskip

\noindent \textbf{Bipolaronic pairing mechanisms and cuprate
superconductivity.} As in the present paper, no ad hoc assumptions, in
particular concerning anisotropy, are made in \cite{articulo2}. In fact, 
\cite{articulo2} proves that unconventional pairing may occur, breaking
spontaneously discrete symmetries of the model, like the $d$-wave pairing,
whose wave function is antisymmetric with respect to $90^{\circ }$%
-rotations. It turns out that electrostatic (screened Coulomb) repulsion is
crucial for such unconventional pairings, which are meanwhile shown to be
concomitant with a strong depletion of superconducting pairs.

Notice that the results of \cite{articulo2} are coherent with experimental
observations on the cuprate LaSr 214: The coherence length at optimal doping
and the $d$-wave pair formation in the pseudogap regime, i.e., at
temperatures much higher than the superconducting transition temperature,
are predicted in good accordance with experimental data. In addition to the $%
d$-wave pairing and the high-temperature pseudogap regime, the model
considered here also captures another very special feature of high-Tc
cuprate superconductors, namely the density waves \cite{denswave}. For more
details, see also Section \ref{Section physics}.

In fact, it is shown in \cite[Section 4.1]{articulo} that three-body
fermion-boson exchange interactions, like the one studied in this paper,
imply an effective fermion-fermion interaction. Then, by considering the
mean-field limit of it, which corresponds to taking couplings (\ref{model
off diag}) that are very localized in momentum space \cite[Section 4.2]%
{articulo}, it was rigorously proven \cite{Antoniothesis} that, below the
critical temperature, the equilibrium states of the (purely fermionic)
associated \emph{many}-body Hamiltonian exhibit periodic modulation in space
of the charge density, even incommensurate with respect to the lattice
spacing.

\subsection{Mathematical Results}

\noindent \textbf{Previous results.} To our knowledge, the model considered
here has not been studied mathematically, apart from our own articles \cite%
{articulo2,articulo} published in recent years. See also the Ph.D. thesis 
\cite{Antoniothesis}. Mathematical studies for explicit exchange interaction
terms are mainly those presented above. As far as we know, concerning its
physical interpretation regarding cuprate superconductivity, our approach
has also never been considered by other physicists and we therefore doubt
that any theoretical results in this direction exist in the literature. For
more details, see the introductory discussions in \cite{articulo2}, which
give a concise overview of theories of high-temperature superconductivity.

Mathematically, the present paper improves \cite{articulo2,articulo} to get
more complete and general rigorous results, including, among other things,
extended Hubbard interactions and scattering properties. While \cite%
{articulo2,articulo} focus only on the ground state energy and the
unconventional pairings in the limit of large Hubbard interactions, here we
provide the full spectral properties of the corresponding Hamiltonian. In
particular we study in depth the effective dispersion relation associated
with \emph{dressed} bound fermion pairs. It confirms that off-diagonal
interactions of the form (\ref{off-diagonal}) produce bounded states by
reducing the energy of the system, similar to \cite{BruZagrebnov3}, possibly
with a spectral gap.

This was already done in \cite{articulo2,articulo}, but only for usual (%
\emph{non}-extended) Hubbard interactions and \emph{one-range} creation /
annihilation operators of fermion pairs. Even in this specific case, the
dispersion relation of dressed bound fermion pairs was analyzed only to a
level of detail enough to deduce unconventional pairings near the ground
state. By contrast, in the present paper other important properties of the
dispersion relation, like its regularity, are studied for the first time and
in a more general framework.

Last but not least, the localization of dressed bound fermion pairs or the
scattering properties of the model have not been studied before.\bigskip

\noindent \textbf{Localized dressed bound fermion pairs.} Using
Combes-Thomas estimates we show, among other things, that the dressed bound
fermion pairs are localized, in the sense that the fermion-fermion
correlation decays very fast in space. Group velocities and tensor masses of
dressed bound fermion pairs are also shown to exist under very natural
conditions on the (absolutely summable function) $\upsilon :\mathbb{Z}%
^{2}\rightarrow \mathbb{R}$ appearing in (\ref{model off diag}).

In fact, our analysis allows one to accurately understand which features of
the exchange strength function $\upsilon $ can strengthen the stability of
the dressed bound fermion pairs. For instance, $\upsilon $ has to be
sufficiently strong and localized in Fourier space in order to get a
sufficiently strong \textquotedblleft gluing effect\textquotedblright .
Additionally, the boson should be heavier than two fermions.

Notice that this second condition is consistent with the physical
interpretation that the boson is a bipolaron, which is known to be
(effectively) much heavier than the fermions (electrons or holes), in
superconducting cuprates. Observe additionally that the very large mass of
bipolarons (and polarons, in general), is one of the main arguments used to
discredit theoretical approaches based on bipolarons, because it is known
from experiments that the charge carriers in superconducting cuprates have
an effective mass comparable to that of electrons and holes.

In fact, we prove that the effective mass of bound pairs mainly depends on
the properties of the function $\upsilon $, that encodes the fermion-boson
exchange processes, but not much on the mass of the boson itself. This issue
is discussed in \cite{articulo2}, in detail. See also the discussion at the
end of Section \ref{sect disp rel bound pairs}. That is why we are
interested in results concerning the mass tensor for bound pairs and we
think we provide here a convincing solution for the \textquotedblleft mass
paradox\textquotedblright\ related to bipolaronic pairing mechanisms in the
microscopic theory of cuprate superconductors.\bigskip

\noindent \textbf{Relationship with the enhanced binding of QFT.} The
formation of dressed bonded fermion pairs as described above is reminiscent
of what is known as \emph{enhanced binding} in Quantum Field Theory (QFT).
For more details on this phenomenon we recommend the lecture notes \cite%
{enhanced-binding}, where it is well explained in the context of
non-relativistic QFT. See also the references therein.

For example, the Pauli-Fierz model, which refers to non-relativistic quantum
charge particles interacting with a massless quantized radiation field
(photons), can have at low energies a dressed particle with an effective
mass bigger than the non-interacting one, leading to the existence of a
ground state for the model. A similar fact occurs in the Nelson model, in
which $N$ quantum particles interact linearly with a field of photons (or
mesons). The formation of such dressed particles is a direct consequence of
the bosonic field acting as mediator of a force.

Indeed, in both cases, the model involves a sum of interaction terms of the
form $\psi _{k}\otimes b_{k}+\bar{\psi}_{k}\otimes b_{k}^{\ast }$, coupling
the $N$-body quantum system with a spinless boson field of momentum $k$ via
annihilation/creation operators $b_{k},b_{k}^{\ast }$. Note that in this
case there is no transformation of particles of one type into another, as in
the exchange interactions described above, but both cases are still similar,
especially as we are carrying out our analysis in the sector with only two
fermions and one boson. This makes the comparison quite relevant, even if
the model and mathematical methods considered here have essential
differences as compared to the previous ones.\bigskip

\noindent \textbf{Scattering properties of the model.} We also study here
scattering properties of the three-body model in two channels, the \emph{%
unbound} and \emph{bound} pair channels:

\begin{itemize}
\item The unbound pair (scattering) channel corresponds to the wave and
scattering operators with respect to fermionic part, respectively defined
via the strong limits 
\begin{equation*}
W^{\pm }\doteq s-{\lim\limits_{t\rightarrow \pm \infty }}\mathrm{e}^{it%
\mathrm{H}}\mathrm{e}^{-it\mathrm{H}_{f}}P_{\mathrm{ac}}\left( \mathrm{H}%
_{f}\right) \text{\qquad and\qquad }S\doteq \left( W^{+}\right) ^{\ast
}W^{-}\ ,
\end{equation*}%
where $\mathrm{H}_{f}$ is a generic, purely fermionic Hamiltonian
representing free fermions that do not interact with any bosonic field, $%
\mathrm{H}$ is the Hamiltonian of the full model and $P_{\mathrm{ac}}(%
\mathrm{H}_{f})$ is the orthogonal projection onto the absolutely continuous
space of $\mathrm{H}_{f}$. It refers to the case in which two fermions start
far apart from each other and only experience a very weak repulsion force
due to the extended Hubbard interaction, while the probability that they
bind together to form a boson is very small. It In this situation, we show
that two (almost) freely propagating fermions in the distant past can come
together and interact with one another, either via the repulsive
electrostatic force or by exchanging a boson, and then propagate away, again
freely in the distant future. In this channel, the scattering matrix can be
explicitly computed via convergent (Dyson) series, making in particular the
study of the scattering effect of the fermion-boson-exchange interaction (%
\ref{off-diagonal}) technically uncomplicated.

\item The bound pair (scattering) channel corresponds to the time evolution $%
\mathrm{e}^{it\mathrm{H}}\mathfrak{P}$, $t\in \mathbb{R}$, where $\mathrm{H}$
is again the Hamiltonian of the full model and $\mathfrak{P}$ is an isometry
from the $L^{2}$-functions on the Brillouin zone to the subspace associated
with the fiber bound states of $\mathrm{H}$. We show in particular that 
\begin{equation*}
\mathrm{e}^{it\mathrm{H}}\mathfrak{P}=\mathfrak{P}\mathrm{e}^{itM_{\mathrm{E}%
\left( \cdot \right) }}\ ,\qquad t\in \mathbb{R}\ ,
\end{equation*}%
with $M_{\mathrm{E}\left( \cdot \right) }$ being some multiplication
operator given by the dispersion relation $k\mapsto \mathrm{E}\left(
k\right) $ characterizing the (fiber) bound states at fixed quasi-momentum
in the normalized Brillouin zone $\mathbb{T}^{2}\doteq \lbrack -\pi ,\pi
)^{2}$. In terms of wave operators, it follows that%
\begin{equation*}
W^{\pm }\doteq s-{\lim\limits_{t\rightarrow \pm \infty }}\mathrm{e}^{it%
\mathrm{H}}\mathfrak{P}\mathrm{e}^{-itM_{\mathrm{E}\left( \cdot \right) }}P_{%
\mathrm{ac}}\left( M_{\mathrm{E}\left( \cdot \right) }\right) =\mathfrak{P}%
P_{\mathrm{a}\mathrm{c}}\left( M_{\mathrm{E}\left( \cdot \right) }\right) \ ,
\end{equation*}%
which gives a scattering operator equal to%
\begin{equation*}
S\doteq \left( W^{+}\right) ^{\ast }W^{-}=P_{\mathrm{a}\mathrm{c}}\left( M_{%
\mathrm{E}\left( \cdot \right) }\right) \ .
\end{equation*}%
It refers to the case in which dressed bound fermion pairs are formed. In
contrast with the first channel, now there is a non-negligible bosonic
component related with the exchanged boson that \textquotedblleft
glues\textquotedblright\ the two fermions together. We prove that those
(spatially localized) dressed bound fermion pairs effectively move like a
free (quantum spinless) particle. In this case, strictly speaking in the
physical sense, there is no scattering and the pairs evolve freely in space,
governed by an effective dispersion relation, the Fourier transform of which
is the effective hopping strength for the (spatially localized) dressed
bound pairs.
\end{itemize}

\noindent \textbf{Composite system at strong on-site Hubbard repulsions.} We
additionally prove that all these properties hold also true in the limit of
large on-site fermionic repulsions, provided that two fermions on two
different lattice sites can interact via the fermion-boson exchange
interaction. It refers to a hard core limit, preventing two fermions from
occupying the same lattice site.

For cuprate superconductors, it is an important issue addressed and answered
here, because of the undeniable experimental evidence of very strong on-site
Coulomb repulsions in cuprates, leading to the universally observed Mott
transition at zero doping \cite{Imada,Nature2015}.

\subsection{Concluding Remarks and Structure of the Paper}

To conclude, the mathematical properties of the model studied in the present
work are well understood and, as a consequence, the model can serve as a
prototypical example of a quantum system including exchange interaction
terms of the form (\ref{off-diagonal}). From a physics viewpoint, it is also
interesting, since dressed bound fermion pairs are good candidates for
superconducting charge carriers in cuprate superconductors, as advocated in 
\cite{articulo2}.

More specifically, our main results are Theorems \ref{Maintheorem1}, \ref%
{Maintheorem2}, \ref{Maintheorem3}, \ref{existence of a dispersion relation
with low energy copy(2)}, \ref{existence of a kato wave operator for the
system} and \ref{kato's wave operator for d-wave pairing copy(1)}. The paper
is organized as follows: Section \ref{Setup of the Problem} explains in
detail the model, while Section \ref{Main Results} gives the main results.
Technical outcomes, along with all their proofs, are gathered in Section \ref%
{Technical Results}. Section \ref{Appendix} is an appendix that gathers
important standard mathematical results used here, an overview of cuprate
physics for non-physicists, as well as the Fock-space formalism, in order to
make the article self-contained and accessible to a wide audience.

\begin{remark}[$d$-dimensional lattices]
\mbox{ }\newline
Our study focuses on two-dimensional lattice systems because of their
application to the superconductivity of cuprates and, in particular, their $%
d $-wave symmetry. However, it can also be done at arbitrary dimension $%
d\geq 1 $ provided the coupling functions used, i.e., $\mathrm{u},\mathfrak{p%
}_{1},\mathfrak{p}_{2},\upsilon :\mathbb{Z}^{d}\rightarrow \mathbb{R}_{0}^{+}
$ below, stays absolutely summable. It is also important that the Fourier
transforms $\hat{\upsilon}$, $\mathfrak{\hat{p}}_{1}$ and $\mathfrak{\hat{p}}%
_{2}$ of the functions $\upsilon $, $\mathfrak{p}_{1}$ and $\mathfrak{p}_{2}$
remain real-valued\footnote{%
A real-valued absolutely summble function $f$ on $\mathbb{Z}^{d}$ has a
real-valued Fourier transform iff $f(-z)=f(z)$. Considering two-dimensional
systems that are invariant under $90^{\circ }$-degree rotations (like we
did, because of cuprates), this property is always true and has not to be
additionally imposed.} continuous functions on the $d$-dimensional torus $%
\mathbb{T}^{d}$.
\end{remark}

\begin{remark}[Notation]
\label{Notation}\mbox{ }\newline
For any normed vector space $\mathcal{X}$ over $\mathbb{C}$, we omit the
subscript $\mathcal{X}$ to denote its norm $\Vert \cdot \Vert \equiv \Vert
\cdot \Vert _{\mathcal{X}}$, unless there is any risk of confusion. Mutatis
mutandis for the scalar product $\langle \cdot ,\cdot \rangle \equiv \langle
\cdot ,\cdot \rangle _{\mathcal{X}}$ in Hilbert spaces. As is usual, $%
\mathcal{B}(\mathcal{X},\mathcal{Y})$ denotes the set of bounded (linear)
operators $\mathcal{X}\rightarrow \mathcal{Y}$ between two normed spaces $%
\mathcal{X}$ to $\mathcal{Y}$. If $\mathcal{X}=\mathcal{Y}$, $\mathcal{B}(%
\mathcal{X})\equiv \mathcal{B}(\mathcal{X},\mathcal{X})$ and its (operator)
norm and its identity are respectively denoted by $\Vert \cdot \Vert _{%
\mathrm{op}}\equiv \Vert \cdot \Vert _{\mathcal{B}(\mathcal{X})}$ and $%
\mathfrak{1}_{\mathcal{X}}\equiv \mathfrak{1}$. $\mathbb{R}_{0}^{+}$ denotes
the set of positive real numbers including zero, whereas $\mathbb{R}%
^{+}\doteq \mathbb{R}_{0}^{+}\backslash \{0\}$ is the set of strictly
positive real numbers.
\end{remark}

\section{Setup of the Problem\label{Setup of the Problem}}

\subsection{Background Lattice}

Copper oxide superconductors have a relatively complex three-dimensional
lattice structure. However, they always contain parallel two-dimensional
layers of copper ($\mathrm{Cu}^{++}$) and oxygen ($\mathrm{O}^{--}$) ions.
These $\mathrm{CuO}_{2}$ layers are essential to understanding
low-temperature superconducting properties, because the (superconducting)
charge transport takes place within the layers. This is explained in \cite%
{Saxena,tinpou1,unitcoherence}. Considering a weak inter-layer interaction
might also help to increase prediction accuracy, but charge transport
between each $\mathrm{CuO}_{2}$ layer or, more generally, in the direction
orthogonal to each layer remains negligible\footnote{%
The superconducting coherence length is much smaller in this orthogonal
direction than in the parallel planes made of copper and oxygen ions.}. Each 
$\mathrm{CuO}_{2}$ layer generally has the symmetries of the square. In
other words, it is invariant under the group $\{0,\pi /2,\pi ,3\pi /2\}$
generated by $90^{\circ }$-degree rotations. See, e.g., \cite[Section 9.1.2]%
{Tsuei2003}, \cite[Section 2.3]{Saxena} and \cite[Section 6.3.1]%
{unitcoherence}. This is an important symmetry property that we keep in mind
throughout our study.

Having in mind these physical observations on cuprates, we consider here
quantum particles on lattices $\mathbb{Z}^{2}$. It means in particular that,
(disregarding internal degrees of freedom of the quantum particles, like
their spin) the (separable) Hilbert space $\ell ^{2}(\mathbb{Z}^{2})$ is the
\textquotedblleft one-particle space\textquotedblright\ associated with the
physical system we are interested in. Its canonical orthonormal basis is $\{%
\mathfrak{e}_{x}\}_{x\in \mathbb{Z}^{2}}$:%
\begin{equation}
\mathfrak{e}_{x}\left( y\right) \doteq \delta _{x,y}\ ,\qquad x,y\in \mathbb{%
Z}^{2}\ ,  \label{e frac}
\end{equation}%
where $\delta _{\mathfrak{i},\mathfrak{j}}$ is the Kronecker delta.

\subsection{Composite of two Fermions and one Boson\label{Hubbard
Hamiltonian with interaction term}}

We consider a system of two fermions (electrons or holes in cuprates) with
opposite spins interacting via the exchange of one boson in a two
dimensional square lattice. Physically, the boson that we have in mind in
cuprate superconductors is a spinless bipolaron, since the very strong
Jahn-Teller (JT) effect associated with copper ions is an important property
of such cuprates \cite{BM86,Jan-teller}. See Section \ref{Section physics}
for more details. However, the exchanged spinless boson could be of any
type, like a phonon or a spin wave, depending on the physical system and
mechanism one has in mind.\bigskip

\noindent \textbf{Hilbert Spaces.} All quantum particles possess an
intrinsic form of angular momentum known as spin, which is characterized by
a quantum number $\mathfrak{s}\in \mathbb{N}/2$ and a finite spin set%
\footnote{$\mathrm{S}$ represents the spectrum of the spin observable.} $%
\mathrm{S}\doteq \{-\mathfrak{s},-\mathfrak{s}+1,\ldots \mathfrak{s}-1,%
\mathfrak{s\}}\subseteq \mathbb{N}$. If $\mathfrak{s}\notin \mathbb{N}$ is
half-integer then the corresponding particles are named \emph{fermions}
while $\mathfrak{s}\in \mathbb{N}$ means by definition that we have \emph{%
bosons}. For example, photons or spinless bipolarons ($\mathfrak{s}=0$) are
bosons, while electrons ($\mathfrak{s}=1/2$) are fermions. In the latter
case, $\mathrm{S}\doteq \{-1/2,1/2\}$ and in physics, the spin set is always
written as $\mathrm{S}\equiv \{\uparrow ,\downarrow \}$ and we thus use this
completely standard notation. By the celebrated spin-statistics theorem,
fermionic wave functions are antisymmetric with respect to permutations of
particles, whereas the bosonic ones are symmetric.

Therefore, the one-particle Hilbert space for the fermions is $\ell ^{2}(%
\mathbb{Z}^{2}\times \{\uparrow ,\downarrow \})$, $\{\uparrow ,\downarrow \}$
being the usual spin set for electrons or holes, and, for two fermions we
hence use the Hilbert space 
\begin{equation*}
\mathfrak{h}_{f}\doteq \bigwedge\nolimits^{2}\ell ^{2}(\mathbb{Z}^{2}\times
\{\uparrow ,\downarrow \})\subseteq \mathfrak{F}_{-}\equiv \mathfrak{F}%
\left( \ell ^{2}(\mathbb{Z}^{2}\times \{\uparrow ,\downarrow \})\right)
\end{equation*}%
of antisymmetric functions\footnote{$\bigwedge\nolimits^{2}\ell ^{2}(\mathbb{%
Z}^{2}\times \{\uparrow ,\downarrow \})$ denotes the $2$-fold antisymmetric
tensor product of $\ell ^{2}(\mathbb{Z}^{2}\times \{\uparrow ,\downarrow \})$%
.}, which is a subspace of the fermionic ($-$) Fock space\footnote{%
I.e., $\mathfrak{F}_{-}\doteq \bigoplus_{n=0}^{\infty
}\bigwedge\nolimits^{n}\ell ^{2}(\mathbb{Z}^{2}\times \{\uparrow ,\downarrow
\})$.} $\mathfrak{F}_{-}$ associated with the one-particle Hilbert space $%
\ell ^{2}(\mathbb{Z}^{2}\times \{\uparrow ,\downarrow \})$. See Equation (%
\ref{Fock f}) below for the precise definition of $\mathfrak{F}_{-}$. The
one-particle Hilbert space of the spinless boson is $\ell ^{2}(\mathbb{Z}%
^{2})$, which can also be seen as a subspace of the bosonic ($+$) Fock space 
\begin{equation*}
\mathfrak{F}_{+}\equiv \mathfrak{F}\left( \ell ^{2}(\mathbb{Z}^{2})\right)
\end{equation*}%
associated with $\ell ^{2}(\mathbb{Z}^{2})$. See Equation (\ref{Fock b})
below for the precise definition of $\mathfrak{F}_{+}$. For a concise review
of bosonic and fermionic Fock spaces, as well as the corresponding
annihilation and creation operators, see Section \ref{Section Fock}.

We study here the effect of processes of annihilation of two fermions of
opposite spins to create a boson, which can conversely be annihilated to
create two new fermions. The Hilbert space associated with this composite
system, made of two fermions and one boson, is the direct sum $\mathfrak{h}%
_{f}\oplus \ell ^{2}(\mathbb{Z}^{2})$, and not the tensor product $\mathfrak{%
h}_{f}\otimes \ell ^{2}(\mathbb{Z}^{2})$. Note indeed that $\mathfrak{h}%
_{f}\oplus \ell ^{2}(\mathbb{Z}^{2})$ can naturally be identified\footnote{%
Denote the vacuum of the Fock space $\mathfrak{F}_{\pm }$ by $\Omega _{\pm }$
and define the mapping $\varsigma $ from $\mathfrak{h}_{f}\oplus \ell ^{2}(%
\mathbb{Z}^{2})$ to $\mathfrak{F}_{-}\otimes \mathfrak{F}_{+}$ by $\varsigma
\left( \varphi \oplus \psi \right) =\varphi \otimes \Omega _{-}+\Omega
_{+}\otimes \psi $. Then, observe that $\varsigma $ is an isometric linear
transformation from $\mathfrak{h}_{f}\oplus \ell ^{2}(\mathbb{Z}^{2})$ to $%
\mathfrak{F}_{-}\otimes \mathfrak{F}$.} with a subspace of $\mathfrak{F}%
_{-}\otimes \mathfrak{F}_{+}$. This fact already unveils the strong
interdependence of the bosonic and fermionic parts. For this reason, from
now on we rather use the term \textquotedblleft composite of two fermions
and one boson\textquotedblright\ instead of \textquotedblleft three-body
system\textquotedblright , in order to avoid any misinterpretation.\bigskip

\noindent \textbf{Fermionic Hamiltonian.} The fermionic part of the
(infinite volume) Hamiltonian of the composite is defined to be the
restriction $H_{f}\in \mathcal{B}(\mathfrak{h}_{f})$ of the formal
expression 
\begin{equation}
-{\frac{\epsilon }{2}}\sum_{s\in \{\uparrow ,\downarrow \},\ x,y\in \mathbb{Z%
}^{2}:|x-y|=1}a_{x,s}^{\ast }a_{y,s}+2\epsilon \sum\limits_{s\in \{\uparrow
,\downarrow \},\,x\in \mathbb{Z}^{2}}a_{x,s}^{\ast }a_{x,s}+\mathrm{U}%
\sum_{x\in \mathbb{Z}^{2}}n_{x,\uparrow }n_{x,\downarrow }+\sum_{x,z\in 
\mathbb{Z}^{2}}\mathrm{u}\left( z\right) n_{x,\uparrow }n_{x+z,\downarrow }
\label{Hamiltonian-f}
\end{equation}%
to the Hilbert space $\mathfrak{h}_{f}$. Here, $a_{x,s}$ ($a_{x,s}^{\ast }$)
denotes the annihilation (creation) operator acting on the fermionic Fock
space $\mathfrak{F}_{-}$ of a fermion at lattice position $x\in \mathbb{Z}%
^{2}$, the spin of which is $s\in \{\uparrow ,\downarrow \}$. As is usual, $%
n_{x,s}\doteq a_{x,s}^{\ast }a_{x,s}$ stands for the number operator of
fermions at lattice position $x\in \mathbb{Z}^{2}$ and spin $s\in \{\uparrow
,\downarrow \}$.

The parameter $\epsilon \in \mathbb{R}_{0}^{+}$ quantifies the hopping
amplitude of fermions. In high-$T_{c}$ superconductors \cite%
{Imada,Nature2015}, $\epsilon $ is expected to be much smaller than the
fermion-fermion interaction energy, more precisely the on-site repulsion
strength $\mathrm{U}\in \mathbb{R}_{0}^{+}$. The function $\mathrm{u}:%
\mathbb{Z}^{2}\rightarrow \mathbb{R}_{0}^{+}$, which represents the
fermion-fermion repulsion at all distances, is absolutely summable and
invariant with respect to $90^{\circ }$-rotations, i.e., 
\begin{equation}
\sum_{z\in \mathbb{Z}^{2}}\left\vert \mathrm{u}\left( z\right) \right\vert
<\infty \qquad \text{and}\qquad \mathrm{u}\left( x,y\right) =\mathrm{u}%
\left( -y,x\right) ,\qquad x,y\in \mathbb{Z}\ .  \label{summable1-U}
\end{equation}%
Clearly, one could set $\mathrm{U}=0$, by redefining the coupling function $%
\mathrm{u}:\mathbb{Z}^{2}\rightarrow \mathbb{R}_{0}^{+}$. It is however
convenient to have a separate parameter $\mathrm{U}\in \mathbb{R}_{0}^{+}$
for the on-site repulsion, because we shall later on consider the
\textquotedblleft hard-core limit\textquotedblright\ $\mathrm{U}\rightarrow
\infty $ for some fixed coupling function $\mathrm{u}$.\bigskip

\noindent \textbf{Extended Hubbard interactions.} The above fermion-fermion
interactions have been extensively studied in condensed matter physics
during the last decade, in particular for two dimensional systems. For
non-zero functions $\mathrm{u}$, they are named \emph{extended} Hubbard
interactions and they can drastically change the behavior of the system, as
compared to the zero-range case (usual Hubbard interaction, $\mathrm{u}=0$).
As one example, they are used in the context of ultracold atoms, ions, and
molecules \cite{extended-hubbard0}. Its bosonic version is also
experimentally investigated. See, e.g., \cite{extended-hubbard1} published
in 2022.

In theoretical studies, frequently, only nearest-neighbor interactions added
to the on-site (zero-range) Hubbard interactions are considered. Here, we do
not need the restriction to one-range (nearest-neighbor) interactions. We
only assume that $\mathrm{u}$ is absolutely summable (see (\ref{summable1-U}%
)), which is physically a very mild restriction, since the effective
two-particle repulsive electrostatic potential in crystals is expected to
decay exponentially fast in space, because of screening effects.

The rotation invariance in Equation (\ref{summable1-U}) refers to the
isotropy of the system under consideration. However, as shown in \cite%
{articulo,articulo2}, the system has low energy states that spontaneously
break the isotropy. This refers to unconventional parings, typically of $d$%
-wave type, of electrons one experimentally observes in many high-$T_{c}$
superconductors \cite{Tsuei,Nature2015,tinpou1}. In fact, to derive the
existence of $d$- and $p$-wave pairings starting from a physically sound
microscopic model was the aim of \cite{articulo,articulo2}. Here, instead,
we keep a broader perspective and do not study this particular
question.\bigskip

\noindent \textbf{Bosonic Hamiltonian.} Similar to the fermionic part, the
bosonic part of the (infinite volume) Hamiltonian of the system is defined
to be the restriction $H_{b}\in \mathcal{B}(\ell ^{2}(\mathbb{Z}^{2}))$ to
the one-boson Hilbert space $\ell ^{2}(\mathbb{Z}^{2})$ of the formal
expression%
\begin{equation}
\epsilon \left( -{\frac{h_{b}}{2}\sum\limits_{x,y\in \mathbb{Z}%
^{2}\,:\,|x-y|=1}}b_{x}^{\ast }\,b_{y}\,+2h_{b}{\sum\limits_{x\in \mathbb{Z}%
^{2}}}b_{x}^{\ast }\,b_{x}\right) \text{ }.  \label{Hamiltonian-b}
\end{equation}%
Here, $b_{x}$ ($b_{x}^{\ast }$) denotes the annihilation (creation) operator
acting on the bosonic Fock space $\mathfrak{F}_{+}$ of a boson at lattice
position $x\in \mathbb{Z}^{2}$. Observe that the bosonic part only contains
a kinetic term. The parameter $h_{b}\in \mathbb{R}_{0}^{+}$ quantifies the
ratio of the effective masses of fermions and bosons: Taking $h_{b}$ smaller
than one physically means that the bosons are heavier than the fermions. As
experimentally found \cite%
{reviewer2-1,reviewer2-2,reviewer2-3,unitmass-bipolaron} for cuprate
superconductors, bipolarons should be much more massive than electrons or
holes and, thus, in the physically relevant regime, $h_{b}$ is to be taken
very small (or even zero, in an idealized situation). See \cite[Section 3.1]%
{articulo2}. In the sequel, we take $h_{b}\in \lbrack 0,1/2]$, meaning that
the boson mass is at least as big as the mass of two fermions, as discussed
in Section \ref{Main Results}.\bigskip\ 

\noindent \textbf{Exchange interactions.} The term of the Hamiltonian that
encodes the decay of a boson into two fermions (i.e., one of the
two-electron(hole)-bipolaron-exchange interaction of the Hamiltonian) refers
to the bounded operator 
\begin{equation}
W_{\mathrm{b\rightarrow f}}:\ell ^{2}\left( \mathbb{Z}^{2}\right)
\rightarrow \mathfrak{h}_{f}\text{ },  \label{Wbf}
\end{equation}%
which is defined to be the restriction of the formal expression \ 
\begin{equation}
{2^{-1/2}\sum\limits_{x,y\in \mathbb{Z}^{2}}}\upsilon \left( x-y\right)
c_{y}^{\ast }\,b_{x}  \label{Hamiltonian-fb}
\end{equation}%
to $\ell ^{2}(\mathbb{Z}^{2})$, where 
\begin{equation}
c_{y}^{\ast }\doteq {\sum\limits_{z\in \mathbb{Z}^{2}}}\,\left( \mathfrak{p}%
_{1}\left( z\right) a_{y+z,\uparrow }^{\ast }\,a_{y,\downarrow }^{\ast }+%
\mathfrak{p}_{2}\left( 2z\right) a_{y+z,\uparrow }^{\ast
}\,a_{y-z,\downarrow }^{\ast }\right)  \label{Hamiltonian-fbbis}
\end{equation}%
for some fixed functions $\mathfrak{p}_{1},\mathfrak{p}_{2}:\mathbb{Z}%
^{2}\rightarrow \mathbb{R}$ that are invariant under $90^{\circ }$-rotations
and exponentially decay in space, i.e., 
\begin{equation}
{\sum_{z\in \mathbb{Z}^{2}}}\,\mathrm{e}^{\alpha _{0}\left\vert z\right\vert
}\left\vert \mathfrak{p}_{\sharp }\left( z\right) \right\vert <\infty \qquad 
\text{and}\qquad \mathfrak{p}_{\sharp }\left( x,y\right) =\mathfrak{p}%
_{\sharp }\left( -y,x\right) ,\qquad x,y\in \mathbb{Z},\ \sharp \in \{1,2\}\
,  \label{ssdsds}
\end{equation}%
for some $\alpha _{0}>0$. In particular, the functions $\mathfrak{p}_{1},%
\mathfrak{p}_{2}$\ are absolutely summable in space.

By definition, we take $\mathfrak{p}_{2}\left( z\right) \doteq 0$ if $z\in 
\mathbb{Z}^{2}\backslash (2\mathbb{Z})^{2}$ and we also assume that 
\begin{equation}
\mathfrak{p}_{1}+\mathfrak{p}_{2}\neq 0\text{\qquad }\mathrm{and}\text{%
\qquad }\mathfrak{p}_{2}\left( x\right) \neq -\mathrm{e}^{i\frac{k}{2}\cdot
x}\mathfrak{p}_{1}\left( x\right) \ ,\qquad x\in \mathbb{Z},\ k\in \lbrack
-\pi ,\pi )^{2}\ .  \label{conditionsdsdsd}
\end{equation}%
The condition $\mathfrak{p}_{1}+\mathfrak{p}_{2}\neq 0$ only ensures the
non-triviality of the exchange interaction, while the second condition
avoids the singular case of a quasi-momentum $k_{0}\in \lbrack -\pi ,\pi
)^{2}$ at which the exchange interaction trivially vanishes, see below (\ref%
{fourier D fract}). This case can easily be analyzed, but it makes the
argumentation cumbersome, so we omit it here, as it is a highly unusual and
irrelevant situation. For example, (\ref{conditionsdsdsd}) is already
satisfied as soon as $\mathfrak{p}_{1}\left( z\right) \neq 0$ for some $z\in 
\mathbb{Z}^{2}\backslash (2\mathbb{Z})^{2}$, since $\mathfrak{p}_{2}\left(
z\right) \doteq 0$ for any $z\notin (2\mathbb{Z})^{2}$. In \cite[Eq. (6)]%
{articulo}, $\mathfrak{p}_{2}=0$ and, given $\kappa >0$, $\mathfrak{p}%
_{1}(z)=\mathrm{e}^{-\kappa |z|}$ for $|z|\leq 1$ and $\mathfrak{p}_{1}(z)=0$
otherwise, while in \cite[Eq. (4)]{articulo2}, $\mathfrak{p}_{2}\left(
2z\right) =\mathfrak{p}_{1}\left( z\right) =1$ when $|z|\leq 1$ and $%
\mathfrak{p}_{1}\left( z\right) =\mathfrak{p}_{2}\left( z\right) =0$
otherwise. This are the typical examples we have in mind, the point here
being the fact two fermions on \emph{different} lattice sites can interact
by exchanging a boson. See also Section \ref{Section physics}.

Physically, $c_{y}^{\ast }$ represents the creation of a fermion pair of
zero total spin, the two components of which are slightly spread around the
lattice position $y\in \mathbb{Z}^{2}$. Such pairs have finite size, because
of (\ref{ssdsds}). In fact, 
\begin{equation}
r_{\mathfrak{p}}\doteq \frac{1}{2}\left( r_{\mathfrak{p}_{1}}+r_{\mathfrak{p}%
_{2}}\right) \ ,  \label{summable2-Urrrrr}
\end{equation}%
where, for any $\sharp \in \{1,2\}$, $r_{\mathfrak{p}_{\sharp }}\doteq 0$ if 
$\mathfrak{p}_{\sharp }=0$, otherwise it is equal to 
\begin{equation}
r_{\mathfrak{p}_{\sharp }}\doteq \frac{{\sum_{z\in \mathbb{Z}^{2}}}%
\,\left\vert z\right\vert \left\vert \mathfrak{p}_{\sharp }\left( z\right)
\right\vert }{{\sum_{z\in \mathbb{Z}^{2}}}\,\left\vert \mathfrak{p}_{\sharp
}\left( z\right) \right\vert }\leq \inf_{\alpha _{0}>0}\alpha _{0}^{-1}\sqrt{%
\frac{{\sum_{z\in \mathbb{Z}^{2}}}\,\mathrm{e}^{\alpha _{0}\left\vert
z\right\vert }\left\vert \mathfrak{p}_{\sharp }\left( z\right) \right\vert }{%
{\sum_{z\in \mathbb{Z}^{2}}}\,\left\vert \mathfrak{p}_{\sharp }\left(
z\right) \right\vert }}<\infty \ ,  \label{summable2-Urrrrr0}
\end{equation}%
is naturally seen as being the actual size of such pairs. Note that the last
inequality is a consequence of the Cauchy-Schwarz inequality, along with the
bound%
\begin{equation*}
\left\vert z\right\vert ^{2}\mathrm{e}^{-\alpha _{0}\left\vert z\right\vert
}\leq \alpha _{0}^{-2},\text{\qquad }\alpha _{0}>0\text{ }.
\end{equation*}%
Allowing two fermions in different lattice site to interact by exchanging a
boson simply means that $r_{\mathfrak{p}}>0$. For the physical significance
of this property for cuprates, see \cite{articulo,articulo2}.

Recall that the exchange strength function $\upsilon :\mathbb{Z}%
^{2}\rightarrow \mathbb{R}$ is only absolutely summable, not necessarily
exponentially decaying as $\mathfrak{p}_{1}$ and $\mathfrak{p}_{2}$, and
invariant under $90^{\circ }$-rotations, i.e., 
\begin{equation}
{\sum_{z\in \mathbb{Z}^{2}}}\,\left\vert \upsilon \left( z\right)
\right\vert <\infty \qquad \text{and}\qquad \upsilon \left( x,y\right)
=\upsilon \left( -y,x\right) ,\qquad x,y\in \mathbb{Z}\ .
\label{summable2-U}
\end{equation}%
Note that the Fourier transforms $\hat{\upsilon}$,$\mathfrak{\hat{p}}_{1}$
and $\mathfrak{\hat{p}}_{2}$ of $\upsilon $, $\mathfrak{p}_{1}$ and $%
\mathfrak{p}_{2}$ are real-valued continuous functions (on the
two-dimensional torus $\mathbb{T}^{2}$) that are again invariant under $%
90^{\circ }$-rotations. Additionally, $\mathfrak{\hat{p}}_{1}$ and $%
\mathfrak{\hat{p}}_{2}$ are real analytic, for $\mathfrak{p}_{1}$ and $%
\mathfrak{p}_{2}$ are exponentially decaying. The reverse process, that is,
the annihilation of two unbound fermions to form a boson, is represented by
the adjoint operator 
\begin{equation}
W_{\mathrm{f\rightarrow b}}\doteq W_{\mathrm{b\rightarrow f}}^{\ast }:%
\mathfrak{h}_{f}\rightarrow \ell ^{2}\left( \mathbb{Z}^{2}\right) \ .
\label{Wfb}
\end{equation}%
\bigskip

\noindent \textbf{Mathematical remarks.} The infinite sums (\ref%
{Hamiltonian-f}), (\ref{Hamiltonian-b}) and (\ref{Hamiltonian-fb}) defining
formally $H_{f}\in \mathcal{B}\left( \mathfrak{h}_{f}\right) $, $H_{b}\in 
\mathcal{B}(\ell ^{2}(\mathbb{Z}^{2}))$ and $W_{\mathrm{b\rightarrow f}}\in 
\mathcal{B}(\ell ^{2}(\mathbb{Z}^{2}),\mathfrak{h}_{f})$ (\ref{Wbf}) are to
be understood as follows: If $\psi \in \mathfrak{h}_{f}$ or $\psi \in \ell
^{2}(\mathbb{Z}^{2})$ is a finitely supported function, then the sum
corresponding to $H_{f}\psi $, $H_{b}\psi $ or $W_{\mathrm{b\rightarrow f}%
}\psi $ is absolutely convergent. Thus, $H_{f}$, $H_{b}$ and $W_{\mathrm{%
b\rightarrow f}}$ are well-defined linear operators acting on the dense
subspace of such functions. One checks that $H_{f}$, $H_{b}$\ and $W_{%
\mathrm{b\rightarrow f}}$ are all bounded on this subspace and they thus
have a unique bounded linear extension to the whole Hilbert space where they
are defined, namely $\mathfrak{h}_{f}$ for $H_{f}$, and $\ell ^{2}(\mathbb{Z}%
^{2})$ for $H_{b}$ and $W_{\mathrm{b\rightarrow f}}$. We denote the
extensions again by $H_{f}$, $H_{b}$\ and $W_{\mathrm{b\rightarrow f}}$.
Note that, being a bounded operator, $W_{\mathrm{b\rightarrow f}}$ has an
adjoint (\ref{Wfb}), while $H_{f}$ and $H_{b}$\ are clearly symmetric and
so, self-adjoint, for they are also bounded.\bigskip

\noindent \textbf{Full model.} Finally, the full Hamiltonian for the
fermion-boson composite is defined, in matrix notation for the direct sum $%
\mathfrak{h}_{f}\oplus \ell ^{2}(\mathbb{Z}^{2})$, as follows: 
\begin{equation}
\begin{pmatrix}
H_{f} & W_{\mathrm{b\rightarrow f}} \\[0.5em] 
W_{\mathrm{f\rightarrow b}} & H_{b}%
\end{pmatrix}%
\in \mathcal{B}\left( \mathfrak{h}_{f}\oplus \ell ^{2}\left( \mathbb{Z}%
^{2}\right) \right) \text{ }.  \label{H}
\end{equation}%
Observe that this\ Hamiltonian is invariant under translations, as well as $%
90^{\circ }$-rotations.

Using the canonical orthonormal basis\footnote{%
I.e., for any $x,y\in \mathbb{Z}^{2}$ and $s,t\in \{\uparrow ,\downarrow \}$%
, $\mathfrak{e}_{\left( x,s\right) }(y,s)=\delta _{s,t}\delta _{x,y}$, where 
$\delta _{x,y}$ is the Kronecker delta.} 
\begin{equation*}
\left\{ \mathfrak{e}_{\left( x,s\right) }:x\in \mathbb{Z}^{2},s\in
\{\uparrow ,\downarrow \}\right\} \subseteq \ell ^{2}\left( \mathbb{Z}%
^{2}\times \{\uparrow ,\downarrow \}\right)
\end{equation*}%
to define the closed subspace 
\begin{equation}
\mathfrak{h}_{0}\doteq \text{\ }\overline{\mathrm{span}}\left\{ \mathfrak{e}%
_{(x,\uparrow )}\wedge \mathfrak{e}_{(y,\downarrow )}\,:\,x,y\in \mathbb{Z}%
^{2}\right\} \subseteq \mathfrak{h}_{f}\ ,  \label{h0}
\end{equation}%
we remark that the zero-spin subspace 
\begin{equation}
\mathfrak{H}\doteq \mathfrak{h}_{0}\oplus \ell ^{2}\left( \mathbb{Z}%
^{2}\right) \subseteq \mathfrak{h}_{f}\oplus \ell ^{2}\left( \mathbb{Z}%
^{2}\right)  \label{fract H}
\end{equation}%
is invariant under the action of the (full) Hamiltonian (\ref{H}). We can
thus consider its restriction 
\begin{equation}
H\doteq \left. 
\begin{pmatrix}
H_{f} & W_{\mathrm{b\rightarrow f}} \\[0.5em] 
W_{\mathrm{f\rightarrow b}} & H_{b}%
\end{pmatrix}%
\right\vert _{\mathfrak{H}}\in \mathcal{B}\left( \mathfrak{H}\right)
\label{H0}
\end{equation}%
to this particular subspace $\mathfrak{H}\subseteq \mathfrak{h}_{f}\oplus
\ell ^{2}(\mathbb{Z}^{2})$.

In fact, as the boson is assumed to be spinless, by the conservation of
angular momentum, we have that the total spin of the fermion pair resulting
from a bosonic decay must be zero. In other words, the physically relevant
(vector) states of the fermion-boson compound system always lie in $%
\mathfrak{H}$. Note finally that $H$ inherits the symmetries of the
Hamiltonian (\ref{H}), i.e., $H$ is invariant under translations and $%
90^{\circ }$-rotations. Note that this last symmetry, i.e., the rotation
invariance, is mainly relevant for the study of unconventional pairings,
which is not done here.

\subsection{The Model in Spaces of Quasi-Momenta\label{Model section}}

We have a composite of two fermions and one boson whose Hamiltonian is
translation invariant. In this case, it is a standard procedure (see, e.g. 
\cite[Chapter XIII.16 ]{ReedSimonIV}) to use the direct integral
decomposition of the Hamiltonian in Fourier space in order to study its
spectral properties.

For the two dimensional lattice $\mathbb{Z}^{2}$, the (Fourier) space of
quasi-momenta is nothing else than the torus%
\begin{equation*}
\mathbb{T}^{2}\doteq \lbrack -\pi ,\pi )^{2}\subseteq \mathbb{R}^{2}\ .
\end{equation*}%
This set is endowed with the metric $d_{\mathbb{T}^{2}}$ defined by%
\begin{equation}
d_{\mathbb{T}^{2}}\left( k,p\right) \doteq \min \left\{ \left\vert
k-p-q\right\vert :q\in 2\pi \mathbb{Z}^{2}\right\} \ ,  \label{metric}
\end{equation}%
where $|k-p-q|$ is the Euclidean distance between $k$ and $p+q$ in $\mathbb{R%
}^{2}$. This defines a compact metric space $(\mathbb{T}^{2},d_{\mathbb{T}%
^{2}})$. Observe also that the usual group operation in $\mathbb{T}^{2}$,
i.e., the sum in $\mathbb{R}^{2}$ modulo $(2\pi ,2\pi )$, is a continuous
operation, while any Borel set in $\mathbb{T}^{2}$ is also a Borel set in $%
\mathbb{R}^{2}$ (endowed with the Euclidean metric).

We also need the normalized Haar measure $\nu $ on $\mathbb{T}^{2}$ defined
for any Borel set $B\subseteq \mathbb{T}^{2}$ by 
\begin{equation}
\nu (B)=(2\pi )^{-2}\mathbf{\lambda }(B)\ ,  \label{Haar mesure}
\end{equation}%
where $\mathbf{\lambda }$ is the Lebesgue measure in $\mathbb{R}^{2}$. This
measure appears in relation with direct integrals of constant Hilbert spaces
on the two-dimensional torus $\mathbb{T}^{2}$, like the Hilbert space%
\begin{equation*}
L^{2}\left( \mathbb{T}^{2}\right) \equiv L^{2}\left( \mathbb{T}^{2},\mathbb{C%
}\right) \equiv L^{2}\left( \mathbb{T}^{2},\mathbb{C},\nu \right) \doteq {%
\int_{\mathbb{T}^{2}}^{\oplus }}\mathbb{C}\,\nu (\mathrm{d}k)
\end{equation*}%
of square-integrable, complex-valued functions on $\mathbb{T}^{2}$. Since
the Haar measure $\nu $ is used in all our direct integrals on $\mathbb{T}%
^{2}$, for simplicity we often remove the symbol $\nu $ from the notation of 
$L^{2}$-spaces, unless this information is important to recall.

The Fourier transform can be applied in the fermionic and bosonic sectors.
In the fermionic one, there is more than one natural way of implementing the
transform, as the corresponding functions have two arguments in $\mathbb{Z}%
^{2}$. It turns out that to be very useful to extract the total
quasi-momentum of fermionic pairs. In fact, we consider the direct integral 
\begin{equation}
L^{2}\left( \mathbb{T}^{2},\mathcal{H}\right) \equiv L^{2}\left( \mathbb{T}%
^{2},\mathcal{H},\nu \right) \doteq {\int_{\mathbb{T}^{2}}^{\oplus }}%
L^{2}\left( \mathbb{T}^{2},\mathbb{C},\nu \right) \oplus \mathbb{C}\,\nu (%
\mathrm{d}k)  \label{Hilbert space good}
\end{equation}%
of the (constant fiber) Hilbert space%
\begin{equation}
\mathcal{H}\doteq L^{2}\left( \mathbb{T}^{2}\right) \oplus \mathbb{C}\equiv
L^{2}\left( \mathbb{T}^{2},\mathbb{C},\nu \right) \oplus \mathbb{C}
\label{cal H}
\end{equation}%
over the torus $\mathbb{T}^{2}$, and choose a unitary transformation 
\begin{equation*}
\mathbb{U}:\mathfrak{H}\longrightarrow L^{2}\left( \mathbb{T}^{2},\mathcal{H}%
\right)
\end{equation*}%
in such a way that $k\in \mathbb{T}^{2}$, the fiber quasi-momentum, is
exactly the total quasi-momentum of the fermion pair.

Recall that $\mathfrak{H}$ defined in (\ref{fract H}) is the Hilbert space
on which $H$ is originally defined. More precisely, 
\begin{equation}
\mathbb{U}\doteq U_{f}\oplus \mathcal{F}\ ,  \label{U}
\end{equation}%
where 
\begin{equation}
\mathcal{F}:\ell ^{2}\left( \mathbb{Z}^{2}\right) \rightarrow L^{2}\left( 
\mathbb{T}^{2}\right)
\end{equation}%
is the Fourier transform on $\ell ^{2}(\mathbb{Z}^{2})$, while the fermionic
part%
\begin{equation}
U_{f}\doteq U_{2}U_{1}:\mathfrak{h}_{0}\rightarrow \int_{\mathbb{T}%
^{2}}^{\oplus }L^{2}\left( \mathbb{T}^{2}\right) \nu \left( \mathrm{d}%
k\right)  \label{Uf}
\end{equation}%
is the composition of two unitary (linear) transformations $U_{1}$ and $%
U_{2} $, whose exact definitions are given as follows:%
\begin{equation}
\begin{array}{cccccccc}
U_{1}: & \mathfrak{h}_{0} & \rightarrow & \ell ^{2}\left( \mathbb{Z}%
^{2}\times \mathbb{Z}^{2}\right) & \rightarrow & \ell ^{2}\left( \mathbb{Z}%
^{2}\times \mathbb{Z}^{2}\right) & \rightarrow & \ell ^{2}\left( \mathbb{Z}%
^{2}\right) \otimes \ell ^{2}\left( \mathbb{Z}^{2}\right) \\ 
& \mathfrak{e}_{(x,\uparrow )}\wedge \mathfrak{e}_{(y,\downarrow )} & \mapsto
& \mathfrak{e}_{(x,y)} & \mapsto & \mathfrak{e}_{(x,x-y)} & \mapsto & 
\mathfrak{e}_{x}\otimes \mathfrak{e}_{x-y}%
\end{array}%
\end{equation}%
and%
\begin{equation}
\begin{array}{cccccc}
U_{2}: & \ell ^{2}\left( \mathbb{Z}^{2}\right) \otimes \ell ^{2}\left( 
\mathbb{Z}^{2}\right) & \rightarrow & L^{2}\left( \mathbb{T}^{2}\right)
\otimes L^{2}\left( \mathbb{T}^{2}\right) & \rightarrow & \int_{\mathbb{T}%
^{2}}^{\oplus }L^{2}\left( \mathbb{T}^{2}\right) \nu \left( \mathrm{d}%
k\right) \\ 
& \mathfrak{e}_{x}\otimes \mathfrak{e}_{x-y} & \mapsto & \mathfrak{\hat{e}}%
_{x}\otimes \mathfrak{\hat{e}}_{x-y} & \mapsto & \mathfrak{\hat{e}}%
_{x}\left( \cdot \right) \mathfrak{\hat{e}}_{x-y}%
\end{array}%
\ .
\end{equation}%
Because $\{\mathfrak{e}_{(x,\uparrow )}\wedge \mathfrak{e}_{(y,\downarrow
)}\}_{x,y\in \mathbb{Z}^{2}}$, $\{\mathfrak{e}_{(x,y)}\}_{x,y\in \mathbb{Z}%
^{2}}$ and $\{\mathfrak{e}_{x}\otimes \mathfrak{e}_{y}\}_{x,y\in \mathbb{Z}%
^{2}}$ are orthonormal bases and $(x,y)\mapsto (x,x-y)$ is a bijection on $%
\mathbb{Z}^{2}\times \mathbb{Z}^{2}$, $U_{1}$ is well-defined as a
composition of three unitary linear transformations. Note also that the last
unitary linear transformation defining $U_{2}$ is defined as in Proposition %
\ref{direct integral as a tensor product}, while the first one defining $%
U_{2}$ is the tensor product $\mathcal{F}\otimes \mathcal{F}$ of the Fourier
transform $\mathcal{F}$ on $\ell ^{2}(\mathbb{Z}^{2})$, defined for any $%
f\in \ell ^{1}(\mathbb{Z}^{2})\subseteq \ell ^{2}(\mathbb{Z}^{2})$ by 
\begin{equation}
\hat{f}\left( k\right) \equiv \mathcal{F}f\left( k\right) ={\sum_{x\in 
\mathbb{Z}^{2}}}\,\mathrm{e}^{ik\cdot x}f\left( x\right) \ ,\qquad k\in 
\mathbb{T}^{2}\ ,  \label{fourier transforms}
\end{equation}%
$k\cdot x$ being the usual scalar product of $k\in \mathbb{T}^{2}$ and $x\in 
\mathbb{Z}^{2}$, seen as vectors of $\mathbb{R}^{2}$. Here we use the symbol 
$\widehat{\left( \cdot \right) }$ to shorten the notation of the Fourier
transform. For instance, for any $x\in \mathbb{Z}^{2}$, we write above $\hat{%
\mathfrak{e}}_{x}$ to denote the function $\mathrm{e}^{i(\cdot )\cdot x}$ on
the torus $\mathbb{T}^{2}$. That is, $\{\hat{\mathfrak{e}}_{x}\}_{x\in 
\mathbb{Z}^{2}}$ is the image under the Fourier transform of the canonical
orthonormal basis$\{\mathfrak{e}_{x}\}_{x\in \mathbb{Z}^{2}}$ (\ref{e frac})
of $\ell ^{2}(\mathbb{Z}^{2})$.

For the reader's convenience and completeness, in Section \ref{Constant
fiber direct integral} we gather key results from the theory of direct
integrals with constant fiber Hilbert spaces. In the next subsection, we
explain how the properties of the Hamiltonian $H\in \mathcal{B}(\mathfrak{H}%
) $ defined by (\ref{H0}) can be studied on the direct integral (\ref%
{Hilbert space good}) over total quasi-momenta.

\subsection{Fiber Decomposition of the Hamiltonian\label{fiber decomposition
of H0}}

By explicit computations, exactly like in \cite{articulo,articulo2}, we show
that the conjugation of the Hamiltonian $H\in \mathcal{B}(\mathfrak{H})$
with the unitary transformation $\mathbb{U}$ of Equation (\ref{U}) is a
decomposable operator on the direct integral $L^{2}(\mathbb{T}^{2},\mathcal{H%
})$. To state this result precisely, we need preliminary definitions
allowing to define the so-called \textquotedblleft fiber
Hamiltonians\textquotedblright , or \textquotedblleft
fibers\textquotedblright\ for short, $A(k)\in \mathcal{B}(\mathcal{H})$\ of $%
\mathbb{U}H\mathbb{U}^{\ast }$ at total quasi-momenta $k\in \mathbb{T}^{2}$.
In fact, the mapping $k\mapsto A(k)$ defines an element of the von Neumann
algebra\footnote{%
The (unique) norm of this $C^{\ast }$-algebra is the essential supremum with
respect to the measure $\nu $ on the torus, see (\ref{norm}).} 
\begin{equation*}
L^{\infty }\left( \mathbb{T}^{2},\mathcal{B}(\mathcal{H})\right) \equiv
L^{\infty }\left( \mathbb{T}^{2},\mathcal{B}(\mathcal{H}),\nu \right)
\end{equation*}%
of (equivalence classes of) strongly measurable functions $\mathbb{T}%
^{2}\rightarrow \mathcal{B}(\mathcal{H})$. See Section \ref{Constant fiber
direct integral} for more details.

Given a total quasi-momentum $k\in \mathbb{T}^{2}$ and the parameters $%
\epsilon ,h_{b}\in \mathbb{R}_{0}^{+}$ tuning the strengths of the two
(fermionic and bosonic) kinetic parts of the model, we define continuous,
real-valued functions $\mathfrak{f}(k),\mathfrak{d}(k),\mathfrak{b}\in
C\left( \mathbb{T}^{2}\right) $ on the torus $\mathbb{T}^{2}$ by 
\begin{eqnarray}
\mathfrak{b}\left( p\right) &\doteq &h_{b}\epsilon \left( 2-\cos \left(
p\right) \right) \ ,  \label{b} \\[0.01in]
\mathfrak{f}\left( k\right) \left( p\right) &\doteq &\epsilon \left\{ 4-\cos
\left( p+k\right) -\cos \left( p\right) \right\} \ ,  \label{f} \\[0.01in]
\mathfrak{d}\left( k\right) \left( p\right) &\doteq &\mathfrak{\hat{p}}%
_{1}\left( k+p\right) +\mathfrak{\hat{p}}_{2}\left( k/2+p\right) \ ,
\label{d}
\end{eqnarray}%
for all $p=(p_{1},p_{2})\in \mathbb{T}^{2}$, where%
\begin{equation}
\cos \left( q\right) \doteq \cos \left( q_{1}\right) +\cos \left(
q_{2}\right) \text{ },\qquad q=(q_{1},q_{2})\in \mathbb{R}^{2}\ .
\label{cosinus}
\end{equation}%
Recall that (the $(2\pi ,2\pi )$-periodic function) $\mathfrak{\hat{p}}_{1}$
and $\mathfrak{\hat{p}}_{2}$ are the Fourier transform of $\mathfrak{p}_{1}$
and $\mathfrak{p}_{2}$, which are the functions defining the operator $%
c_{y}^{\ast }$ in (\ref{Hamiltonian-fbbis}), representing the creation of
fermion pairs in the model. Note also from (\ref{conditionsdsdsd}) that $%
\mathfrak{d}(k)\neq 0$ for all $k\in \mathbb{T}^{2}$. Indeed, using $%
\mathfrak{p}_{2}(z)\doteq 0$ for $z\notin 2\mathbb{Z}$ as well as (\ref%
{fourier transforms}) and (\ref{d}), 
\begin{equation}
\mathfrak{d}\left( k\right) =\mathcal{F}\left[ \mathrm{e}^{ik\cdot x}%
\mathfrak{p}_{1}\left( x\right) +\mathrm{e}^{i\frac{k}{2}\cdot x}\mathfrak{p}%
_{2}\left( x\right) \right] \ ,  \label{fourier D fract}
\end{equation}%
where $\mathrm{e}^{ik\cdot x}\mathfrak{p}_{\sharp }(x)$ stands for the
function $x\mapsto \mathrm{e}^{ik\cdot x}\mathfrak{p}_{\sharp }(x)$ with $%
\sharp \in \{1,2\}$.

Then, at any quasi-momentum $k\in \mathbb{T}^{2}$ and on-site repulsion
strength $\mathrm{U}\in \mathbb{R}_{0}^{+}$, we define the bounded operators 
$B_{1,1}\left( k\right) $ and $A_{1,1}(\mathrm{U},k)$ acting on the Hilbert
space $L^{2}\left( \mathbb{T}^{2}\right) $ by 
\begin{eqnarray}
B_{1,1}\left( k\right) &\doteq &M_{\mathfrak{f}\left( k\right) }+{%
\sum\limits_{x\in \mathbb{Z}^{2}}}\,\mathrm{u}\left( x\right) P_{x}\ ,
\label{A11} \\
A_{1,1}\left( \mathrm{U},k\right) &\doteq &B_{1,1}\left( k\right) +\mathrm{U}%
P_{0}\ ,  \label{A11-U0}
\end{eqnarray}%
where $M_{\mathfrak{f}(k)}$ stands for the multiplication operator by $%
\mathfrak{f}(k)\in C(\mathbb{T}^{2})$ and $P_{x}$ is the orthogonal
projection onto the one-dimensional subspace $\mathbb{C}\mathfrak{\hat{e}}%
_{x}\subseteq L^{2}(\mathbb{T}^{2})$. Note that the infinite sum defining
the bounded operator $B_{1,1}(k)$ is absolutely convergent, for the function 
$\mathrm{u}:\mathbb{Z}^{2}\rightarrow \mathbb{R}$ is, by assumption,
absolutely summable. See (\ref{summable1-U}).

We define next 
\begin{eqnarray}
&&%
\begin{array}{cccl}
A_{2,1}\left( k\right) : & L^{2}\left( \mathbb{T}^{2}\right) & \rightarrow & 
\mathbb{C} \\ 
& \varphi & \mapsto & \hat{\upsilon}\left( k\right) \left\langle \mathfrak{d}%
\left( k\right) ,\varphi \right\rangle \ ,%
\end{array}
\label{A21} \\
&&  \notag \\
&&%
\begin{array}{cccl}
A_{1,2}\left( k\right) : & \mathbb{C} & \rightarrow & L^{2}\left( \mathbb{T}%
^{2}\right) \\ 
& z & \mapsto & \hat{\upsilon}\left( k\right) \mathfrak{d}\left( k\right) z%
\end{array}
\label{A12}
\end{eqnarray}%
as well as%
\begin{equation}
\begin{array}{cccl}
A_{2,2}\left( k\right) : & \mathbb{C} & \rightarrow & \mathbb{C} \\ 
& z & \mapsto & \mathfrak{b}\left( k\right) z%
\end{array}
\label{A22}
\end{equation}%
for any fixed $k\in \mathbb{T}^{2}$. By compactness of $\mathbb{T}^{2}$ and
continuity (in operator norm) of the mappings $k\mapsto A_{i,j}(k)$ for all $%
i,j\in \{1,2\}$, we have 
\begin{equation}
A\left( \cdot \right) \equiv A\left( \mathrm{U},\cdot \right) \doteq 
\begin{pmatrix}
A_{1,1}\left( \mathrm{U},\cdot \right) & A_{1,2}\left( \cdot \right) \\%
[0.5em] 
A_{2,1}\left( \cdot \right) & A_{2,2}\left( \cdot \right)%
\end{pmatrix}%
\in L^{\infty }\left( \mathbb{T}^{2},\mathcal{B}(\mathcal{H})\right)
\label{fiber hamiltonians}
\end{equation}%
(see Lemma \ref{maincoro1 copy(1)}), which is meanwhile the fiber
decomposition of the operator $\mathbb{U}H\mathbb{U}^{\ast }$:

\begin{proposition}[Fiber decomposition of the quantum model]
\label{direct integral decomposition of the hamiltonian}\mbox{ }\newline
The conjugation of $H$ by $\mathbb{U}$ (\ref{U}) is decomposable and has $%
A(\cdot )$ as its fibers, that is, 
\begin{equation*}
\mathbb{U}H\mathbb{U}^{\ast }={\int_{\mathbb{T}^{2}}^{\oplus }}A(k)\,\nu (%
\mathrm{d}k)\ .
\end{equation*}
\end{proposition}

\begin{proof}
This is proven from explicit computations which are almost the same as those
done in \cite{articulo,articulo2}. We postpone the details of this
calculation to Section \ref{Unitary transformation copy(1)}.
\end{proof}

The fiber decomposition given by Proposition \ref{direct integral
decomposition of the hamiltonian} is useful because it gives access to
spectral properties of $H$. In fact, for an operator that is decomposable on 
$L^{2}\left( \mathbb{T}^{2},\mathcal{H}\right) $, i.e., an operator
unitarily equivalent to an element of the von Neumann algebra $L^{2}\left( 
\mathbb{T}^{2},\mathcal{B}(\mathcal{H})\right) $, like the Hamiltonian $H$,
the fibers $A(k)$ of which are all self-adjoint, it is known that $\lambda
\in \sigma (H)$ if, and only if, for all $\varepsilon >0$,%
\begin{equation*}
\nu \left( \left\{ k\in \mathbb{T}^{2}:\sigma \left( A\left( k\right)
\right) \cap \left( \lambda -\varepsilon ,\lambda +\varepsilon \right) \neq
\emptyset \right\} \right) >0\text{ }.
\end{equation*}%
See Theorem \ref{borelian functional calculus of a direct integral of
operators}. As is usual, here, $\sigma (X)$ denotes the spectrum of any
operator $X$ acting on some Hilbert space.

\section{Main Results\label{Main Results}}

In this section we state our main results, starting with general spectral
properties of the Hamiltonian $H$ to finish with results related with
scattering.

Recall that the model has parameters $\epsilon ,\mathrm{U},h_{b}\in \mathbb{R%
}_{0}^{+}$ and $\alpha _{0}\in \mathbb{R}^{+}$, and it depends on the choice
of functions 
\begin{equation*}
\mathrm{u}:\mathbb{Z}^{2}\rightarrow \mathbb{R}_{0}^{+}\ ,\quad \mathfrak{p}%
_{1}:\mathbb{Z}^{2}\rightarrow \mathbb{R}\ \ ,\quad \mathfrak{p}_{2}:\mathbb{%
Z}^{2}\rightarrow \mathbb{R}\quad \text{and}\quad \upsilon :\mathbb{Z}%
^{2}\rightarrow \mathbb{R}
\end{equation*}%
(with $\mathfrak{p}_{2}(z)\doteq 0$ for $z\notin 2\mathbb{Z}$) that are
absolutely summable and invariant with respect to $90^{\circ }$-rotations.
Observe additionally that the functions $\mathfrak{p}_{1}$ and $\mathfrak{p}%
_{2}$ are required to be exponentially decaying, that is, $\mathrm{e}%
^{\alpha _{0}|\cdot |}\mathfrak{p}_{1}$ and $\mathrm{e}^{\alpha _{0}|\cdot |}%
\mathfrak{p}_{2}$ are absolutely summable for some $\alpha _{0}>0$. See
Equations (\ref{summable1-U}), (\ref{ssdsds}) and (\ref{summable2-U}). All
details of the Hamiltonian, like the precise choice of its parameters and
functions, are not explicitly mentioned in our discussions or statements
below, unless it is important for clearness. There is however one important
condition to clarify:

While some of our results can be obtained without any other restriction,
frequently we fix the parameter $h_{b}$ in the interval $[0,1/2]$. This
choice physically means that the boson is heavier than two fermions. As
already discussed above, the assumption is perfectly justified when one
views the two fermions and the boson of the model as being electrons or
holes and a bipolaron, respectively, in a cuprate. In fact, polarons (and
thus bipolarons) are charge carriers that are self-trapped inside a strong
and local lattice deformation that surrounds them, caused by electrostatic
interactions between the carriers and the lattice. A priori, such (strong
and local) lattice deformations can barely move, that is, their effective
mass is huge. See, e.g., \cite{reviewer2-1,reviewer2-2,reviewer2-3}. This is
coherent with the assumption of a large mass of JT bipolarons in copper
oxides \cite{unitmass-bipolaron}, similar to JT polarons \cite{HNT83}. See
also Section \ref{Section physics} for more details.

We show that the condition $h_{b}\in \lbrack 0,1/2]$ is crucial to obtain
dressed bound fermion pairs, which are expected to represent the charge
carriers below the pseudogap temperature \cite{articulo2}.

\subsection{Spectral Properties}

Having in mind Proposition \ref{direct integral decomposition of the
hamiltonian} and Theorem \ref{borelian functional calculus of a direct
integral of operators} we start with the spectral properties of fiber
Hamiltonians (\ref{fiber hamiltonians}) at any quasi-momentum $k\in \mathbb{T%
}^{2}$. This refers to the following theorem:

\begin{theorem}[Spectral properties of fiber Hamiltonians]
\label{Maintheorem1}\mbox{ }\newline
Fix $\epsilon ,\mathrm{U}\in \mathbb{R}_{0}^{+}$, $h_{b}\in \lbrack 0,1/2]$
and $k\in \mathbb{T}^{2}$.

\begin{enumerate}
\item[i.)] Essential spectrum $\sigma _{\mathrm{ess}}\left( \cdot \right) $
of the fiber Hamiltonian:%
\begin{equation*}
\sigma _{\mathrm{ess}}\left( A\left( \mathrm{U},k\right) \right) =\,%
\mathfrak{f}\left( k\right) \left( \mathbb{T}^{2}\right) =2\epsilon \cos
\left( k/2\right) \left[ -1,1\right] +4\epsilon \ .
\end{equation*}

\item[ii.)] Ground state energy: There is a unique, non-degenerate
eigenvalue $\mathrm{E}(\mathrm{U},k)\leq \mathfrak{b}(k)$ of $A(\mathrm{U}%
,k) $ below the essential spectrum, with associated eigenvector%
\begin{equation*}
\Psi (\mathrm{U},k)\doteq (\hat{\psi}_{k}\left( \mathrm{U}\right) ,-1)\ ,%
\text{\quad where\quad }\hat{\psi}_{\mathrm{U},k}\doteq \hat{\upsilon}\left(
k\right) \left( A_{1,1}\left( \mathrm{U},k\right) -\mathrm{E}\left( \mathrm{U%
},k\right) \mathfrak{1}\right) ^{-1}\mathfrak{d}\left( k\right) \in
L^{2}\left( \mathbb{T}^{2}\right) \ .
\end{equation*}%
In addition, $\mathrm{E}(\mathrm{U},k)=\mathfrak{b}(k)$ iff $\hat{\upsilon}%
(k)=0$. Recall that $\mathfrak{b}(k)$\ is defined by Equation (\ref{b}).

\item[iii.)] Spectral gap and Anderson localization: If $\hat{\upsilon}%
(0)\neq 0$ and $r_{\mathfrak{p}}>0$ then%
\begin{equation*}
\inf_{\mathrm{U}\in \mathbb{R}_{0}^{+}}\min_{k\in \mathbb{T}^{2}}\left\{
\min \sigma _{\mathrm{ess}}\left( A\left( \mathrm{U},k\right) \right) -%
\mathrm{E}\left( \mathrm{U},k\right) \right\} >0
\end{equation*}%
and there are $C,\alpha \in \mathbb{R}^{+}$ such that, for all $k\in \mathbb{%
T}^{2}$ and $\mathrm{U}\in \mathbb{R}_{0}^{+}$, 
\begin{equation*}
\left\vert \mathcal{F}^{-1}[\hat{\psi}_{\mathrm{U},k}]\left( x\right)
\right\vert \leq C\mathrm{e}^{-\alpha |x|}\ ,\qquad x\in \mathbb{Z}^{2}\ .
\end{equation*}

\item[iv.)] $\mathrm{E}\left( \mathrm{U},\cdot \right) :\mathbb{T}%
^{2}\rightarrow \mathbb{R}$ is a continuous function and if $\hat{\upsilon}$
is of class\footnote{%
Given $n,d\in \mathbb{N}$ and an open set $\Omega \subseteq \mathbb{R}^{n}$, 
$C^{d}(\Omega )$ denotes the set of $d$ times continuously differentiable,
complex-valued functions on $\Omega $, while $C^{\omega }(\Omega )$ and $%
C^{a}(\Omega )$ refer to the space of smooth and real analytic functions on $%
\Omega $, respectively.} $C^{d}$ on $(-\pi ,\pi )^{2}\backslash
\{0\}\subseteq \mathbb{R}^{2}$ with $d\in \mathbb{N}\cup \{\omega ,a\}$ then
so does $\mathrm{E}\left( \mathrm{U},\cdot \right) $ on $(-\pi ,\pi
)^{2}\backslash \{0\}$.
\end{enumerate}
\end{theorem}

\begin{proof}
The theorem is a combination of Theorems \ref{existence of eigenvalue for
each fiber}, \ref{regularity of E}, \ref{hardcore limit of the eigenspace}
and \ref{Exponentially localized bound pairs} together with Propositions \ref%
{essential spectrum of a fiber}, \ref{spectral gap} and Corollary \ref%
{eigenspace of a fiber} (see (\ref{eigenvector})).
\end{proof}

\begin{remark}
\label{remark regularity}\mbox{ }\newline
Recall that if, for some natural number $d\geq 1$, 
\begin{equation*}
\sum_{x\in \mathbb{Z}^{2}}\left\vert x\right\vert ^{d}\left\vert \upsilon
\left( x\right) \right\vert <\infty
\end{equation*}%
then the Fourier transform $\hat{\upsilon}$ of the function $\upsilon :%
\mathbb{Z}^{2}\rightarrow \mathbb{R}$, as defined by (\ref{fourier
transforms}), is of class $C^{d}$ on the whole torus $\mathbb{T}^{2}$.
\end{remark}

\begin{remark}
\label{remark add}\mbox{ }\newline
If $\mathfrak{p}_{1}=\mathfrak{p}_{2}\in \mathbb{C}\mathfrak{e}_{0}$,i.e., $%
r_{\mathfrak{p}}=0$, then Theorem \ref{Maintheorem1} (iii) remains true, but
not uniformly in $\mathrm{U}\in \mathbb{R}_{0}^{+}$. That is, in this case,
one only has: 
\begin{equation*}
\min_{k\in \mathbb{T}^{2}}\left\{ \min \sigma _{\mathrm{ess}}\left( A\left( 
\mathrm{U},k\right) \right) -\mathrm{E}\left( \mathrm{U},k\right) \right\} >0
\end{equation*}%
and there are $C_{\mathrm{U}},\alpha _{\mathrm{U}}\in \mathbb{R}^{+}$ such
that, for all $k\in \mathbb{T}^{2}$, 
\begin{equation*}
\left\vert \mathcal{F}^{-1}[\hat{\psi}_{\mathrm{U},k}]\left( x\right)
\right\vert \leq C_{\mathrm{U}}\mathrm{e}^{-\alpha _{\mathrm{U}}|x|}\
,\qquad x\in \mathbb{Z}^{2}\ .
\end{equation*}
\end{remark}

Assertion (i) of Theorem \ref{Maintheorem1} holds true for all $h_{b}\in 
\mathbb{R}_{0}^{+}$, but the other assertions need the restriction $h_{b}\in
\lbrack 0,1/2]$ to ensure that the eigenvalues are below the essential
spectrum, as stated in Assertion (ii). In fact, $h_{b}\in \lbrack 0,1/2]$ iff%
\begin{equation}
\mathfrak{b}\left( k\right) \leq \mathfrak{z}\left( k\right) \doteq
4\epsilon -2\epsilon \cos (k/2)=\min \sigma _{\mathrm{ess}}\left( A\left( 
\mathrm{U},k\right) \right)  \label{z}
\end{equation}%
for all $k\in \mathbb{T}^{2}$, with equality \emph{only} at $k=0$. See\
Equation (\ref{b}). Therefore, by\ Assertion (ii), $\mathrm{E}(\mathrm{U},k)$
belongs to the essential spectrum iff $k=0$ and $\hat{\upsilon}(0)=0$.
Otherwise, we have a uniform spectral gap, as stated in Assertion (iii).

As is explained in \cite{articulo,articulo2}, the eigenvalue $\mathrm{E}%
\left( \mathrm{U},k\right) $ given by Theorem \ref{Maintheorem1} is
associated with the formation of \emph{dressed bound fermion pairs} with
total quasi-momentum $k\in \mathbb{T}^{2}$. These pairs are generally
exponentially localized, thanks to Theorem \ref{Maintheorem1} (iii), which
basically implies that the two fermions move together confined within some
small ball, that is, they are tightly bound in space, provided $\hat{\upsilon%
}(0)\neq 0$. When\ $\mathfrak{p}_{1}(z)\neq 0$ or $\mathfrak{p}_{2}(z)\neq 0$
for some $z\neq 0,$ or, equivalently, $r_{\mathfrak{p}}>0$, the size of the
small does not depend upon the Hubbard coupling constant $\mathrm{U}$ and a
very large $\mathrm{U}\gg 1$ only prevents two fermions from occupying the
same lattice site. The condition $\mathfrak{p}_{1}(z)\neq 0$ or $\mathfrak{p}%
_{2}(z)\neq 0$ for some $z\neq 0$ or, equivalently, $r_{\mathfrak{p}}>0$, is
therefore \textbf{pivotal} to get (Cooper) fermion pairs, the natural
candidates for superconducting charge carriers, in presence of strong
on-site Coulomb repulsions, like in cuprates \cite{Imada,Nature2015}.

For the usual (i.e., non-extended) Hubbard interaction ($\mathrm{u}=0$) and
one-range\footnote{%
It means here that $c_{y}^{\ast }\doteq {\sum_{\left\vert z\right\vert \leq
1}}\,\big(\mathfrak{p}_{1}\left( z\right) a_{y+z,\uparrow }^{\ast
}\,a_{y,\downarrow }^{\ast }+\mathfrak{p}_{2}\left( 2z\right)
a_{y+z,\uparrow }^{\ast }\,a_{y-z,\downarrow }^{\ast }\big)$.} creation
operators $c_{y}^{\ast }$ of fermion pairs (in this case (\ref{ssdsds})
holds true for all $\alpha _{0}\in \mathbb{R}_{0}^{+}$), note that a weak
form of pair localization was previously shown in the ground state. See for
instance \cite[Theorem 3 and Proposition 13]{articulo}. In this particular
case, estimates of $\mathrm{E}\left( \mathrm{U},k\right) $ and $\Psi (%
\mathrm{U},k)$ are known for large $\mathrm{U}\gg 1$. See, e.g., \cite[%
Theorem 4, Corollary 5, Theorem 16]{articulo}. Recall that the aim in \cite%
{articulo,articulo2} was to show the existence of $d$- and $p$-wave pairings
in the ground state for some physically sound model, and not the systematic
study of a general class of models. In \cite{articulo2} we conjecture that
such dressed bound fermion pairs represent the charge carriers below the
pseudogap temperature in cuprates.

Theorem \ref{Maintheorem1} combined with Proposition \ref{direct integral
decomposition of the hamiltonian} and the theory of direct integrals (cf.
Theorem \ref{borelian functional calculus of a direct integral of operators}%
) has direct consequences for the spectrum of the full Hamiltonian $H\in 
\mathcal{B}(\mathfrak{H})$, which is defined by Equation (\ref{H0}). Among
other things we obtain the following corollary:\ 

\begin{corollary}[Spectral properties of $H$]
\label{maincoro1}\mbox{ }\newline
Fix $\epsilon ,\mathrm{U}\in \mathbb{R}_{0}^{+}$ and $h_{b}\in \lbrack
0,1/2] $. Then, 
\begin{equation*}
\sigma \left( H\right) \cap \left( -\infty ,8\epsilon \right] =\left\{ 
\mathrm{E}\left( \mathrm{U},k\right) :k\in \mathbb{T}^{2}\right\} \cup
\left( 0,8\epsilon \right) \ ,
\end{equation*}%
where $\sigma \left( H\right) $ denotes, as is usual, the spectrum of $H$,
and%
\begin{equation*}
\min \sigma \left( H\right) =E\left( \mathrm{U}\right) \doteq \min_{k\in 
\mathbb{T}^{2}}\mathrm{E}\left( \mathrm{U},k\right) \leq 0\ .
\end{equation*}%
If additionally $\hat{\upsilon}(0)\neq 0$ and $r_{\mathfrak{p}}>0$, then 
\begin{equation*}
\sup_{\mathrm{U}\in \mathbb{R}_{0}^{+}}E\left( \mathrm{U}\right) <0\ .
\end{equation*}
\end{corollary}

\begin{proof}
To prove the first assertion, it suffices to combine Proposition \ref{direct
integral decomposition of the hamiltonian} and Theorem \ref{Maintheorem1}
with Theorem \ref{borelian functional calculus of a direct integral of
operators}. The second one can be proven like in \cite{articulo} by using
Kato's perturbation theory \cite{Kato}. In Proposition \ref{determining the
fundamental energy} we give an alternative and more direct proof of it.
Finally, the last assertion is a consequence of the inequalities 
\begin{equation*}
\min_{k\in \mathbb{T}^{2}}\mathrm{E}\left( \mathrm{U},k\right) \leq \mathrm{E%
}\left( \mathrm{U},0\right) =\mathrm{E}\left( \mathrm{U},0\right) -\min
\sigma _{\mathrm{ess}}\left( A\left( \mathrm{U},0\right) \right)
\end{equation*}%
and Theorem \ref{Maintheorem1} (iii).
\end{proof}

Physically speaking, the spectral values of $H$ represent the energy levels
that are available to the composite of two fermions and one boson, in
particular for a fermion pair exchanging a boson. As expected, the minimum
energy $E$, also well-known as the ground state energy, is given by
minimizing the eigenvalues $\mathrm{E}(\mathrm{U},k)$ over the torus $%
\mathbb{T}^{2}$.

We now study the model at very large on-site repulsion $\mathrm{U}\gg 1$. In
fact, quoting \cite{articulo2}, \textquotedblleft in all cuprates, there is
undeniable experimental evidence of strong on-site Coulomb repulsions,
leading to the universally observed Mott transition at zero doping \cite%
{Imada,Nature2015}. This phase is characterized by a periodic distribution
of fermions (electrons or holes) with exactly one particle per lattice site.
Doping copper oxides with holes or electrons can prevent this situation.
Instead, at sufficiently small temperatures a superconducting phase is
achieved, as first discovered in 1986 for the copper oxide perovskite $%
\mathrm{La}_{2-x}\mathrm{Ba}_{x}\mathrm{CuO}_{4}$ \cite{BM86}%
.\textquotedblright\ However, instead of the usual $s$-wave
superconductivity, one experimentally observes $d$-wave superconductivity 
\cite{Tsuei,Nature2015,tinpou1}. The fact that \emph{only} the $s$-wave
pairing is suppressed also advocates for a very local (i.e., on-site) and
strong effective repulsion of fermions. For this reason we consider the
limit $\mathrm{U}\rightarrow \infty $ in our model. It corresponds to a hard
core limit because it prevents two fermions from being on the same lattice
site.

In the limit $\mathrm{U}\rightarrow \infty $, it is easy to see that the
ground state energy $E(\mathrm{U})$ of Corollary \ref{maincoro1} defines an
increasing function of $\mathrm{U}\in \mathbb{R}_{0}^{+}$, which is bounded
from above by $0$. Hence, 
\begin{equation}
E\left( \infty \right) \doteq \lim\limits_{\mathrm{U}\rightarrow \infty
}E\left( \mathrm{U}\right) ={\sup\limits_{\mathrm{U}\in \mathbb{R}_{0}^{+}}}%
E\left( \mathrm{U}\right) \leq 0\ .  \label{hard-core ground state energy}
\end{equation}%
For more details, see Lemma \ref{formula for hard-core ground state energy}.
The limit $\mathrm{U}\rightarrow \infty $ of the eigenvalue and eigenvector
of each fiber, given by Theorem \ref{Maintheorem1}, is less trivial to
obtain and is the object of the next theorem:

\begin{theorem}[Spectral properties of fiber Hamiltonians -- Hard-core limit]

\label{Maintheorem2}\mbox{ }\newline
Fix $\epsilon ,\mathrm{U}\in \mathbb{R}_{0}^{+}$ and $h_{b}\in \lbrack
0,1/2] $. The following limits exist: 
\begin{eqnarray*}
\mathrm{E}\left( \infty ,k\right) &\doteq &{\lim\limits_{\mathrm{U}%
\rightarrow \infty }}\mathrm{E}\left( \mathrm{U},k\right) ={\sup\limits_{%
\mathrm{U}\in \mathbb{R}_{0}^{+}}}\mathrm{E}\left( \mathrm{U},k\right) \leq 
\mathfrak{b}\left( k\right) \ ,\qquad k\in \mathbb{T}^{2}\ . \\
\Psi \left( \infty ,k\right) &\doteq &{\lim_{\mathrm{U}\rightarrow \infty }}%
\,\Psi \left( \mathrm{U},k\right) \in \mathcal{H}\backslash \{0\}\ ,\qquad
k\in \mathbb{T}^{2}\backslash \{0\}\ .
\end{eqnarray*}%
Assertion (iv) of Theorem \ref{Maintheorem1} also holds true for $\mathrm{U}%
=\infty $. In addition, when $r_{\mathfrak{p}}>0$, $\mathrm{E}(\infty ,k)=%
\mathfrak{b}(k)$ iff $\hat{\upsilon}(k)=0$. If $r_{\mathfrak{p}}>0$ and $%
\hat{\upsilon}(0)\neq 0$, then $\Psi (\infty ,0)$ exists.
\end{theorem}

\begin{proof}
See Theorems \ref{properties of effective dispersion relation with low
energy} and \ref{hardcore limit of the eigenspace}.
\end{proof}

Note that the eigenvalues given by Theorem \ref{Maintheorem1} are not
explicitly known. The same is of course true in the hard-core limit $\mathrm{%
U}\rightarrow \infty $. For applications it is important to have a
sufficiently good control on these objects to be able to compute them,
either analytically or numerically. This is done in \cite{articulo,articulo2}
for the special case of the usual Hubbard interaction ($\mathrm{u}=0$) and
one-range creation operators $c_{y}^{\ast }$ of fermion pairs, by providing
estimates for $\mathrm{E}\left( \mathrm{U},k\right) $ and $\Psi \left( 
\mathrm{U},k\right) $ at large $\mathrm{U}$. See \cite[Theorem 4, Corollary
5, Theorem 16]{articulo}.

Recall that $\hat{\upsilon}$ is the Fourier transform of $\upsilon $, which
is the function appearing in Equation (\ref{Hamiltonian-fb}), encoding the
(exchange) interaction between fermion pairs\ and bosons. By Theorem \ref%
{Maintheorem1}, if $\hat{\upsilon}(k)=0$ then $\mathrm{E}(\mathrm{U},k)$ is
nothing else than the explicit function $\mathfrak{b}(k)$ (\ref{b}). Hence,
we focus on the physically more relevant case $\hat{\upsilon}(k)\neq 0$.
Using the Birman-Schwinger principle (Theorem \ref{birman-schwinger's
theorem}), we show in this case that the eigenvalue $\mathrm{E}(\mathrm{U}%
,k) $ is the unique solution to a relatively simple equation for real
numbers, similar to the characteristic equation used to compute eigenvalues
of matrices.

To this end, we define a function $\mathfrak{T}:\mathcal{D}\rightarrow 
\mathbb{R}$ on the set 
\begin{equation*}
\mathcal{D}\doteq \left\{ \left( \mathrm{U},k,x\right) \in \left[ 0,\infty %
\right] \times \mathbb{T}^{2}\times \mathbb{R}:x<\mathfrak{z}\left( k\right)
\right\} \subseteq \mathbb{R}^{3}
\end{equation*}%
by%
\begin{equation}
\mathfrak{T}\left( \mathrm{U},k,x\right) \doteq \left\langle \mathfrak{d}%
\left( k\right) ,\left( A_{1,1}\left( \mathrm{U},k\right) -x\mathfrak{1}%
\right) ^{-1}\mathfrak{d}\left( k\right) \right\rangle  \label{fract T}
\end{equation}%
for any finite $\mathrm{U}\in \mathbb{R}_{0}^{+}$, $k\in \mathbb{T}^{2}$ and 
$x\in \left( 0,\mathfrak{z}\left( k\right) \right) $, while for the infinite
on-site repulsion, $k\in \mathbb{T}^{2}$ and $x\in \left( 0,\mathfrak{z}%
\left( k\right) \right) $,%
\begin{equation*}
\mathfrak{T}\left( \infty ,k,x\right) \doteq \lim_{\mathrm{U}\rightarrow
\infty }\mathfrak{T}\left( \mathrm{U},k,x\right) \text{ },
\end{equation*}%
the above limit existing by virtue of Corollary \ref{limit of I(U,k,x) when
U goes to infinity}. Recall that $\mathfrak{z}\left( k\right) $\ is defined
by Equation (\ref{equation sur bk}). In fact, for any $k\in \mathbb{T}^{2}$
and $x\in \left( 0,\mathfrak{z}\left( k\right) \right) $, 
\begin{equation*}
\mathfrak{T}\left( \infty ,k,x\right) =R_{\mathfrak{s},\mathfrak{s}%
}^{-1}\left( {R_{\mathfrak{d},\mathfrak{d}}R_{\mathfrak{s},\mathfrak{s}}-}%
\left\vert R_{\mathfrak{s},\mathfrak{d}}\right\vert ^{2}\right) >0\ ,
\end{equation*}%
where $R_{\mathfrak{s},\mathfrak{s}}$, $R_{\mathfrak{s},\mathfrak{d}}$, $R_{%
\mathfrak{d},\mathfrak{s}}$ and $R_{\mathfrak{d},\mathfrak{d}}$ are four
constants defined by Equations (\ref{R1})--(\ref{R4}) with $\lambda =x$.
When $\mathrm{u}=0$ these constants are given by explicit integrals on the
torus $\mathbb{T}^{2}$ \cite{articulo,articulo2}. Then, the eigenvalues, the
existence of which is stated in Theorem \ref{Maintheorem1}, as well as their
limits (Theorem \ref{Maintheorem2}), can be studied via the following
characteristic equation:

\begin{theorem}[Characteristic equation for the fiber ground states]
\label{Maintheorem3}\mbox{ }\newline
Fix $\epsilon \in \mathbb{R}_{0}^{+}$, $h_{b}\in \lbrack 0,1/2]$ and $k\in 
\mathbb{T}^{2}$ such that $\hat{\upsilon}(k)\neq 0$. Then, for any $\mathrm{U%
}\in \left[ 0,\infty \right] $, $\mathrm{E}(\mathrm{U},k)$ is the unique
solution to the equation 
\begin{equation*}
\hat{\upsilon}\left( k\right) ^{2}\mathfrak{T}\left( \mathrm{U},k,x\right)
+x-\mathfrak{b}\left( k\right) =0\text{ },\qquad x<\mathfrak{z}\left(
k\right) \ .
\end{equation*}
\end{theorem}

\begin{proof}
For $\mathrm{U}\in \mathbb{R}_{0}^{+}$, combine Theorem \ref{eigenvalue as a
root of a non linear equation} with Theorem \ref{existence of eigenvalue for
each fiber}, while for $\mathrm{U}=\infty $ use Theorem \ref{properties of
effective dispersion relation with low energy}.
\end{proof}

Notice that, more generally, for any fixed $\mathrm{U}\in \mathbb{R}_{0}^{+}$%
, the same characteristic equation determines \emph{all} eigenvalues of the
fiber lying in the resolvent set $\rho (A_{1,1}(\mathrm{U},k))$ of the
operator $A_{1,1}(\mathrm{U},k)$. See Theorem \ref{eigenvalue as a root of a
non linear equation}. Also the associated eigenspaces can be explicitly
characterized, thanks to Corollary \ref{eigenspace of a fiber}. In this
context, Corollary \ref{there is at most one eigenvalue on each connected
component} shows that, for any $h_{b}\in \lbrack 0,1/2]$ and total
quasi-momentum $k\in \mathbb{T}^{2}$, there is at most one eigenvalue of $A(%
\mathrm{U},k)$ in each connected component of $\rho (A_{1,1}(\mathrm{U}%
,k))\cap \mathbb{R}$.

\subsection{Dispersion Relation of Dressed Bound Fermion Pairs\label{sect
disp rel bound pairs}}

By Theorem \ref{Maintheorem1}, $\mathrm{E}:\mathbb{T}^{2}\rightarrow \mathbb{%
R}$ is a continuous family of non-degenerate eigenvalues, generally (at
least for $k\neq 0$) associated with exponentially localized eigenvectors.
Note that the case $k=0$ is particular when $\hat{\upsilon}(0)=0$, since $%
\mathrm{E}(0)$ is not an isolated eigenvalue of $A(\mathrm{U},0)$. However,
the family $(\mathrm{E}(k))_{k\in \mathbb{T}^{2}}$ is still continuous. The
peculiar behavior at $k=0$ leads us to only consider total quasi-momenta in
the subset%
\begin{equation}
\mathbb{S}^{2}\doteq \left( -\pi ,\pi \right) ^{2}\backslash \left\{
0\right\} \subseteq \mathbb{T}^{2}\ ,  \label{sdsdsdssdssdd}
\end{equation}%
as, for instance, in Theorem \ref{Maintheorem1} (iv).

Because of Proposition \ref{direct integral decomposition of the hamiltonian}%
, the family $(\mathrm{E}(k))_{k\in \mathbb{T}^{2}}$ can thus be seen as the
effective dispersion relation of dressed bound fermion pairs. It is expected
to determine transport properties of the quantum system at low temperatures.
We now define in mathematical terms what a dispersion relation is.

First, a dispersion relation $\varkappa :\mathbb{T}^{2}\rightarrow \mathbb{R}
$ should be a functions mapping quasi-momenta $k\in \mathbb{T}^{2}$ on the
torus to spectral values of the corresponding fibers. More precisely, $%
\varkappa (k)$ should be an isolated eingenvalue of the fiber associated
with the total quasi-momentum $k$. Recall that the dispersion relation of a
(non-relativistic) particle in the $d$-dimensional continuum (that is, the
particle moves in the continuum $d$-dimensional space $\mathbb{R}^{d}$),
whose (isotropic) mass is $m$, is$\ k^{2}/2m$ and velocity $v(k)=k/m$, $k\in 
\mathbb{R}^{d}$. Having this standard example in mind, we would like also to
derive from a dispersion relation a group velocity and a mass tensor, at any
fixed quasi-momentum $k\in \mathbb{T}^{2}$, as is usual. These are key
objects, for instance in the study of transport properties. Notice that they
require sufficient regularity of the dispersion relation to be defined.

Keeping in mind that all our quantities are parametrized by the on-site
repulsion $\mathrm{U}\in \left[ 0,\infty \right] $, we define a family of
dispersion relations associated with the fiber Hamiltonians $A(\mathrm{U},k)$
as follows:

\begin{definition}[Family of dispersion relations]
\label{dispersion relation}\label{dispersion relation copy(3)}\mbox{ }%
\newline
A function $\varkappa :\left[ 0,\infty \right] \times \mathbb{T}%
^{2}\rightarrow \mathbb{R}$ is said to be a family of dispersion relations $%
\varkappa (\mathrm{U},\cdot )$ if the following properties are satisfied for
all $\mathrm{U}\in \left[ 0,\infty \right] $:

\begin{enumerate}
\item[i.)] For any $k\in \mathbb{T}^{2}$ and $\mathrm{U}\in \mathbb{R}%
_{0}^{+}$, $\varkappa (\mathrm{U},k)$ is an eigenvalue of $A(\mathrm{U},k)$
and 
\begin{equation*}
\varkappa \left( \infty ,k\right) =\lim\limits_{\mathrm{U}\rightarrow \infty
}\varkappa \left( \mathrm{U},k\right) \ .
\end{equation*}

\item[ii.)] For all $\mathrm{U}\in \left[ 0,\infty \right] $, $\varkappa (%
\mathrm{U},\cdot )\in C(\mathbb{T}^{2})$ and is of class $C^{2}$ on the open
set $\mathbb{S}^{2}\subseteq \mathbb{R}^{2}$.
\end{enumerate}
\end{definition}

The first property is a very natural property, having in mind Proposition %
\ref{direct integral decomposition of the hamiltonian} and the theory of
direct integrals (Theorem \ref{borelian functional calculus of a direct
integral of operators}). The second property of Definition \ref{dispersion
relation} is needed to define \emph{group velocities} and \emph{mass tensors}%
.

To explain these two concepts, we need the Hessian of functions $f\in C^{2}(%
\mathbb{S}^{2})$ at fixed $k$, which is denoted by 
\begin{equation}
\mathrm{Hess}\left( f\right) \left( k\right) \doteq \left( 
\begin{array}{cc}
\partial _{k_{1}}^{2}f & \partial _{k_{1}}\partial _{k_{2}}f \\ 
\partial _{k_{2}}\partial _{k_{1}}f & \partial _{k_{2}}^{2}f%
\end{array}%
\right) \left( k\right) \in \mathsf{M}_{2}\left( \mathbb{R}\right) \ ,\qquad
k\in \mathbb{S}^{2}\ ,  \label{def hessian}
\end{equation}%
where $\mathsf{M}_{2}(\mathbb{R})$ is the set of $2\times 2$ matrices with
real coefficients. It is a straightforward consequence of the regularity of $%
f\in C^{2}(\mathbb{S}^{2})$ that 
\begin{equation*}
\mathrm{Hess}\left( f\right) :\mathbb{S}^{2}\longrightarrow \mathsf{M}%
_{2}\left( \mathbb{R}\right)
\end{equation*}%
is continuous. For any $f\in C^{2}(\mathbb{S}^{2})$, we consider the set%
\begin{equation}
\mathfrak{M}_{f}\doteq \left\{ k\in \mathbb{S}^{2}\,:\,\mathrm{Hess}\left(
f\right) \left( k\right) \in \mathsf{GL}_{2}\left( \mathbb{R}\right)
\right\} \subseteq \mathbb{S}^{2}  \label{sdsdsdsds}
\end{equation}%
with $\mathsf{GL}_{2}\left( \mathbb{R}\right) \subseteq \mathsf{M}_{2}\left( 
\mathbb{R}\right) $ being the set of invertible $2\times 2$ matrices with
real coefficients. As $\mathsf{GL}_{2}(\mathbb{R})\subseteq \mathsf{M}_{2}(%
\mathbb{R})$ is an open set (see \cite[Theorem 1.4]{Folland}), it then
follows that%
\begin{equation*}
\mathfrak{M}_{f}=\mathrm{Hess}\left( f\right) ^{-1}\left( \mathsf{GL}%
_{2}\left( \mathbb{R}\right) \right)
\end{equation*}%
is also an open set.

We are now in a position to define group velocities and mass tensors of a
family of dispersion relations.

\begin{definition}[Group velocities and mass tensors]
\label{group velocity and mass tensor copy(2)}\mbox{ }\newline
At any $\mathrm{U}\in \left[ 0,\infty \right] $, the group velocity $\mathbf{%
v}_{\varkappa ,\mathrm{U}}:\mathbb{S}^{2}\rightarrow \mathbb{R}$ and the
mass tensor $\mathbf{m}_{\varkappa ,\mathrm{U}}:\mathfrak{M}_{\varkappa
\left( \mathrm{U},\cdot \right) }\rightarrow \mathsf{M}_{2}\left( \mathbb{R}%
\right) $ associated with a family $\varkappa :\left[ 0,\infty \right]
\times \mathbb{T}^{2}\rightarrow \mathbb{R}$ of dispersion relations are
respectively defined by%
\begin{equation*}
\mathbf{v}_{\varkappa ,\mathrm{U}}\left( k\right) \doteq \vec{\nabla}%
_{k}\varkappa \left( \mathrm{U},k\right) \qquad \text{and}\qquad \mathbf{m}%
_{\varkappa ,\mathrm{U}}\left( k\right) \doteq \mathrm{Hess}\left( \varkappa
\left( \mathrm{U},\cdot \right) \right) \left( k\right) ^{-1}\ .
\end{equation*}
\end{definition}

We deduce from Theorem \ref{regularity of E} that $\mathrm{E}$ is a
dispersion relation when the function $\upsilon :\mathbb{Z}^{2}\rightarrow 
\mathbb{R}$ is at least $2$ times continuously differentiable and, in this
case, we can even compute the group velocity via the characteristic equation
(Theorem \ref{Maintheorem3}).

\begin{theorem}[Dispersion relations of dressed bound fermion pairs]
\label{existence of a dispersion relation with low energy copy(2)}\mbox{ }%
\newline
Fix $\epsilon \in \mathbb{R}_{0}^{+}$ and $h_{b}\in \lbrack 0,1/2]$. Assume
that $\hat{\upsilon}\in C^{2}(\mathbb{S}^{2})$.

\begin{enumerate}
\item[i.)] Then, $\mathrm{E}:\left[ 0,\infty \right] \times \mathbb{T}%
^{2}\rightarrow \mathbb{R}$ given by Theorems \ref{Maintheorem1} and \ref%
{Maintheorem2} is a family of dispersion relations.

\item[ii.)] The associated group velocities are equal to 
\begin{equation*}
\mathbf{v}_{\mathrm{E},\mathrm{U}}\left( k\right) =\left. \left( \hat{%
\upsilon}\left( k\right) ^{2}\partial _{x}\mathfrak{T}\left( \mathrm{U}%
,k,x\right) +1\right) ^{-1}\vec{\nabla}\left( \hat{\upsilon}\left( k\right)
^{2}\mathfrak{T}\left( \mathrm{U},k,x\right) -\mathfrak{b}\left( k\right)
\right) \right\vert _{x={\mathrm{E}\left( \mathrm{U},k\right) }}
\end{equation*}%
for any $\mathrm{U}\in \left[ 0,\infty \right] $ and $k\in \mathbb{S}^{2}$,
with 
\begin{equation*}
\mathbf{v}_{\mathrm{E},\infty }\left( k\right) =\lim_{\mathrm{U}\rightarrow
\infty }\mathbf{v}_{\mathrm{E},\mathrm{U}}\left( k\right) \ ,\qquad k\in 
\mathbb{S}^{2}\ .
\end{equation*}

\item[iii.)] If $\hat{\upsilon}$ is real analytic on $\mathbb{S}^{2}$, then,
for any $\mathrm{U}\in \left[ 0,\infty \right] $, either $\mathfrak{M}_{%
\mathrm{E}\left( \mathrm{U},\cdot \right) }$ has full measure or $\mathfrak{M%
}_{\mathrm{E}}=\emptyset $. In particular, the tensor masses $\mathbf{m}_{%
\mathrm{E},\mathrm{U}}$\ are either defined almost everywhere in $\mathbb{S}%
^{2}$ or not defined at all.
\end{enumerate}
\end{theorem}

\begin{proof}
Use Corollaries \ref{existence of a dispersion relation with low energy} and %
\ref{Hard-core dispersion relation}.
\end{proof}

Similar to Remark \ref{remark regularity}, if for some strictly positive
constant $\gamma >0$, 
\begin{equation*}
\sum_{x\in \mathbb{Z}^{2}}\mathrm{e}^{\gamma \left\vert x\right\vert
}\left\vert \upsilon \left( x\right) \right\vert <\infty
\end{equation*}%
then the Fourier transform $\hat{\upsilon}$ of the function $\upsilon :%
\mathbb{Z}^{2}\rightarrow \mathbb{R}$, as defined via (\ref{fourier
transforms}), is real analytic on the whole torus $\mathbb{T}^{2}$. It is
very natural to expect a local interaction between fermion pairs\ and bosons
in (\ref{Hamiltonian-fb}), meaning here that the function $\upsilon $ should
even have finite support. In particular, all conditions of Theorem \ref%
{existence of a dispersion relation with low energy copy(2)}, including the
ones of the third assertion, should hold true in the application to
superconducting cuprates.

In fact, as shown in \cite{articulo2}, the dispersion relation of Theorem %
\ref{existence of a dispersion relation with low energy copy(2)} yields the
formation of $d$-wave pairs when one adjusts the parameters of the model
(with $\mathrm{u}=0$) to fit those of cuprate superconductors, in particular
the ones of the cuprate $\mathrm{La}_{2-x}\mathrm{Sr}_{x}\mathrm{CuO}_{4}$
(LaSr 214) near optimal doping. When considering the usual Hubbard model,
that is, the case where there is no other fermionic repulsion than the
on-site one (i.e., $\mathrm{u}=0$) and no fermion-boson exchange (i.e., $%
\upsilon =0$), $\mathrm{E}$ turns out to be the function $\mathfrak{b}:%
\mathbb{T}^{2}\rightarrow \mathbb{R}$, defined by (\ref{b}), which is
nothing else than the dispersion relation%
\begin{equation*}
\mathfrak{b}(k)\doteq h_{b}\epsilon \left( 2-\cos \left( k\right) \right) \
,\qquad k\in \mathbb{T}^{2}\ ,
\end{equation*}%
of free bosons (bipolarons for cuprates).

By turning on the fermion-boson-exchange interaction, the dispersion
relation of dressed bound fermion pairs with lowest energy can strongly
deviate from $\mathfrak{b}$, the unperturbed one. Recall for instance that $%
\mathfrak{b}$ describes bosons with a very large mass as compared to the
effective mass of electrons or holes in cuprates. However, as shown in \cite%
{articulo2}, for typical parameters of the cuprate LaSr 214, the effective
mass of the bound pair (with dispersion relation $\mathrm{E}$) is comparable
to the mass of electrons or holes. This is a consequence of the mass of
charge carriers calculated in \cite{polaronsize}, and the fact that a large
effective mass of dressed bound fermion pairs and a high fermion-pair
depletion\footnote{%
I.e., the fermionic component of the dressed bound fermion pairs is very
small in comparison with the bipolaronic component.}, close to $90\%$ as
measured in \cite{Nature2016}, is not compatible with our model. This solves
the so-called \textquotedblleft large mass paradox\textquotedblright\ of the
microscopic theory of cuprate superconductors, based on some kind mechanism
involving bipolarons. For more details, see \cite{articulo2} and references
therein.

In fact, the effective mass of dressed bound fermion pairs, or more
generally its (effective) mass tensor, depends strongly on the coupling
function $\hat{\upsilon}$ near its maximum. Bearing in mind Definitions \ref%
{dispersion relation} and \ref{group velocity and mass tensor copy(2)}, one
can therefore provide via\ Theorems \ref{Maintheorem3} and \ref{group
velocity and mass tensor copy(2)} not only qualitative but also quantitative
information, which is important for describing the physical behaviour of
fermionic pairs formed in this way by means of a bosonic field. A natural
question is then to study its scattering properties and this is precisely
what we propose to do in the next section.

\subsection{Quantum Scattering}

Scattering in quantum mechanics constitutes a well-established mathematical
theory aiming at analyzing the behavior of quantum systems at large times.
To this end, a reference (or free) Hamiltonian $Y$ is chosen and the
dynamics $(\mathrm{e}^{itX})_{t\in \mathbb{R}}$ of the quantum system driven
by the (full) Hamiltonian $X$ is compared at large (negative and positive)
times to $(\mathrm{e}^{itY})_{t\in \mathbb{R}}$. In fact, scattering theory
can be viewed as a kind of perturbation theory for the absolutely continuous
spectrum of $X$. See, e.g., \cite[Chapter X]{Kato}. For standard textbooks
explaining in detail the scattering theory, we recommend \cite%
{ReedSimonIII,Newton,Yafaev}. Below, for the reader's convenience, we
shortly recall definitions that are relevant here.

Take two bounded self-adjoint operators $X$ and $Y$ acting on two Hilbert
spaces $\mathcal{X}$ and $\mathcal{Y}$, respectively. Let $P_{\mathrm{ac}%
}(Y) $ be the orthogonal projection onto the absolutely continuous space of $%
Y$, which is defined as follows: 
\begin{eqnarray}
\mathrm{ran}\left( P_{\mathrm{ac}}(Y)\right) &\doteq &%
\big\{%
\psi \in \mathcal{Y}:\left\langle \psi ,\chi _{(\cdot )}(Y)\psi
\right\rangle _{\mathcal{Y}}\text{ is absolutely continuous}  \notag \\
&&\qquad \qquad \qquad \qquad \qquad \qquad \text{with respect to the
Lebesgue measure}%
\big\}%
\ ,  \label{abs cont space}
\end{eqnarray}%
where $\chi _{\Omega }$ is its characteristic function\footnote{$\chi
_{\Omega }\left( x\right) =1$ for $x\in \Omega $ and $\chi _{\Omega }\left(
0\right) =1$ otherwise.} of any Borel set $\Omega \subseteq \mathbb{R}$. The
so-called \emph{wave operators}\ for the pair $(X,Y)$ with \emph{%
identification operator}\ $J\in \mathcal{B}\left( \mathcal{Y},\mathcal{X}%
\right) $ is, by definition, the strong limit%
\begin{equation}
W^{\pm }\left( X,Y;J\right) \doteq s-{\lim\limits_{t\rightarrow \pm \infty }}%
\mathrm{e}^{itX}J\mathrm{e}^{-itY}P_{\mathrm{ac}}\left( Y\right) \text{ },
\label{scattering operator1}
\end{equation}%
when it exists. See for instance \cite[Definition 1.3]{Yafaev}. When $%
\mathcal{Y}=\mathcal{X}$ and $J=\mathfrak{1}$, like in \cite[Definition 1.1]%
{Yafaev}, we use the shorter notation 
\begin{equation}
W^{\pm }\left( X,Y\right) \equiv W^{\pm }\left( X,Y;\mathfrak{1}\right)
\doteq s-{\lim\limits_{t\rightarrow \pm \infty }}\mathrm{e}^{itX}\mathrm{e}%
^{-itY}P_{\mathrm{ac}}\left( Y\right) \text{ }.  \label{scattering operator2}
\end{equation}%
In case the above wave operators exist, they are partial isometries \cite[%
Proposition 1,\ Sect. XI.3]{ReedSimonIII}. They are said to be \emph{complete%
} when 
\begin{equation*}
\mathrm{ran}\left( W^{+}\left( X,Y\right) \right) =\mathrm{ran}\left(
W^{-}\left( X,Y\right) \right) =\mathrm{ran}\left( P_{\mathrm{ac}}(X)\right)
\ .
\end{equation*}%
See \cite[p. 19,\ Sect. XI.3]{ReedSimonIII}.

Similarly, in the general case, $W^{\pm }\left( X,Y;J\right) $ are said to
be complete whenever 
\begin{equation*}
\overline{\mathrm{ran}\left( W^{+}\left( X,Y;J\right) \right) }=\overline{%
\mathrm{ran}\left( W^{-}\left( X,Y;J\right) \right) }=\mathrm{ran}\left( P_{%
\mathrm{ac}}(X)\right) \ .
\end{equation*}%
See \cite[p. 35,\ Sect. XI.3]{ReedSimonIII}. The corresponding scattering
operator is equal to 
\begin{equation}
S\left( X,Y;J\right) \doteq W^{+}\left( X,Y;J\right) ^{\ast }W^{-}\left(
X,Y;J\right) \in \mathcal{B}\left( \mathcal{Y}\right) \ .
\label{scaterring operator}
\end{equation}%
It leads to the scattering matrix (or simply $S$-matrix) in a representation
where $Y$ is diagonal, because the scattering operator commutes with $Y$.
See \cite[Equation (1.12)]{Yafaev}.

\begin{remark}
\label{direct sum of absolutely continuous space}\mbox{ }\newline
For two bounded self-adjoint operators $X$ and $Y$ acting on two Hilbert
spaces $\mathcal{X}$ and $\mathcal{Y}$, note\footnote{%
To show this property, take any Borel set $\Omega \subseteq \mathbb{R}$ and
observe that $\langle (\varphi ,\psi ),\chi _{\Omega }(X\oplus Y)(\varphi
,\psi )\rangle _{\mathcal{X}\oplus \mathcal{Y}}=\langle \varphi ,\chi
_{\Omega }(X)\varphi \rangle _{\mathcal{X}}+\langle \psi ,\chi _{\Omega
}(Y)\psi \rangle _{\mathcal{Y}}.$ Since $\chi _{\Omega }(X)$ and $\chi
_{\Omega }(Y)$ are positive operators, the left-hand side of the last
expression is zero iff each term in the right-hand side is zero.} that 
\begin{equation*}
\mathrm{ran}\left( P_{\mathrm{ac}}(X\oplus Y)\right) =\mathrm{ran}\left( P_{%
\mathrm{ac}}(X)\right) \oplus \mathrm{ran}\left( P_{\mathrm{ac}}(Y)\right) \
.
\end{equation*}%
This is an elementary observation used to study the scattering channels in\
Section \ref{scattering channels}.
\end{remark}

In our framework, the Hamiltonian $X$ is the bounded self-adjoint operator 
\begin{equation*}
\mathbb{U}H\mathbb{U}^{\ast }={\int_{\mathbb{T}^{2}}^{\oplus }}A(k)\,\nu (%
\mathrm{d}k)
\end{equation*}%
of Proposition \ref{direct integral decomposition of the hamiltonian}, which
acts on the Hilbert space $L^{2}(\mathbb{T}^{2},\mathcal{H})$. Below, two
different (reference) Hamiltonians $Y$ are taken into account, corresponding
to two scattering channels: the \emph{unbound} and \emph{bound} pair
channels. For cuprates, the first channel should be\ associated with the
high temperature regime, while the second one is related to sufficiently low
temperatures.

\subsubsection{Unbound pair scattering channel\label{Unbound Pair Channel}}

Far apart from each other, two fermions only experience a very weak
repulsion force due to the extended Hubbard interaction while the
probability that they bind together to form a boson is also very small.
Thus, in this situation, one expects that the dynamics of such a pair is
governed by the fermionic part, and even by the hopping term alone. During
intermediate times they could of course interact, as they may get close to
each other, and they could even be bound together via the effective
attraction caused by fermion-boson exchange processes. The lifetime of bound
fermions should however be finite in this situation and they are expected to
be released at some point and behave again as two free fermions that go far
apart from each other for large times. See Fig. \ref{figure1}. We show below
that this heuristics can be put in precise mathematical terms. 
%
%
\begin{figure}[!hbtp]
\begin{center}
\begin{tikzpicture}
\begin{feynman}

\vertex[blob, label={right:$S_{k}$}, minimum height=3cm,minimum width=2cm](m) at ( 0, 0) {};
\vertex (a) at (-5,-1) {$k-q$};
\vertex (b) at ( 5,-3) {$k-p$};
\vertex (c) at (-5, 1) {$q$};
\vertex (d) at ( 5, 3) {$p$};
\diagram* {
(a) -- [fermion] (m) -- [anti fermion] (c),
(b) -- [anti fermion] (m) -- [fermion] (d),
};
\end{feynman}

\end{tikzpicture}
\end{center}
\caption{Illustration of the unbound pair scattering channel: Two free fermions of (quasi-) momentum $k-p$ and $q$ respectively (i.e.
the full momentum of the fermionic pair is $k$) at time $t=-\infty $
interact in finite time with the composite system, in particular with the
bosonic field, to be asymptotically free again at time $t=+\infty $, thanks
to Theorem \ref{existence of a kato wave operator for the system}. Here $S_{k}=S\left( A\left( k\right) ,\left( M_{\mathfrak{f}\left( k\right)
}+R\left( \mathrm{V},\mathrm{v}\right) \right) \oplus A_{2,2}\left( k\right)
\right) $ is the scattering operator of this process in each fiber $k$,
which depends explicitly on $\hat{\upsilon}\left( k\right) $. See Theorem \ref{existence of a kato wave operator for the system copy(1)} and the
example given by Equations (\ref{ex1})--(\ref{ex2}).}
\label{figure1}
\end{figure}%

To this end, define the Hilbert space 
\begin{equation}
\mathfrak{H}_{f}\doteq L^{2}\left( \mathbb{T}^{2},L^{2}\left( \mathbb{T}%
^{2}\right) ,\nu \right) \doteq \int_{\mathbb{T}^{2}}^{\oplus }L^{2}\left( 
\mathbb{T}^{2},\mathbb{C},\nu \right) \,\nu \left( \mathrm{d}k\right)
\label{H0-HIlbert space}
\end{equation}%
as well as the Hamiltonian 
\begin{equation}
\mathrm{H}_{f}\equiv \mathrm{H}_{f}\left( \mathrm{V},\mathrm{v}\right)
\doteq {\int_{\mathbb{T}^{2}}^{\oplus }}\left( M_{\mathfrak{f}\left(
k\right) }+R\left( \mathrm{V},\mathrm{v}\right) \right) \,\nu (\mathrm{d}%
k)\in \mathcal{B}\left( \mathfrak{H}_{f}\right)  \label{H0H0}
\end{equation}%
for any $\mathrm{V}\in \mathbb{R}_{0}^{+}$ and absolutely summable function $%
\mathrm{v}:\mathbb{Z}^{2}\rightarrow \mathbb{R}_{0}^{+}$, where 
\begin{equation}
R\left( \mathrm{V},\mathrm{v}\right) \doteq {\sum\limits_{x\in \mathbb{Z}%
^{2}}}\,\mathrm{v}\left( x\right) P_{x}+\mathrm{V}P_{0}\in \mathcal{B}\left(
L^{2}(\mathbb{T}^{2})\right) \ ,  \label{repulsion}
\end{equation}%
$M_{\mathfrak{f}\left( k\right) }$ being the fiber Hamiltonian defined as
the multiplication operator by $\mathfrak{f}(k)\in C(\mathbb{T}^{2})$ (see (%
\ref{f})) while $P_{x}$ is the orthogonal projection onto the
one-dimensional subspace $\mathbb{C}\mathfrak{\hat{e}}_{x}\subseteq L^{2}(%
\mathbb{T}^{2})$. Observe then that 
\begin{equation*}
\mathbb{U}H\mathbb{U}^{\ast }-%
\begin{pmatrix}
\mathrm{H}_{f} & 0 \\[0.5em] 
0 & 0%
\end{pmatrix}%
={\int_{\mathbb{T}^{2}}^{\oplus }}%
\begin{pmatrix}
{\sum\limits_{x\in \mathbb{Z}^{2}}}\,\left( \mathrm{u}\left( x\right) -%
\mathrm{v}\left( x\right) \right) P_{x}+\left( \mathrm{U}-\mathrm{V}\right)
P_{0} & A_{1,2}\left( k\right) \\[0.5em] 
A_{2,1}\left( k\right) & A_{2,2}\left( k\right)%
\end{pmatrix}%
\,\nu (\mathrm{d}k)\ .
\end{equation*}%
Compare indeed (\ref{H0H0}) with Equations (\ref{fiber hamiltonians})--(\ref%
{A11}) and Proposition \ref{direct integral decomposition of the hamiltonian}%
. By Lemma \ref{lemma for existence of non-bound channel}, note that $P_{%
\mathrm{ac}}(\mathrm{H}_{f})=\mathfrak{1}$.

Let us consider the identification operator $\mathfrak{U}:\mathfrak{H}%
_{f}\rightarrow L^{2}\left( \mathbb{T}^{2},\mathcal{H}\right) $ defined for
any purely fermionic state $\psi \in \mathfrak{H}_{f}$ by 
\begin{equation}
\begin{array}{cccl}
\mathfrak{U}\psi : & \mathbb{T}^{2} & \rightarrow & \mathcal{H}\doteq
L^{2}\left( \mathbb{T}^{2}\right) \oplus \mathbb{C} \\ 
& k & \mapsto & \left( \psi \left( k\right) ,0\right)%
\end{array}%
\ .  \label{identification operator1}
\end{equation}%
See Equation (\ref{cal H}). Note that $\mathfrak{U}$ is an isometry, i.e., a
norm preserving linear transformation. In fact, it is the canonical
fiberwise inclusion of $\mathfrak{H}_{f}$ into $L^{2}\left( \mathbb{T}^{2},%
\mathcal{H}\right) $. Recall from Proposition \ref{direct integral
decomposition of the hamiltonian} that $A\left( \mathrm{U},\cdot \right) $,
defined by (\ref{A11})--(\ref{fiber hamiltonians}), is the fiber
decomposition of the operator $\mathbb{U}H\mathbb{U}^{\ast }$. Then, we
obtain wave and scattering operators with respect to fermionic parts:

\begin{theorem}[Unbound pair (scattering) channel]
\label{existence of a kato wave operator for the system}\mbox{ }\newline
Let $\mathrm{V}\in \mathbb{R}_{0}^{+}$ and $\mathrm{v}:\mathbb{Z}%
^{2}\rightarrow \mathbb{R}_{0}^{+}$ be any absolutely summable function.

\begin{enumerate}
\item[i.)] \textit{The wave operators, as defined by (\ref{scattering
operator1}) for }$X=\mathbb{U}H\mathbb{U}^{\ast }$, $Y=\mathrm{H}_{f}$ and $%
J=\mathfrak{U}$\textit{, satisfy}%
\begin{equation}
W^{\pm }\left( \mathbb{U}H\mathbb{U}^{\ast },\mathrm{H}_{f};\mathfrak{U}%
\right) =\left( {\int_{\mathbb{T}^{2}}^{\oplus }}W^{\pm }\left( A\left(
k\right) ,\left( M_{\mathfrak{f}\left( k\right) }+R\left( \mathrm{V},\mathrm{%
v}\right) \right) \oplus A_{2,2}\left( k\right) \right) \nu \left( \mathrm{d}%
k\right) \right) \mathfrak{U}  \label{ran0}
\end{equation}%
with range equal to 
\begin{equation}
\mathrm{ran}\left( W^{\pm }\left( \mathbb{U}H\mathbb{U}^{\ast },\mathrm{H}%
_{f};\mathfrak{U}\right) \right) =\int_{\mathbb{T}^{2}}^{\oplus }L^{2}\left( 
\mathbb{T}^{2}\right) \oplus \{0\}\,\nu \left( \mathrm{d}k\right) \ .
\label{ran1}
\end{equation}

\item[ii.)] \textit{The scattering operator, as defined by (\ref{scaterring
operator}) for }$X=\mathbb{U}H\mathbb{U}^{\ast }$, $Y=\mathrm{H}_{f}$ and $J=%
\mathfrak{U}$\textit{, equals} 
\begin{equation*}
S\left( \mathbb{U}H\mathbb{U}^{\ast },\mathrm{H}_{f};\mathfrak{U}\right) =%
\mathfrak{U}^{\ast }\left( {\int_{\mathbb{T}^{2}}^{\oplus }}S\left( A\left(
k\right) ,\left( M_{\mathfrak{f}\left( k\right) }+R\left( \mathrm{V},\mathrm{%
v}\right) \right) \oplus A_{2,2}\left( k\right) \right) \nu \left( \mathrm{d}%
k\right) \right) \mathfrak{U}\ .
\end{equation*}
\end{enumerate}
\end{theorem}

\begin{proof}
Observe that the operator difference $(\mathbb{U}H\mathbb{U}^{\ast }-\mathrm{%
H}_{f})$ is not trace-class (it is not even compact) and, thus, the
existence of this scattering channel is not a direct consequence of the
well-known Kato-Rosenblum theorem \cite[Theorem XI.8]{ReedSimonIII}. In
fact, one of the main steps of the proof is to show that this difference is
the direct integral of a strongly measurable family of trace-class
operators. By this means we are then able to apply the Kato-Rosenblum
theorem \textquotedblleft fiberwise\textquotedblright\ to deduce the first
assertion. See Section \ref{non-bound electronic pair channel}\ for more
details, in particular Theorem \ref{existence of a kato wave operator for
the systembis}. Assertion (ii) is a direct consequence of Assertion (i)
together with the theory of direct integrals.
\end{proof}

\begin{remark}
\label{remark spacial space}\mbox{ }\newline
If one would like to go back to the original Hilbert space $\mathfrak{h}_{0}$
(\ref{h0}) for fermion pairs with opposite spins, that is, if one wishes to
use space coordinates, instead of the quasi-momenta, then one employs
Theorem \ref{existence of a kato wave operator for the system}, along with
the observation that 
\begin{equation*}
W^{\pm }\left( H,U_{f}^{\ast }\mathrm{H}_{f}U_{f};\mathbb{U}^{\ast }%
\mathfrak{U}U_{f}\right) =U^{\ast }W^{\pm }\left( \mathbb{U}H\mathbb{U}%
^{\ast },\mathrm{H}_{f};\mathfrak{U}\right) U_{f}\ ,
\end{equation*}%
where $U_{f}$ is defined by (\ref{Uf}). See also Equation (\ref{U}) and
Proposition \ref{direct integral decomposition of the hamiltonian}.
\end{remark}

Theorem \ref{existence of a kato wave operator for the system} refers to the
unbound pair (scattering) channel. The subspace $\mathfrak{H}_{f}\subseteq 
\mathfrak{H}$ corresponds to the \textquotedblleft
incoming\textquotedblright\ ($+$) and \textquotedblleft
outcoming\textquotedblright\ ($-$) scattering states of the quantum system,
in this particular scattering channel. Physically, this theorem shows, among
other things, that the bosonic component of \textrm{e}$^{itH}$ vanishes on
this channel, as $t\rightarrow \pm \infty $. This is a direct consequence of
Equation (\ref{ran1}).

In addition, Equation (\ref{ran0}) gives an explicit fiber decomposition of
wave operators with respect to the purely fermionic Hamiltonian in terms of $%
k$-dependent wave operators defined naturally from the fiber decomposition
of the operator $\mathbb{U}H\mathbb{U}^{\ast }$. Mutatis mutandis for the
scattering operator, thanks to Theorem \ref{existence of a kato wave
operator for the system} (ii). In other words, the knowledge of scattering
properties of each fiber, almost everywhere, entirely determines the
scattering properties of the composite system, made of two fermions and one
boson. We can now use this property, i.e. Theorem \ref{existence of a kato
wave operator for the system}, to obtain a more computable expression for
the wave and scattering operators in each given fiber. This can be done via
infinite series (perturbative expansions), thanks to Corollary \ref%
{finite-time scattering and wave operators copy(2)}.

Below we give an example of such a computation by taking $\mathrm{U}=\mathrm{%
V}\in \mathbb{R}_{0}^{+}$ and $\mathrm{v}=\mathrm{u}:\mathbb{Z}%
^{2}\rightarrow \mathbb{R}_{0}^{+}$ in Theorem \ref{existence of a kato wave
operator for the system}. With this particular choice, the unbound pair
channel allows one to isolate the fermion-boson exchange mechanism, which in
terms of Hamiltonians refers to the use of off-diagonal operators 
\begin{equation}
B^{(t)}\left( k\right) \doteq 
\begin{pmatrix}
0 & B_{1,2}^{(t)}\left( k\right) \\[0.5em] 
B_{2,1}^{(t)}\left( k\right) & 0%
\end{pmatrix}%
\in \mathcal{B}\left( \mathcal{H}\right) \ ,\qquad t\in \mathbb{R},\ k\in 
\mathbb{T}^{2}\ ,  \label{B0}
\end{equation}%
in the fibers, where, for any $t\in \mathbb{R}$ and $k\in \mathbb{T}^{2}$, 
\begin{equation}
B_{1,2}^{(t)}\left( k\right) \doteq \mathrm{e}^{itA_{1,1}\left( \mathrm{U}%
,k\right) }A_{1,2}\left( k\right) \mathrm{e}^{-itA_{2,2}\left( k\right)
}\qquad \text{and}\qquad B_{2,1}^{(t)}\left( k\right) \doteq \mathrm{e}%
^{itA_{2,2}\left( k\right) }A_{2,1}\left( k\right) \mathrm{e}%
^{-itA_{1,1}\left( \mathrm{U},k\right) }\ .  \label{B0bis}
\end{equation}%
Recall that, for $m,n\in \{1,2\}$, $A_{m,n}(k)$ is defined by (\ref{A11})--(%
\ref{A22}). Below, $B_{m,n}(k)$, $m\neq n$, stands for the norm-continuous
family of operators $(B_{m,n}^{(t)}(k))_{t\in \mathbb{R}}$.

To shorten the notation, for any $s,t\in \mathbb{R}$, as well as two
norm-continuous families $X\equiv (X_{t})_{t\in \mathbb{R}}$ and $Y\equiv
(Y_{t})_{t\in \mathbb{R}}$ of bounded operators $X_{t}:\mathcal{X}%
\rightarrow \mathcal{Y}$ and $Y_{t}:\mathcal{Y}\rightarrow \mathcal{X}$ on
two Hilbert spaces $\mathcal{X}$ and $\mathcal{Y}$, respectively, we define
the bounded operators:%
\begin{eqnarray}
\cos _{\succ }\left( XY;s,t\right) &\doteq &\mathbf{1}+\sum_{p=1}^{\infty
}\left( -1\right) ^{p}\int_{s}^{t}\mathrm{d}\tau _{1}\cdots \int_{s}^{\tau
_{2p-1}}\mathrm{d}\tau _{2p}(X_{\tau _{1}}Y_{\tau _{2}})\cdots (X_{\tau
_{2p-1}}Y_{\tau _{2p}})\ ,  \label{cosinus-operator} \\
\sin _{\succ }\left( XY;s,t\right) &\doteq &\int_{s}^{t}\mathrm{d}\tau
X_{\tau }+\sum_{p=1}^{\infty }\left( -1\right) ^{p}\int_{s}^{t}\mathrm{d}%
\tau _{1}\cdots \int_{s}^{\tau _{2p}}\mathrm{d}\tau _{2p+1}X_{\tau
_{1}}\left( (Y_{\tau _{2}}X_{\tau _{3}})\cdots (Y_{\tau _{2p}}X_{\tau
_{2p+1}})\right) \ .  \notag \\
&&  \label{sinus-operator}
\end{eqnarray}%
The integrals above are Riemann ones, noting that $(X_{t})_{t\in \mathbb{R}}$
and $(Y_{t})_{t\in \mathbb{R}}$ are continuous families in Banach spaces,
namely $\mathcal{B}\left( \mathcal{X};\mathcal{Y}\right) $ and $\mathcal{B}%
\left( \mathcal{Y};\mathcal{X}\right) $, respectively. Note that $\cos
_{\succ }\left( XY;s,t\right) \in \mathcal{B}\left( \mathcal{Y}\right) $ and 
$\sin _{\succ }\left( XY;s,t\right) \in \mathcal{B}\left( \mathcal{X},%
\mathcal{Y}\right) $ are always absolutely summable series in the operator
norm. Then, we obtain the following results:

\begin{theorem}[Scattering operators as pertubative series]
\label{existence of a kato wave operator for the system copy(1)}\mbox{ }%
\newline
Let $\varepsilon \in \mathbb{R}^{+}$ and $\mathrm{H}_{f}\equiv \mathrm{H}%
_{f}\left( \mathrm{U},\mathrm{u}\right) $. Then, for any $\varphi \in 
\mathfrak{H}_{f}$, there is $T>0$ such that 
\begin{eqnarray*}
T &<&t\implies \left\Vert \left( W^{+}\left( \mathbb{U}H\mathbb{U}^{\ast },%
\mathrm{H}_{f};\mathfrak{U}\right) -V_{0,t}\mathfrak{U}\right) \varphi
\right\Vert _{\mathcal{X}}\leq \varepsilon \ , \\
t &<&-T\implies \left\Vert \left( W^{-}\left( \mathbb{U}H\mathbb{U}^{\ast },%
\mathrm{H}_{f};\mathfrak{U}\right) -V_{0,t}\mathfrak{U}\right) \varphi
\right\Vert _{\mathcal{X}}\leq \varepsilon \ ,
\end{eqnarray*}%
Moreover, for any $\varphi ,\psi \in \mathfrak{H}_{f}$, there is $T>0$ such
that%
\begin{equation*}
s<-T<T<t\implies \left\langle \psi ,S\left( \mathbb{U}H\mathbb{U}^{\ast },%
\mathrm{H}_{f};\mathfrak{U}\right) \varphi \right\rangle _{\mathcal{X}%
}=\left\langle \mathfrak{U}\psi ,V_{t,s}\mathfrak{U}\varphi \right\rangle _{%
\mathcal{X}}+\mathcal{O}\left( \varepsilon \right) \ ,
\end{equation*}%
where, for all $s,t\in \mathbb{R}$,%
\begin{equation*}
V_{t,s}\doteq {\int_{\mathbb{T}^{2}}^{\oplus }}\left( 
\begin{array}{cc}
\cos _{\succ }\left( B_{1,2}\left( k\right) B_{2,1}\left( k\right)
;s,t\right) & -i\sin _{\succ }\left( B_{1,2}\left( k\right) B_{2,1}\left(
k\right) ;s,t\right) \\ 
-i\sin _{\succ }\left( B_{2,1}\left( k\right) B_{1,2}\left( k\right)
;s,t\right) & \cos _{\succ }\left( B_{2,1}\left( k\right) B_{1,2}\left(
k\right) ;s,t\right)%
\end{array}%
\right) \nu \left( \mathrm{d}k\right) \ .
\end{equation*}
\end{theorem}

\begin{proof}
It suffices to combine Lemma \ref{finite-time scattering and wave operators
copy(3)} with Equation (\ref{interwining good}) and Theorem \ref{existence
of a kato wave operator for the system}, similar to Corollary \ref%
{finite-time scattering and wave operators copy(2)}.
\end{proof}

Theorem \ref{existence of a kato wave operator for the system copy(1)}
provides a way to approximate the scattering matrix associated with the
fermion-boson-exchange interaction. Note for instance from (\ref{A11})--(\ref%
{A22}) that the operator $B_{1,2}^{(t)}\left( k\right) B_{2,1}^{(s)}\left(
k\right) $ and $B_{2,1}^{(t)}\left( k\right) B_{1,2}^{(s)}\left( k\right) $
have a relatively simple form for any $s,t\in \mathbb{R}$ and $k\in \mathbb{T%
}^{2}$: 
\begin{eqnarray}
B_{1,2}^{(t)}\left( k\right) B_{2,1}^{(s)}\left( k\right) &=&\left( \hat{%
\upsilon}\left( k\right) ^{2}\mathrm{e}^{i\left( s-t\right) \mathfrak{b}%
\left( k\right) }\right) \mathrm{e}^{itA_{1,1}\left( \mathrm{U},k\right) }P_{%
\mathfrak{d}\left( k\right) }\mathrm{e}^{-isA_{1,1}\left( \mathrm{U}%
,k\right) }\ ,  \label{ex1} \\
B_{2,1}^{(t)}\left( k\right) B_{1,2}^{(s)}\left( k\right) &=&\left( \hat{%
\upsilon}\left( k\right) ^{2}\mathrm{e}^{i\left( t-s\right) \mathfrak{b}%
\left( k\right) }\right) \left\langle \mathfrak{d}\left( k\right) ,\mathrm{e}%
^{i(s-t)A_{1,1}\left( \mathrm{U},k\right) }\mathfrak{d}\left( k\right)
\right\rangle \ ,  \label{ex2}
\end{eqnarray}%
where $P_{\mathfrak{d}\left( k\right) }$ is the orthogonal projection onto
the one-dimensional subspace $\mathbb{C}\mathfrak{d}\left( k\right)
\subseteq L^{2}(\mathbb{T}^{2})$. Similar computations can be done for other
choices of $\mathrm{V}\in \mathbb{R}_{0}^{+}$ and $\mathrm{v}:\mathbb{Z}%
^{2}\rightarrow \mathbb{R}_{0}^{+}$ in Theorem \ref{existence of a kato wave
operator for the system}, like $\mathrm{V}=0=\mathrm{v}$ (non-interacting
fermion systems). See again Corollary \ref{finite-time scattering and wave
operators copy(2)}.

The understanding of this kind of scattering, regarding free fermion
(electron) collisions, is relevant in physics, because it can allow the
exchange function $\upsilon $ of a real system to be studied. It is
therefore important to have a model from which not only qualitative, but
also quantitative, information can be obtained. This is the purpose of this
section, in particular of Theorem \ref{existence of a kato wave operator for
the system copy(1)}, which gives the explicit dependency of scattering in
terms of $\upsilon $.

\subsubsection{Bound pair scattering channel}

Similarly, we also prove the existence of a scattering channel for dressed
bound pairs. As dressed bound pairs are space-localized objects (see, e.g.,
Theorem \ref{Maintheorem1} (iii)), the fermions forming the pair efficiently
exchange a boson, at a non-negligible rate, via the terms $W_{\mathrm{%
b\rightarrow f}}$ (\ref{Wbf}) and $W_{\mathrm{f\rightarrow b}}$ (\ref{Wfb})
in the Hamiltonian $H$. In particular, such quantum states must have some
non-negligible bosonic component representing the exchanged boson that
\textquotedblleft glues\textquotedblright\ the two fermions together. This
dressed bound pair is however expected to move like a free (quantum
spinless) particle in the real space. See Fig. \ref{figure2}. We translate
this physical heuristics in precise mathematical terms by considering the
effective dispersion relations\ 
\begin{equation*}
\mathrm{E}:\left[ 0,\infty \right] \times \mathbb{T}^{2}\rightarrow \mathbb{R%
}
\end{equation*}%
given by Theorems \ref{Maintheorem1}, \ref{Maintheorem2} and \ref{existence
of a dispersion relation with low energy copy(2)}. 
%
%
\begin{figure}[!hbtp]
\begin{center}
\begin{tikzpicture}
\begin{feynman}

\vertex[blob,minimum height=2cm,minimum width=2cm](m) at ( 0, 0) {};
\vertex (a) at (-5,-1) {$k-q$};
\vertex (b) at ( 5,-1) {$k-p$};
\vertex (c) at (-5, 1) {$q$};
\vertex (d) at ( 5, 1) {$p$};
\diagram* {
(a) -- [fermion] (m) -- [anti fermion] (c),
(b) -- [anti fermion] (m) -- [fermion] (d),
(a) -- [photon, edge label=$\hat{\upsilon}\left( k\right) $] (c),
(b) -- [photon, edge label=$\hat{\upsilon}\left( k\right) $] (d),
};
\end{feynman}

\end{tikzpicture}
\end{center}
\caption{Illustration of the bound pair scattering channel. Here $k$ is the full
(quasi-)momentum of the (exponentially localized) dressed bound fermion pairs. The oscillating vertical lines between 
the two fermions (e.g. electrons) before the scattering process and afterwards characterise their bound via a bosonic (e.g., bipolaronic) particle transfer with coupling function $\hat{\upsilon}\left( k\right) $, see Fig. \ref{figure0}. It illustrates the stability of these pairs of fermions in time, as
expressed by Theorem \ref{kato's wave operator for d-wave pairing copy(1)},
i.e., the pairs cannot decay into an (even only asymptotically) unbound pair
of fermions.}
\label{figure2}
\end{figure}%

For any $\mathrm{U}\in \mathbb{R}_{0}^{+}\cup \{\infty \}$, we consider the
identification operator 
\begin{equation*}
\mathfrak{P}_{\mathrm{U}}:L^{2}\left( \mathbb{T}^{2}\right) \rightarrow
L^{2}\left( \mathbb{T}^{2},\mathcal{H}\right)
\end{equation*}%
defined for any $\varphi \in L^{2}\left( \mathbb{T}^{2}\right) $ by\footnote{%
Despite the fact that $\mathrm{E}(0)$ might not be in the resolvent set of $%
A_{1,1}(\mathrm{U},0)$, we can simply ignore it for $\{k=0\}$ has null
Lebesgue measure.} 
\begin{equation}
\begin{array}{cccl}
\mathfrak{P}_{\mathrm{U}}\varphi : & \mathbb{T}^{2}\backslash \{0\} & 
\rightarrow & \mathcal{H}\doteq L^{2}\left( \mathbb{T}^{2}\right) \oplus 
\mathbb{C} \\ 
& k & \mapsto & \varphi \left( k\right) \left\Vert \Psi \left( \mathrm{U}%
,k\right) \right\Vert ^{-1}\Psi \left( \mathrm{U},k\right)%
\end{array}%
\ ,  \label{frac P}
\end{equation}%
where $\Psi \left( \mathrm{U},k\right) $ is the eigenvector associated with
the (non-degenerate) eigenvalue $\mathrm{E}\left( \mathrm{U},k\right) $, as
given by Theorems \ref{Maintheorem1}, \ref{Maintheorem2} and \ref{existence
of a dispersion relation with low energy copy(2)}. Note from Theorem \ref%
{Maintheorem1} that the mapping 
\begin{equation*}
k\mapsto \left\Vert \Psi \left( \mathrm{U},k\right) \right\Vert ^{-1}\Psi
\left( \mathrm{U},k\right)
\end{equation*}%
is continuous on $\mathbb{T}^{2}$ for any $\mathrm{U}\in \mathbb{R}_{0}^{+}$
and its pointwise limit 
\begin{equation*}
k\mapsto \left\Vert \Psi \left( \infty ,k\right) \right\Vert ^{-1}\Psi
\left( \infty ,k\right)
\end{equation*}%
(cf. Theorem \ref{Maintheorem2}) is therefore measurable. In particular, the
linear transformation $\mathfrak{P}_{\mathrm{U}}$ is well defined for any $%
\mathrm{U}\in \mathbb{R}_{0}^{+}\cup \{\infty \}$. Moreover, one checks that
it is norm-preserving.

Since $\mathrm{E}\left( \mathrm{U},\cdot \right) \in C(\mathbb{T}^{2};%
\mathbb{R})\subseteq L^{\infty }(\mathbb{T}^{2},\nu )$ (see Theorems \ref%
{Maintheorem1} (iv) and \ref{Maintheorem2}), we can consider the
multiplication operator by $\mathrm{E}\left( \mathrm{U},\cdot \right) $ on $%
L^{2}(\mathbb{T}^{2})$, which is denoted by 
\begin{equation}
M_{\mathrm{E}\left( \mathrm{U},\cdot \right) }\doteq {\int_{\mathbb{T}%
^{2}}^{\oplus }}\mathrm{E}\left( \mathrm{U},k\right) \,\nu \left( \mathrm{d}%
k\right) \ ,\qquad \mathrm{U}\in \mathbb{R}_{0}^{+}\cup \{\infty \}\ .
\label{ME}
\end{equation}%
Remark also from Theorems \ref{Maintheorem1} (iv) and \ref{Maintheorem2}
together with Corollary \ref{example of absolutely continuous space} that $%
P_{\mathrm{ac}}(M_{\mathrm{E}\left( \mathrm{U},\cdot \right) })=\mathfrak{1}$
whenever $\hat{\upsilon}$ is real analytic on $\mathbb{S}^{2}$. We then
study now the (dressed) bound pair scattering channel, which is much simpler
than in the unbound pair channel:

\begin{theorem}[Bound pair (scattering) channel]
\label{kato's wave operator for d-wave pairing copy(1)}\mbox{ }\newline
Let $h_{b}\in \lbrack 0,1/2]$. Then the following assertions hold true:

\begin{enumerate}
\item[i.)] Dynamics and wave operators at finite $\mathrm{U}\in \mathbb{R}%
_{0}^{+}$: 
\begin{equation*}
\mathrm{e}^{it\mathbb{U}H\mathbb{U}^{\ast }}\mathfrak{P}_{\mathrm{U}}=%
\mathfrak{P}_{\mathrm{U}}\mathrm{e}^{itM_{\mathrm{E}\left( \mathrm{U},\cdot
\right) }}\ ,\qquad t\in \mathbb{R}\ .
\end{equation*}

\item[ii.)] Dynamics in the hard-core limit $\mathrm{U}\rightarrow \infty $:%
\begin{equation*}
s-{\lim\limits_{\mathrm{U}\rightarrow \infty }}\mathfrak{P}_{\mathrm{U}}=%
\mathfrak{P}_{\infty }\qquad \text{and}\qquad s-{\lim\limits_{\mathrm{U}%
\rightarrow \infty }}\,\mathrm{e}^{it\mathbb{U}H\mathbb{U}^{\ast }}\mathfrak{%
P}_{\mathrm{U}}=\mathfrak{P}_{\infty }\mathrm{e}^{itM_{\mathrm{E}\left( 
\mathrm{\infty },\cdot \right) }}\ ,\qquad t\in \mathbb{R}\ .
\end{equation*}
\end{enumerate}
\end{theorem}

\begin{proof}
Assertion (i) is Proposition \ref{kato's wave operator for d-wave pairing}.
Assertion (ii) results from Proposition \ref{some strong limits}.
\end{proof}

From Theorem \ref{kato's wave operator for d-wave pairing copy(1)} (i) the
scattering channel is time independent. For instance, for any $\mathrm{U}\in 
\mathbb{R}_{0}^{+}$, one trivially checks that 
\begin{equation*}
W^{\pm }\left( \mathbb{U}H\mathbb{U}^{\ast },M_{\mathrm{E}\left( \mathrm{U}%
,\cdot \right) };\mathfrak{P}_{\mathrm{U}}\right) =\mathfrak{P}_{\mathrm{U}%
}P_{\mathrm{a}\mathrm{c}}\left( M_{\mathrm{E}\left( \mathrm{U},\cdot \right)
}\right)
\end{equation*}%
and, since $\mathfrak{P}_{\mathrm{U}}^{\ast }\mathfrak{P}_{\mathrm{U}}=%
\mathfrak{1}$ and $P_{\mathrm{ac}}$ is a projection, its scattering operator
is equal to 
\begin{equation*}
S\left( \mathbb{U}H\mathbb{U}^{\ast },M_{\mathrm{E}\left( \mathrm{U},\cdot
\right) };\mathfrak{P}_{\mathrm{U}}\right) =P_{\mathrm{a}\mathrm{c}}\left(
M_{\mathrm{E}\left( \mathrm{U},\cdot \right) }\right) \ .
\end{equation*}%
If $\hat{\upsilon}$ is additionally real analytic on $\mathbb{S}^{2}$, then $%
P_{\mathrm{ac}}(M_{\mathrm{E}\left( \mathrm{U},\cdot \right) })=\mathfrak{1}$%
, thanks to Theorem \ref{Maintheorem1} and Corollary \ref{example of
absolutely continuous space}. In this case, the wave and scattering
operators are given by 
\begin{equation*}
W^{\pm }\left( \mathbb{U}H\mathbb{U}^{\ast },M_{\mathrm{E}\left( \mathrm{U}%
,\cdot \right) };\mathfrak{P}_{\mathrm{U}}\right) =\mathfrak{P}_{\mathrm{U}%
}\qquad \text{and}\qquad S\left( \mathbb{U}H\mathbb{U}^{\ast },M_{\mathrm{E}%
\left( \mathrm{U},\cdot \right) };\mathfrak{P}_{\mathrm{U}}\right) =%
\mathfrak{1}
\end{equation*}%
for any $\mathrm{U}\in \mathbb{R}_{0}^{+}$. Their hard-core limit are then
also trivial, thanks to Theorem \ref{kato's wave operator for d-wave pairing
copy(1)} (ii).

This scattering channel is therefore easy to study. In particular, similar
to Remark \ref{remark spacial space}, we can easily go back to the original
Hilbert space $\mathfrak{h}_{0}$ (\ref{h0}), referring to space coordinates
instead of the quasi-momenta. With this aim, we first observe that, for any $%
\mathrm{U}\in \mathbb{R}_{0}^{+}$,%
\begin{equation}
\mathrm{e}^{itH}\mathcal{P}_{\mathrm{U}}=\mathcal{P}_{\mathrm{U}}\mathrm{e}%
^{itU_{f}^{\ast }M_{\mathrm{E}\left( \mathrm{U},\cdot \right) }U_{f}}\
,\qquad t\in \mathbb{R}\ ,  \label{tutyuitui}
\end{equation}%
where $\mathcal{P}_{\mathrm{U}}\in \mathcal{B}\left( \mathfrak{h}_{0},%
\mathfrak{H}\right) $ is the new identification operator%
\begin{equation*}
\mathcal{P}_{\mathrm{U}}\doteq \mathbb{U}^{\ast }\mathfrak{P}_{\mathrm{U}%
}U_{f}\ ,\qquad \mathrm{U}\in \mathbb{R}_{0}^{+}\ ,
\end{equation*}%
and $U_{f}\doteq U_{2}U_{1}$. See Equations (\ref{U})--(\ref{Uf}) and
Proposition \ref{direct integral decomposition of the hamiltonian}.

On the one hand, Equation (\ref{tutyuitui}) together with Theorem \ref%
{existence of a dispersion relation with low energy copy(2)} shows that $%
\mathrm{E}\left( \mathrm{U},\cdot \right) $ defines (a family of) dispersion
relations, in the sense of Definition \ref{dispersion relation}. The Fourier
transform of $\mathrm{E}\left( \mathrm{U},\cdot \right) $ is the (effective)
hopping strength for the (spatially localized) dressed bound pairs. On the
other hand, the new identification operator $\mathcal{P}_{\mathrm{U}}$ is
translation invariant, i.e., 
\begin{equation*}
\mathcal{P}_{\mathrm{U}}\theta _{x}=\Theta _{x}\mathcal{P}_{\mathrm{U}}\
,\qquad x\in \mathbb{Z}^{2}\ ,
\end{equation*}%
where, for any fixed $x\in \mathbb{Z}^{2}$, $\theta _{x}\in \mathcal{B}(%
\mathfrak{h}_{0})$ and $\Theta _{x}\in \mathcal{B}(\mathfrak{H})$ are the
unique unitary operators respectively satisfying%
\begin{equation*}
\theta _{x}\left( \mathfrak{e}_{(y,\uparrow )}\wedge \mathfrak{e}%
_{(z,\downarrow )}\right) =\mathfrak{e}_{(y+x,\uparrow )}\wedge \mathfrak{e}%
_{(z+x,\downarrow )}\ ,\qquad y,z\in \mathbb{Z}^{2}\ ,
\end{equation*}%
(see Equations (\ref{h0})) and%
\begin{equation*}
\Theta _{x}\left( \mathfrak{\psi }\oplus \varphi \right) =\left( \theta _{x}%
\mathfrak{\psi }\right) \oplus \varphi \left( x+\cdot \right) \ ,\qquad 
\mathfrak{\psi \in h}_{0},\ \varphi \in \ell ^{2}\left( \mathbb{Z}%
^{2}\right) \ .
\end{equation*}%
See Equations (\ref{fract H}). In addition, when $\hat{\upsilon}(0)\neq 0$,
Theorem \ref{Maintheorem1} (iii) shows that the (dressed) fermion pair in
the bound pair channel is exponentially localized in space, that is, the
associate fermion-fermion correlation function decays exponentially fast
with respect to the distance between the fermions, uniformly in time. Note
that it is not required that the range of $\mathcal{P}_{\mathrm{U}}$ has a
vanishing bosonic component, because of the expected presence of
\textquotedblleft gluing bosons\textquotedblright\ in the \emph{dressed}
bound fermionic pair.

As a consequence, the bound channel describes an effective system of free
localized, spinless quasi-particles which minimize the energy at any fixed
total quasi-momentum. In particular, such quasi-particles of lowest energy,
or dressed fermion pairs, are stable in time, that is, they cannot decay
into an (even only asymptotically) unbound pair of fermions. Conversely, we
also show in Section \ref{Unbound Pair Channel} that a pair of fermions that
is asymptotically unbound far in the past is not able to bind together to
form a stable bound pair in the distant future.

Nevertheless, these quasi-particles should only be stable with respect to 
\emph{external} perturbations as soon as their states are related to
quasi-momenta $k$ such that $\mathrm{E}\left( \mathrm{U},k\right) <0$. If
the (dressed) quasi-particle is in a state whose support contains fibers $k$
such that $\mathrm{E}\left( \mathrm{U},k\right) \geq 0$, it is not in the
most energetically favorable state, since%
\begin{equation*}
\min \sigma _{\mathrm{ess}}\left( A\left( \mathrm{U},0\right) \right) =0
\end{equation*}%
(see Theorem \ref{Maintheorem1}). In fact, if the component corresponding to
quasi-momenta $k$ such that $\mathrm{E}\left( \mathrm{U},k\right) \geq 0$
has non-vanishing Lebesgue measure, then the quasi-particle should be
instable with respect to \emph{external} perturbations, by possibly creating
unbounded fermions with small quasi-momenta to decrease its total energy.
This situation clearly appears for quasi-momenta $k\in \mathbb{T}^{2}$ such
that $\hat{\upsilon}(k)=0$ or sufficiently small $\left\vert \hat{\upsilon}%
(k)\right\vert \ll 1$ when $k\neq 0$, since in these two cases, either $%
\mathrm{E}\left( \mathrm{U},k\right) =\mathfrak{b}\left( k\right) $\ ($\hat{%
\upsilon}(k)=0$) or $\mathrm{E}\left( \mathrm{U},k\right) \simeq \mathfrak{b}%
\left( k\right) $ ($\left\vert \hat{\upsilon}(k)\right\vert \ll 1$) with $%
\mathfrak{b}\left( k\right) \doteq h_{b}\epsilon \left( 2-\cos \left(
k\right) \right) $ (see (\ref{b})). If such a decay process really occurs,
one should see critical quasi-momenta, like in physical superconductors.

To prevent from this situation, one needs sufficiently strong $\left\vert 
\hat{\upsilon}\left( k\right) \right\vert \gg 1$ to have $\mathrm{E}\left( 
\mathrm{U},k\right) <0$ for all $k\in \mathbb{T}^{2}$. In the position
space, this means that the exchange strength between the two fermions and
the boson, represented by the function $\upsilon :\mathbb{Z}^{2}\rightarrow 
\mathbb{R}$ appearing in (\ref{Wbf})--(\ref{Wfb}), has to be sufficiently
strong and localized, like in Remark \ref{remark regularity}, in order to
get a sufficiently strong \textquotedblleft gluing effect\textquotedblright\
for dressed pairs, at all quasi-momenta. Recall also that the boson should
be heavier than the two fermions, i.e., $h_{b}\in \lbrack 0,1/2]$.

Last but not least, all this discussion can be extended to the hard-core
limit $\mathrm{U}\rightarrow \infty $, in view of Theorems \ref{existence of
a dispersion relation with low energy copy(2)} and \ref{kato's wave operator
for d-wave pairing copy(1)} (ii).

\section{Technical Results\label{Technical Results}}

\subsection{Notation\label{notation}}

The purpose of this section is to fix (or recall) the notation and
terminology that is used throughout the rest of the article. Let $\mathcal{X}
$ be any complex Hilbert space. We denote its scalar product by $\langle
\cdot ,\cdot \rangle _{\mathcal{X}}$, with the convention that it is
antilinear in the first argument and linear in the second one. The norm of $%
\mathcal{X}$ is thus 
\begin{equation*}
\left\Vert \varphi \right\Vert _{\mathcal{X}}\doteq \sqrt{\left\langle
\varphi ,\varphi \right\rangle _{\mathcal{X}}}\ ,\qquad \varphi \in \mathcal{%
X}\ .
\end{equation*}%
When there is no danger of confusion, as already said in Remark \ref%
{Notation}, we usually omit the subscript referring to the Hilbert space and
write $\Vert \cdot \Vert $ for $\Vert \cdot \Vert _{\mathcal{X}}$ and $%
\langle \cdot ,\cdot \rangle $ for $\langle \cdot ,\cdot \rangle _{\mathcal{X%
}}$.

Recall that $\mathcal{B}(\mathcal{X})$ denotes the set of bounded (linear)
operators on $\mathcal{X}$. $\mathfrak{1}\equiv \mathfrak{1}_{\mathcal{X}%
}\in \mathcal{B}(\mathcal{X})$ is the identity operator. Given $T\in 
\mathcal{B}(\mathcal{X})$, $T^{\ast }$ denotes its adjoint operator. The
(full) spectrum, essential spectrum and resolvent set of any $T\in \mathcal{B%
}(\mathcal{X})$ are denoted by $\sigma (T)$, $\sigma _{\mathrm{ess}}(T)$ and 
$\rho (T)$, respectively. The operator norm of $\mathcal{B}(\mathcal{X})$ is 
\begin{equation*}
\Vert T\Vert _{\mathrm{op}}\doteq \sup \left\{ \left\Vert T\varphi
\right\Vert _{\mathcal{X}}:\varphi \in \mathcal{X}\ \text{with }\left\Vert
\varphi \right\Vert _{\mathcal{X}}=1\right\} .
\end{equation*}%
Given $T\in \mathcal{B}(\mathcal{X})$ , $\mathcal{E}_{T}(\lambda )$ stands
for the eigenspace associated with an eigenvalue $\lambda \in \sigma (T)$ of 
$T$.

In all the Section \ref{Technical Results}, we study properties of the
Hamiltonian $H\in \mathcal{B}(\mathfrak{H})$ defined by (\ref{H0}). As one
can see from (\ref{H})--(\ref{H0}) combined with (\ref{Hamiltonian-f}), (\ref%
{Hamiltonian-b})--(\ref{Hamiltonian-fbbis}) and (\ref{Wfb}), it depends on
several parameters. More precisely, $\epsilon ,\mathrm{U},h_{b}\in \mathbb{R}%
_{0}^{+}$ and $\alpha _{0}\in \mathbb{R}^{+}$, while%
\begin{equation*}
\mathrm{u}:\mathbb{Z}^{2}\rightarrow \mathbb{R}_{0}^{+}\ ,\quad \mathrm{e}%
^{\alpha _{0}\left\vert \cdot \right\vert }\mathfrak{p}_{1}:\mathbb{Z}%
^{2}\rightarrow \mathbb{R}\ \ ,\quad \mathrm{e}^{\alpha _{0}\left\vert \cdot
\right\vert }\mathfrak{p}_{2}:\mathbb{Z}^{2}\rightarrow \mathbb{R}\quad 
\text{and}\quad \upsilon :\mathbb{Z}^{2}\rightarrow \mathbb{R}
\end{equation*}%
(with $\mathfrak{p}_{2}(z)\doteq 0$ for $z\notin 2\mathbb{Z}$) are all
absolutely summable functions that are invariant with respect to $90^{\circ
} $-rotations. See Equations (\ref{summable1-U}), (\ref{ssdsds}) and (\ref%
{summable2-U}). The parameters of the operator $H$ are always fixed and
arbitrary, unless we need to specify them to clarify some particular
statement. Recall that the invariance under $90^{\circ }$-rotation is not
that important here. In fact, here, the only important point concerning this
symmetry is that it implies that the Fourier transforms $\hat{\upsilon}$, $%
\hat{\mathfrak{p}}_{1}$ and $\hat{\mathfrak{p}}_{2}$ are real-valued,
because 
\begin{equation*}
\upsilon (-x)=\upsilon (x)=\overline{\upsilon (x)}\ ,\text{\quad }\mathfrak{p%
}_{1}(-x)=\mathfrak{p}_{1}(x)=\overline{\mathfrak{p}_{1}(x)}\text{\quad
and\quad }\mathfrak{p}_{2}(-x)=\mathfrak{p}_{2}(x)=\overline{\mathfrak{p}%
_{2}(x)}\text{ },
\end{equation*}%
i.e., the real valued functions $\upsilon $, $\mathfrak{p}_{1}$ and $%
\mathfrak{p}_{2}$ are reflection invariant, as a consequence of their $%
90^{\circ }$-rotation invariance. Apart of this technical point, it is
mainly relevant for the study of unconventional pairings, which is not done
in the present work.

Note additionally that the on-site repulsion $\mathrm{U}\in \mathbb{R}%
_{0}^{+}$ appears explicitly in all the quantities defined in Sections \ref%
{Setup of the Problem}--\ref{Main Results}. However, in Section \ref%
{Technical Results}, this parameter is only important for the Subsections %
\ref{effective behavior at hard-core limit of the electronic repulsion} and %
\ref{scattering channels}. Therefore, unless the parameter $\mathrm{U}$ is
important for our discussions or statements, we omit it in order to shorten
the notation, by writing 
\begin{equation*}
f\left( k\right) \equiv f\left( \mathrm{U},k\right)
\end{equation*}%
for any function $f(\mathrm{U},k)$ of the parameters $\mathrm{U}$ and $k$.

\subsection{Computation of the Fiber Decomposition of the Hamiltonian \label%
{Unitary transformation copy(1)}}

For completeness, we first proof in a simple lemma that the fiber
Hamiltonians defined by (\ref{fiber hamiltonians}) yield an element of $%
L^{\infty }\left( \mathbb{T}^{2},\mathcal{B}(\mathcal{H})\right) $. Then, we
prove Proposition \ref{direct integral decomposition of the hamiltonian}.

\begin{lemma}[Elementary properties of fiber Hamiltonians]
\label{maincoro1 copy(1)}\mbox{ }\newline
Fix $h_{b},\epsilon ,\mathrm{U}\in \mathbb{R}_{0}^{+}$. Then, $A:\mathbb{T}%
^{2}\rightarrow \mathcal{B}(\mathcal{H})$, as defined by (\ref{fiber
hamiltonians}), is continuous and, in particular, $A\left( \cdot \right) \in
L^{\infty }\left( \mathbb{T}^{2},\mathcal{B}(\mathcal{H})\right) $.
\end{lemma}

\begin{proof}
Since $\cos :\mathbb{R}^{2}\rightarrow \mathbb{R}$, as defined by (\ref%
{cosinus}), is a non-expansive mapping with period $2\pi $, given $%
k,k^{\prime },p\in \mathbb{T}^{2}$, the quantity%
\begin{equation*}
\mathfrak{f}(k^{\prime })\left( p\right) -\mathfrak{f}\left( k\right) \left(
p\right) =\epsilon \left\{ \cos \left( p+k\right) -\cos \left( p+k^{\prime
}\right) \right\} =\epsilon \left\{ \cos \left( p+k\right) -\cos \left(
p+k^{\prime }+2\pi q\right) \right\}
\end{equation*}%
(see (\ref{f})) is bounded for any $q\in \mathbb{Z}^{2}$ by%
\begin{equation*}
\left\vert \mathfrak{f}(k^{\prime })\left( p\right) -\mathfrak{f}\left(
k\right) \left( p\right) \right\vert \leq \epsilon \left\vert \left(
p+k\right) -\left( p+k^{\prime }+2\pi q\right) \right\vert =\epsilon
\left\vert k-k^{\prime }+2\pi q\right\vert \ .
\end{equation*}%
Hence, taking the minimum over all $q\in \mathbb{Z}^{2}$ and the supremum
over all $p\in \mathbb{T}^{2}$, we obtain from (\ref{metric}) and (\ref%
{A11-U0}) that%
\begin{eqnarray*}
\left\Vert A_{1,1}(k^{\prime })-A_{1,1}(k)\right\Vert _{\mathrm{o}\mathrm{p}%
} &=&\left\Vert M_{\mathfrak{f}(k^{\prime })}-M_{\mathfrak{f}\left( k\right)
}\right\Vert _{\mathrm{o}\mathrm{p}}=\sup_{p\in \mathbb{T}^{2}}\left\vert 
\mathfrak{f}(k^{\prime })\left( p\right) -\mathfrak{f}\left( k\right) \left(
p\right) \right\vert \\
&\leq &\epsilon \min_{q\in \mathbb{Z}^{2}}\left\vert k-k^{\prime }+2\pi
q\right\vert =\epsilon d_{\mathbb{T}^{2}}(k,k^{\prime })
\end{eqnarray*}%
for all $k,k^{\prime }\in \mathbb{T}^{2}$. In other words, the mapping 
\begin{equation*}
A_{1,1}\left( \cdot \right) :\mathbb{T}^{2}\rightarrow \mathcal{B}\left(
L^{2}\left( \mathbb{T}^{2}\right) \right)
\end{equation*}%
is ($\epsilon $-Lipschitz) continuous with respect to the metric $d_{\mathbb{%
T}^{2}}$. Similarly, we see that $\mathfrak{b}:\mathbb{T}^{2}\rightarrow 
\mathbb{R}$, defined by (\ref{b}), is continuous with respect to $d_{\mathbb{%
T}^{2}}$, and hence so is $A_{2,2}:\mathbb{T}^{2}\rightarrow \mathcal{L}(%
\mathbb{C})$ (see (\ref{A22})). In addition, by the triangle and
Cauchy-Schwarz inequalities, for any $k,k^{\prime }\in \mathbb{T}^{2}$ and $%
\varphi \in L^{2}(\mathbb{T}^{2})$, 
\begin{equation*}
\left\vert \hat{\upsilon}(k^{\prime })\left\langle \mathfrak{d}(k^{\prime
}),\varphi \right\rangle -\hat{\upsilon}(k)\left\langle \mathfrak{d}%
(k),\varphi \right\rangle \right\vert \leq \left\vert \hat{\upsilon}%
(k^{\prime })-\hat{\upsilon}(k)\right\vert \left\Vert \mathfrak{d}(k^{\prime
})\right\Vert \left\Vert \varphi \right\Vert +\left\vert \hat{\upsilon}%
(k)\right\vert \left\Vert \mathfrak{d}(k^{\prime })-\mathfrak{d}%
(k)\right\Vert \left\Vert \varphi \right\Vert .
\end{equation*}%
Because of (\ref{ssdsds}) and (\ref{summable2-U}), $\mathfrak{d}(k),\hat{%
\upsilon}\in C\left( \mathbb{T}^{2}\right) $. So, since $\mathbb{T}^{2}$ is (%
$d_{\mathbb{T}^{2}}$-)compact and%
\begin{equation*}
\left\Vert A_{2,1}(k^{\prime })-A_{2,1}(k)\right\Vert _{\mathrm{o}\mathrm{p}%
}=\sup_{\varphi \in L^{2}(\mathbb{T}^{2}),||\varphi ||_{2}=1}\left\vert \hat{%
\upsilon}(k^{\prime })\left\langle \mathfrak{d}(k^{\prime }),\varphi
\right\rangle -\hat{\upsilon}(k)\left\langle \mathfrak{d}(k),\varphi
\right\rangle \right\vert ,
\end{equation*}%
we deduce from the last inequality and (\ref{A21}) that $A_{2,1}:\mathbb{T}%
^{2}\rightarrow L^{2}(\mathbb{T}^{2})^{\ast }$ is continuous. As $%
A_{1,2}(k)=A_{2,1}(k)^{\ast }$ for all $k\in \mathbb{T}^{2}$, we conclude
that the mapping $A:\mathbb{T}^{2}\rightarrow \mathcal{B}(\mathcal{H})$ is
continuous, and hence bounded on the $d_{\mathbb{T}^{2}}$-compact set $%
\mathbb{T}^{2}$.
\end{proof}

We now compute the following unitary transformation of the Hamiltonian $H$
(see (\ref{H})):%
\begin{equation}
\mathbb{U}H\mathbb{U}^{\ast }=\left( 
\begin{array}{cc}
U_{f} & 0 \\ 
0 & \mathcal{F}%
\end{array}%
\right) 
\begin{pmatrix}
H_{f} & W_{\mathrm{b\rightarrow f}} \\[0.5em] 
W_{\mathrm{f\rightarrow b}} & H_{b}%
\end{pmatrix}%
\left( 
\begin{array}{cc}
U_{f}^{\ast } & 0 \\ 
0 & \mathcal{F}^{\ast }%
\end{array}%
\right) =%
\begin{pmatrix}
U_{f}H_{f}U_{f}^{\ast } & U_{f}W_{\mathrm{b\rightarrow f}}\mathcal{F}^{\ast }
\\[0.5em] 
\mathcal{F}W_{\mathrm{f\rightarrow b}}U_{f}^{\ast } & \mathcal{F}H_{b}%
\mathcal{F}^{\ast }%
\end{pmatrix}
\label{asdsdasdasd}
\end{equation}%
with $\mathbb{U}$ defined by (\ref{U})--(\ref{fourier transforms}). In fact,
the remaining part of this section is devoted to the computations leading to
Proposition \ref{direct integral decomposition of the hamiltonian}.

To begin with, we observe that, for any lattice site $x\in \mathbb{Z}^{2}$
and spin $s\in \{\uparrow ,\downarrow \}$, $b_{x}\doteq b(\mathfrak{e}_{x})$
and $a_{x,s}\doteq a(\mathfrak{e}_{(x,s)})$, where $\{\mathfrak{e}_{x}\doteq
\delta _{x,\cdot }\}_{x\in \mathbb{Z}^{2}}$ is the canonical orthonormal
basis (\ref{e frac}) of $\ell ^{2}(\mathbb{Z}^{2})$ and $a_{x,s}$ ($%
b_{x}^{\ast }$) denotes the annihilation operator acting on the fermionic
(bosonic) Fock space $\mathfrak{F}_{-}$ ($\mathfrak{F}_{+}$) of a fermion
(boson). In both cases, $\Omega $ denotes the vacuum state. We compute each
term of the the right-hand side (\ref{asdsdasdasd}) separately:\medskip

\noindent \underline{Computation of $U_{f}H_{f}U_{f}^{\ast }$ in relation to 
$A_{1,1}$.} We first note from (\ref{fermion operator}) that, for any $%
x,y,u\in \mathbb{Z}^{2}$ and $s\in \{\uparrow ,\downarrow \}$,%
\begin{equation*}
a_{x,s}\left( \mathfrak{e}_{(y,\uparrow )}\wedge \mathfrak{e}_{(u,\downarrow
)}\right) =\frac{1}{\sqrt{2}}\left( \langle \mathfrak{e}_{(x,s)},\mathfrak{e}%
_{(y,\uparrow )}\rangle \mathfrak{e}_{(u,\downarrow )}-\langle \mathfrak{e}%
_{(x,s)},\mathfrak{e}_{(u,\downarrow )}\rangle \mathfrak{e}_{(y,\uparrow
)}\right)
\end{equation*}%
vanishes whenever $(x,s)\notin \{(y,\uparrow ),(u,\downarrow )\}$. Using
this observation and (\ref{fermion operator creat}), one concludes that, for
any $y,u\in \mathbb{Z}^{2}$,%
\begin{eqnarray}
&&\sum_{s\in \{\uparrow ,\downarrow \},\ x,z\in \mathbb{Z}%
^{2}\,:\,|z|=1}a_{x,s}^{\ast }a_{x+z,s}\,\left( \mathfrak{e}_{(y,\uparrow
)}\wedge \mathfrak{e}_{(u,\downarrow )}\right)  \notag \\
&=&\sum_{z\in \mathbb{Z}^{2}\,:\,|z|=1}\left( a_{y+z,\uparrow }^{\ast
}a_{y,\uparrow }\,\left( \mathfrak{e}_{(y,\uparrow )}\wedge \mathfrak{e}%
_{(u,\downarrow )}\right) +a_{u+z,\downarrow }^{\ast }a_{u,\downarrow
}\,\left( \mathfrak{e}_{(y,\uparrow )}\wedge \mathfrak{e}_{(u,\downarrow
)}\right) \right)  \notag \\
&=&\frac{1}{\sqrt{2}}\sum_{z\in \mathbb{Z}^{2}\,:\,|z|=1}\left(
a_{y+z,\uparrow }^{\ast }\left( \mathfrak{e}_{(u,\downarrow )}\right)
-a_{u+z,\downarrow }^{\ast }\left( \mathfrak{e}_{(y,\uparrow )}\right)
\right)  \notag \\
&=&\sum_{z\in \mathbb{Z}^{2}\,:\,|z|=1}\left( \mathfrak{e}_{(y+z,\uparrow
)}\wedge \mathfrak{e}_{(u,\downarrow )}+\mathfrak{e}_{(y,\uparrow )}\wedge 
\mathfrak{e}_{(u+z,\downarrow )}\right) \ .  \label{calcul1}
\end{eqnarray}%
Likewise, we see that, for any $y,u\in \mathbb{Z}^{2}$,%
\begin{equation}
\sum_{s\in \{\uparrow ,\downarrow \},\ x\in \mathbb{Z}^{2}}a_{x,s}^{\ast
}a_{x,s}\left( \mathfrak{e}_{(y,\uparrow )}\wedge \mathfrak{e}%
_{(u,\downarrow )}\right) =2\mathfrak{e}_{(y,\uparrow )}\wedge \mathfrak{e}%
_{(u,\downarrow )}\ .  \label{calcul2}
\end{equation}%
Moreover, as $\mathrm{u}:\mathbb{Z}^{2}\rightarrow \mathbb{R}$ is absolutely
summable and invariant with respect to $180^{\circ }$-rotations (cf. (\ref%
{summable1-U})), we also get that%
\begin{eqnarray}
\sum_{x,z\in \mathbb{Z}^{2}}\mathrm{u}\left( z\right) n_{x,\uparrow
}n_{x+z,\downarrow }\left( \mathfrak{e}_{(y,\uparrow )}\wedge \mathfrak{e}%
_{(u,\downarrow )}\right) &=&\sum_{z\in \mathbb{Z}^{2}}\mathrm{u}\left(
z\right) n_{u-z,\uparrow }\left( \mathfrak{e}_{(y,\uparrow )}\wedge 
\mathfrak{e}_{(u,\downarrow )}\right) =\mathrm{u}\left( u-y\right) \left( 
\mathfrak{e}_{(y,\uparrow )}\wedge \mathfrak{e}_{(u,\downarrow )}\right) 
\notag \\
&=&\mathrm{u}\left( y-u\right) \left( \mathfrak{e}_{(y,\uparrow )}\wedge 
\mathfrak{e}_{(u,\downarrow )}\right) \ ,  \label{calcul3}
\end{eqnarray}%
for any $y,u\in \mathbb{Z}^{2}$, which, for $\mathrm{u}\left( z\right)
=\delta _{z,0}$, is equal to 
\begin{equation}
\sum_{x\in \mathbb{Z}^{2}}n_{x,\uparrow }n_{x,\downarrow }\left( \mathfrak{e}%
_{(y,\uparrow )}\wedge \mathfrak{e}_{(u,\downarrow )}\right) =\delta
_{y,u}\left( \mathfrak{e}_{(y,\uparrow )}\wedge \mathfrak{e}_{(u,\downarrow
)}\right) \ .  \label{calcul4}
\end{equation}%
We thus infer from (\ref{Hamiltonian-f}) combined with (\ref{calcul1})--(\ref%
{calcul4}) that%
\begin{eqnarray*}
H_{f}\left( \mathfrak{e}_{(y,\uparrow )}\wedge \mathfrak{e}_{(u,\downarrow
)}\right) &=&-{\frac{\epsilon }{2}}\sum_{z\in \mathbb{Z}^{2}\,:\,|z|=1}%
\left( \mathfrak{e}_{(y+z,\uparrow )}\wedge \mathfrak{e}_{(u,\downarrow )}+%
\mathfrak{e}_{(y,\uparrow )}\wedge \mathfrak{e}_{(u+z,\downarrow )}\right) \\
&&+\left( 4\epsilon +\mathrm{U}\delta _{y,u}+\mathrm{u}\left( y-u\right)
\right) \left( \mathfrak{e}_{(y,\uparrow )}\wedge \mathfrak{e}%
_{(u,\downarrow )}\right) \ ,
\end{eqnarray*}%
for any $y,u\in \mathbb{Z}^{2}$. Then, conjugating $H_{f}$ by the unitary
operator $U_{f}$ (\ref{Uf})--(\ref{fourier transforms}) gives the equality%
\begin{eqnarray*}
U_{f}H_{f}U_{f}^{\ast }\left( \mathfrak{\hat{e}}_{y}\left( \cdot \right) 
\mathfrak{\hat{e}}_{y-u}\right) &=&U_{f}H_{f}\left( \mathfrak{e}%
_{(y,\uparrow )}\wedge \mathfrak{e}_{(u,\downarrow )}\right) \\
&=&-{\frac{\epsilon }{2}}\sum_{z\in \mathbb{Z}^{2}\,:\,|z|=1}\left( 
\mathfrak{\hat{e}}_{y+z}\left( \cdot \right) \mathfrak{\hat{e}}_{y+z-u}+%
\mathfrak{\hat{e}}_{y}\left( \cdot \right) \mathfrak{\hat{e}}%
_{y-(u+z)}\right) \\
&&+\left( 4\epsilon +\mathrm{U}\delta _{y,u}+\mathrm{u}\left( y-u\right)
\right) \mathfrak{\hat{e}}_{y}\left( \cdot \right) \mathfrak{\hat{e}}_{y-u}
\end{eqnarray*}%
for any $y,u\in \mathbb{Z}^{2}$. By first evaluating the above expression at 
$k\in \mathbb{T}^{2}$, and then at $p\in \mathbb{T}^{2}$, and using (\ref{f}%
) and (\ref{cosinus}), we obtain that%
\begin{eqnarray*}
\left( U_{f}H_{f}U_{f}^{\ast }\left( \mathfrak{\hat{e}}_{y}\left( \cdot
\right) \mathfrak{\hat{e}}_{y-u}\right) (k)\right) \left( p\right) &=&%
\mathfrak{\hat{e}}_{y}\left( k\right) \mathfrak{\hat{e}}_{y-u}\left(
p\right) \left( \mathrm{U}\delta _{y,u}+\mathrm{u}\left( y-u\right) \right)
\\
&&+\mathfrak{\hat{e}}_{y}\left( k\right) \mathfrak{\hat{e}}_{y-u}\left(
p\right) \epsilon \left( 4-{\frac{1}{2}}\sum_{z\in \mathbb{Z}%
^{2}\,:\,|z|=1}\left( \mathrm{e}^{i\left( k+p\right) \cdot z}+\mathrm{e}%
^{ip\cdot z}\right) \right) \\
&=&\mathfrak{\hat{e}}_{y}\left( k\right) \mathfrak{\hat{e}}_{y-u}\left(
p\right) \left( \mathrm{U}\delta _{y,u}+\mathrm{u}\left( y-u\right) +%
\mathfrak{f}\left( k\right) \left( p\right) \right) \\
&=&\mathfrak{\hat{e}}_{y}\left( k\right) \left( \mathrm{U}P_{0}+{%
\sum\limits_{x\in \mathbb{Z}^{2}}}\,\mathrm{u}\left( x\right) P_{x}+M_{%
\mathfrak{f}(k)}\left( p\right) \right) \left( \mathfrak{\hat{e}}%
_{y-u}\right) \left( p\right) \ ,
\end{eqnarray*}%
for any $y,u\in \mathbb{Z}^{2}$. By (\ref{A11})--(\ref{A11-U0}), it follows
that%
\begin{equation}
U_{f}H_{f}U_{f}^{\ast }\left( \mathfrak{\hat{e}}_{y}\left( \cdot \right) 
\mathfrak{\hat{e}}_{y-u}\right) =\left( {\int_{\mathbb{T}^{2}}^{\oplus }}%
A_{1,1}\left( p\right) \,\nu (\mathrm{d}p)\right) \mathfrak{\hat{e}}%
_{y}\left( \cdot \right) \mathfrak{\hat{e}}_{y-u}\ .  \label{gfjhhgjg}
\end{equation}%
As $\{\mathfrak{\hat{e}}_{y}\left( \cdot \right) \mathfrak{\hat{e}}%
_{y-u}\}_{y,u\in \mathbb{Z}^{2}}$ is an orthonormal basis for the Hilbert
space 
\begin{equation*}
\int_{\mathbb{T}^{2}}^{\oplus }L^{2}\left( \mathbb{T}^{2}\right) \nu \left( 
\mathrm{d}k\right) \ ,
\end{equation*}%
we deduce from (\ref{gfjhhgjg}) that 
\begin{equation*}
U_{f}H_{f}U_{f}^{\ast }={\int_{\mathbb{T}^{2}}^{\oplus }}A_{1,1}\left(
k\right) \,\nu (\mathrm{d}k)\ .
\end{equation*}

\noindent \underline{Computation of $\mathcal{F}H_{b}\mathcal{F}^{\ast }$ in
relation to $A_{2,2}$.} Using (\ref{boson operator}) and (\ref{boson
operatorcreat}) we conclude that, for any $y\in \mathbb{Z}^{2}$,%
\begin{equation*}
H_{b}\left( \mathfrak{e}_{y}\right) =\epsilon h_{b}\left( 2{%
\sum\limits_{x\in \mathbb{Z}^{2}}}\langle \mathfrak{e}_{x},\mathfrak{e}%
_{y}\rangle b_{x}^{\ast }\Omega -{\frac{1}{2}\sum\limits_{z\in \mathbb{Z}%
^{2}\,:\,|z|=1}}\left\langle \mathfrak{e}_{x+z},\mathfrak{e}%
_{y}\right\rangle b_{x}^{\ast }\Omega \right) =\epsilon h_{b}\left( 2%
\mathfrak{e}_{y}-{\frac{1}{2}\sum\limits_{z\in \mathbb{Z}^{2}\,:\,|z|=1}}%
\mathfrak{e}_{y+z}\right)
\end{equation*}%
so that 
\begin{equation*}
\mathcal{F}H_{b}\left( \mathfrak{e}_{y}\right) =\epsilon h_{b}\left( 2%
\mathfrak{\hat{e}}_{y}-{\frac{1}{2}\sum\limits_{z\in \mathbb{Z}%
^{2}\,:\,|z|=1}}\mathfrak{\hat{e}}_{y+z}\right) \ ,\qquad y\in \mathbb{Z}%
^{2}\ .
\end{equation*}%
Therefore, using that $\mathfrak{\hat{e}}_{y}\equiv \mathcal{F(\mathfrak{e}}%
_{y}\mathcal{)=}\,\mathrm{e}^{ik\cdot y}$ (see (\ref{fourier transforms}))
as well as (\ref{b}), (\ref{cosinus}) and (\ref{A22}), we arrive at the
result 
\begin{eqnarray*}
\mathcal{F}H_{b}\mathcal{F}^{\ast }(\mathfrak{\hat{e}}_{y})\left( k\right)
&=&\epsilon h_{b}\mathrm{e}^{ik\cdot y}\left( 2-{\frac{1}{2}%
\sum\limits_{z\in \mathbb{Z}^{2}\,:\,|z|=1}}\mathrm{e}^{ik\cdot z}\right)
=\epsilon h_{b}\left( 2-\cos \left( k\right) \right) \mathrm{e}^{ik\cdot y}
\\
&=&\mathfrak{b}\left( k\right) \mathrm{e}^{ik\cdot y}=A_{2,2}\left( k\right) 
\mathfrak{\hat{e}}_{y}\left( k\right)
\end{eqnarray*}%
for all $y\in \mathbb{Z}^{2}$ and $k\in \mathbb{T}^{2}$. As $\{\hat{%
\mathfrak{e}}_{y}\}_{y\in \mathbb{Z}^{2}}$ is an orthonormal basis for $%
L^{2}(\mathbb{T}^{2})$, it follows that 
\begin{equation*}
\mathcal{F}H_{b}\mathcal{F}^{\ast }={\int_{\mathbb{T}^{2}}^{\oplus }}%
A_{2,2}(k)\,\nu (\mathrm{d}k)\ .
\end{equation*}%
\noindent \underline{Computation of $\mathcal{F}W_{\mathrm{f\rightarrow b}%
}U_{f}^{\ast }$ and $U_{f}W_{\mathrm{b\rightarrow f}}\mathcal{F}^{\ast }$ in
relation to $A_{2,1}$ and $A_{1,2}$.} Observe from (\ref{Hamiltonian-fbbis})
and (\ref{fermion operator creat}) that, for all $y\in \mathbb{Z}^{2}$,%
\begin{equation*}
c_{y}^{\ast }\Omega =\sqrt{2}{\sum\limits_{z\in \mathbb{Z}^{2}}}\,\left( 
\mathfrak{p}_{1}\left( z\right) \mathfrak{e}_{(y+z,\uparrow )}\wedge 
\mathfrak{e}_{(y,\downarrow )}+\mathfrak{p}_{2}\left( 2z\right) \mathfrak{e}%
_{(y+z,\uparrow )}\wedge \mathfrak{e}_{(y-z,\downarrow )}\right)
\end{equation*}%
and, as a consequence, using (\ref{Wbf})--(\ref{Hamiltonian-fb}) as well as (%
\ref{boson operator}), we get that, for any $u\in \mathbb{Z}^{2}$, 
\begin{equation*}
W_{\mathrm{b}\rightarrow \mathrm{f}}\left( \mathfrak{e}_{u}\right) ={%
\sum\limits_{y\in \mathbb{Z}^{2}}}\upsilon \left( u-y\right) {%
\sum\limits_{z\in \mathbb{Z}^{2}}}\,\left( \mathfrak{p}_{1}\left( z\right) 
\mathfrak{e}_{(y+z,\uparrow )}\wedge \mathfrak{e}_{(y,\downarrow )}+%
\mathfrak{p}_{2}\left( 2z\right) \mathfrak{e}_{(y+z,\uparrow )}\wedge 
\mathfrak{e}_{(y-z,\downarrow )}\right) \ .
\end{equation*}%
Therefore, by (\ref{Uf})--(\ref{fourier transforms}), for any $u\in \mathbb{Z%
}^{2}$, 
\begin{equation*}
U_{f}W_{\mathrm{b\rightarrow f}}\mathcal{F}^{\ast }\left( \mathfrak{\hat{e}}%
_{u}\right) ={\sum\limits_{y\in \mathbb{Z}^{2}}}\upsilon \left( u-y\right) {%
\sum\limits_{z\in \mathbb{Z}^{2}}}\,\left( \mathfrak{p}_{1}\left( z\right) 
\mathfrak{\hat{e}}_{y+z}\left( \cdot \right) \mathfrak{\hat{e}}_{z}+%
\mathfrak{p}_{2}\left( 2z\right) \mathfrak{\hat{e}}_{y+z}\left( \cdot
\right) \mathfrak{\hat{e}}_{2z}\right) \ .
\end{equation*}%
In particular, as $\upsilon ,\mathfrak{p}_{1},\mathfrak{p}_{2}:\mathbb{Z}%
^{2}\rightarrow \mathbb{R}$ are absolutely summable and invariant with
respect to $180^{\circ }$-rotations (cf. (\ref{ssdsds}) and (\ref%
{summable2-U})), we deduce from the last equality that, for any $u\in 
\mathbb{Z}^{2}$ and $k,p\in \mathbb{T}^{2}$,%
\begin{eqnarray*}
\left( U_{f}W_{\mathrm{b\rightarrow f}}\mathcal{F}^{\ast }\left( \mathfrak{%
\hat{e}}_{u}\right) \left( k\right) \right) \left( p\right) &=&{%
\sum\limits_{y\in \mathbb{Z}^{2}}}\upsilon \left( u-y\right) {%
\sum\limits_{z\in \mathbb{Z}^{2}}}\,\left( \mathfrak{p}_{1}\left( z\right) 
\mathrm{e}^{ik\cdot (y+z)}\mathrm{e}^{ip\cdot z}+\mathfrak{p}_{2}\left(
2z\right) \mathrm{e}^{ik\cdot (y+z)}\mathrm{e}^{i2p\cdot z}\right) \\
&=&{\sum\limits_{z\in \mathbb{Z}^{2}}}\,\left( \mathfrak{p}_{1}\left(
z\right) \mathrm{e}^{i\left( k+p\right) \cdot z}+\mathfrak{p}_{2}\left(
2z\right) \mathrm{e}^{i\left( k+2p\right) \cdot z)}\right) \mathrm{e}%
^{ik\cdot u}{\sum\limits_{y\in \mathbb{Z}^{2}}}\upsilon \left( y-u\right) 
\mathrm{e}^{ik\cdot \left( y-u\right) }
\end{eqnarray*}%
Using now that 
\begin{equation*}
{\sum\limits_{z\in \mathbb{Z}^{2}}}\mathfrak{p}_{1}\left( z\right) \mathrm{e}%
^{i\left( k+p\right) \cdot z}=\mathfrak{\hat{p}}_{1}\left( k+p\right) \qquad 
\text{and}\qquad {\sum\limits_{z\in \mathbb{Z}^{2}}}\mathfrak{p}_{2}\left(
2z\right) \mathrm{e}^{i\left( k+2p\right) \cdot z)}=\mathfrak{\hat{p}}%
_{2}\left( k/2+p\right)
\end{equation*}%
($\mathfrak{p}_{2}\left( z\right) \doteq 0$ for $z\notin (2\mathbb{Z})^{2}$)
we arrive at the equalities%
\begin{eqnarray*}
\left( U_{f}W_{\mathrm{b\rightarrow f}}\mathcal{F}^{\ast }\left( \mathfrak{%
\hat{e}}_{u}\right) \left( k\right) \right) \left( p\right) &=&\left( 
\mathfrak{\hat{p}}_{1}\left( k+p\right) +\mathfrak{\hat{p}}_{2}\left(
k/2+p\right) \right) \mathrm{e}^{ik\cdot u}{\sum\limits_{z\in \mathbb{Z}^{2}}%
}\upsilon \left( z\right) \mathrm{e}^{ik\cdot z} \\
&=&\hat{\upsilon}\left( k\right) \mathfrak{d}\left( k\right) \left( p\right) 
\hat{\mathfrak{e}}_{u}(k)=\left[ A_{1,2}\left( k\right) \hat{\mathfrak{e}}%
_{u}(k)\right] \left( p\right) \ ,
\end{eqnarray*}%
with $\mathfrak{d}(k)(p)$ and $A_{1,2}(k)$ defined by (\ref{d}) and (\ref%
{A12}), respectively. As $u\in \mathbb{Z}^{2}$ and $k,p\in \mathbb{T}^{2}$
are arbitrary in the above equations, this shows that $U_{f}W_{\mathrm{b}%
\rightarrow \mathrm{f}}\mathcal{F}^{\ast }$ coincides with the bounded
linear transformation%
\begin{equation*}
\begin{array}{cccc}
J: & L^{2}\left( \mathbb{T}^{2}\right) & \rightarrow & \int_{\mathbb{T}%
^{2}}^{\oplus }L^{2}\left( \mathbb{T}^{2}\right) \nu \left( \mathrm{d}%
k\right) \\ 
&  &  &  \\ 
& \varphi & \mapsto & A_{1,2}\left( \cdot \right) \hat{\mathfrak{e}}%
_{u}(\cdot )%
\end{array}%
\end{equation*}%
on the orthonormal basis $\{\hat{\mathfrak{e}}_{u}\}_{u\in \mathbb{Z}^{2}}$
and therefore $U_{f}W_{\mathrm{b}\rightarrow \mathrm{f}}\mathcal{F}^{\ast
}=J $. By taking adjoints on both sides, we also obtain $\mathcal{F}W_{%
\mathrm{f}\rightarrow \mathrm{b}}U_{f}^{\ast }=J^{\ast }$. Finally, one can
easily check from (\ref{A21}) that 
\begin{equation*}
\left( J^{\ast }\psi \right) \left( k\right) =A_{2,1}\left( k\right) \psi
\left( k\right) \ ,\qquad k\in \mathbb{T}^{2},\ \psi \in \int_{\mathbb{T}%
^{2}}^{\oplus }L^{2}\left( \mathbb{T}^{2}\right) \nu \left( \mathrm{d}%
k\right) \ .
\end{equation*}%
This completes the proof of Proposition \ref{direct integral decomposition
of the hamiltonian}.

\subsection{Spectrum of the Fiber Hamiltonians\label{spectrum of Hamiltonian
fibers}}

We start with the study of the essential spectrum of fiber Hamiltonians (\ref%
{fiber hamiltonians}) at any total quasi-momentum $k\in \mathbb{T}^{2}$,
before considering afterwards the discrete one in Section \ref{discrete
spectrum of fiber hamiltonians}. Then, in Section\ \ref{Bottom of the
spectrum} we study the bottom of the spectrum ($k\in \mathbb{T}^{2}$ being
fixed). The whole study leads to important spectral properties of $H$, as
previously explained, via Proposition \ref{direct integral decomposition of
the hamiltonian} combined with Theorem \ref{borelian functional calculus of
a direct integral of operators}.

\subsubsection{Essential spectrum\label{essential spectrum of the fiber
hamiltonians}}

The essential spectrum $\sigma _{\mathrm{ess}}(A(k))$ of the fiber
Hamiltonian 
\begin{equation*}
A\left( k\right) \equiv A\left( \mathrm{U},k\right) \ ,
\end{equation*}%
defined by (\ref{fiber hamiltonians}) at fixed $\mathrm{U}\in \mathbb{R}%
_{0}^{+}$ and total quasi-momentum $k\in \mathbb{T}^{2}$, is completely
determined by the following proposition:

\begin{proposition}[Essential spectrum of fiber Hamiltonians]
\label{essential spectrum of a fiber}\mbox{ }\newline
For any $k\in \mathbb{T}^{2}$ and $h_{b},\epsilon ,\mathrm{U}\in \mathbb{R}%
_{0}^{+}$, one has%
\begin{equation*}
\sigma _{\mathrm{ess}}\left( A\left( k\right) \right) =\sigma _{\mathrm{ess}%
}\left( A_{1,1}\left( k\right) \right) =\sigma _{\mathrm{ess}}\left(
B_{1,1}\left( k\right) \right) =\sigma \left( M_{\mathfrak{f}\left( k\right)
}\right) =2\epsilon \cos \left( k/2\right) \left[ -1,1\right] +4\epsilon 
\text{ },
\end{equation*}%
where\ $M_{\mathfrak{f}\left( k\right) }$ stands for the multiplication
operator associated with the function $\mathfrak{f}(k)$ (\ref{f}), while $%
B_{1,1}(k)$ and $A_{1,1}(k)\equiv A_{1,1}(\mathrm{U},k)$ are defined by (\ref%
{A11}) and (\ref{A11-U0}), respectively.
\end{proposition}

\begin{proof}
Recall that $\nu $ is the normalized Haar measure defined by (\ref{Haar
mesure}) on $\mathbb{T}^{2}$. Fix $k\in \mathbb{T}^{2}$. If $\lambda $ is an
eigenvalue of $M_{\mathfrak{f}\left( k\right) }$ with associated eigenvector 
$\varphi \in L^{2}(\mathbb{T}^{2})$, then%
\begin{equation*}
M_{\mathfrak{f}\left( k\right) }\varphi \left( p\right) \doteq \mathfrak{f}%
\left( k\right) \left( p\right) \varphi \left( p\right) =\lambda \varphi
\left( p\right)
\end{equation*}%
for almost every $p\in \mathbb{T}^{2}$. As $\varphi \neq 0$, there exists $%
\Omega \subseteq \mathbb{T}^{2}$ with strictly positive measure $\nu (\Omega
)>0$ such that the above equality holds true with $\varphi (p)\neq 0$ for
every $p\in \Omega $. Thus, $\mathfrak{f}(k)(p)=\lambda $ for all $p\in
\Omega $. Because%
\begin{equation*}
\nu ([-\pi ,\pi )^{2}\backslash (-\pi ,\pi )^{2})=0\ ,
\end{equation*}%
we can assume without loss of generality that $\Omega \subseteq (-\pi ,\pi
)^{2}$. Since $\mathfrak{f}(k)-\lambda $ is real analytic on the open domain 
$(-\pi ,\pi )^{2}$ in $\mathbb{R}^{2}$ and the zeros of any non-constant
real analytic function have null Lebesgue measure \cite{analytic}, we would
have $\nu (\Omega )=0$, which contradicts our choice of the set $\Omega $.
Recall indeed that $\nu $ is the Lebesgue measure, up to a normalization
constant (see (\ref{Haar mesure})).\ Hence, $M_{\mathfrak{f}\left( k\right)
} $ has no eigenvalues and, thus,%
\begin{equation*}
\sigma _{\mathrm{ess}}\left( M_{\mathfrak{f}\left( k\right) }\right) =\sigma
\left( M_{\mathfrak{f}\left( k\right) }\right) =\,\mathfrak{f}\left(
k\right) (\mathbb{T}^{2})\text{ }.
\end{equation*}%
The last equality holds true, for $\mathfrak{f}$ is a continuous function on
a compact domain, namely the torus $\mathbb{T}^{2}$. Clearly, 
\begin{equation*}
\mathfrak{f}\left( k\right) (\mathbb{T}^{2})=2\epsilon \cos \left(
k/2\right) \left[ -1,1\right] +4\epsilon \text{ }.
\end{equation*}%
Observing that $A_{1,2}(k)$, $A_{2,1}(k)$, $A_{2,2}(k)$ and $P_{x}$ are all
rank-one linear transformations, we can apply the stability of the essential
spectrum under compact perturbations (see \cite[Corollary 8.16]{schmudgen})
to conclude that 
\begin{equation}
\sigma _{\mathrm{ess}}\left( A\left( k\right) \right) =\sigma _{\mathrm{ess}%
}\left( M_{\mathfrak{f}\left( k\right) }\right) =\,\mathfrak{f}\left(
k\right) (\mathbb{T}^{2})\text{ }.  \label{ess1}
\end{equation}%
In fact, from the absolute summability of the function $\mathrm{u}:\mathbb{Z}%
^{2}\rightarrow \mathbb{R}_{0}^{+}$ (see (\ref{summable1-U})), along with
the closedness of the subspace of compact operators in the Banach space of
all bounded operators, the operator defined by the infinite sum%
\begin{equation*}
{\sum\limits_{x\in \mathbb{Z}^{2}}}\,\mathrm{u}\left( x\right) P_{x}
\end{equation*}%
is not only bounded, but even compact on the Hilbert space $L^{2}(\mathbb{T}%
^{2})$. Recall that $P_{x}$ denotes the orthogonal projection onto the
one-dimensional subspace $\mathbb{C}\hat{\mathfrak{e}}_{x}\subseteq L^{2}(%
\mathbb{T}^{2})$ for any $x\in \mathbb{Z}^{2}$. For the same reasons, 
\begin{equation*}
\sigma _{\mathrm{ess}}\left( A_{1,1}\left( k\right) \right) =\sigma _{%
\mathrm{ess}}\left( B_{1,1}\left( k\right) \right) =\sigma _{\mathrm{ess}%
}\left( M_{\mathfrak{f}\left( k\right) }+\mathrm{U}P_{0}\right) =\sigma _{%
\mathrm{ess}}\left( M_{\mathfrak{f}\left( k\right) }\right) \ .
\end{equation*}
\end{proof}

\begin{corollary}[Bottom of the spectrum of $A_{1,1}(k)$ and $B_{1,1}(k)$]
\label{bottom of the spectrum of A11}\mbox{ }\newline
For any $k\in \mathbb{T}^{2}$, one has that 
\begin{equation*}
\min \sigma \left( A_{1,1}\left( k\right) \right) =\min \sigma \left(
B_{1,1}\left( k\right) \right) =\min \sigma \left( M_{\mathfrak{f}\left(
k\right) }\right) =4\epsilon -2\epsilon \cos \left( k/2\right) \doteq 
\mathfrak{z}\left( k\right) \ .
\end{equation*}
\end{corollary}

\begin{proof}
Fix $k\in \mathbb{T}^{2}$. Since $\mathrm{U}\in \mathbb{R}_{0}^{+}$, one has
the operator inequalities 
\begin{equation*}
M_{\mathfrak{f}\left( k\right) }\leq M_{\mathfrak{f}\left( k\right) }+{%
\sum\limits_{x\in \mathbb{Z}^{2}}}\,\mathrm{u}\left( x\right) P_{x}\doteq
B_{1,1}\left( k\right) \leq B_{1,1}\left( k\right) +\mathrm{U}P_{0}\doteq
A_{1,1}(k)\ ,
\end{equation*}%
for the set of positive operators on a Hilbert space forms a norm-closed
convex cone. By combining the last inequalities with Proposition \ref%
{essential spectrum of a fiber}, one arrives at the assertion.
\end{proof}

\subsubsection{Discrete spectrum\label{discrete spectrum of fiber
hamiltonians}}

In the following it is technically convenient to assume that $\mathfrak{b}%
(k) $, which is the kinetic energy of a boson with quasi-momentum $k$, is
below the bottom of the spectrum of $A_{1,1}(k)$, that is, the minimum
energy of the fermion pair for the same total quasi-momentum. In other
words, we assume from now on that 
\begin{equation}
\mathfrak{b}(k)\doteq h_{b}\epsilon \left( 2-\cos \left( k\right) \right)
\leq \mathfrak{z}(k)\doteq 4\epsilon -2\epsilon \cos (k/2)
\label{equation sur bk}
\end{equation}%
for all $k\in \mathbb{T}^{2}$, with equality \emph{only} at $k=0$. See\
Equation (\ref{b}) and Corollary \ref{bottom of the spectrum of A11}. By
direct computations\footnote{%
(\ref{equation sur bk}) is clearly true for $h_{b}=0$. Take $h_{b}>0$. Using 
$\cos \left( \theta \right) =2\cos ^{2}\left( \theta /2\right) -1$ and (\ref%
{cosinus}), one verifies that (\ref{equation sur bk}) is equivalent to $%
h_{b}\left( 4-2x^{2}-2y^{2}\right) \leq 4-2\left( x+y\right) $ for $x,y\in %
\left[ 0,1\right] $. Since $\inf_{x\in \left[ 0,1\right] }\left\{
h_{b}x^{2}-x\right\} =-1/(4h_{b})$ for $h_{b}\geq 1/2$ and $\inf_{x\in \left[
0,1\right] }\left\{ h_{b}x^{2}-x\right\} =h_{b}-1$ for $h_{b}\in (0,1/2]$,
we deduce that (\ref{equation sur bk}) holds true iff $h_{b}\in \lbrack
0,1/2]$.}, one verifies that this amounts to take $h_{b}$ in the interval $%
[0,1/2]$. This means that we consider a regime where the boson mass is at
least the mass of the two fermions, as physically expected for cuprate
superconductors, see \cite[Section 3.1]{articulo2}.

\begin{proposition}[Eigenvalues of fiber Hamiltonians -- I]
\label{eigenvalue of a fiber}\mbox{ }\newline
Take any $k\in \mathbb{T}^{2}$ and $h_{b}\in \lbrack 0,1/2]$.

\begin{enumerate}
\item[i.)] $\lambda \neq \mathfrak{b}(k)$ is an eigenvalue of $A(k)$ iff
there is a non-zero vector $\varphi \in L^{2}(\mathbb{T}^{2})$ in the kernel
of the bounded operator%
\begin{equation*}
A_{1,1}\left( k\right) -\lambda \mathfrak{1}-\left( \mathfrak{b}\left(
k\right) -\lambda \right) ^{-1}A_{1,2}\left( k\right) A_{2,1}\left( k\right)
\in \mathcal{B}\left( L^{2}(\mathbb{T}^{2})\right) \text{ }.
\end{equation*}

In this case, $\lambda $ is an eigenvalue of $A(k)$ with associated
eigenvector%
\begin{equation*}
\left( \varphi ,-\left( \mathfrak{b}\left( k\right) -\lambda \right)
^{-1}A_{2,1}\left( k\right) \varphi \right) \in \mathcal{H}\doteq
L^{2}\left( \mathbb{T}^{2}\right) \oplus \mathbb{C}\text{ }.
\end{equation*}

\item[ii.)] $\mathfrak{b}(k)$ is an eigenvalue of $A(k)$ iff $\hat{\upsilon}%
(k)=0$.
\end{enumerate}
\end{proposition}

\begin{proof}
Fix $k\in \mathbb{T}^{2}$ and $h_{b}\in \lbrack 0,1/2]$. We start with the
proof of Assertion (i): If $\lambda \neq \mathfrak{b}(k)$ is an eigenvalue
of $A(k)$ with associated eigenvector $(\varphi ,z)\in \mathcal{H}\backslash
\{0\}$, then we directly deduce from (\ref{fiber hamiltonians}) that%
\begin{eqnarray}
\left( A_{1,1}\left( k\right) -\lambda \mathfrak{1}\right) \varphi
+A_{1,2}\left( k\right) z &=&0\text{ },  \label{eq1} \\[0.01in]
\,A_{2,1}\left( k\right) \varphi +\left( \mathfrak{b}\left( k\right)
-\lambda \right) z &=&0\text{ }.  \label{eq2}
\end{eqnarray}%
By combining these two equations, we obtain 
\begin{equation}
z=-\left( \mathfrak{b}\left( k\right) -\lambda \right) ^{-1}A_{2,1}\left(
k\right) \varphi  \label{sdsdsdsd}
\end{equation}%
and, thus, 
\begin{equation*}
\left[ A_{1,1}\left( k\right) -\lambda \mathfrak{1}-\left( \mathfrak{b}%
\left( k\right) -\lambda \right) ^{-1}A_{1,2}\left( k\right) A_{2,1}\left(
k\right) \right] \varphi =0\text{ }.
\end{equation*}%
We have that $\varphi \neq 0$, for otherwise $z$ would also be zero, by (\ref%
{sdsdsdsd}), and this would contradict the fact that $(\varphi ,z)$ is a
non-zero vector. The converse is obvious and Assertion (i) holds true.

We now prove Assertion (ii): It is easy to check from (\ref{fiber
hamiltonians}) that $\hat{\upsilon}(k)=0$ implies that 
\begin{equation}
A\left( k\right) \left( 0,1\right) =\mathfrak{b}\left( k\right) \left(
0,1\right) \text{ }.  \label{sdfsdfsdf}
\end{equation}%
Conversely, suppose that $\mathfrak{b}(k)$ is an eigenvalue of $A(k)$ with
associated eigenvector $(\varphi ,z)\in \mathcal{H}\backslash \{0\}$, but $%
\hat{\upsilon}(k)\neq 0$. Then, by (\ref{A21}) and (\ref{fiber hamiltonians}%
),%
\begin{equation}
\left( A_{1,1}\left( k\right) -\mathfrak{b}\left( k\right) \mathfrak{1}%
\right) \varphi +A_{1,2}\left( k\right) z=0\qquad \text{and}\qquad
A_{2,1}\left( k\right) \varphi \doteq \hat{\upsilon}\left( k\right)
\left\langle \mathfrak{d}\left( k\right) ,\varphi \right\rangle =0\text{ }.
\label{ssdssd}
\end{equation}%
Remark that the second equality says that $\varphi \,\bot \,\mathfrak{d}(k)$%
, since we assume $\hat{\upsilon}(k)\neq 0$. Considering the scalar product
of $\varphi $ with both sides of the first equation, we then get that 
\begin{equation}
\left\langle \varphi ,\left( A_{1,1}\left( k\right) -\mathfrak{b}\left(
k\right) \mathfrak{1}\right) \varphi \right\rangle +z\hat{\upsilon}\left(
k\right) \left\langle \varphi ,\mathfrak{d}\left( k\right) \right\rangle
=\left\langle \varphi ,\left( A_{1,1}\left( k\right) -\mathfrak{b}\left(
k\right) \mathfrak{1}\right) \varphi \right\rangle =0\text{ },  \label{dfg}
\end{equation}%
see (\ref{A12}). Because $h_{b}\in \lbrack 0,1/2]$, if $k\neq 0$ then (\ref%
{equation sur bk}) holds true with a strict inequality and therefore, 
\begin{equation*}
\mathfrak{b}\left( k\right) <4\epsilon -2\epsilon \cos \left( k/2\right)
=\min \sigma \left( A_{1,1}\left( k\right) \right) \ ,
\end{equation*}%
thanks to Corollary \ref{bottom of the spectrum of A11}. Hence, 
\begin{equation*}
A_{1,1}\left( k\right) -\mathfrak{b}\left( k\right) \mathfrak{1}\geq c%
\mathfrak{1}\ ,
\end{equation*}%
for some constant $c>0$, which, combined with (\ref{dfg}), in turn implies
that $\varphi =0$. If now $k=0$, then $\mathfrak{b}(0)=0$ (see (\ref{b}))
and we obtain from (\ref{dfg}) that%
\begin{equation}
{\int_{\mathbb{T}^{2}}}\left\vert \varphi \left( p\right) \right\vert ^{2}%
\mathfrak{f}\left( 0\right) \left( p\right) \,\nu \left( \mathrm{d}p\right)
=\langle \varphi ,M_{\mathfrak{f}\left( 0\right) }\varphi \rangle \leq
\langle \varphi ,A_{1,1}\left( 0\right) \varphi \rangle =0\text{ },
\label{sd}
\end{equation}%
since $M_{\mathfrak{f}\left( 0\right) }\leq A_{1,1}(\mathrm{U},0)$ (see (\ref%
{A11})--(\ref{A11-U0})). As 
\begin{equation*}
\mathfrak{f}(0)(p)\doteq \epsilon \{4-2\cos (p)\}\ ,\qquad p\in \mathbb{T}%
^{2}\ ,
\end{equation*}%
(see (\ref{f}) and (\ref{cosinus})) defines a positive and continuous
function that vanishes at $p=0$ only, one deduces from (\ref{sd}) that $%
\varphi =0$ also when $k=0$. In any case, $\varphi =0$ and so, (\ref{ssdssd}%
) combined with (\ref{A12}) yields 
\begin{equation*}
A_{1,2}(k)z\doteq \hat{\upsilon}(k)\mathfrak{d}(k)z=0\text{ }.
\end{equation*}%
Since $\mathfrak{d}(k)\neq 0$ and $\hat{\upsilon}(k)\neq 0$, we must have
that $z=0$.Thus, we arrive at $(\varphi ,z)=(0,0)$, which contradicts the
fact that $(\varphi ,z)$ is a non-zero vector. Therefore, if $\mathfrak{b}%
(k) $ is an eigenvalue of $A(k)$ then we must have $\hat{\upsilon}(k)=0$.
\end{proof}

The Birman-Schwinger principle (Theorem \ref{birman-schwinger's theorem})
allows us to transform the eigenvalue problem for the fiber Hamiltonian $%
A(k) $ into a non-linear equation on the resolvent set $\rho (A_{1,1}(%
\mathrm{U},k))$ of the operator $A_{1,1}(\mathrm{U},k)$, which is the
resolvent set for a fermion pair with total quasi-momentum $k\in \mathbb{T}%
^{2}$:

\begin{theorem}[Characteristic equation for eigenvalues]
\label{eigenvalue as a root of a non linear equation}\mbox{ }\newline
Fix $h_{b}\in \lbrack 0,1/2]$ and $k\in \mathbb{T}^{2}$. Then, $\lambda \in
\rho (A_{1,1}(k))$ is an eigenvalue of $A(k)$ iff it is a solution to the
equation 
\begin{equation}
\hat{\upsilon}\left( k\right) ^{2}\mathfrak{T}\left( k,z\right) +z-\mathfrak{%
b}\left( k\right) =0,\qquad z\in \rho \left( A_{1,1}\left( k\right) \right) 
\text{ },  \label{non lin eq ev}
\end{equation}%
where $\mathfrak{T}$ is the function defined by (\ref{fract T}), that is, 
\begin{equation}
\mathfrak{T}\left( k,z\right) \equiv \mathfrak{T}\left( \mathrm{U}%
,k,z\right) \doteq \left\langle \mathfrak{d}\left( k\right) ,\left(
A_{1,1}\left( k\right) -z\mathfrak{1}\right) ^{-1}\mathfrak{d}\left(
k\right) \right\rangle \ ,  \label{cal T}
\end{equation}
\end{theorem}

\begin{proof}
Fix $h_{b}\in \lbrack 0,1/2]$ and $k\in \mathbb{T}^{2}$. We divide the proof
in several cases:\ \medskip

\noindent \underline{Case 1:} We first consider the case $\hat{\upsilon}%
(k)=0 $ and $k\neq 0$. In that situation, $\mathfrak{b}(k)$ is trivially the
only solution to (\ref{non lin eq ev}). On the other hand, we already know
from Proposition \ref{eigenvalue of a fiber} (ii) that $\mathfrak{b}(k)$ is
an eigenvalue of $A(k)$. We must therefore prove that there is no other
eigenvalue $\lambda $ of $A(k)$ in $\rho (A_{1,1}(k))$ but $\mathfrak{b}(k)$%
. In fact, if such a $\lambda \in \rho (A_{1,1}(k))$ exists then, by
Proposition \ref{eigenvalue of a fiber} (i) with $\hat{\upsilon}(k)=0$, $%
A_{1,1}(k)-\lambda \mathfrak{1}$ would have a non-trivial kernel, which is
not possible, for $\lambda $ is in the resolvent set of $A_{1,1}(k)$.
\medskip

\noindent \underline{Case 2:} Suppose that $k=0$ and $\hat{\upsilon}(0)=0$.
We observe that (\ref{non lin eq ev}) has no solution because 
\begin{equation*}
0\in \lbrack 0,8\epsilon ]=\sigma _{\mathrm{ess}}(A_{1,1}(0))\ ,
\end{equation*}%
thanks to Proposition \ref{essential spectrum of a fiber}. In addition, by
applying Proposition \ref{eigenvalue of a fiber} (i) and noting that $%
\mathfrak{b}(0)=0$, we see that $A(0)$ has no eigenvalues in $\rho
(A_{1,1}(0))$. \medskip

\noindent \underline{Case 3:} Finally, assume that $\hat{\upsilon}(k)\neq 0$
and take $\lambda \in \rho (A_{1,1}(k))$. Observe from Proposition \ref%
{eigenvalue of a fiber} (ii) that $\mathfrak{b}(k)$ cannot be an eigenvalue
of $A(k)$. Additionally, $\mathfrak{b}(k)$ cannot be a solution to Equation (%
\ref{non lin eq ev}). This last observation is proven as follows: When $k=0$
this is clear because $\mathfrak{b}(0)=0$ is not even in the domain of the
equation to be solved in (\ref{non lin eq ev}). For $k\neq 0$, if $\mathfrak{%
b}(k)$ is a solution to (\ref{non lin eq ev}), then $\mathfrak{T}(k,%
\mathfrak{b}(k))=0$, but we know from Corollary \ref{bottom of the spectrum
of A11} and $h_{b}\in \lbrack 0,1/2]$ that 
\begin{equation*}
A_{1,1}(k)-\mathfrak{b}(k)\mathfrak{1}\geq c\mathfrak{1}
\end{equation*}%
for some constant $c>0$. Therefore, $\mathfrak{T}(k,\mathfrak{b}(k))=0$
would yield $\mathfrak{d}(k)=0$, which is obviously wrong, by (\ref{d}).
Therefore, in all cases, $\mathfrak{b}(k)$ cannot be a solution to Equation (%
\ref{non lin eq ev}) and we can assume that $\lambda \neq \mathfrak{b}(k)$.
Now, the remaining part of the proof is essentially the same as the one of 
\cite[Proposition 10]{articulo}, but we reproduce it for completeness. By (%
\ref{A21})--(\ref{A12}), the orthogonal projection $S$ onto the subspace $%
\mathbb{C}\mathfrak{d}(k)\subseteq L^{2}(\mathbb{T}^{2})$ can be written as 
\begin{equation*}
S\varphi =\Vert \mathfrak{d}(k)\Vert ^{-2}\langle \mathfrak{d}(k),\varphi
\rangle \mathfrak{d}(k)=\hat{\upsilon}(k)^{-2}\Vert \mathfrak{d}(k)\Vert
^{-2}A_{1,2}(k)A_{2,1}(k)\varphi ,\qquad \varphi \in L^{2}\left( \mathbb{T}%
^{2}\right) \ .
\end{equation*}%
Then, observe from Proposition \ref{eigenvalue of a fiber} (i) that $\lambda 
$ is an eigenvalue of $A(k)$ iff $\lambda $ is an eigenvalue of $T-V^{2}$
with 
\begin{eqnarray}
V &\doteq &\hat{\upsilon}(k)(\mathfrak{b}(k)-\lambda )^{-1/2}\Vert \mathfrak{%
d}(k)\Vert S\ ,  \label{V} \\[0.01in]
T &\doteq &A_{1,1}(k)\text{ },  \label{T}
\end{eqnarray}%
Thus, by applying Theorem \ref{birman-schwinger's theorem}, we deduce that $%
\lambda $ is an eigenvalue of $A(k)$ iff $1$ is an eigenvalue of the
corresponding Birman-Schwinger operator, which, with the above operators $T$
and $V$, is equal to%
\begin{equation*}
\mathrm{B}(\lambda )=\hat{\upsilon}(k)^{2}(\mathfrak{b}(k)-\lambda
)^{-1}\Vert \mathfrak{d}(k)\Vert ^{2}S(A_{1,1}(k)-\lambda \mathfrak{1})^{-1}S%
\text{ }.
\end{equation*}%
Remark in this case that 
\begin{equation}
\mathcal{E}_{\mathrm{B}(\lambda )}(1)=\mathbb{C}\mathfrak{d}(k)\qquad \text{%
and}\qquad \dim \mathcal{E}_{T-V^{2}}(\lambda )=\dim \mathcal{E}_{\mathrm{B}%
(\lambda )}(1)=1\text{ },  \label{dimension equal}
\end{equation}%
since, obviously, 
\begin{equation*}
\mathrm{B}(\lambda )L^{2}\left( \mathbb{T}^{2}\right) \subseteq SL^{2}\left( 
\mathbb{T}^{2}\right) =\mathbb{C}\mathfrak{d}(k)\text{ }.
\end{equation*}%
We thus conclude that $\lambda $ is an eigenvalue of $A(k)$ iff%
\begin{align*}
\mathrm{B}(\lambda )\mathfrak{d}(k)=\mathfrak{d}(k)& \Leftrightarrow \langle 
\mathfrak{d}(k),\mathrm{B}(\lambda )\mathfrak{d}(k)-\mathfrak{d}(k)\rangle =0
\\[1em]
& \Leftrightarrow \langle \mathfrak{d}(k),\mathrm{B}(\lambda )\mathfrak{d}%
(k)\rangle =\Vert \mathfrak{d}(k)\Vert ^{2} \\[1em]
& \Leftrightarrow \hat{\upsilon}(k)^{2}(\mathfrak{b}(k)-\lambda )^{-1}\Vert 
\mathfrak{d}(k)\Vert ^{2}\langle \mathfrak{d}(k),S(A_{1,1}(k)-\lambda 
\mathfrak{1})^{-1}S\mathfrak{d}(k)\rangle =\Vert \mathfrak{d}(k)\Vert ^{2} \\%
[1em]
& \Leftrightarrow \hat{\upsilon}(k)^{2}\mathfrak{T}(k,\lambda )=\mathfrak{b}%
(k)-\lambda \text{ }.
\end{align*}%
This completes the proof.
\end{proof}

\begin{corollary}[Eigenspaces of fiber Hamiltonians]
\label{eigenspace of a fiber}\mbox{ }\newline
Fix $h_{b}\in \lbrack 0,1/2]$ and $k\in \mathbb{T}^{2}$. If $\lambda \in
\rho (A_{1,1}(k))$ is an eigenvalue of the fiber Hamiltonian $A(k)$ then the
associated eigenspace is 
\begin{equation*}
\mathcal{E}_{A\left( k\right) }\left( \lambda \right) =\mathbb{C}g\left(
k,\lambda \right) \text{ },
\end{equation*}%
where%
\begin{equation*}
g\left( k,\lambda \right) \doteq \left( \hat{\upsilon}\left( k\right) \left(
A_{1,1}\left( k\right) -\lambda \mathfrak{1}\right) ^{-1}\mathfrak{d}\left(
k\right) ,-1\right) \in \mathcal{H}\ .
\end{equation*}%
In particular, $\lambda $ is a non-degenerated eigenvalue of $A(k)$.
\end{corollary}

\begin{proof}
Assume that $\hat{\upsilon}(k)\neq 0$. Recall from the proof of Theorem \ref%
{eigenvalue as a root of a non linear equation} that in this case $\mathfrak{%
b}(k)$ is not an eigenvalue of $A(k)$ and so, we assume without loss of
generality that $\lambda \neq \mathfrak{b}(k)$. In this case, observe that
we have (\ref{dimension equal}) and a close look at the proof of Theorem \ref%
{birman-schwinger's theorem}, in particular Lemma \ref{Lemma dfddfdfdf},
leads us to%
\begin{equation*}
\mathcal{E}_{T-V^{2}}(\lambda )\doteq \ker (T-V^{2}-\lambda \mathfrak{1})=%
\mathbb{C}\varphi _{0}\ ,
\end{equation*}%
where%
\begin{equation}
\varphi _{0}\doteq (T-\lambda \mathfrak{1})^{-1}V\mathfrak{d}(k)=\hat{%
\upsilon}(k)(\mathfrak{b}(k)-\lambda )^{-1/2}\Vert \mathfrak{d}(k)\Vert
(A_{1,1}(k)-\lambda \mathfrak{1})^{-1}\mathfrak{d}(k)\ ,  \label{phi0}
\end{equation}%
by (\ref{V}) and (\ref{T}). From Proposition \ref{eigenvalue of a fiber} (i)
($\hat{\upsilon}(k)\neq 0$), one then obtains that%
\begin{eqnarray*}
\mathcal{E}_{A(k)}(\lambda ) &=&\left\{ (\varphi ,-(\mathfrak{b}(k)-\lambda
)^{-1}A_{2,1}(k)\varphi )\in \mathcal{H}\,:\,\varphi \in \ker
(T-V^{2}-\lambda \mathfrak{1})\right\} \\
&=&\mathbb{C}\ (\varphi _{0},-(\mathfrak{b}(k)-\lambda
)^{-1}A_{2,1}(k)\varphi _{0})\ .
\end{eqnarray*}%
In view of Equation (\ref{A21}), (\ref{phi0}) and Theorem \ref{eigenvalue as
a root of a non linear equation}, the last vector can be rewritten as
follows: 
\begin{align*}
(\varphi _{0},-(\mathfrak{b}(k)-\lambda )^{-1}A_{2,1}(k)\varphi _{0})&
=(\varphi _{0},-(\mathfrak{b}(k)-\lambda )^{-1}\hat{\upsilon}(k)\langle 
\mathfrak{d}(k),\varphi _{0}\rangle ) \\[1em]
& =(\varphi _{0},-(\mathfrak{b}(k)-\lambda )^{-3/2}\hat{\upsilon}%
(k)^{2}\Vert \mathfrak{d}(k)\Vert \mathfrak{T}(k,\lambda )) \\[1em]
& =(\varphi _{0},-(\mathfrak{b}(k)-\lambda )^{-1/2}\Vert \mathfrak{d}%
(k)\Vert ) \\
& =(\mathfrak{b}(k)-\lambda )^{-1/2}\Vert \mathfrak{d}(k)\Vert g(k,\lambda
)\ ,
\end{align*}%
whenever $\hat{\upsilon}(k)\neq 0$. Finally, if $\hat{\upsilon}(k)=0$ and $%
\lambda \in \rho (A_{1,1}(k))$ is an eigenvalue of $A(k)$ then it is
straightforward to check that $\lambda =\mathfrak{b}(k)$ with eigenspace
generated by the vector $(0,1)\in \mathcal{H}$, see for instance (\ref{eq1}%
)--(\ref{eq2}) and (\ref{sdfsdfsdf}).
\end{proof}

\begin{corollary}[Eigenvalues of fiber Hamiltonians -- II]
\label{there is at most one eigenvalue on each connected component}\mbox{ }%
\newline
Fix $h_{b}\in \lbrack 0,1/2]$ and $k\in \mathbb{T}^{2}$. There is at most
one eigenvalue of $A(k)$ in each connected component of $\rho
(A_{1,1}(k))\cap \mathbb{R}$.
\end{corollary}

\begin{proof}
In view of Theorem \ref{eigenvalue as a root of a non linear equation}, it
suffices to show that the derivative of the mapping%
\begin{equation*}
\rho (A_{1,1}(k))\cap \mathbb{R}\ni x\longmapsto \hat{\upsilon}(k)^{2}%
\mathfrak{T}(k,x)+x-\mathfrak{b}(k)\in \mathbb{R}
\end{equation*}%
is strictly positive. For any $x_{0}\in \rho (A_{1,1}(k))\cap \mathbb{R}$,
we have that 
\begin{equation}
\left. \partial _{x}\left\{ \hat{\upsilon}(k)^{2}\mathfrak{T}(k,x)+x-%
\mathfrak{b}(k)\right\} \right\vert _{x=x_{0}}=\hat{\upsilon}(k)^{2}\Vert
(A_{1,1}(k)-x_{0}\mathfrak{1})^{-1}\mathfrak{d}(k)\Vert ^{2}+1>1\ .
\label{sdfsdfsdfsdf}
\end{equation}
\end{proof}

\subsubsection{Bottom of the spectrum\label{Bottom of the spectrum}}

As is well-known, physical properties of quantum systems at very low
temperatures are essentially determined by the bottom of the spectrum of the
corresponding Hamiltonian. In our case, having in mind the application to
superconductivity in cuprates, we would like to study the bottom of the
spectrum of the Hamiltonian $H\in \mathcal{B}(\mathfrak{H})$ defined by (\ref%
{H0}). By Proposition \ref{direct integral decomposition of the hamiltonian}
and Theorem \ref{borelian functional calculus of a direct integral of
operators}, we thus study the bottom of the spectrum of the fiber
Hamiltonian $A(k)$ (\ref{fiber hamiltonians}) at fixed total quasi-momentum $%
k\in \mathbb{T}^{2}$, similar to \cite{articulo,articulo2}.

\begin{theorem}[Bottom of the spectrum of $A(k)$]
\label{existence of eigenvalue for each fiber}\mbox{ }\newline
Fix $h_{b}\in \lbrack 0,1/2]$. If $k \neq 0$ then there is exactly one
eigenvalue $\mathrm{E}\left( k\right) \equiv \mathrm{E}\left( \mathrm{U}%
,k\right)$ of the Hamiltonian $A(k)$ strictly below $\sigma _{\mathrm{ess}%
}\left( A\left( k\right) \right)$. In this case, the eigenvalue is
non-degenerated and $\mathrm{E}(k) < \mathfrak{b}(k)$ when $\hat{\upsilon}%
(k) \neq 0$, whereas $\mathrm{E}(k) = \mathfrak{b}(k)$ if $\hat{\upsilon}(k)
= 0$. This statement remains valid for $k = 0$ provided $\hat{\upsilon}(0)
\neq 0$.
\end{theorem}

\begin{proof}
Assume that $k\neq 0$. Recall that $\mathfrak{z}(k)$ is defined in Corollary %
\ref{bottom of the spectrum of A11}. From Corollaries \ref{bottom of the
spectrum of A11} and \ref{there is at most one eigenvalue on each connected
component}, the interval $(-\infty ,\mathfrak{z}(k))$ contains at most one
eigenvalue of $A(k)$. By Corollary \ref{eigenspace of a fiber}, the
eigenvalue is non-degenerate, if it exists. If $\hat{\upsilon}(k)=0$, we
know from Proposition \ref{eigenvalue of a fiber} (ii) that $\mathfrak{b}(k)$
is such an eigenvalue. Recall that for $k\neq 0$ one has that $\mathfrak{b}%
(k)<\mathfrak{z}(k)$, because $h_{b}\in \lbrack 0,1/2]$. Now, suppose that $%
\hat{\upsilon}(k)\neq 0$. When $\mathfrak{b}(k)\leq x<\mathfrak{z}(k)$, we
have 
\begin{equation*}
\left( A_{1,1}\left( k\right) -x\mathfrak{1}\right) ^{-1}\geq c\mathfrak{1}
\end{equation*}%
for some constant $c>0$ and hence, $\mathfrak{T}(k,x)>0$, see (\ref{cal T}).
Consequently,%
\begin{equation*}
\hat{\upsilon}(k)^{2}\mathfrak{T}(k,x)+x-\mathfrak{b}(k)>0\ ,
\end{equation*}%
which means that $A(k)$ has no eigenvalues in the interval $[\mathfrak{b}(k),%
\mathfrak{z}(k))$, by Theorem \ref{eigenvalue as a root of a non linear
equation}. We shall now look for an eigenvalue in the interval $(-\infty ,%
\mathfrak{b}(k))$. On the one hand, using Corollary \ref{bottom of the
spectrum of A11}, observe that 
\begin{equation*}
\Vert \mathfrak{d}(k)\Vert ^{-2}|\mathfrak{T}(k,x)|\leq \Vert (A_{1,1}(k)-x%
\mathfrak{1})^{-1}\Vert _{\mathrm{o}\mathrm{p}}=(\mathfrak{z}(k)-x)^{-1}\leq
(\mathfrak{b}(k)-x)^{-1},
\end{equation*}%
whenever $x<\mathfrak{b}(k)$. Taking $x\rightarrow -\infty $, $\mathfrak{T}%
(k,x)$ tends to zero and, hence, 
\begin{equation*}
{\lim\limits_{x\rightarrow -\infty }}\left\{ \hat{\upsilon}(k)^{2}\mathfrak{T%
}(k,x)+x-\mathfrak{b}(k)\right\} =-\infty \text{ }.
\end{equation*}%
On the other hand, the continuity of the mapping 
\begin{equation*}
\mathfrak{T}(k,\cdot ):\rho (A_{1,1}(k))\rightarrow \mathbb{R}
\end{equation*}%
on $(-\infty ,\mathfrak{b}(k)]$ gives us 
\begin{equation*}
{\lim\limits_{x\rightarrow \mathfrak{b}(k)}}\left\{ \hat{\upsilon}(k)^{2}%
\mathfrak{T}(k,x)+x-\mathfrak{b}(k)\right\} =\hat{\upsilon}(k)^{2}\mathfrak{T%
}(k,\mathfrak{b}(k))>0\text{ }.
\end{equation*}%
By the intermediate value theorem, there is $\mathrm{E}(k)\in (-\infty ,%
\mathfrak{b}(k))$ such that 
\begin{equation*}
\hat{\upsilon}(k)^{2}\mathfrak{T}(k,\mathrm{E}(k))+\mathrm{E}(k)-\mathfrak{b}%
(k)=0\text{ }.
\end{equation*}%
By Theorem \ref{eigenvalue as a root of a non linear equation}, $\mathrm{E}%
(k)$ must be an eigenvalue of $A(k)$.

The proof for $k=0$ is done in a similar way. Basically, the only difference
is that, in this case, $\mathfrak{b}(0)=\mathfrak{z}(0)=0$ and 
\begin{equation*}
{\lim\limits_{x\rightarrow 0^{-}}}\left\{ \hat{\upsilon}(0)^{2}\mathfrak{T}%
(0,x)+x\right\} \in (0,\infty]
\end{equation*}%
occurs due to other reasons. Indeed, from Corollary \ref{bottom of the
spectrum of A11} we deduce that $\mathfrak{T}(0,\cdot)$ is strictly positive
on the interval $(-\infty,0)$. Because of (\ref{sdfsdfsdfsdf}) we also have
that $\partial_x\mathfrak{T}(0,x)|_{x = x_0} \geq 0$ whenever $x_0 < 0$.
Thus the limit of $\mathfrak{T}(0,x)$ as $x \to 0^-$ exists, being possibly
infinite.
\end{proof}

If $\hat{\upsilon}(0)=0$ then, by Theorem \ref{eigenvalue as a root of a non
linear equation}, the fiber Hamiltonian $A(0)$ has no negative eigenvalues.
In this case we set $\mathrm{E}(0)=0$, which is obviously an eigenvalue of $%
A(0)$ with associated eigenvector $(0,1)$. (Note that $\sigma _{\mathrm{ess}%
}(A(0))=[0,8\epsilon ]$, by Proposition \ref{essential spectrum of a fiber}%
.) With this definition, observe that, for all $k\in \mathbb{T}^{2}$, $%
\mathrm{E}(k)$ is the minimum spectral value of $A(k)$: 
\begin{equation}
\mathrm{E}\left( k\right) =\min \sigma \left( A\left( k\right) \right) \leq 
\mathfrak{z}\left( k\right) =\min \sigma _{\mathrm{ess}}\left( A\left(
k\right) \right) \ .  \label{minimum2}
\end{equation}%
The lowest eigenvalue $\mathrm{E}(k)$ of $A(k)$, when $\mathrm{E}(k)<0$, is
related to the formation of dressed bound fermion pairs of fermions with
total quasi-momentum $k\in \mathbb{T}^{2}$. In Theorem \ref{Exponentially
localized bound pairs}, we make this claim more precise and prove the
spatial localization of such bounded pairs. Before doing that we study the
regularity of the real-valued function 
\begin{equation*}
\mathrm{E}\equiv \mathrm{E}\left( \mathrm{U},\cdot \right) :\mathbb{T}%
^{2}\rightarrow \mathbb{R}
\end{equation*}%
on the two-dimensional torus.

To this end, we rewrite the characteristic equation given by Theorem \ref%
{eigenvalue as a root of a non linear equation} via the function $\Phi :%
\mathcal{O}\rightarrow \mathbb{R}$ defined by 
\begin{equation}
\Phi \left( k,x\right) \equiv \Phi \left( \mathrm{U},k,x\right) \doteq \hat{%
\upsilon}\left( k\right) ^{2}\mathfrak{T}\left( k,x\right) +x-\mathfrak{b}%
\left( k\right) \ ,  \label{Phi}
\end{equation}%
where $\mathcal{O}$ is the open set%
\begin{equation}
\mathcal{O}\doteq \left\{ \left( k,x\right) \in \mathbb{S}^{2}\times \mathbb{%
R}:x<\mathfrak{z}\left( k\right) \right\} \subseteq \mathbb{R}^{3}\ .
\label{sdsdsd1}
\end{equation}%
Observe from Equation (\ref{sdfsdfsdfsdf}) that $\partial _{x}\Phi >0$ over
the whole domain of $\Phi $.

We now study the continuity of the function $\mathrm{E}:\mathbb{T}%
^{2}\rightarrow \mathbb{R}$ and give a sufficient condition for $\mathrm{E}$%
\ to be of class $C^{d}$ on $\mathbb{S}^{2}$ for every $d\in \mathbb{N}\cup
\{\omega ,a\}$, where $C^{d}(\Omega )$, $d\in \mathbb{N}$, stands for the
space of $d\ $times continuously differentiable functions on $\Omega $,
while $C^{\omega }(\Omega )$ and $C^{a}(\Omega )$ refer to the space of
smooth and real analytic functions on $\Omega $, respectively.

\begin{theorem}[Regularity of the function $\mathrm{E}$]
\label{regularity of E}\mbox{ }\newline
Let $h_{b}\in \lbrack 0,1/2]$.

\begin{enumerate}
\item[i.)] The family $\{\mathrm{E}(\mathrm{U},\cdot )\}_{\mathrm{U}\in 
\mathbb{R}_{0}^{+}}$ of real-valued functions on $\mathbb{T}^{2}$ is
equicontinuous with respect to the metric\footnote{%
Note that as $d_{\mathbb{T}^{2}}$ is smaller than the Euclidean metric for $%
\mathbb{T}^{2}$ as a subset of $\mathbb{R}^{2}$, the equicontinuity also
holds true for the Euclidean metric.} $d_{\mathbb{T}^{2}}$.

\item[ii.)] If $\hat{\upsilon}\in C^{d}(\mathbb{S}^{2})$ ($\mathbb{S}%
^{2}\subseteq \mathbb{R}^{2}$) for some $d\in \mathbb{N}\cup \{\omega ,a\}$,
then $\mathrm{E}\equiv \mathrm{E}(\mathrm{U},\cdot )\in C^{d}(\mathbb{S}%
^{2}) $ for all $\mathrm{U}\in \mathbb{R}_{0}^{+}$. In this case, 
\begin{equation}
\partial _{k_{j}}\mathrm{E}\left( k\right) =-\left( \partial _{x}\Phi \left( 
{k,\mathrm{E}\left( k\right) }\right) \right) ^{-1}\partial _{k_{j}}\Phi
\left( {k,\mathrm{E}\left( k\right) }\right) \ ,\qquad k\in \mathbb{S}^{2},\
j\in \{1,2\}\ .  \label{sdsdsdsdsd}
\end{equation}
\end{enumerate}
\end{theorem}

\begin{proof}
By the spectral theorem, we deduce from (\ref{minimum2}) that, for any $%
\mathrm{U}\in \mathbb{R}_{0}^{+}$, 
\begin{equation}
\mathrm{E}\left( \mathrm{U},k\right) =\min \sigma \left( A\left( \mathrm{U}%
,k\right) \right) ={\inf_{\psi \in \mathcal{H},\Vert \psi \Vert =1}}%
\left\langle \psi ,A\left( \mathrm{U},k\right) \psi \right\rangle ,\qquad
k\in \mathbb{T}^{2}\ .  \label{minimum2bis}
\end{equation}%
Given any $\varepsilon >0$ and $k_{0}\in \mathbb{T}^{2}$, by the (operator
norm) continuity of the mapping $A\left( 0,\cdot \right) :\mathbb{T}%
^{2}\rightarrow \mathcal{B}(\mathcal{H})$ at the point $k_{0}$ we can find $%
\delta >0$ such that%
\begin{align*}
\sup_{\mathrm{U}\in \mathbb{R}_{0}^{+}}{\sup_{\psi \in \mathcal{H},\Vert
\psi \Vert =1}}\left\vert \left\langle \psi ,A\left( \mathrm{U},k\right)
\psi \right\rangle -\left\langle \psi ,A\left( \mathrm{U},k_{0}\right) \psi
\right\rangle \right\vert & \leq \sup_{\mathrm{U}\in \mathbb{R}%
_{0}^{+}}\left\Vert A\left( \mathrm{U},k\right) -A\left( \mathrm{U}%
,k_{0}\right) \right\Vert _{\mathrm{op}} \\
& =\sup_{\mathrm{U}\in \mathbb{R}_{0}^{+}}\left\Vert A\left( 0,k\right)
-A\left( 0,k_{0}\right) \right\Vert _{\mathrm{op}}<\varepsilon
\end{align*}%
for every $k\in \mathbb{T}^{2}$ with $d_{\mathbb{T}^{2}}(k,k_{0})<\delta $.
Recall that $\mathcal{H}$ stands for the (fiber) Hilbert space $L^{2}(%
\mathbb{T}^{2})\oplus \mathbb{C}$, see (\ref{cal H}), while $d_{\mathbb{T}%
^{2}}$ is the metric (\ref{metric}) on the torus $\mathbb{T}^{2}$.
Therefore, $\mathrm{E}:\mathbb{T}^{2}\rightarrow \mathbb{R}$ can be
expressed as the infimum over the equicontinuous family $\{\langle \psi ,A(%
\mathrm{U},\cdot )\psi \rangle \}_{\mathrm{U}\in \mathbb{R}_{0}^{+},\psi \in 
\mathcal{H},\Vert \psi \Vert =1}$ of (continuous) functions and Assertion
(i) follows.

Take again some $k_{0}\in \mathbb{S}^{2}$. Assume that $\hat{\upsilon}$ is
of class $C^{d}$ on $\mathbb{S}^{2}\subseteq \mathbb{R}^{2}$ with $d\in 
\mathbb{N}\cup \{\omega ,a\}$. Let $\vartheta =(k_{0},\mathrm{E}(k_{0}))\in 
\mathcal{O}$. Using Theorems \ref{eigenvalue as a root of a non linear
equation} and \ref{existence of eigenvalue for each fiber} as well as
Equation (\ref{sdfsdfsdfsdf}) and the (operator norm) continuity of the
mapping 
\begin{equation}
A_{1,1}\left( \cdot \right) :\mathbb{T}^{2}\rightarrow \mathcal{B}\left(
L^{2}(\mathbb{T}^{2})\right) \ ,  \label{mapping utilse}
\end{equation}%
one checks that $\Phi \in C^{d}(\mathcal{O})$ with $d\geq 1$, $\Phi
(\vartheta )=0$ and $\partial _{x}\Phi (\vartheta )\neq 0$. See for instance
(\ref{sdfsdfsdfsdf}). We can thus apply the implicit function theorem (see,
e.g., \cite{Implicittheorem1} for an ordinary version and \cite%
{Implicittheorem2} for an analytic version) to obtain open subsets $%
U\subseteq \mathbb{R}^{2}$ and $J\subseteq \mathbb{R}$ such that $\vartheta
\in U\times J\subseteq \mathcal{O}$ and, for each $k\in U$, there is a
unique real number $\xi (k)\in J$ satisfying $\Phi (k,\xi (k))=0$. Moreover,
the mapping $\xi :U\rightarrow J$ defined in this way is of class $C^{d}$
and its partial derivatives are given by%
\begin{equation}
{\partial }_{k_{j}}{\xi \left( {k}\right) }=-\left( \partial _{x}\Phi {%
\left( {k,\xi \left( k\right) }\right) }\right) ^{-1}{{\partial }%
_{k_{j}}\Phi \left( {k,\xi \left( k\right) }\right) }\ ,\qquad k\in U,\ j\in
\{1,2\}\ .  \label{dfdfdfdf}
\end{equation}%
As $\xi (k_{0})=\mathrm{E}(k_{0})<\mathfrak{z}(k_{0})$ (see (\ref{minimum2}%
)), by continuity there exists a neighborhood $V\subseteq U$ of $k_{0}$ such
that $\xi (k)<\mathfrak{z}(k)$ for every $k\in V$. It follows that, for all $%
k\in V$, $\xi (k)$ and $\mathrm{E}(k)$ are in the same connected component $%
(-\infty ,\mathfrak{z}(k))$ and from $\partial _{x}\Phi >0$ we conclude that 
$\mathrm{E}\upharpoonright V=\xi \upharpoonright V$. So, $\mathrm{E}$ is of
class $C^{d}$ near $k_{0}$ and (\ref{dfdfdfdf}) yields (\ref{sdsdsdsdsd})
for any $k\in V$, in particular for $k=k_{0}$. As $k_{0}$ is arbitrary,
Assertion (ii) follows.
\end{proof}

We can now deduce from Theorem \ref{regularity of E} that $\mathrm{E}$ is a
dispersion relation (see Definition \ref{dispersion relation}) when the
function $\hat{\upsilon}:\mathbb{S}^{2}\rightarrow \mathbb{R}$ is at least $%
2 $ times continuously differentiable and in this case, we can even compute
the group velocity. To see this, recall that, for any $f\in C^{2}(\mathbb{S}%
^{2})$, we define in (\ref{sdsdsdsds}) the subset%
\begin{equation*}
\mathfrak{M}_{f}\doteq \left\{ k\in \mathbb{S}^{2}\,:\,\mathrm{Hess}\left(
f\right) \left( k\right) \in \mathsf{GL}_{2}\left( \mathbb{R}\right)
\right\} \subseteq \mathbb{S}^{2}
\end{equation*}%
with $\mathsf{GL}_{2}\left( \mathbb{R}\right) $ being the set of invertible $%
2\times 2$ matrices with real coefficients.

\begin{corollary}[$\mathrm{E}$ as a dispersion relation and group velocity]
\label{existence of a dispersion relation with low energy}\mbox{ }\newline
Let $h_{b}\in \lbrack 0,1/2]$ and $\mathrm{U}\in \mathbb{R}_{0}^{+}$. Then, $%
\mathrm{E}\equiv \mathrm{E}(\mathrm{U},\cdot )\in C(\mathbb{T}^{2})$ and is
of class $C^{2}$ on the open set $\mathbb{S}^{2}\subseteq \mathbb{R}^{2}$
whenever $\hat{\upsilon}$ is of class $C^{2}$ on $\mathbb{S}^{2}$. In this
case, the corresponding group velocity is 
\begin{equation*}
\mathbf{v}_{\mathrm{E}}\left( k\right) \doteq \vec{\nabla}_{k}\mathrm{E}%
\left( k\right) =-\left( \partial _{x}\Phi \left( {k,\mathrm{E}\left(
k\right) }\right) \right) ^{-1}\vec{\nabla}_{k}\Phi \left( {k,\mathrm{E}%
\left( k\right) }\right) \ ,\qquad k\in \mathbb{S}^{2}\ .
\end{equation*}%
Moreover, if $\hat{\upsilon}$ is real analytic (i.e., of class $C^{a}$ in
the above terminology) on $\mathbb{S}^{2}$, then either $\mathfrak{M}_{%
\mathrm{E}}$ has full measure or is empty.
\end{corollary}

\begin{proof}
The first part of the assertion is a direct application of Theorem \ref%
{regularity of E}. It remains to study the set $\mathfrak{M}_{\mathrm{E}}$.
If $\hat{\upsilon}$ is real analytic then, from Equation (\ref{def hessian})
and Theorem \ref{regularity of E}, the function $f:\mathbb{S}%
^{2}\longrightarrow \mathbb{R}$ defined by%
\begin{equation*}
f(k)\doteq \det \left( \mathrm{Hess}{\left( \mathrm{E}\right) \left(
k\right) }\right) \ ,\qquad k\in \mathbb{S}^{2}\ ,
\end{equation*}%
is real analytic and satisfies $f^{-1}(\{0\})=\mathbb{S}^{2}\backslash 
\mathfrak{M}_{\mathrm{E}}$. Since the zeros of any non-constant real
analytic function have null Lebesgue measure (see, e.g., \cite{analytic}),
either $\mathfrak{M}_{\mathrm{E}}$ has full measure or is empty.
\end{proof}

It is natural to derive now the ground state energy of the Hamiltonian $H\in 
\mathcal{B}(\mathfrak{H})$ defined by (\ref{H0}), which is related to the
ground state energy of fiber Hamiltonians, thanks to Proposition \ref{direct
integral decomposition of the hamiltonian} and Theorem \ref{borelian
functional calculus of a direct integral of operators}. As expected, one has
the following equality for the ground state energy 
\begin{equation*}
E{\left( \mathrm{U}\right) }\doteq \min \sigma \left( H\right) =\min \mathrm{%
E}\left( \mathbb{T}^{2}\right) \text{ }.
\end{equation*}%
A\ proof can be done like \cite[Lemma 8]{articulo} by using Kato's
perturbation theory. In the sequel, we provide an alternative way of proving
the equality, which is much more direct.

\begin{proposition}[Bottom of the spectrum of $H$]
\label{determining the fundamental energy}\mbox{ }\newline
We have that 
\begin{equation*}
E{\left( \mathrm{U}\right) }={\min\limits_{k\in \mathbb{T}^{2}}}\,\min
\sigma \left( A\left( k\right) \right) =\min \mathrm{E}\left( \mathbb{T}%
^{2}\right) \leq 0\ .
\end{equation*}
\end{proposition}

\begin{proof}
We first remark that the union 
\begin{equation*}
\mathcal{K}\doteq {\bigcup }\left\{ \sigma \left( A\left( k\right) \right)
\,:\,k\in \mathbb{T}^{2}\right\} \subseteq \mathbb{R}
\end{equation*}%
of the spectra of all fiber Hamiltonians is closed. To see this, let $%
(\lambda _{n})_{n\in \mathbb{N}}$ be a sequence of real numbers converging
to $\lambda \in \mathbb{R}$ with $\lambda _{n}\in \sigma (A(k_{n}))$ for
some $k_{n}\in \mathbb{T}^{2}$ at $n\in \mathbb{N}$. By compactness of $%
\mathbb{T}^{2}$, we can assume without loss of generality that $%
(k_{n})_{n\in \mathbb{N}}$ converges to some point $k_{0}\in \mathbb{T}^{2}$%
. By the (operator norm) continuity of the mapping $A:\mathbb{T}%
^{2}\rightarrow \mathcal{B}(\mathcal{H})$, it follows that%
\begin{equation*}
{\lim\limits_{n\rightarrow \infty }}\,(A(k_{n})-\lambda _{n}\mathfrak{1)}%
=A(k_{0})-\lambda \mathfrak{1}
\end{equation*}%
(in operator norm). Hence, $\lambda \in \sigma (A(k_{0}))$, for otherwise $%
A(k_{n})-\lambda _{n}\mathfrak{1}$ would be invertible\footnote{%
Recall that the set of invertible operators on a Banach space $X$ is an open
subset of $\mathcal{B}(X)$, with respect to the operator norm. See, e.g., 
\cite[Theorem 1.4]{Folland}.} for sufficiently large $n$. Thus, $\mathcal{K}$
is a closed set and, as a consequence, for any $s\notin \mathcal{K}$, there
is $\varepsilon >0$ such that%
\begin{equation*}
\left( s-\varepsilon ,s+\varepsilon \right) \cap \sigma \left( A\left(
k\right) \right) =\emptyset \text{ },\qquad k\in \mathbb{T}^{2}\ .
\end{equation*}%
With this property, we infer from Proposition \ref{direct integral
decomposition of the hamiltonian} and Theorem \ref{borelian functional
calculus of a direct integral of operators} that $\sigma (H)\subseteq 
\mathcal{K}$, which yields the inequality 
\begin{equation}
E{\left( \mathrm{U}\right) }\geq {\min\limits_{k\in \mathbb{T}^{2}}}\,\min
\sigma \left( A\left( k\right) \right) =\min \mathrm{E}\left( \mathbb{T}%
^{2}\right) \text{ }.  \label{ineq}
\end{equation}%
Note that the last equality results from (\ref{minimum2}). By Theorem \ref%
{regularity of E} (i), $\mathrm{E}:\mathbb{T}^{2}\rightarrow \mathbb{R}$ is
continuous. From the compactness of the torus $\mathbb{T}^{2}$ and the
Weierstrass extreme value theorem, $\mathrm{E}$ has a minimizer in $\mathbb{T%
}^{2}$, say $k_{0}\in \mathbb{T}^{2}$. The continuity of $\mathrm{E}$ at $%
k_{0}$ implies that, for every $\varepsilon >0$, there is $\delta >0$ such
that, for all $k\in \mathbb{T}^{2}$ satisfying $d_{\mathbb{T}%
^{2}}(k,k_{0})<\delta $, 
\begin{equation*}
\mathrm{E}\left( k\right) \in \left( \mathrm{E}\left( k_{0}\right)
-\varepsilon ,\mathrm{E}\left( k_{0}\right) +\varepsilon \right) \ ,
\end{equation*}%
and, as a consequence,%
\begin{equation*}
\nu \left( \left\{ k\in \mathbb{T}^{2}:\sigma \left( A\left( k\right)
\right) \cap \left( \mathrm{E}\left( k_{0}\right) -\varepsilon ,\mathrm{E}%
\left( k_{0}\right) +\varepsilon \right) \neq \emptyset \right\} \right)
\geq \nu \left( \mathbb{T}^{2}\cap B_{\delta }\left( k_{0}\right) \right) >0%
\text{ },
\end{equation*}%
where $B_{\delta }(k_{0})$ is the open ball (for the metric $d_{\mathbb{T}%
^{2}}$) centered at $k_{0}\in \mathbb{T}^{2}$ of radius $\delta \in \mathbb{R%
}^{+}$. By Theorem \ref{borelian functional calculus of a direct integral of
operators}, this implies that $\mathrm{E}(k_{0})\in \sigma (H)$, which,
combined with (\ref{ineq}), yields the equalities 
\begin{equation*}
E{\left( \mathrm{U}\right) }={\min\limits_{k\in \mathbb{T}^{2}}}\,\min
\sigma \left( A\left( k\right) \right) =\min \mathrm{E}\left( \mathbb{T}%
^{2}\right) \text{ }.
\end{equation*}%
Using Equations (\ref{equation sur bk}) and (\ref{minimum2}), we note that 
\begin{equation*}
E{\left( \mathrm{U}\right) }=\min \mathrm{E}\left( \mathbb{T}^{2}\right)
\leq {\min\limits_{k\in \mathbb{T}^{2}}\ }\mathfrak{z}(k)=4\epsilon
-2\epsilon \max_{k\in \mathbb{T}^{2}}\cos (k/2)=\mathfrak{z}(0)=0\ .
\end{equation*}
\end{proof}

\subsection{Spectral Properties in the Hard-Core Limit\label{effective
behavior at hard-core limit of the electronic repulsion}}

We now study the spectral properties of fiber Hamiltonians $A(k)$ (\ref%
{fiber hamiltonians}) in the hard-core limit. It refers to the limit $%
\mathrm{U}\rightarrow \infty $. In fact, a very strong on-site repulsion $%
\mathrm{U}$ (see Equation (\ref{Hamiltonian-f})) prevents two fermions of
opposite spins from occupying the same lattice site. We study in particular
the continuous function $\mathrm{E}:\mathbb{T}^{2}\rightarrow \mathbb{R}$
defined by Theorem \ref{existence of eigenvalue for each fiber}, which
corresponds to the continuous family of non-degenerate eigenvalues at lowest
energies in each fiber, in this limit.

An important result in this context is the characterization of such
eigenvalues via the Birman-Schwinger principle, given by Theorem \ref%
{eigenvalue as a root of a non linear equation}. In particular, we need
first to study the hard-core limit of the characteristic equation, which
amounts to determine the limit $\mathrm{U}\rightarrow \infty $ of the
quantity (\ref{cal T}), i.e., 
\begin{equation}
\mathfrak{T}\left( \mathrm{U},k,\lambda \right) \equiv \mathfrak{T}\left(
k,\lambda \right) \doteq \left\langle \mathfrak{d}\left( k\right) ,\left(
A_{1,1}\left( \mathrm{U},k\right) -\lambda \mathfrak{1}\right) ^{-1}%
\mathfrak{d}\left( k\right) \right\rangle \text{ },\qquad \lambda \in \rho
\left( A_{1,1}\left( \mathrm{U},k\right) \right) \ ,  \label{cal Tbis}
\end{equation}%
for $h_{b}\in \lbrack 0,1/2]$ and $k\in \mathbb{T}^{2}$. We start with this
point, which allows us to study afterwards the limit of the lowest
eigenvalues and all the derived quantities, like for instance the group
velocity.

\subsubsection{The characteristic equation in the hard-core limit}

Let $\mathfrak{s}=\mathfrak{\hat{e}}_{0}\in L^{2}(\mathbb{T}^{2})$ denote
the constant function$\ 1$ on the torus $\mathbb{T}^{2}$. For any fixed $%
k\in \mathbb{T}^{2}$ and $\lambda \in \rho (B_{1,1}(k))$, define the
following four constants: 
\begin{align}
R_{\mathfrak{s},\mathfrak{s}}& \equiv R_{\mathfrak{s},\mathfrak{s}}\left(
k,\lambda \right) \doteq \left\langle \mathfrak{s},\left( B_{1,1}\left(
k\right) -\lambda \mathfrak{1}\right) ^{-1}\mathfrak{s}\right\rangle \ ,
\label{R1} \\[0.02in]
R_{\mathfrak{s},\mathfrak{d}}& \equiv R_{\mathfrak{s},\mathfrak{d}}\left(
k,\lambda \right) \doteq \left\langle \mathfrak{s},\left( B_{1,1}\left(
k\right) -\lambda \mathfrak{1}\right) ^{-1}\mathfrak{d}\left( k\right)
\right\rangle \ ,  \label{R2} \\[0.02in]
R_{\mathfrak{d},\mathfrak{s}}& \equiv R_{\mathfrak{d},\mathfrak{s}}\left(
k,\lambda \right) \doteq \left\langle \mathfrak{d}\left( k\right) ,\left(
B_{1,1}\left( k\right) -\lambda \mathfrak{1}\right) ^{-1}\mathfrak{s}%
\right\rangle \ ,  \label{R3} \\
R_{\mathfrak{d},\mathfrak{d}}& \equiv R_{\mathfrak{d},\mathfrak{d}}\left(
k,\lambda \right) \doteq \left\langle \mathfrak{d}\left( k\right) ,\left(
B_{1,1}\left( k\right) -\lambda \mathfrak{1}\right) ^{-1}\mathfrak{d}\left(
k\right) \right\rangle \ ,  \label{R4}
\end{align}%
where we recall that $B_{1,1}\left( k\right) $ is defined by (\ref{A11}). In
the following lemma we write $\mathfrak{T}(\mathrm{U},k,\lambda )$, which is
defined by (\ref{cal Tbis}), in terms of these four quantities.

In the following it is technically convenient to assume that $\mathfrak{d}%
(k)\notin \mathbb{C}\mathfrak{s}$ for all $k\in \mathbb{T}^{2}$. Notice that
this holds true iff $\mathfrak{p}_{1}\notin \mathbb{C}\mathfrak{e}_{0}$ or $%
\mathfrak{p}_{2}\notin \mathbb{C}\mathfrak{e}_{0}$, i.e., $r_{\mathfrak{p}%
}>0 $. Indeed, recall (\ref{fourier D fract}), that is, 
\begin{equation*}
\mathfrak{d}\left( k\right) =\mathcal{F}\left[ \mathrm{e}^{ik\cdot x}%
\mathfrak{p}_{1}\left( x\right) +\mathrm{e}^{i\frac{k}{2}\cdot x}\mathfrak{p}%
_{2}\left( x\right) \right] \ ,
\end{equation*}%
where $\mathrm{e}^{ik\cdot x}\mathfrak{p}_{\sharp }(x)$ stands for the
function $x\mapsto \mathrm{e}^{ik\cdot x}\mathfrak{p}_{\sharp }(x)$ with $%
\sharp \in \{1,2\}$. See also discussions around Equations (\ref%
{conditionsdsdsd})--(\ref{summable2-Urrrrr}).

\begin{lemma}
\label{representing I(k,x) by explicit integrals}\mbox{ }\newline
Let $k\in \mathbb{T}^{2}$, $\mathrm{U}\in \mathbb{R}_{0}^{+}$ and $\lambda <%
\mathfrak{z}(k)$, with $\mathfrak{z}(k)\in \mathbb{R}$ defined in Corollary %
\ref{bottom of the spectrum of A11}. Then,%
\begin{equation*}
\mathfrak{T}\left( \mathrm{U},k,\lambda \right) =\frac{R_{\mathfrak{d},%
\mathfrak{d}}}{\mathrm{U}R_{\mathfrak{s},\mathfrak{s}}+1}+\mathrm{U}\frac{R_{%
\mathfrak{d},\mathfrak{d}}R_{\mathfrak{s},\mathfrak{s}}-\left\vert R_{%
\mathfrak{s},\mathfrak{d}}\right\vert ^{2}}{\mathrm{U}R_{\mathfrak{s},%
\mathfrak{s}}+1}\text{ },
\end{equation*}%
with $R_{\mathfrak{d},\mathfrak{d}}R_{\mathfrak{s},\mathfrak{s}}-\left\vert
R_{\mathfrak{s},\mathfrak{d}}\right\vert ^{2}\geq 0$. Moreover, if $r_{%
\mathfrak{p}}>0$ (see (\ref{summable2-Urrrrr})--(\ref{summable2-Urrrrr0})),
then the inequality is strict.
\end{lemma}

\begin{proof}
The proof of the first part is a slightly more complicated version of the
one of \cite[Lemma 14]{articulo}. Fix $k\in \mathbb{T}^{2}$, $\mathrm{U}\in 
\mathbb{R}_{0}^{+}$ and $\lambda <\mathfrak{z}(k)$. Define the complex
numbers%
\begin{eqnarray*}
Q_{\mathfrak{s},\mathfrak{s}} &\doteq &\left\langle \mathfrak{s},\left(
A_{1,1}\left( \mathrm{U},k\right) -\lambda \mathfrak{1}\right) ^{-1}%
\mathfrak{s}\right\rangle \ , \\
Q_{\mathfrak{s},\mathfrak{d}} &\doteq &\left\langle \mathfrak{s},\left(
A_{1,1}\left( \mathrm{U},k\right) -\lambda \mathfrak{1}\right) ^{-1}%
\mathfrak{d}\left( k\right) \right\rangle \ , \\
Q_{\mathfrak{d},\mathfrak{s}} &\doteq &\left\langle \mathfrak{d}\left(
k\right) ,\left( A_{1,1}\left( \mathrm{U},k\right) -\lambda \mathfrak{1}%
\right) ^{-1}\mathfrak{s}\right\rangle \ , \\
Q_{\mathfrak{d},\mathfrak{d}} &\doteq &\left\langle \mathfrak{d}\left(
k\right) ,\left( A_{1,1}\left( \mathrm{U},k\right) -\lambda \mathfrak{1}%
\right) ^{-1}\mathfrak{d}\left( k\right) \right\rangle =\mathfrak{T}\left( 
\mathrm{U},k,\lambda \right) \ ,
\end{eqnarray*}%
where we recall that 
\begin{equation}
A_{1,1}\left( \mathrm{U},k\right) \doteq B_{1,1}\left( k\right) +\mathrm{U}%
P_{0}\geq B_{1,1}\left( k\right) \ ,  \label{ghghghh}
\end{equation}%
by Equation (\ref{A11-U0}). Using Corollary \ref{bottom of the spectrum of
A11} note at this point that 
\begin{equation}
\mathfrak{z}\left( k\right) =\min \sigma \left( A_{1,1}\left( \mathrm{U}%
,k\right) \right) =\min \sigma \left( B_{1,1}\left( k\right) \right) \geq 0\
.  \label{gjhhjh0}
\end{equation}%
In particular, 
\begin{equation}
\lambda \in \left( -\infty ,\mathfrak{z}\left( k\right) \right) \subseteq
\rho \left( A_{1,1}\left( \mathrm{U},k\right) \right) \cap \rho \left(
B_{1,1}\left( k\right) \right)  \label{gjhhjh}
\end{equation}%
and the resolvent operators $(A_{1,1}(\mathrm{U},k)-\lambda \mathfrak{1}%
)^{-1}$ and $(B_{1,1}(k)-\lambda \mathfrak{1})^{-1}$ are strictly positive.
By using the second resolvent identity together with Equation (\ref{ghghghh}%
), we compute that 
\begin{equation*}
\left( A_{1,1}\left( \mathrm{U},k\right) -\lambda \mathfrak{1}\right)
^{-1}=\left( B_{1,1}\left( k\right) -\lambda \mathfrak{1}\right) ^{-1}-%
\mathrm{U}\left( A_{1,1}\left( \mathrm{U},k\right) -\lambda \mathfrak{1}%
\right) ^{-1}P_{0}\left( B_{1,1}\left( k\right) -\lambda \mathfrak{1}\right)
^{-1}\ .
\end{equation*}%
Recalling that $P_{0}$ is the orthogonal projection onto $\mathbb{C}%
\mathfrak{s=\mathbb{C}\hat{e}}_{0}$, we have that%
\begin{align*}
Q_{\mathfrak{s},\mathfrak{s}}& =R_{\mathfrak{s},\mathfrak{s}}-\mathrm{U}%
\left\langle \mathfrak{s},\left( A_{1,1}\left( \mathrm{U},k\right) -\lambda 
\mathfrak{1}\right) ^{-1}P_{0}\left( B_{1,1}\left( k\right) -\lambda 
\mathfrak{1}\right) ^{-1}\mathfrak{s}\right\rangle \\
& =R_{\mathfrak{s},\mathfrak{s}}-\mathrm{U}Q_{\mathfrak{s},\mathfrak{s}}R_{%
\mathfrak{s},\mathfrak{s}}\ , \\
Q_{\mathfrak{d},\mathfrak{d}}& =R_{\mathfrak{d},\mathfrak{d}}-\mathrm{U}%
\left\langle \mathfrak{d}\left( k\right) ,\left( A_{1,1}\left( \mathrm{U}%
,k\right) -\lambda \mathfrak{1}\right) ^{-1}P_{0}\left( B_{1,1}\left(
k\right) -\lambda \mathfrak{1}\right) ^{-1}\mathfrak{d}\left( k\right)
\right\rangle \\
& =R_{\mathfrak{d},\mathfrak{d}}-\mathrm{U}Q_{\mathfrak{d},\mathfrak{s}}R_{%
\mathfrak{s},\mathfrak{d}}\ , \\
Q_{\mathfrak{s},\mathfrak{d}}& =R_{\mathfrak{s},\mathfrak{d}}-\mathrm{U}%
\left\langle \mathfrak{s},\left( A_{1,1}\left( \mathrm{U},k\right) -\lambda 
\mathfrak{1}\right) ^{-1}P_{0}\left( B_{1,1}\left( k\right) -\lambda 
\mathfrak{1}\right) ^{-1}\mathfrak{d}\left( k\right) \right\rangle \\
& =R_{\mathfrak{s},\mathfrak{d}}-\mathrm{U}Q_{\mathfrak{s},\mathfrak{s}}R_{%
\mathfrak{s},\mathfrak{d}}\ , \\
Q_{\mathfrak{d},\mathfrak{s}}& =R_{\mathfrak{d},\mathfrak{s}}-\mathrm{U}%
\left\langle \mathfrak{d}\left( k\right) ,\left( A_{1,1}\left( \mathrm{U}%
,k\right) -\lambda \mathfrak{1}\right) ^{-1}P_{0}\left( B_{1,1}\left(
k\right) -\lambda \mathfrak{1}\right) ^{-1}\mathfrak{s}\right\rangle \\
& =R_{\mathfrak{d},\mathfrak{s}}-\mathrm{U}Q_{\mathfrak{d},\mathfrak{s}}R_{%
\mathfrak{s},\mathfrak{s}}\ .
\end{align*}%
In matrix notation, the above equations can be rewritten as 
\begin{equation*}
\begin{pmatrix}
R_{\mathfrak{s},\mathfrak{s}} & R_{\mathfrak{d},\mathfrak{s}} \\[0.5em] 
R_{\mathfrak{s},\mathfrak{d}} & R_{\mathfrak{d},\mathfrak{d}}%
\end{pmatrix}%
=\mathrm{U}%
\begin{pmatrix}
Q_{\mathfrak{s},\mathfrak{s}}R_{\mathfrak{s},\mathfrak{s}} & Q_{\mathfrak{d},%
\mathfrak{s}}R_{\mathfrak{s},\mathfrak{s}} \\ 
Q_{\mathfrak{s},\mathfrak{s}}R_{\mathfrak{s},\mathfrak{d}} & Q_{\mathfrak{d},%
\mathfrak{s}}R_{\mathfrak{s},\mathfrak{d}}%
\end{pmatrix}%
+%
\begin{pmatrix}
Q_{\mathfrak{s},\mathfrak{s}} & Q_{\mathfrak{d},\mathfrak{s}} \\ 
Q_{\mathfrak{s},\mathfrak{d}} & Q_{\mathfrak{d},\mathfrak{d}}%
\end{pmatrix}%
=%
\begin{pmatrix}
\mathrm{U}R_{\mathfrak{s},\mathfrak{s}}+1 & 0 \\[0.5em] 
\mathrm{U}R_{\mathfrak{s},\mathfrak{d}} & 1%
\end{pmatrix}%
\begin{pmatrix}
Q_{\mathfrak{s},\mathfrak{s}} & Q_{\mathfrak{d},\mathfrak{s}} \\ 
Q_{\mathfrak{s},\mathfrak{d}} & Q_{\mathfrak{d},\mathfrak{d}}%
\end{pmatrix}%
\ .
\end{equation*}%
As $(B_{1,1}(k)-\lambda \mathfrak{1})^{-1}\geq 0$ (because of (\ref{gjhhjh}%
)) and $\mathrm{U}R_{\mathfrak{s},\mathfrak{s}}\geq 0$,%
\begin{equation*}
\det 
\begin{pmatrix}
\mathrm{U}R_{\mathfrak{s},\mathfrak{s}}+1 & 0 \\[0.5em] 
\mathrm{U}R_{\mathfrak{s},\mathfrak{d}} & 1%
\end{pmatrix}%
=\mathrm{U}R_{\mathfrak{s},\mathfrak{s}}+1>0\ ,
\end{equation*}%
which means that the matrix appearing in the above determinant is
invertible. From this we conclude that%
\begin{equation*}
\begin{pmatrix}
Q_{\mathfrak{s},\mathfrak{s}} & Q_{\mathfrak{d},\mathfrak{s}} \\ 
Q_{\mathfrak{s},\mathfrak{d}} & Q_{\mathfrak{d},\mathfrak{d}}%
\end{pmatrix}%
=\frac{1}{\mathrm{U}R_{\mathfrak{s},\mathfrak{s}}+1}%
\begin{pmatrix}
1 & 0 \\ 
-\mathrm{U}R_{\mathfrak{s},\mathfrak{d}} & \mathrm{U}R_{\mathfrak{s},%
\mathfrak{s}}+1%
\end{pmatrix}%
\begin{pmatrix}
R_{\mathfrak{s},\mathfrak{s}} & R_{\mathfrak{d},\mathfrak{s}} \\[0.5em] 
R_{\mathfrak{s},\mathfrak{d}} & R_{\mathfrak{d},\mathfrak{d}}%
\end{pmatrix}%
\ .
\end{equation*}%
In particular, since $R_{\mathfrak{s},\mathfrak{d}}=\overline{R_{\mathfrak{d}%
,\mathfrak{s}}}$, 
\begin{equation*}
\mathfrak{T}\left( \mathrm{U},k,\lambda \right) =Q_{\mathfrak{d},\mathfrak{d}%
}=R_{\mathfrak{d},\mathfrak{d}}-\frac{\mathrm{U}}{\mathrm{U}R_{\mathfrak{s},%
\mathfrak{s}}+1}\left\vert R_{\mathfrak{s},\mathfrak{d}}\right\vert ^{2}=%
\frac{R_{\mathfrak{d},\mathfrak{d}}}{\mathrm{U}R_{\mathfrak{s},\mathfrak{s}%
}+1}+\mathrm{U}\frac{R_{\mathfrak{d},\mathfrak{d}}R_{\mathfrak{s},\mathfrak{s%
}}-\left\vert R_{\mathfrak{s},\mathfrak{d}}\right\vert ^{2}}{\mathrm{U}R_{%
\mathfrak{s},\mathfrak{s}}+1}\ .
\end{equation*}%
Because $\lambda <\mathfrak{z}(k)\leq 0$ and $(B_{1,1}(k)-\lambda \mathfrak{1%
})^{-1}\geq \left\vert \mathfrak{z}(k)-\lambda \right\vert ^{-1}\mathfrak{1}$
(see (\ref{gjhhjh})), the sesquilinear form 
\begin{equation*}
\left( \varphi ,\psi \right) \mapsto \langle \varphi ,(B_{1,1}(k)-\lambda 
\mathfrak{1})^{-1}\psi \rangle
\end{equation*}%
is a scalar product and using the Cauchy-Schwarz inequality\footnote{%
Recall that the Cauchy-Schwarz inequality applied to a scalar product is an
equality iff the vectors are linearly dependent.}, we deduce that 
\begin{equation*}
R_{\mathfrak{d},\mathfrak{d}}R_{\mathfrak{s},\mathfrak{s}}-\left\vert R_{%
\mathfrak{s},\mathfrak{d}}\right\vert ^{2}\geq 0\ .
\end{equation*}%
When $r_{\mathfrak{p}}>0$, the set $\{\mathfrak{d}(k),\mathfrak{s}\}$ is
linearly independent for every $k\in \mathbb{T}^{2}$.
\end{proof}

The last lemma is useful to deduce the behavior of the quantity $\mathfrak{T}%
(\mathrm{U},k,\lambda )$ at large Hubbard coupling constant $\mathrm{U}\gg 1$%
:

\begin{corollary}[$\mathfrak{T}(\mathrm{U}_{0},k,\protect\lambda )$ at large
on-site repulsions]
\label{limit of I(U,k,x) when U goes to infinity}\mbox{ }\newline
Let $k\in \mathbb{T}^{2}$ and $\lambda <\mathfrak{z}(k)$, with $\mathfrak{z}%
(k)\in \mathbb{R}$ defined in Corollary \ref{bottom of the spectrum of A11}.
Then, for all $\mathrm{U}\in \mathbb{R}_{0}^{+}$,%
\begin{equation*}
0 \le \mathfrak{T}\left( \mathrm{U},k,\lambda \right) -\mathfrak{T}\left(
\infty ,k,\lambda \right) \le\frac{R_{\mathfrak{d},\mathfrak{d}}}{1+\mathrm{U%
}R_{\mathfrak{s},\mathfrak{s}}}\ ,
\end{equation*}%
where 
\begin{equation}
\mathfrak{T}\left( \infty ,k,\lambda \right) \doteq R_{\mathfrak{s},%
\mathfrak{s}}^{-1}\left( {R_{\mathfrak{d},\mathfrak{d}}R_{\mathfrak{s},%
\mathfrak{s}}-}\left\vert R_{\mathfrak{s},\mathfrak{d}}\right\vert
^{2}\right) \geq 0{\ .}  \label{Tinfinity}
\end{equation}
\end{corollary}

\begin{proof}
By Lemma \ref{representing I(k,x) by explicit integrals}, we have $R_{%
\mathfrak{s},\mathfrak{s}}>0$ and $R_{\mathfrak{d},\mathfrak{d}}R_{\mathfrak{%
s},\mathfrak{s}}\geq\left\vert R_{\mathfrak{s},\mathfrak{d}}\right\vert ^{2}$
while 
\begin{equation*}
\mathfrak{T}(\mathrm{U},k,\lambda )-{\frac{R_{\mathfrak{d},\mathfrak{d}}R_{%
\mathfrak{s},\mathfrak{s}}-\left\vert R_{\mathfrak{s},\mathfrak{d}%
}\right\vert ^{2}}{R_{\mathfrak{s},\mathfrak{s}}}} =\frac{\left\vert R_{%
\mathfrak{s},\mathfrak{d}}\right\vert ^{2}}{\left( 1+\mathrm{U}R_{\mathfrak{s%
},\mathfrak{s}}\right) R_{\mathfrak{s},\mathfrak{s}}}\le\frac{R_{\mathfrak{d}%
,\mathfrak{d}}}{1+\mathrm{U}R_{\mathfrak{s},\mathfrak{s}}}\ .
\end{equation*}
\end{proof}

\subsubsection{Hard-core dispersion relation of bound pairs of lowest energy}

We are now in a position to study the spectral properties of the model $H\in 
\mathcal{B}(\mathfrak{H})$ defined by (\ref{H0}) in the hard-core limit. We
study in particular its ground state energy $E(\mathrm{U})$ (see Corollary %
\ref{maincoro1}) and the limit of the continuous function $\mathrm{E}:%
\mathbb{T}^{2}\rightarrow \mathbb{R}$ defined by Theorem \ref{existence of
eigenvalue for each fiber}, which corresponds to the continuous family of
non-degenerate eigenvalues at lowest energies in the fibers.

We start with the hard-core ground state energy, which is a well-defined
quantity that even stays negative:

\begin{lemma}[Existence of the hard-core ground state energy]
\label{formula for hard-core ground state energy}\mbox{ }\newline
In the hard-core limit $\mathrm{U}\rightarrow \infty $, the hard-core ground
state energy (\ref{hard-core ground state energy}) is well-defined and is
equal to 
\begin{equation*}
E\left( \infty \right) ={\sup\limits_{\mathrm{U}\in \mathbb{R}_{0}^{+}}}E(%
\mathrm{U})\leq 0.
\end{equation*}
\end{lemma}

\begin{proof}
When $0\leq \mathrm{U}\leq \mathrm{V}$, one obviously has 
\begin{equation*}
A(\mathrm{V},k)-A(\mathrm{U},k)=(\mathrm{V}-\mathrm{U})%
\begin{pmatrix}
P_{0} & 0 \\ 
0 & 0%
\end{pmatrix}%
\geq 0\ ,
\end{equation*}%
see Equation (\ref{fiber hamiltonians}). In other words, $A(\mathrm{U},k)$, $%
\mathrm{U}\in \mathbb{R}_{0}^{+}$, defines an increasing family of bounded
operators and by Proposition \ref{essential spectrum of a fiber} and
Equation (\ref{equation sur bk}), 
\begin{equation}
\min \sigma \left( A\left( \mathrm{U},k\right) \right) \leq \min \sigma
\left( A\left( \mathrm{V},k\right) \right) \leq \mathfrak{z}\left( k\right)
\label{sdsdsdssds}
\end{equation}%
whenever $0\leq \mathrm{U}\leq \mathrm{V}$. In particular, by taking the
minimum over $k\in \mathbb{T}^{2}$, if $0\leq \mathrm{U}\leq \mathrm{V}$\
then 
\begin{equation*}
E(\mathrm{U})\leq E(\mathrm{V})\leq 0\ .
\end{equation*}%
See Proposition \ref{determining the fundamental energy}. This shows that $E$
is an increasing function of $\mathrm{U}\in \mathbb{R}_{0}^{+}$, which is
bounded from above by $0$. This yields the assertion, thanks to the monotone
convergence theorem.
\end{proof}

We give in the next theorem a hard-core limit version of Theorems \ref%
{eigenvalue as a root of a non linear equation}, \ref{existence of
eigenvalue for each fiber} and \ref{regularity of E}. To this end, recall
that $\mathrm{E}\left( k\right) $\textrm{, }$k\in \mathbb{T}^{2}$\textrm{, }%
are given by Theorem \ref{existence of eigenvalue for each fiber} as a
family of non-degenerate eigenvalues. This family depends upon the parameter 
$\mathrm{U}\in \mathbb{R}_{0}^{+}$ and we thus use here the notation $%
\mathrm{E}(\mathrm{U},k)\equiv \mathrm{E}(k)$. This defines a function $%
\mathrm{E}:\mathbb{R}_{0}^{+}\times \mathbb{T}^{2}\rightarrow \mathbb{R}$.

\begin{theorem}[Dispersion relation in the hard-core limit]
\label{properties of effective dispersion relation with low energy}\mbox{ }%
\newline
Let $h_{b}\in \lbrack 0,1/2]$. Recall that $\mathfrak{T}(\infty ,k,\lambda
)\geq 0$ is defined by (\ref{Tinfinity}).

\begin{enumerate}
\item[i.)] For every $k\in \mathbb{T}^{2}$, the following limit exists: 
\begin{equation*}
\mathrm{E}\left( \infty ,k\right) ={\lim\limits_{\mathrm{U}\rightarrow
\infty }}\mathrm{E}\left( \mathrm{U},k\right) ={\sup\limits_{\mathrm{U}\in 
\mathbb{R}_{0}^{+}}}\mathrm{E}\left( \mathrm{U},k\right) \ ;
\end{equation*}

\item[ii.)] $\mathrm{E}\left( \infty ,\cdot \right) :\mathbb{T}%
^{2}\rightarrow \mathbb{R}$ is a continuous function;

\item[iii.)] For $k\neq 0$, $\mathrm{E}(\infty ,k)$ is the unique solution
to the equation 
\begin{equation}
\hat{\upsilon}\left( k\right) ^{2}\mathfrak{T}\left( \infty ,k,z\right) +z-%
\mathfrak{b}\left( k\right) =0\text{ },\qquad z<\mathfrak{z}(k)\ .
\label{jkljklkjl}
\end{equation}

\item[iv.)] If $\hat{\upsilon}$ is of class $C^{d}$ on $\mathbb{S}%
^{2}\subseteq \mathbb{R}^{2}$ with $d\in \mathbb{N}\cup \{\omega ,a\}$ then
so does $\mathrm{E}(\infty ,\cdot )$.
\end{enumerate}

If, in addition, $r_{\mathfrak{p}}>0$ then:

\begin{enumerate}
\item[v.)] For every $k\in \mathbb{T}^{2}$, $\mathrm{E}(\infty ,k)\leq 
\mathfrak{b}(k)$ with equality iff $\hat{\upsilon}(k)=0$.
\end{enumerate}
\end{theorem}

\begin{proof}
Fix $h_{b}\in \lbrack 0,1/2]$. By (\ref{minimum2}) and (\ref{sdsdsdssds})
together with Corollary \ref{bottom of the spectrum of A11} and Theorem \ref%
{existence of eigenvalue for each fiber}, 
\begin{equation}
\mathrm{E}\left( \mathrm{U},k\right) \leq \mathrm{E}\left( \mathrm{V}%
,k\right) \leq \mathfrak{b}\left( k\right) \leq \mathfrak{z}\left( k\right)
=\min \sigma \left( B_{1,1}\left( k\right) \right) \ ,\qquad k\in \mathbb{T}%
^{2}\ ,  \label{sssss}
\end{equation}%
whenever $0\leq \mathrm{U}\leq \mathrm{V}$. This shows that, at any fixed $%
k\in \mathbb{T}^{2}$, the function $\mathrm{U}\mapsto \mathrm{E}\left( 
\mathrm{U},k\right) $ from $\mathbb{R}_{0}^{+}$ to $\mathbb{R}$ is
increasing and bounded. Therefore, for any $k\in \mathbb{T}^{2}$, 
\begin{equation}
\mathrm{E}\left( \infty ,k\right) ={\lim\limits_{\mathrm{U}\rightarrow
\infty }}\mathrm{E}\left( \mathrm{U},k\right) ={\sup\limits_{\mathrm{U}\in 
\mathbb{R}_{0}^{+}}}\mathrm{E}\left( \mathrm{U},k\right) \leq \mathfrak{b}%
\left( k\right) \ ,  \label{limit1}
\end{equation}%
thanks to the monotone convergence theorem. In particular, Assertion (i)
holds true.

By Combining Equation (\ref{limit1}) with (\ref{minimum2bis}), note that%
\begin{equation*}
\mathrm{E}\left( \infty ,k\right) ={\sup\limits_{\mathrm{U}\in \mathbb{R}%
_{0}^{+}}\inf_{\psi \in \mathcal{H},\Vert \psi \Vert =1}}\left\langle \psi
,A\left( \mathrm{U},k\right) \psi \right\rangle \ ,\qquad k\in \mathbb{T}%
^{2}\ .
\end{equation*}%
Meanwhile, for any $\varepsilon >0$ and $k_{0}\in \mathbb{T}^{2}$, there is $%
\delta >0$ such that 
\begin{equation*}
\left\Vert A\left( \mathrm{U},k\right) -A\left( \mathrm{U},k_{0}\right)
\right\Vert _{\mathrm{op}}=\left\Vert A\left( 0,k\right) -A\left(
0,k_{0}\right) \right\Vert _{\mathrm{op}}\leq \delta \ .
\end{equation*}%
As a consequence, similar to what is done after (\ref{minimum2bis}), the set 
\begin{equation*}
\{\left\langle \psi ,A(\mathrm{U},\cdot )\psi \right\rangle :\mathrm{U}\in 
\mathbb{R}_{0}^{+},\ \psi \in \mathcal{H}\text{ with }\left\Vert \psi
\right\Vert =1\}
\end{equation*}%
is a family of equicontinuous functions on $\mathbb{T}^{2}$. Therefore, $%
\mathrm{E}(\infty ,\cdot )$ is continuous and Assertion (ii) holds true.

Fix $k\neq 0$. If $\hat{\upsilon}(k)\neq 0$ and $\mathrm{U}\in \mathbb{R}%
^{+} $, then we deduce from Corollary \ref{bottom of the spectrum of A11},
Theorem \ref{eigenvalue as a root of a non linear equation} and Equations (%
\ref{sssss})--(\ref{limit1}) that 
\begin{equation*}
\left\{ \mathrm{E}\left( \mathrm{U},k\right) :\mathrm{U}\in \mathbb{R}%
_{0}^{+}\cup \left\{ \infty \right\} \right\} \subseteq \rho \left(
B_{1,1}\left( k\right) \right)
\end{equation*}%
and%
\begin{equation*}
\hat{\upsilon}\left( k\right) ^{-2}\left( \mathfrak{b}\left( k\right) -%
\mathrm{E}\left( \mathrm{U},k\right) \right) -\mathfrak{T}\left( \infty ,k,%
\mathrm{E}\left( \mathrm{U},k\right) \right) =\mathfrak{T}\left( k,\mathrm{E}%
\left( \mathrm{U},k\right) \right) -\mathfrak{T}\left( \infty ,k,\mathrm{E}%
\left( \mathrm{U},k\right) \right) \ .
\end{equation*}%
Invoking\ next Corollary \ref{limit of I(U,k,x) when U goes to infinity}, we
obtain the inequality 
\begin{equation*}
\left\vert \mathfrak{b}\left( k\right) -\mathrm{E}\left( \mathrm{U},k\right)
-\hat{\upsilon}\left( k\right) ^{2}\mathfrak{T}\left( \infty ,k,\mathrm{E}%
\left( \mathrm{U},k\right) \right) \right\vert <{\frac{\hat{\upsilon}\left(
k\right) ^{2}R_{\mathfrak{d},\mathfrak{d}}}{\mathrm{U}R_{\mathfrak{s},%
\mathfrak{s}}}}\text{ }.
\end{equation*}%
We can take the limit $\mathrm{U}\rightarrow \infty $ in this last
inequality by using (\ref{limit1}) and the continuity of the mapping 
\begin{equation*}
\mathfrak{T}\left( \infty ,k,\cdot \right) :\rho \left( B_{1,1}\left(
k\right) \right) \rightarrow \mathbb{R}^{+}
\end{equation*}%
at the point $\mathrm{E}(\infty ,k)$, to arrive at the equality%
\begin{equation}
\hat{\upsilon}(k)^{2}\mathfrak{T}\left( \infty ,k,\mathrm{E}\left( \infty
,k\right) \right) +\mathrm{E}\left( \infty ,k\right) -\mathfrak{b}\left(
k\right) =0\ .  \label{hjhjj}
\end{equation}%
This proves that $\mathrm{E}(\infty ,k)$ is a solution to (\ref{jkljklkjl}).
There is no other solution below $\mathfrak{z}(k)$ because of the following
arguments: Given any $\mathrm{U}\in \lbrack 0,\infty ]$, let%
\begin{equation*}
f_{\mathrm{U}}\left( x\right) \doteq \hat{\upsilon}(k)^{2}\mathfrak{T}(%
\mathrm{U},k,x)+x\ ,\quad x\in (-\infty ,\mathfrak{z}(k))\ .
\end{equation*}%
By Corollary \ref{limit of I(U,k,x) when U goes to infinity}, $(f_{\mathrm{U}%
})_{\mathrm{U}\in \mathbb{R}_{0}^{+}}$ converges pointwise to $f_{\infty }$,
as $\mathrm{U}\rightarrow \infty $. Since the pointwise limit of
monotonically increasing function is again monotonically increasing, it
follows that $f_{\infty }$ is monotonically increasing. Given any $x<y<%
\mathfrak{z}(k)$, take any $r>0$ with $r\geq f_{\infty }(y)-f_{\infty
}(x)\geq 0$. Then, for some $\mathrm{U}_{0}\in \mathbb{R}_{0}^{+}$
sufficiently large, one has 
\begin{equation*}
-r<f_{\mathrm{U}_{0}}(y)-f_{\infty }(y)<r\qquad \text{and}\qquad -r<f_{%
\mathrm{U}_{0}}(x)-f_{\infty }(x)<r,
\end{equation*}%
so that 
\begin{align*}
2r& >(f_{\mathrm{U}_{0}}(y)-f_{\infty }(y))-(f_{\mathrm{U}_{0}}(x)-f_{\infty
}(x)) \\
& =(f_{\mathrm{U}_{0}}(y)-f_{\mathrm{U}_{0}}(x))-(f_{\infty }(y)-f_{\infty
}(x)) \\
& \geq f_{\mathrm{U}_{0}}(y)-f_{\mathrm{U}_{0}}(x)-r.
\end{align*}%
Then, the mean value theorem combined with (\ref{sdfsdfsdfsdf}) implies that 
\begin{equation}
3r\geq f_{\mathrm{U}_{0}}(y)-f_{\mathrm{U}_{0}}(x)=f_{\mathrm{U}%
_{0}}^{\prime }(c)(y-x)\geq y-x>0\   \label{finfinity}
\end{equation}%
for some $c\in (x,y)$. This implies that the function $f_{\infty }$ is
strictly increasing on $(-\infty ,\mathfrak{z}(k))$ and, hence, there is a
unique solution, $\mathrm{E}\left( \infty ,k\right) $, to (\ref{jkljklkjl}).
Meanwhile, if $\hat{\upsilon}(k)=0$ then Theorem \ref{existence of
eigenvalue for each fiber} implies that $\mathrm{E}(\mathrm{U},k)=\mathfrak{b%
}(k)$ for all $\mathrm{U}\in \mathbb{R}_{0}^{+}$ and, obviously, $\mathrm{E}%
(\infty ,k)=\mathfrak{b}(k)$ is the unique solution to (\ref{jkljklkjl}).

Consider now the open set $\mathcal{O}$ defined by (\ref{sdsdsd1}). Let $%
\Phi (\infty ,\cdot ):\mathcal{O}\rightarrow \mathbb{R}$ be defined by 
\begin{equation}
\Phi \left( \infty ,k,x\right) \doteq \hat{\upsilon}\left( k\right) ^{2}%
\mathfrak{T}\left( \infty ,k,x\right) +x-\mathfrak{b}\left( k\right) \
,\qquad \left( k,x\right) \in \mathcal{O}\ .  \label{sdsdsd20}
\end{equation}%
By Corollary \ref{limit of I(U,k,x) when U goes to infinity}, note that $%
\Phi (\infty ,\cdot ):\mathcal{O}\rightarrow \mathbb{R}$ is nothing else
than the pointwise limit of the function $\Phi (\mathrm{U},\cdot ):\mathcal{O%
}\rightarrow \mathbb{R}$ defined by (\ref{Phi}):%
\begin{equation}
\lim_{\mathrm{U}\rightarrow \infty }\Phi \left( \mathrm{U},k,x\right) =\Phi
\left( \infty ,k,x\right) \ .  \label{sdsdsd2}
\end{equation}%
The function $\Phi (\infty ,\cdot ,\cdot )$ is a continuously differentiable
function satisfying 
\begin{equation*}
\partial _{x}\Phi \left( \infty ,k,x\right) \geq 1/3>0\ ,\qquad \left(
k,x\right) \in \mathcal{O}\ ,
\end{equation*}%
thanks to Inequality (\ref{finfinity}). Observe also that $\Phi (\infty
,\cdot )\in C^{d}(\mathcal{O})$\ if $\hat{\upsilon}$ is of class $C^{d}$ on $%
\mathbb{S}^{2}\subseteq \mathbb{R}^{2}$ with $d\in \mathbb{N}\cup \{\omega
,a\}$. Therefore, by repeating essentially the same argument used in the
proof of Theorem \ref{regularity of E}, one concludes that $\mathrm{E}%
(\infty ,\cdot )\in C^{d}(\mathbb{S}^{2})$ whenever $\hat{\upsilon}$ is of
class $C^{d}$ on $\mathbb{S}^{2}\subseteq \mathbb{R}^{2}$ with $d\in \mathbb{%
N}\cup \{\omega ,a\}$.

Finally, assume that $r_{\mathfrak{p}}>0$, that is, $\mathfrak{p}_{1}\notin 
\mathbb{C}\mathfrak{e}_{0}$ or $\mathfrak{p}_{2}\notin \mathbb{C}\mathfrak{e}%
_{0}$.For $k\neq 0$, we deduce from (\ref{hjhjj}) and Lemma \ref%
{representing I(k,x) by explicit integrals} that 
\begin{equation*}
\mathrm{E}(\infty ,k)=\mathfrak{b}(k)\Leftrightarrow \hat{\upsilon}(k)^{2}%
\mathfrak{T}\left( \infty ,k,\mathrm{E}\left( \infty ,k\right) \right)
=0\Leftrightarrow \hat{\upsilon}(k)=0\ .
\end{equation*}%
To conclude the proof of Assertion (v), it remains to show that $\hat{%
\upsilon}(0)\neq 0$ implies $\mathrm{E}(\infty ,0)<\mathfrak{b}(0)$. Assume
on the contrary that 
\begin{equation}
\lim_{\mathrm{U}\rightarrow \infty }\mathrm{E}(\mathrm{U},0)=\mathfrak{b}%
\left( 0\right) \ ,  \label{sdfsfsdfsfsdfsdfsdf}
\end{equation}%
keeping in mind that $\mathfrak{b}\left( 0\right) =\mathfrak{z}(0)=0$. Then,
we infer from Theorems \ref{eigenvalue as a root of a non linear equation}
and \ref{existence of eigenvalue for each fiber} that the following equality
must be true:%
\begin{equation*}
\lim_{\mathrm{U}\rightarrow \infty }\mathfrak{T}\left( \mathrm{U},0,\mathrm{E%
}(\mathrm{U},0)\right) =\lim_{\mathrm{U}\rightarrow \infty }\hat{\upsilon}%
(0)^{-2}\left( \mathfrak{b}\left( 0\right) -\mathrm{E}(\mathrm{U},0)\right)
=0\ .
\end{equation*}%
By Lemma \ref{representing I(k,x) by explicit integrals}, it follows that%
\begin{equation}
\lim_{\mathrm{U}\rightarrow \infty }\frac{R_{\mathfrak{d},\mathfrak{d}%
}\left( \mathrm{U}\right) }{\mathrm{U}R_{\mathfrak{s},\mathfrak{s}}\left( 
\mathrm{U}\right) +1}=0\ ,  \label{solve1}
\end{equation}%
where, by a slight abuse of notation, 
\begin{eqnarray*}
R_{\mathfrak{s},\mathfrak{s}}\left( \mathrm{U}\right) &\doteq &R_{\mathfrak{s%
},\mathfrak{s}}\left( 0,\mathrm{E}(\mathrm{U},0)\right) \ , \\
R_{\mathfrak{d},\mathfrak{d}}\left( \mathrm{U}\right) &\doteq &R_{\mathfrak{d%
},\mathfrak{d}}\left( 0,\mathrm{E}(\mathrm{U},0)\right) \ , \\
R_{\mathfrak{s},\mathfrak{d}}\left( \mathrm{U}\right) &\doteq &R_{\mathfrak{s%
},\mathfrak{d}}\left( 0,\mathrm{E}(\mathrm{U},0)\right) \ .
\end{eqnarray*}%
Therefore, we deduce from Corollary \ref{limit of I(U,k,x) when U goes to
infinity} that 
\begin{equation}
\lim_{\mathrm{U}\rightarrow \infty }\frac{R_{\mathfrak{d},\mathfrak{d}%
}\left( \mathrm{U}\right) R_{\mathfrak{s},\mathfrak{s}}\left( \mathrm{U}%
\right) -\left\vert R_{\mathfrak{s},\mathfrak{d}}\left( \mathrm{U}\right)
\right\vert ^{2}}{R_{\mathfrak{s},\mathfrak{s}}\left( \mathrm{U}\right) }%
=\lim_{\mathrm{U}\rightarrow \infty }\mathfrak{T}(\infty ,0,\mathrm{E}(%
\mathrm{U},0))=0\ .  \label{sssssss}
\end{equation}%
Now, observe that%
\begin{eqnarray}
&&\frac{R_{\mathfrak{d},\mathfrak{d}}\left( \mathrm{U}\right) R_{\mathfrak{s}%
,\mathfrak{s}}\left( \mathrm{U}\right) -\left\vert R_{\mathfrak{s},\mathfrak{%
d}}\left( \mathrm{U}\right) \right\vert ^{2}}{R_{\mathfrak{s},\mathfrak{s}%
}\left( \mathrm{U}\right) }  \notag \\
&=&\frac{1}{R_{\mathfrak{s},\mathfrak{s}}\left( \mathrm{U}\right) }%
\left\langle \left( R_{\mathfrak{s},\mathfrak{s}}\left( \mathrm{U}\right) 
\mathfrak{d}\left( 0\right) -R_{\mathfrak{s},\mathfrak{d}}\left( \mathrm{U}%
\right) \mathfrak{s}\right) ,(B_{1,1}(0)-\mathrm{E}(\mathrm{U},0)\mathfrak{1}%
)^{-1}\left( R_{\mathfrak{s},\mathfrak{s}}\left( \mathrm{U}\right) \mathfrak{%
d}\left( 0\right) -R_{\mathfrak{s},\mathfrak{d}}\left( \mathrm{U}\right) 
\mathfrak{s}\right) \right\rangle  \notag \\
&=&\left\langle \mathfrak{d}\left( 0\right) -\mathfrak{\alpha \left( \mathrm{%
U}\right) s},(B_{1,1}(0)-\mathrm{E}(\mathrm{U},0)\mathfrak{1})^{-1}\left( 
\mathfrak{d}\left( 0\right) -\mathfrak{\alpha \left( \mathrm{U}\right) s}%
\right) \right\rangle  \label{fhjgjj}
\end{eqnarray}%
where 
\begin{equation*}
\alpha \left( \mathrm{U}\right) \doteq R_{\mathfrak{s},\mathfrak{s}}\left( 
\mathrm{U}\right) ^{-1}R_{\mathfrak{s},\mathfrak{d}}\left( \mathrm{U}\right)
\ .
\end{equation*}%
Let $\varphi _{\mathrm{U}}=\mathfrak{d}(0)-\alpha (\mathrm{U})\mathfrak{s}%
\neq 0$ and $W_{\mathrm{U}}=B_{1,1}(0)-\mathrm{E}(\mathrm{U},0)\mathfrak{1}$%
. Since $W_{\mathrm{U}}$ is strictly positive, it has an unique square root,
which is also strictly positive. Then 
\begin{multline*}
||\varphi _{\mathrm{U}}||^{4}=|\langle W_{\mathrm{U}}^{-1/2}\varphi _{%
\mathrm{U}},W_{\mathrm{U}}^{1/2}\varphi _{\mathrm{U}}\rangle |^{2}\leq ||W_{%
\mathrm{U}}^{-1/2}\varphi _{\mathrm{U}}||^{2}||W_{\mathrm{U}}^{1/2}\varphi _{%
\mathrm{U}}||^{2}\leq \\[1em]
\leq ||W_{\mathrm{U}}||_{\mathrm{op}}||\varphi _{\mathrm{U}}||^{2}\langle
\varphi _{\mathrm{U}},W_{\mathrm{U}}^{-1}\varphi _{\mathrm{U}}\rangle \leq
||\varphi _{\mathrm{U}}||^{2}\langle \varphi _{\mathrm{U}},W_{\mathrm{U}%
}^{-1}\varphi _{\mathrm{U}}\rangle \big(||B_{1,1}(0)||_{\mathrm{op}}+|%
\mathrm{E}(0,0)|\big).
\end{multline*}%
As $\langle \varphi _{\mathrm{U}},W_{\mathrm{U}}^{-1}\varphi _{\mathrm{U}%
}\rangle $ tends to $0$ when $\mathrm{U}\rightarrow \infty $, it follows
that also $\varphi _{\mathrm{U}}$ vanishes in this limit, that is, 
\begin{equation}
\lim_{\mathrm{U}\rightarrow \infty }\mathfrak{\alpha \left( \mathrm{U}%
\right) s}=\mathfrak{d}\left( 0\right) \ .  \label{sdsdsdsdsdsd}
\end{equation}%
As $\mathbb{C}\mathfrak{s}$ is a closed subspace of $L^{2}(\mathbb{T}^{2})$,
we get that $\mathfrak{d}(0)\in \mathbb{C}\mathfrak{s}$, which is a
contradiction. Therefore, Assertion (v) holds true.
\end{proof}

\begin{remark}
\label{case k = 0}\mbox{ }\newline
Under the additional conditions that $r_{\mathfrak{p}}>0$ (i.e., $\mathfrak{p%
}_{1}\notin \mathbb{C}\mathfrak{e}_{0}$ or $\mathfrak{p}_{2}\notin \mathbb{C}%
\mathfrak{e}_{0}$) and $\hat{\upsilon}(0)\neq 0$, Assertion (iv) remains
valid with $(-\pi ,\pi )^{2}$ instead of $\mathbb{S}^{2}$, i.e., by
including the zero quasi-momentum case $k=0$. Also, Assertion (iii) holds
true for $k=0$. In fact, under these further conditions, we know from
Assertion (v) that $\mathrm{E}(\infty ,0)<\mathfrak{z}(0)=0$.
\end{remark}

\begin{corollary}[Hard-core dispersion relation]
\label{Hard-core dispersion relation}\mbox{ }\newline
Let $h_{b}\in \lbrack 0,1/2]$ and $\hat{\upsilon}$ be of class $C^{2}$ on $%
\mathbb{S}^{2}$. Then, $\mathrm{E}(\infty ,\cdot )\in C(\mathbb{T}^{2})$ and
is of class $C^{2}$ on $\mathbb{S}^{2}$. In this case, for any $k\in \mathbb{%
S}^{2}$,%
\begin{equation*}
\mathbf{v}_{\mathrm{E},\infty }\left( k\right) \doteq \vec{\nabla}_{k}%
\mathrm{E}\left( \infty ,k\right) =-\left( \partial _{x}\Phi \left( \infty ,{%
k,\mathrm{E}\left( \infty ,k\right) }\right) \right) ^{-1}\vec{\nabla}%
_{k}\Phi \left( \infty ,{k,\mathrm{E}\left( \infty ,k\right) }\right) ={%
\lim_{\mathrm{U}\rightarrow \infty }}\mathbf{v}_{\mathrm{E},\mathrm{U}%
}\left( k\right) \ .
\end{equation*}%
Moreover, if $\hat{\upsilon}$ is real analytic (of class $C^{a}$, in our
terminology) on $\mathbb{S}^{2}$, then either $\mathfrak{M}_{\mathrm{E}%
\left( \infty ,\cdot \right) }$ has full measure or is empty.
\end{corollary}

\begin{proof}
Recall that $\mathcal{O}$ is the open set (\ref{sdsdsd1}) and $\Phi (\cdot
,\cdot ):\mathbb{R}_{0}^{+}\times \mathcal{O}\rightarrow \mathbb{R}$ is the
real-valued function defined by (\ref{Phi}). Note that, for any $\mathrm{U}%
\in \mathbb{R}_{0}^{+}$, the function $\Phi (\mathrm{U},\cdot )$ is a smooth
function on $\mathcal{O}$ and we estimate its derivatives with respect to
the parameter $k$, at fixed $\mathrm{U}\in \mathbb{R}_{0}^{+}$ and $x$,
where $(k,x)\in \mathcal{O}$: For any $(k,x)\in \mathcal{O}$ and $\mathrm{U}%
\in \mathbb{R}_{0}^{+}$, Equations (\ref{A11})--(\ref{A11-U0}) together with
the second resolvent formula yield the derivative 
\begin{equation*}
\partial _{k_{j}}\left( A_{1,1}\left( \mathrm{U},k\right) -x\mathfrak{1}%
\right) ^{-1}=-\left( A_{1,1}\left( \mathrm{U},k\right) -x\mathfrak{1}%
\right) ^{-1}\left\{ \partial _{k_{j}}M_{\mathfrak{f}\left( k\right)
}\right\} \left( A_{1,1}\left( \mathrm{U},k\right) -x\mathfrak{1}\right)
^{-1}
\end{equation*}%
for any $j\in \{1,2\}$, where $k=(k_{1},k_{2})\in \mathbb{S}^{2}$.
Therefore, by (\ref{f}) and (\ref{Phi}), for any $(k,x)\in \mathcal{O}$ and $%
\mathrm{U}\in \mathbb{R}_{0}^{+}$, 
\begin{equation}
\left\vert \partial _{k_{j}}\Phi \left( \mathrm{U},k,x\right) \right\vert
\leq 8\epsilon \hat{\upsilon}\left( k\right) ^{2}\left\vert \mathfrak{z}%
\left( k\right) -x\right\vert ^{-2}  \label{combined1}
\end{equation}%
for any $j\in \{1,2\}$, where $k=(k_{1},k_{2})\in \mathbb{S}^{2}$. Taking
the second derivative, one can easily check that 
\begin{equation}
\left\vert \partial _{k_{j}}^{2}\Phi \left( \mathrm{U},k,x\right)
\right\vert \leq 8\epsilon \hat{\upsilon}\left( k\right) ^{2}\left\vert 
\mathfrak{z}\left( k\right) -x\right\vert ^{-3}  \label{combined2}
\end{equation}%
for $j\in \{1,2\}$, where $k=(k_{1},k_{2})\in \mathbb{S}^{2}$. In the same
way, we deduce from the identities%
\begin{eqnarray*}
\partial _{x}\Phi \left( \mathrm{U},k,x\right) &=&\hat{\upsilon}\left(
k\right) ^{2}\left\langle \mathfrak{d}\left( k\right) ,\left( A_{1,1}\left( 
\mathrm{U},k\right) -x\mathfrak{1}\right) ^{-2}\mathfrak{d}\left( k\right)
\right\rangle +1 \\
\partial _{x}^{2}\Phi \left( \mathrm{U},k,x\right) &=&2\hat{\upsilon}\left(
k\right) ^{2}\left\langle \mathfrak{d}\left( k\right) ,\left( A_{1,1}\left( 
\mathrm{U},k\right) -x\mathfrak{1}\right) ^{-3}\mathfrak{d}\left( k\right)
\right\rangle
\end{eqnarray*}%
the following inequalities: 
\begin{eqnarray}
\left\vert \partial _{x}\Phi \left( \mathrm{U},k,x\right) \right\vert &\leq &%
\hat{\upsilon}\left( k\right) ^{2}\left\vert \mathfrak{z}\left( k\right)
-x\right\vert ^{-2}+1\ ,  \label{combined3} \\
\left\vert \partial _{x}^{2}\Phi \left( \mathrm{U},k,x\right) \right\vert
&\leq &2\hat{\upsilon}\left( k\right) ^{2}\left\vert \mathfrak{z}\left(
k\right) -x\right\vert ^{-3}\ ,  \label{combined4}
\end{eqnarray}%
for any $(k,x)\in \mathcal{O}$ and $\mathrm{U}\in \mathbb{R}_{0}^{+}$.

Now, fix $(k_{0},x_{0})\in \mathcal{O}$. The function $\mathfrak{z}$ of
Corollary \ref{bottom of the spectrum of A11} is continuous with respect to $%
k_{1},k_{2}$, where $k=(k_{1},k_{2})\in \mathbb{S}^{2}$. Then, there is a
closed cube centered at $(k_{0},x_{0})$ with side length $\delta \in \mathbb{%
R}^{+}$ contained in $\mathcal{O}$. Suppose that $\partial _{x}\Phi (\mathrm{%
U},k_{0},x_{0})$ does not converge to $\partial _{x}\Phi (\infty
,k_{0},x_{0})$. Then we can find $r_{0}\in \mathbb{R}^{+}$ and a sequence $(%
\mathrm{U}_{n})_{n\in \mathbb{N}}$ of positive numbers such that $\mathrm{U}%
_{n}\rightarrow \infty $ as $n\rightarrow \infty $ and, for every $n\in 
\mathbb{N}$, 
\begin{equation*}
|\partial _{x}\Phi (\mathrm{U}_{n},k_{0},x_{0})-\partial _{x}\Phi (\infty
,k_{0},x_{0})|\geq r_{0}\ .
\end{equation*}%
Let 
\begin{equation*}
\mathfrak{F}=\{\partial _{x}\Phi (\mathrm{U}_{n},k_{0},\cdot
)\upharpoonright \lbrack x_{0}-\delta ,x_{0}+\delta ]\,:\,n\in \mathbb{N}\}\
.
\end{equation*}%
By combining Equation (\ref{combined4}) with the mean value theorem, we see
that $\mathfrak{F}$ is Lipschitz equicontinuous. In particular this family
of functions is equicontinuous. Moreover, it follows from (\ref{combined3})
that $\mathfrak{F}$ is bounded in the supremum norm. We can hence apply the
(Arzel\`{a}-) Ascoli theorem \cite[Theorem A5]{Rudin}, according to which $%
\partial _{x}\Phi (\mathrm{U}_{n},k_{0},\cdot )$, when restricted to the
compact interval $[x_{0}-\delta ,x_{0}+\delta ]$, converges uniformly along
some subsequence. Assume, for simplicity and without loss of generality,
that the (full) sequence of functions converges itself. By Equation (\ref%
{sdsdsd2}), recall that $\Phi (\mathrm{U},\cdot ,\cdot )$ converges
pointwise to the function $\Phi (\infty ,\cdot ,\cdot )$ on $\mathcal{O}$,
which is is the real-valued function defined by (\ref{sdsdsd20}). Then, by 
\cite[Theorem 7.17]{Rudin2}, 
\begin{equation*}
\partial _{x}\Phi \left( \infty ,k_{0},x\right) =\lim_{n\rightarrow \infty
}\partial _{x}\Phi \left( \mathrm{U}_{n},k_{0},x\right) \ ,\qquad x\in
\lbrack x_{0}-\delta ,x_{0}+\delta ]\ .
\end{equation*}%
For $x=x_{0}$, this lead us to a contradiction. Thus 
\begin{equation*}
\partial _{x}\Phi \left( \infty ,k,x\right) =\lim_{\mathrm{U}\rightarrow
\infty }\partial _{x}\Phi \left( \mathrm{U},k,x\right) \ 
\end{equation*}%
for every $(k,x)\in \mathcal{O}$. In the same way, we invoke Equations (\ref%
{combined1}) and (\ref{combined2}) together with the mean value theorem and
the (Arzel\`{a}-) Ascoli theorem \cite[Theorem A5]{Rudin} to deduce that 
\begin{equation*}
\partial _{k_{j}}\Phi \left( \infty ,k,x\right) =\lim_{\mathrm{U}\rightarrow
\infty }\partial _{k_{j}}\Phi \left( \mathrm{U},k,x\right) \ ,\qquad j=1,2\ .
\end{equation*}%
for every $(k,x)\in \mathcal{O}$. To prove the corollary, we eventually use
these observations together with Theorem \ref{properties of effective
dispersion relation with low energy} (i), Corollary \ref{existence of a
dispersion relation with low energy} and the equicontinuity of 
\begin{equation*}
\{\partial _{\mu }\Phi \left( \mathrm{U},k_{0},\cdot \right) \upharpoonright
(-\infty ,\mathfrak{b}(k_{0})]\,:\,\mathrm{U}\in \mathbb{R}^{+}\}\ ,
\end{equation*}%
with $\mu $ standing for the variables $k_{1}$, $k_{2}$ or $x$. Note that
the last assertion concerning $\mathfrak{M}_{\mathrm{E}\left( \infty ,\cdot
\right) }$ is a direct consequence of the fact that the zeros of any
non-constant real analytic function have null Lebesgue measure \cite%
{analytic}, see the proof of the same assertion for $\mathrm{U}<\infty $ in
Corollary \ref{existence of a dispersion relation with low energy}.
\end{proof}

Note that no additional condition is required for $\mathrm{E}$ to have a
well-defined hard-core limit. Compare Corollary \ref{existence of a
dispersion relation with low energy} with Theorem \ref{properties of
effective dispersion relation with low energy}. Moreover, by Corollary \ref%
{Hard-core dispersion relation}, $\mathrm{E}(\infty ,\cdot )$ can be viewed
as the (effective) dispersion relation of the dressed bound fermion pairs,
with lowest energy, in the hard-core limit.

We close this section by showing the convergence of the low-energy
eigenvector of $A(\mathrm{U},k)$ for large Hubbard couplings. This refers to
the (hard-core)\ limit $\mathrm{U}\rightarrow \infty $ of the vector 
\begin{equation*}
\Psi \left( \mathrm{U},k\right) \equiv \Psi \left( k\right) \doteq \left( 
\hat{\upsilon}\left( k\right) \left( A_{1,1}\left( \mathrm{U},k\right) -%
\mathrm{E}\left( \mathrm{U},k\right) \mathfrak{1}\right) ^{-1}\mathfrak{d}%
\left( k\right) ,-1\right) \in \mathcal{H}\ ,
\end{equation*}%
see Equation (\ref{eigenvector}).

\begin{proposition}[Hard-core limit of eigenvectors]
\label{hardcore limit of the eigenspace}\mbox{ }\newline
Let $h_{b}\in \lbrack 0,1/2]$. Fix $k\in \mathbb{T}^{2}\backslash \{0\}$.
The following limit exits:%
\begin{equation*}
\Psi \left( \infty ,k\right) \doteq {\lim_{\mathrm{U}\rightarrow \infty }}%
\,\Psi \left( \mathrm{U},k\right) \in \mathcal{H}\backslash \{0\}\ .
\end{equation*}%
This statement remains valid for $k=0$ provided that $\hat{\upsilon}(0)\neq
0 $ and $r_{\mathfrak{p}}>0$ (i.e., $\mathfrak{p}_{1}\notin \mathbb{C}%
\mathfrak{e}_{0}$ or $\mathfrak{p}_{2}\notin \mathbb{C}\mathfrak{e}_{0}$).
\end{proposition}

\begin{proof}
Fix $k\in \mathbb{T}^{2}$ with $k\neq 0$. By using the first resolvent
formula together with Theorem \ref{existence of eigenvalue for each fiber},
Theorem \ref{properties of effective dispersion relation with low energy}
(i) and (v), we find that%
\begin{equation*}
\Vert (A_{1,1}(\mathrm{U},k)-\mathrm{E}\left( \infty ,k\right) \mathfrak{1}%
)^{-1}-(A_{1,1}(\mathrm{U},k)-\mathrm{E}(\mathrm{U},k)\mathfrak{1}%
)^{-1}\Vert _{\mathrm{o}\mathrm{p}}\leq {\frac{|\mathrm{E}(\mathrm{U},k)-%
\mathrm{E}(\infty ,k)|}{|\mathfrak{z}(k)-\mathrm{E}(\infty ,k)|^{2}}}%
\rightarrow 0\ .
\end{equation*}%
Note that $\mathrm{E}(\infty ,k)\leq \mathfrak{b}\left( k\right) <\mathfrak{z%
}\left( k\right) $ for any $k\neq 0$. (When $\hat{\upsilon}(0)\neq 0$ and $%
\mathfrak{p}_{1}\notin \mathbb{C}\mathfrak{e}_{0}$ or $\mathfrak{p}%
_{2}\notin \mathbb{C}\mathfrak{e}_{0}$, we also have that $\mathrm{E}(\infty
,0)<\mathfrak{b}(0)=\mathfrak{z}(0)$.) On the other hand, by Proposition \ref%
{inversion is monotone decreasing}, $\{(A_{1,1}(\mathrm{U},k)-\mathrm{E}%
(\infty ,k)\mathfrak{1})^{-1}\}_{\mathrm{U}\geq 0}$ is a decreasing family
of positive operators and, by Proposition \ref{prop weak lim monotone}, it
converges strongly as $\mathrm{U}\rightarrow \infty $. Consequently, $%
(A_{1,1}(\mathrm{U},k)-\mathrm{E}(\mathrm{U},k)\mathfrak{1})^{-1}$ also
converges strongly.
\end{proof}

\subsection{Spectral Gap and Anderson Localization}

By Equation (\ref{minimum2}), the spectral gap of fiber Hamiltonians is
equal to%
\begin{equation}
\mathfrak{g}\left( \mathrm{U},k\right) \doteq \min \sigma _{\mathrm{ess}%
}\left( A\left( \mathrm{U},k\right) \right) -\mathrm{E}\left( \mathrm{U}%
,k\right) \geq 0,\qquad k\in \mathbb{T}^{2}\ ,  \label{gap function}
\end{equation}%
for any Hubbard coupling constant $\mathrm{U}\in \mathbb{R}_{0}^{+}$. When $%
r_{\mathfrak{p}}>0$ (i.e., $\mathfrak{p}_{1}\notin \mathbb{C}\mathfrak{e}%
_{0} $ or $\mathfrak{p}_{2}\notin \mathbb{C}\mathfrak{e}_{0}$) and $\hat{%
\upsilon}(0)\neq 0$, this quantity turns out to be strictly positive,
uniformly with respect to the parameter $\mathrm{U}$:

\begin{proposition}[Uniform spectral gap of fiber Hamiltonians]
\label{spectral gap}\mbox{ }\newline
Fix $h_{b}\in \lbrack 0,1/2]$. If $r_{\mathfrak{p}}>0$ (i.e., $\mathfrak{p}%
_{1}\notin \mathbb{C}\mathfrak{e}_{0}$ or $\mathfrak{p}_{2}\notin \mathbb{C}%
\mathfrak{e}_{0}$) and $\hat{\upsilon}(0)\neq 0$ then%
\begin{equation*}
\inf_{\mathrm{U}\in \mathbb{R}_{0}^{+}}\min_{k\in \mathbb{T}^{2}}\mathfrak{g}%
\left( \mathrm{U},k\right) >0\ .
\end{equation*}
\end{proposition}

\begin{proof}
The family $\{\mathrm{E}(\mathrm{U},\cdot )\}_{\mathrm{U}\in \mathbb{R}%
_{0}^{+}}$ of real-valued functions on $\mathbb{T}^{2}$ is equicontinuous,
thanks to Theorem \ref{regularity of E} (i). Since Proposition \ref%
{essential spectrum of a fiber} says that, for any $k\in \mathbb{T}^{2}$, 
\begin{equation}
\min \sigma _{\mathrm{ess}}\left( A\left( \mathrm{U},k\right) \right) =%
\mathfrak{z}\left( k\right) \doteq 4\epsilon -2\epsilon \cos \left(
k/2\right) \ ,  \label{sdsdssdsdsdsdsd}
\end{equation}%
we thus deduce from (\ref{gap function}) that the family $\{\mathfrak{g}(%
\mathrm{U},\cdot )\}_{\mathrm{U}\in \mathbb{R}_{0}^{+}}$ of real-valued
functions on $\mathbb{T}^{2}$ is equicontinuous. It follows that the
function 
\begin{equation}
\mathbb{R}_{0}^{+}\ni \mathrm{U}\longmapsto \min_{k\in \mathbb{T}^{2}}%
\mathfrak{g}\left( \mathrm{U},k\right) \in \mathbb{R}  \label{hj}
\end{equation}%
is continuous. Moreover, from the compactness of $\mathbb{T}^{2}$, $%
\mathfrak{g}(\mathrm{U},\cdot )$ has a global minimizer, say $k_{\mathrm{U}%
}\in \mathbb{T}^{2}$ for all $\mathrm{U}\in \mathbb{R}_{0}^{+}$. Since $\hat{%
\upsilon}(0)\neq 0$, by Theorem \ref{existence of eigenvalue for each fiber}%
, 
\begin{equation*}
\mathrm{E}\left( \mathrm{U},k_{\mathrm{U}}\right) <\min \sigma _{\mathrm{ess}%
}\left( A\left( \mathrm{U},k_{\mathrm{U}}\right) \right) =0
\end{equation*}%
when $k_{\mathrm{U}}=0$, while in the case $k_{\mathrm{U}}\neq 0$, 
\begin{equation*}
\mathrm{E}\left( \mathrm{U},k_{\mathrm{U}}\right) \leq \mathfrak{b}(k_{%
\mathrm{U}})<\mathfrak{z}\left( k\right) =\min \sigma _{\mathrm{ess}}\left(
A\left( \mathrm{U},k_{\mathrm{U}}\right) \right) \ .
\end{equation*}%
In particular, 
\begin{equation*}
\min_{k\in \mathbb{T}^{2}}\mathfrak{g}\left( \mathrm{U},k\right) = \mathfrak{%
g}\left( \mathrm{U},k_\mathrm{U}\right) >0\ ,\qquad \mathrm{U}\in \mathbb{R}%
_{0}^{+}\ .
\end{equation*}%
Using this together with the continuity of the function (\ref{hj}) we arrive
at the inequality%
\begin{equation}
\inf_{\mathrm{U}\in \left[ 0,c\right] }\min_{k\in \mathbb{T}^{2}}\mathfrak{g}%
\left( \mathrm{U},k\right) >0  \label{c1}
\end{equation}%
for any positive parameter $c\in \mathbb{R}_{0}^{+}$. Now, we perform the
limit $\mathrm{U}\rightarrow \infty $. Since $\mathbb{T}^{2}$ is compact,
the net$\ (k_{\mathrm{U}})_{\mathrm{U}\in \mathbb{R}_{0}^{+}}$ converges
along subnets (in fact subsequences). Assume without loss of generality that 
$(k_{\mathrm{U}})_{\mathrm{U}\in \mathbb{R}_{0}^{+}}$ converges to some $%
k_{\infty }\in \mathbb{T}^{2}$ (otherwise one uses all the following
arguments on subsequences). If $k_{\infty }\neq 0$ then 
\begin{equation}
\lim_{\mathrm{U}\rightarrow \infty }\min_{k\in \mathbb{T}^{2}}\mathfrak{g}%
\left( \mathrm{U},k\right) \geq \lim_{\mathrm{U}\rightarrow \infty }%
\mathfrak{z}\left( k_{\mathrm{U}}\right) -\mathfrak{b}(k_{\mathrm{U}})=%
\mathfrak{z}\left( k_{\infty }\right) -\mathfrak{b}(k_{\infty })> 0\ ,
\label{c2}
\end{equation}%
thanks to Theorem \ref{existence of eigenvalue for each fiber} and the
continuity of the functions $\mathfrak{z}$ and $\mathfrak{b}$. Assume now
that $k_{\infty }=0$. Since, for all $\mathrm{U}\in \mathbb{R}_{0}^{+}$, 
\begin{equation*}
\left\vert \mathrm{E}\left( \mathrm{U},k_{\mathrm{U}}\right) -\mathrm{E}%
\left( \infty ,0\right) \right\vert \leq \left\vert \mathrm{E}\left( \mathrm{%
U},k_{\mathrm{U}}\right) -\mathrm{E}\left( \mathrm{U},0\right) \right\vert
+\left\vert \mathrm{E}\left( \mathrm{U},0\right) -\mathrm{E}\left( \infty
,0\right) \right\vert ,
\end{equation*}%
we infer from the equicontinuity of the family $\{\mathrm{E}(\mathrm{U}%
,\cdot )\}_{\mathrm{U}\in \mathbb{R}_{0}^{+}}$ (Theorem \ref{regularity of E}
(i)) and Theorem \ref{properties of effective dispersion relation with low
energy} (i) that 
\begin{equation*}
\lim_{\mathrm{U}\rightarrow \infty }\mathrm{E}\left( \mathrm{U},k_{\mathrm{U}%
}\right) =\mathrm{E}\left( \infty ,0\right) \ .
\end{equation*}%
Combined with Theorem \ref{properties of effective dispersion relation with
low energy} (v) and $\hat{\upsilon}(0)\neq 0$, this last limit in turn
implies that 
\begin{equation}
\lim_{\mathrm{U}\rightarrow \infty }\min_{k\in \mathbb{T}^{2}}\mathfrak{g}%
\left( \mathrm{U},k\right) = \lim_{\mathrm{U}\rightarrow \infty }\left\{%
\mathfrak{z}(k_\mathrm{U}) - \mathrm{E}\left( \mathrm{U},k_{\mathrm{U}%
}\right)\right\} =\mathfrak{z}\left( 0\right) -\mathrm{E}\left( \infty
,0\right) =-\mathrm{E}\left( \infty ,0\right) > 0\ .  \label{c3}
\end{equation}%
The assertion is therefore a combination of Inequalities (\ref{c1}), (\ref%
{c2}) and (\ref{c3}).
\end{proof}

We study now the space localization of the (dressed) bound pair with total
quasi-momentum $k\in \mathbb{T}^{2}$ and energy $\mathrm{E}\left( \mathrm{U}%
,k\right) $. Assume that $\hat{\upsilon}(0)\neq 0$. By Corollary \ref%
{eigenspace of a fiber}, for any fixed $k\in \mathbb{T}^{2}$, it corresponds
to study the fermionic part of the eigenvector 
\begin{equation}
\Psi \left( \mathrm{U},k\right) \doteq g\left( k,\mathrm{E}\left( \mathrm{U}%
,k\right) \right) =\left( \hat{\upsilon}\left( k\right) \left( A_{1,1}\left( 
\mathrm{U},k\right) -\mathrm{E}\left( \mathrm{U},k\right) \mathfrak{1}%
\right) ^{-1}\mathfrak{d}\left( k\right) ,-1\right) \in \mathcal{H}\ ,
\label{eigenvector}
\end{equation}%
written in the real space $\mathbb{Z}^{2}$ via the inverse Fourier transform 
$\mathcal{F}^{-1}$ (see (\ref{fourier transforms})). This function is
denoted by 
\begin{equation}
\psi _{\mathrm{U},k}\doteq \mathcal{F}^{-1}\left[ \hat{\upsilon}\left(
k\right) \left( A_{1,1}\left( \mathrm{U},k\right) -\mathrm{E}\left( \mathrm{U%
},k\right) \mathfrak{1}\right) ^{-1}\mathfrak{d}\left( k\right) \right] \in
\ell ^{2}\left( \mathbb{Z}^{2}\right)  \label{eigenvector2}
\end{equation}%
for any fixed $k\in \mathbb{T}^{2}$. One should not be confused here by the
parameter $k$. Recall for instance that, given $k\in \mathbb{T}^{2}$, $%
\mathfrak{d}(k)\in C\left( \mathbb{T}^{2}\right) $ is itself a function on
the torus $\mathbb{T}^{2}$, defined by 
\begin{equation}
\mathfrak{d}\left( k\right) \left( p\right) \doteq \mathfrak{\hat{p}}%
_{1}\left( k+p\right) +\mathfrak{\hat{p}}_{2}\left( k/2+p\right) \ ,\qquad
p\in \mathbb{T}^{2}\ ,  \label{dbis}
\end{equation}%
see Equation (\ref{d}). In particular, observe that 
\begin{equation}
\psi _{\mathrm{U},k}=\hat{\upsilon}\left( k\right) \mathcal{F}^{-1}\left[
\left( A_{1,1}\left( \mathrm{U},k\right) -\mathrm{E}\left( \mathrm{U}%
,k\right) \mathfrak{1}\right) ^{-1}\mathfrak{d}\left( k\right) \right] \ .
\label{dfdfdfdfd}
\end{equation}%
We now show that this function is exponentially localized in the real space:

\begin{theorem}[Exponentially localized dressed bound fermion pairs]
\label{Exponentially localized bound pairs}\mbox{ }\newline
Fix $h_{b}\in \lbrack 0,1/2]$, $k\in \mathbb{T}^{2}$ and suppose that $r_{%
\mathfrak{p}}>0$ (i.e., $\mathfrak{p}\notin \mathbb{C}\mathfrak{e}_{0}$) and 
$\hat{\upsilon}(0)\neq 0$. There exist positive constants $C,\alpha >0$ such
that, for all $k\in \mathbb{T}^{2}$ and $\mathrm{U}\in \mathbb{R}_{0}^{+}$,%
\begin{equation*}
\left\vert \psi _{\mathrm{U},k}(x)\right\vert \leq C\mathrm{e}^{-\alpha
|x|}\ ,\qquad x\in \mathbb{Z}^{2}\ .
\end{equation*}
\end{theorem}

\begin{proof}
By (\ref{fourier D fract}), we compute that%
\begin{align}
& \mathcal{F}^{-1}\left[ (A_{1,1}(\mathrm{U},k)-\mathrm{E}(\mathrm{U},k)%
\mathfrak{1})^{-1}\mathfrak{d}\left( k\right) \right]  \notag \\[0.01in]
& =(\mathcal{F}^{-1}A_{1,1}(\mathrm{U},k)\mathcal{F}-\mathrm{E}(\mathrm{U},k)%
\mathfrak{1})^{-1}\mathcal{F}^{-1}\left[ \mathfrak{d}\left( k\right) \right]
\notag \\
& ={\sum_{y\in \mathbb{Z}^{2}}}\,\left( \mathrm{e}^{ik\cdot y}\mathfrak{p}%
_{1}(y)+\mathrm{e}^{i\frac{k}{2}\cdot y}\mathfrak{p}_{2}\left( y\right)
\right) (\mathcal{F}^{-1}A_{1,1}(\mathrm{U},k)\mathcal{F}-\mathrm{E}(\mathrm{%
U},k)\mathfrak{1})^{-1}\mathfrak{e}_{y}\ .  \label{sdsdsdsdd}
\end{align}%
By Equation (\ref{dfdfdfdfd}), it suffices to estimate the exponential decay
of this particular function. This is done by using the celebrated
Combes-Thomas estimates, which correspond here to Theorem \ref{combes-thomas
theorem}. To this end, several quantities, one of them being related to the
spectral gap $\mathfrak{g}(\mathrm{U},k)$ (\ref{gap function}), have to be
controlled and, as in Section \ref{Combes-Thomas estimates}, we use the
notation (\ref{def combes10bis}), i.e., 
\begin{equation}
\Delta (\lambda ;T)\doteq \min \left\{ \left\vert \lambda -a\right\vert
:a\in \sigma (T)\right\}  \label{def combes10}
\end{equation}%
for the distance between a complex number $\lambda \in \mathbb{C}$ and the
spectrum $\sigma (T)$ of an operator $T\in \mathcal{B}(\ell ^{2}(\mathbb{Z}%
^{2}))$, as well as (\ref{def combes1}), which, in the present case, refers
to the quantity 
\begin{equation}
\mathbf{S}(T,\mu )\doteq {\sup_{x\in \mathbb{Z}^{2}}\,\sum_{y\in \mathbb{Z}%
^{2}}}\,\left( \mathrm{e}^{\mu \left\vert x-y\right\vert }-1\right)
\left\vert \left\langle \mathfrak{e}_{x},T\mathfrak{e}_{y}\right\rangle
\right\vert \in \left[ 0,\infty \right]  \label{def combes100}
\end{equation}%
for any $T\in \mathcal{B}(\ell ^{2}(\mathbb{Z}^{2}))$ and $\mu \in \mathbb{R}%
_{0}^{+}$. We do it in three steps: The first one controls the spectral gap $%
\mathfrak{g}(\mathrm{U},k)$ (\ref{gap function}) and a quantity like (\ref%
{def combes10}) for $\lambda =\mathrm{E}(\mathrm{U},k)$, while the second
step is an analysis of quantities like (\ref{def combes100}). These two
steps allow us to apply, in the last step, Theorem \ref{combes-thomas
theorem} in order to get the desired result. \medskip

\noindent \underline{Step 1:} Observe from Equation (\ref{gap function}) and
Proposition \ref{spectral gap} that we can find $\alpha >0$ such that, for
all $k\in \mathbb{T}^{2}$ and $\mathrm{U}\in \mathbb{R}_{0}^{+}$,%
\begin{equation*}
0<4\epsilon (\mathrm{e}^{\alpha }-1)<\inf_{\mathrm{U}\in \mathbb{R}%
_{0}^{+}}\min_{k\in \mathbb{T}^{2}}\mathfrak{g}\left( \mathrm{U},k\right)
\leq \mathfrak{g}\left( \mathrm{U},k\right) \doteq \min \sigma _{\mathrm{ess}%
}\left( A\left( \mathrm{U},k\right) \right) -\mathrm{E}\left( \mathrm{U}%
,k\right) \ .
\end{equation*}%
Using now Proposition \ref{essential spectrum of a fiber} and the fact that 
\begin{equation*}
\min \sigma _{\mathrm{ess}}(A_{1,1}(\mathrm{U},k))=\min \sigma (A_{1,1}(%
\mathrm{U},k))\ ,
\end{equation*}%
for all $k\in \mathbb{T}^{2}$ and $\mathrm{U}\in \mathbb{R}_{0}^{+}$, we
deduce from the last inequalities that%
\begin{equation*}
0<4\epsilon (\mathrm{e}^{\alpha }-1)<\inf_{\mathrm{U}\in \mathbb{R}%
_{0}^{+}}\min_{k\in \mathbb{T}^{2}}\mathfrak{g}\left( \mathrm{U},k\right)
\leq \Delta \left( \mathrm{E}(\mathrm{U},k);A_{1,1}(\mathrm{U},k)\right)
=\min \sigma (A_{1,1}(\mathrm{U},k))-\mathrm{E}(\mathrm{U},k)\ ,
\end{equation*}%
see also Equation (\ref{def combes10}). Since $\mathcal{F}$ is a unitary
transformation, $A_{1,1}(\mathrm{U},k)$ and $\mathcal{F}^{\ast }A_{1,1}(%
\mathrm{U},k)\mathcal{F}$ have the same spectrum and it follows that%
\begin{equation}
0<4\epsilon (\mathrm{e}^{\alpha }-1)<\inf_{\mathrm{U}\in \mathbb{R}%
_{0}^{+}}\min_{k\in \mathbb{T}^{2}}\mathfrak{g}\left( \mathrm{U},k\right)
\leq \Delta \left( \mathrm{E}(\mathrm{U},k);\mathcal{F}^{\ast }A_{1,1}(%
\mathrm{U},k)\mathcal{F}\right)  \label{sssdsdsd}
\end{equation}%
for all $k\in \mathbb{T}^{2}$ and $\mathrm{U}\in \mathbb{R}_{0}^{+}$.
\medskip

\noindent \underline{Step 2:} By Equations (\ref{e frac}) and (\ref{fourier
transforms}), one easily checks that 
\begin{equation*}
\mathfrak{\hat{e}}_{y}\left( p\right) \doteq {\sum_{x\in \mathbb{Z}^{2}}}\,%
\mathrm{e}^{ip\cdot x}\mathfrak{e}_{y}\left( x\right) =\mathrm{e}^{ip\cdot
y}\ ,\qquad p\in \mathbb{T}^{2},\ y\in \mathbb{Z}^{2}\ ,
\end{equation*}%
while, for any fixed $k\in \mathbb{T}^{2}$, the real-valued functions $%
\mathfrak{f}(k)$, defined by (\ref{f}) and (\ref{cosinus}) on the torus $%
\mathbb{T}^{2}$, can be rewritten as 
\begin{equation*}
\mathfrak{f}\left( k\right) \left( p\right) \doteq \epsilon \left\{ 4-\cos
\left( p+k\right) -\cos \left( p\right) \right\} =4\epsilon -{\frac{\epsilon 
}{2}\sum_{z\in \mathbb{Z}^{2}\,,\,|z|=1}}\left( \mathrm{e}^{i\left(
p+k\right) \cdot z}+\mathrm{e}^{ip\cdot z}\right) \ ,\qquad p\in \mathbb{T}%
^{2}\ .
\end{equation*}%
Therefore, since $M_{\mathfrak{f}\left( k\right) }$ stands for the
multiplication operator by $\mathfrak{f}(k)\in C(\mathbb{T}^{2})$, for every 
$p,k\in \mathbb{T}^{2}$ and $y\in \mathbb{Z}^{2}$, 
\begin{equation*}
M_{\mathfrak{f}\left( k\right) }\mathfrak{\hat{e}}_{y}(p)=4\mathrm{e}%
^{ip\cdot y}\epsilon -{\frac{\epsilon }{2}\sum_{z\in \mathbb{Z}%
^{2}\,,\,|z|=1}}\left( \mathrm{e}^{ik\cdot z}+1\right) \mathrm{e}^{ip\cdot
(y+z)}=4\epsilon \mathfrak{\hat{e}}_{y}(p)-{\frac{\epsilon }{2}\sum_{z\in 
\mathbb{Z}^{2}\,,\,|z|=1}}\left( \mathrm{e}^{ik\cdot z}+1\right) \mathfrak{%
\hat{e}}_{y+z}(p)\ ,
\end{equation*}%
which, by (\ref{A11})--(\ref{A11-U0}), in turn implies that 
\begin{align*}
\mathcal{F}^{\ast }A_{1,1}(\mathrm{U},k)\mathcal{F}\mathfrak{e}_{y}& =%
\mathcal{F}^{\ast }\left( M_{\mathfrak{f}\left( k\right) }\mathfrak{\hat{e}}%
_{y}+\mathrm{U}P_{0}\mathfrak{\hat{e}}_{y}+{\sum_{z\in \mathbb{Z}^{2}}}\,%
\mathrm{u}(z)P_{z}\mathfrak{\hat{e}}_{y}\right) \\[1em]
& =\mathcal{F}^{\ast }\left( M_{\mathfrak{f}\left( k\right) }\mathfrak{\hat{e%
}}_{y}+\left( \mathrm{U}\delta _{y,0}+\mathrm{u}(y)\right) \mathfrak{\hat{e}}%
_{y}\right) \\[1em]
& =4\epsilon \mathfrak{e}_{y}-{\frac{\epsilon }{2}\sum_{z\in \mathbb{Z}%
^{2}\,,\,|z|=1}}\left( \mathrm{e}^{ik\cdot z}+1\right) \mathfrak{e}%
_{y+z}+\left( \mathrm{U}\delta _{y,0}+\mathrm{u}(y)\right) \mathfrak{e}_{y}\
,
\end{align*}%
keeping in mind that $P_{x}$ is the orthogonal projection onto the
one-dimensional subspace $\mathbb{C}\hat{\mathfrak{e}}_{x}\subseteq L^{2}(%
\mathbb{T}^{2})$. Recall that $\mathcal{F}^{\ast }=\mathcal{F}^{-1}$, the
Fourier transform being unitary. Thus, since $\alpha >0$, for each $x\in 
\mathbb{Z}^{2}$, we obtain that%
\begin{equation*}
{\sum_{y\in \mathbb{Z}^{2}}}\,\left\vert \mathrm{e}^{\alpha \left\vert
x-y\right\vert }-1\right\vert \left\vert \left\langle \mathfrak{e}_{x},%
\mathcal{F}^{\ast }A_{1,1}\left( \mathrm{U},k\right) \mathcal{F}\mathfrak{e}%
_{y}\right\rangle \right\vert =\frac{\epsilon }{2}\left( \mathrm{e}^{\alpha
}-1\right) {\sum_{y\in \mathbb{Z}^{2},\,|x-y|=1}}\left\vert \mathrm{e}%
^{ik\cdot (x-y)}+1\right\vert \leq 4\epsilon \left( \mathrm{e}^{\alpha
}-1\right) \ .
\end{equation*}%
Hence, taking the supremum over all $x\in \mathbb{Z}^{2}$ in this equation
and using the notation given by (\ref{def combes100}) as well as (\ref%
{sssdsdsd}), we arrive at%
\begin{equation}
\mathbf{S}\left( \mathcal{F}^{\ast }A_{1,1}\left( \mathrm{U},k\right) 
\mathcal{F},\alpha \right) \leq 4\epsilon (\mathrm{e}^{\alpha }-1)<\inf_{%
\mathrm{U}\in \mathbb{R}_{0}^{+}}\min_{k\in \mathbb{T}^{2}}\mathfrak{g}%
\left( \mathrm{U},k\right) \leq \Delta \left( \mathrm{E}\left( \mathrm{U}%
,k\right) ;\mathcal{F}^{\ast }A_{1,1}\left( \mathrm{U},k\right) \mathcal{F}%
\right)  \label{combes estimate hyp}
\end{equation}%
for any fixed $k\in \mathbb{T}^{2}$ and $\mathrm{U}\in \mathbb{R}_{0}^{+}$%
.\medskip

\noindent \underline{Step 3:} Thanks to (\ref{combes estimate hyp}), we are
now in a position to apply Theorem \ref{combes-thomas theorem} for $H=%
\mathcal{F}^{\ast }A_{1,1}\left( \mathrm{U},k\right) \mathcal{F}$ and $\mu
=\alpha $ to obtain that, for any $x,y\in \mathbb{Z}^{2}$,%
\begin{eqnarray*}
&&\left\vert \left\langle \mathfrak{e}_{x},\left( \mathcal{F}^{\ast
}A_{1,1}\left( \mathrm{U},k\right) \mathcal{F-}\mathrm{E}(\mathrm{U},k)%
\mathfrak{1}\right) ^{-1}\mathfrak{e}_{y}\right\rangle \right\vert \\
&\leq &\frac{\mathrm{e}^{-\alpha \left\vert x-y\right\vert }}{\Delta \left( 
\mathrm{E}\left( \mathrm{U},k\right) ;\mathcal{F}^{\ast }A_{1,1}\left( 
\mathrm{U},k\right) \mathcal{F}\right) -\mathbf{S}\left( \mathcal{F}^{\ast
}A_{1,1}\left( \mathrm{U},k\right) \mathcal{F},\alpha \right) } \\
&\leq &\frac{\mathrm{e}^{-\alpha \left\vert x-y\right\vert }}{\inf_{\mathrm{U%
}\in \mathbb{R}_{0}^{+}}\min_{k\in \mathbb{T}^{2}}\mathfrak{g}\left( \mathrm{%
U},k\right) -4\epsilon (\mathrm{e}^{\alpha }-1)}\ .
\end{eqnarray*}%
Combined with Equations (\ref{dfdfdfdfd})--(\ref{sdsdsdsdd}) and the
triangle inequality as well as the reverse one $|x-y|\geq |x|-|y|$, we then
arrive at%
\begin{eqnarray}
\left\vert \psi _{\mathrm{U},k}\left( x\right) \right\vert &=&\left\vert
\left\langle \mathfrak{e}_{x},\psi _{\mathrm{U},k}\right\rangle \right\vert 
\notag \\
&\leq &\left\vert \hat{\upsilon}\left( k\right) \right\vert {\sum_{y\in 
\mathbb{Z}^{2}}}\left( \left\vert \mathfrak{p}_{1}\left( y\right)
\right\vert +\left\vert \mathfrak{p}_{2}\left( y\right) \right\vert \right)
\left\vert \left\langle \mathfrak{e}_{x},\left( \mathcal{F}%
^{-1}A_{1,1}\left( \mathrm{U},k\right) \mathcal{F}-\mathrm{E}(\mathrm{U},k)%
\mathfrak{1}\right) ^{-1}\mathfrak{e}_{y}\right\rangle \right\vert  \notag \\
&\leq &\frac{\left\vert \hat{\upsilon}\left( k\right) \right\vert }{\inf_{%
\mathrm{U}\in \mathbb{R}_{0}^{+}}\min_{k\in \mathbb{T}^{2}}\mathfrak{g}%
\left( \mathrm{U},k\right) -4\epsilon (\mathrm{e}^{\alpha }-1)}{\sum_{y\in 
\mathbb{Z}^{2}}}\left( \left\vert \mathfrak{p}_{1}\left( y\right)
\right\vert +\left\vert \mathfrak{p}_{2}\left( y\right) \right\vert \right) 
\mathrm{e}^{-\alpha \left\vert x-y\right\vert }  \notag \\
&\leq &\frac{\left\vert \hat{\upsilon}\left( k\right) \right\vert \mathrm{e}%
^{-\alpha \left\vert x\right\vert }}{\inf_{\mathrm{U}\in \mathbb{R}%
_{0}^{+}}\min_{k\in \mathbb{T}^{2}}\mathfrak{g}\left( \mathrm{U},k\right)
-4\epsilon (\mathrm{e}^{\alpha }-1)}{\sum_{y\in \mathbb{Z}^{2}}}\left(
\left\vert \mathfrak{p}_{1}\left( y\right) \right\vert +\left\vert \mathfrak{%
p}_{2}\left( y\right) \right\vert \right) \mathrm{e}^{\alpha \left\vert
y\right\vert }  \label{fgjghjghjkghj}
\end{eqnarray}%
for all $x\in \mathbb{Z}^{2}$, $k\in \mathbb{T}^{2}$ and $\mathrm{U}\in 
\mathbb{R}_{0}^{+}$. By choosing $\alpha $ sufficiently small (more
precisely $\alpha \leq \alpha _{0}$), we can assume, without loss of
generality, that the above sum is finite, see (\ref{ssdsds}). This completes
the proof, because the Fourier transform $\hat{\upsilon}$ of $\upsilon $ is
a continuous function on the torus $\mathbb{T}^{2}$, which is compact, and
is consequently bounded.
\end{proof}

\subsection{Scattering Channels\label{scattering channels}}

\subsubsection{Unbound pair scattering channel\label{non-bound electronic
pair channel}}

Recall that $\mathfrak{H}_{f}$ is defined by (\ref{H0-HIlbert space}) and $%
\mathrm{H}_{f}$ is the operator defined by (\ref{H0H0}) for any $\mathrm{V}%
\in \mathbb{R}_{0}^{+}$ and absolutely summable function $\mathrm{v}:\mathbb{%
Z}^{2}\rightarrow \mathbb{R}_{0}^{+}$.

For any operator $Y$ acting on a Hilbert space $\mathcal{Y}$, $P_{\mathrm{ac}%
}(Y)$ denotes the orthogonal projection on the absolutely continuous space
of $Y$, defined by (\ref{abs cont space}). In order to show the existence of
a unbound pair scattering channel, we need the following technical lemma:

\begin{lemma}[Absolute continuous space of fermionic Hamiltonians]
\label{lemma for existence of non-bound channel}\mbox{ }\newline
For any $\mathrm{V}\in \mathbb{R}_{0}^{+}$ and every absolutely summable
function $\mathrm{v}:\mathbb{Z}^{2}\rightarrow \mathbb{R}_{0}^{+}$, the
orthogonal projection $P_{\mathrm{ac}}(\mathrm{H}_{f})$ on the absolutely
continuous space of $\mathrm{H}_{f}$, defined by (\ref{abs cont space}), is
equal to $\mathfrak{1}$.
\end{lemma}

\begin{proof}
Take any $\psi \in \mathfrak{H}_{f}$ and observe from Corollary \ref{example
of absolutely continuous space} that%
\begin{equation*}
\psi \left( k\right) \in \mathrm{ran}\left( P_{\mathrm{ac}}\left( M_{%
\mathfrak{f}\left( k\right) }\right) \right) ,\qquad k\in \mathbb{T}^{2}\ ,
\end{equation*}%
where $M_{\mathfrak{f}\left( k\right) }$ is the fiber Hamiltonian defined as
the multiplication operator associated with the continuous function $%
\mathfrak{f}(k)\in C(\mathbb{T}^{2})$ (see (\ref{f})). For any $\mathrm{V}%
\in \mathbb{R}_{0}^{+}$ and absolutely summable function $\mathrm{v}:\mathbb{%
Z}^{2}\rightarrow \mathbb{R}_{0}^{+}$, the operator defined by (\ref%
{repulsion}), that is,%
\begin{equation*}
R\left( \mathrm{V},\mathrm{v}\right) \doteq {\sum\limits_{x\in \mathbb{Z}%
^{2}}}\,\mathrm{v}\left( x\right) P_{x}+\mathrm{V}P_{0}\in \mathcal{B}\left(
L^{2}(\mathbb{T}^{2})\right) \ ,
\end{equation*}%
is a trace-class operator, where we recall that $P_{x}$ is the orthogonal
projection on the one-dimensional subspace $\mathbb{C}\mathfrak{\hat{e}}%
_{x}\subseteq L^{2}(\mathbb{T}^{2})$. By \cite[Theorem 4.4, Chapter X]{Kato}%
, it follows in this case that 
\begin{equation*}
\psi \left( k\right) \in \mathrm{ran}\left( P_{\mathrm{ac}}\left( M_{%
\mathfrak{f}\left( k\right) }+R\left( \mathrm{V},\mathrm{v}\right) \right)
\right) \ ,\qquad k\in \mathbb{T}^{2}\ .
\end{equation*}%
Let $B\subseteq \mathbb{R}$ be an arbitrary Borel set with zero Lebesgue
measure. By using (\ref{H0H0}), we deduce that 
\begin{align*}
\left\langle \psi ,\chi _{B}\left( M_{\mathfrak{f}}\right) \psi
\right\rangle _{\mathfrak{H}_{f}}& =\left\langle \psi ,\left( {\int_{\mathbb{%
T}^{2}}^{\oplus }}\chi _{B}\left( M_{\mathfrak{f}\left( k\right) }+R\left( 
\mathrm{V},\mathrm{v}\right) \right) \,\nu \left( \mathrm{d}k\right) \right)
\psi \right\rangle _{\mathfrak{H}_{f}} \\[1em]
& ={\int_{\mathbb{T}^{2}}}\left\langle \psi \left( k\right) ,\chi _{B}\left(
M_{\mathfrak{f}\left( k\right) }+R\left( \mathrm{V},\mathrm{v}\right)
\right) \psi \left( k\right) \right\rangle _{L^{2}\left( \mathbb{T}%
^{2}\right) }\,\nu \left( \mathrm{d}k\right) =0\text{ }.
\end{align*}%
For the first equality note that we apply Theorem \ref{borelian functional
calculus of a direct integral of operators} (iii).
\end{proof}

The following results imply that the dynamic generated by the Hamiltonian $H$
(i.e., included the exchange interaction and extended Hubbard repulsions)
asymptotically far in the past or future approaches the purely fermionic
dynamics for two unbound fermions. This is of course physically expected,
since all interaction strengths get weak as the distance between the
fermions increases. This is a consequence of the next assertions.

To shorten the notation, for any $\mathrm{V}\in \mathbb{R}_{0}^{+}$ and
every absolutely summable function $\mathrm{v}:\mathbb{Z}^{2}\rightarrow 
\mathbb{R}_{0}^{+}$, we define the Hamiltonian 
\begin{equation*}
H^{(1)}\equiv H^{(1)}\left( \mathrm{V},\mathrm{v}\right) \doteq {\int_{%
\mathbb{T}^{2}}^{\oplus }}H^{(1)}\left( k\right) \oplus A_{2,2}\left(
k\right) \nu \left( \mathrm{d}k\right) \in \mathcal{B}\left( L^{2}(\mathbb{T}%
^{2},\mathcal{H})\right)
\end{equation*}%
with 
\begin{equation}
H^{(1)}\left( k\right) \doteq M_{\mathfrak{f}\left( k\right) }+R\left( 
\mathrm{V},\mathrm{v}\right) \in \mathcal{B}\left( \mathcal{H}\right) \
,\qquad k\in \mathbb{T}^{2}\ .  \label{H1(k)}
\end{equation}%
Here, $M_{\mathfrak{f}\left( k\right) }$, $R\left( \mathrm{V},\mathrm{v}%
\right) $ and $A_{2,2}(k)$ are respectively the multiplication operator
associated with the continuous function $\mathfrak{f}(k)\in C(\mathbb{T}%
^{2}) $ (see (\ref{f})), the trace-class operator (\ref{repulsion}) and the
operator defined on $\mathbb{C}$ by (\ref{A22}). We start with the unbounded
pair scattering channel in each fiber:

\begin{lemma}[Fiberwise unbound pair (scattering) channel]
\label{lemma for existence of non-bound channel copy(1)}\mbox{ }\newline
For any $\mathrm{V}\in \mathbb{R}_{0}^{+}$, every absolutely summable
function $\mathrm{v}:\mathbb{Z}^{2}\rightarrow \mathbb{R}_{0}^{+}$ and all $%
k\in \mathbb{T}^{2}$, the wave operators 
\begin{equation*}
W^{\pm }\left( A\left( k\right) ,H^{(1)}\left( k\right) \oplus A_{2,2}\left(
k\right) \right) =s-{\lim\limits_{t\rightarrow \pm \infty }}\mathrm{e}%
^{itA\left( k\right) }\mathrm{e}^{-itH^{(1)}\left( k\right) \oplus
A_{2,2}\left( k\right) }\ ,
\end{equation*}%
as defined by Equation (\ref{scattering operator2}), exist and are complete
with 
\begin{equation*}
\mathrm{ran}\left( W^{\pm }\left( A\left( k\right) ,H^{(1)}\left( k\right)
\oplus A_{2,2}\left( k\right) \right) \right) =\mathrm{ran}\left( P_{\mathrm{%
ac}}\left( H^{(1)}\left( k\right) \oplus A_{2,2}\left( k\right) \right)
\right) \ .
\end{equation*}
\end{lemma}

\begin{proof}
Note that $P_{\mathrm{ac}}(\mathrm{H}_{f})=\mathfrak{1}$, thanks to Lemma %
\ref{lemma for existence of non-bound channel}. By (\ref{scattering
operator2}), it justifies the strong limit given in the lemma. As we discuss
in the proof of Proposition \ref{essential spectrum of a fiber}, for any $%
k\in \mathbb{T}^{2}$, the operator $A(k)$ is the sum of $M_{\mathfrak{f}%
\left( k\right) }\oplus A_{2,2}(k)$ and a compact operator $T$. In fact, as
the function $\mathrm{u}:\mathbb{Z}^{2}\rightarrow \mathbb{R}$ (defining the
fiber Hamiltonian $A(k)$) is, absolutely summable (see (\ref{summable1-U})),
the operator difference $T$ is even trace-class. As explained in Lemma \ref%
{lemma for existence of non-bound channel}, $R\left( \mathrm{V},\mathrm{v}%
\right) $ is also a trace-class operator, because $\mathrm{v}:\mathbb{Z}%
^{2}\rightarrow \mathbb{R}_{0}^{+}$ is absolutely summable, again by
assumption. By using the Kato-Rosenblum theorem \cite[Theorem XI.8]%
{ReedSimonIII}, it thus follows that the wave operators 
\begin{equation*}
W^{\pm }\left( A\left( k\right) ,H^{(1)}\left( k\right) \oplus A_{2,2}\left(
k\right) \right) =W^{\pm }\left( A\left( k\right) ,\left( M_{\mathfrak{f}%
\left( k\right) }+R\left( \mathrm{V},\mathrm{v}\right) \right) \oplus
A_{2,2}\left( k\right) \right)
\end{equation*}%
exist and are complete for every $k\in \mathbb{T}^{2}$.
\end{proof}

We are now in a position to prove Theorem \ref{existence of a kato wave
operator for the system}. Recall that $\mathfrak{U}:\mathfrak{H}%
_{f}\rightarrow L^{2}\left( \mathbb{T}^{2},\mathcal{H}\right) $ is the
operator defined by (\ref{identification operator1}), while $A_{2,2}(k)$ and 
$H^{(1)}\left( k\right) $ are respectively defined by (\ref{A22}) and (\ref%
{H1(k)}). The definition of wave operators $W^{\pm }$ are given by Equations
(\ref{scattering operator1})--(\ref{scattering operator2}).

\begin{theorem}[Unbound pair (scattering) channel]
\label{existence of a kato wave operator for the systembis}\mbox{ }\newline
For any $\mathrm{V}\in \mathbb{R}_{0}^{+}$ and every absolutely summable
function $\mathrm{v}:\mathbb{Z}^{2}\rightarrow \mathbb{R}_{0}^{+}$,%
\begin{equation*}
W^{\pm }\left( \mathbb{U}H\mathbb{U}^{\ast },\mathrm{H}_{f};\mathfrak{U}%
\right) =\left( {\int_{\mathbb{T}^{2}}^{\oplus }}W^{\pm }\left( A\left(
k\right) ,H^{(1)}\left( k\right) \oplus A_{2,2}\left( k\right) \right) \nu
\left( \mathrm{d}k\right) \right) \mathfrak{U}
\end{equation*}%
with 
\begin{equation*}
\mathrm{ran}\left( W^{\pm }\left( \mathbb{U}H\mathbb{U}^{\ast },\mathrm{H}%
_{f};\mathfrak{U}\right) \right) =\int_{\mathbb{T}^{2}}^{\oplus }L^{2}\left( 
\mathbb{T}^{2}\right) \oplus \{0\}\,\nu \left( \mathrm{d}k\right) \ .
\end{equation*}
\end{theorem}

\begin{proof}
For almost every $k\in \mathbb{T}^{2}$ and every $\psi \in \mathfrak{H}_{f}$%
, 
\begin{equation*}
\left( H^{(1)}\mathfrak{U}\psi \right) \left( k\right) =H^{(1)}\left(
k\right) \oplus A_{2,2}\left( k\right) \left( \mathfrak{U}\psi \right)
\left( k\right) =(H^{(1)}\left( k\right) \psi \left( k\right) ,0)=\left( 
\mathfrak{U}\mathrm{H}_{f}\psi \right) \left( k\right) \ .
\end{equation*}%
In other words, $\mathfrak{U}$ is an intertwining operator for $\mathrm{H}%
_{f}$ and $H^{(1)}$ and hence, for their respective complex exponential:%
\begin{equation}
\mathfrak{U}\mathrm{e}^{-it\mathrm{H}_{f}}=\mathrm{e}^{-itH^{(1)}}\mathfrak{U%
}\ ,\qquad t\in \mathbb{R}\ .  \label{interwining good}
\end{equation}%
We also observe that, for any $z\in \mathbb{C}$ and any Borel set $%
B\subseteq \mathbb{R}$ containing $\mathfrak{b}(k)\in \mathbb{R}$ (see (\ref%
{b})), 
\begin{equation*}
\left\langle z,\chi _{B}\left( A_{2,2}\left( k\right) \right) z\right\rangle
_{\mathbb{C}}=\left\vert z\right\vert ^{2}\chi _{B}\left( \mathfrak{b}\left(
k\right) \right) \neq 0
\end{equation*}%
even if the Lebesgue measure of $B$ is zero. This last observation, together
with Remark \ref{direct sum of absolutely continuous space} and Lemma \ref%
{lemma for existence of non-bound channel}, yields%
\begin{equation}
\mathrm{ran}\left( P_{\mathrm{ac}}\left( H^{(1)}\left( k\right) \oplus
A_{2,2}\left( k\right) \right) \right) =L^{2}\left( \mathbb{T}^{2}\right)
\oplus \left\{ 0\right\} \ ,\qquad k\in \mathbb{T}^{2}\ .  \label{sddsadasd}
\end{equation}%
In particular, 
\begin{equation}
{\int_{\mathbb{T}^{2}}^{\oplus }}P_{\mathrm{ac}}\left( H^{(1)}\left(
k\right) \oplus A_{2,2}\left( k\right) \right) \mathfrak{U=U}\ .
\label{sdsdsds2}
\end{equation}%
We can then apply Proposition \ref{pointwise convergence theorem for direct
integral} together with Lemmata \ref{lemma for existence of non-bound
channel}, \ref{lemma for existence of non-bound channel copy(1)}, Theorem %
\ref{borelian functional calculus of a direct integral of operators} and
Equations (\ref{interwining good})--(\ref{sdsdsds2}) to arrive at%
\begin{align*}
W^{\pm }\left( \mathbb{U}H\mathbb{U}^{\ast },M_{\mathfrak{f}};\mathfrak{U}%
\right) & \doteq s-\lim_{t\rightarrow \mp \infty }\mathrm{e}^{it\mathbb{U}H%
\mathbb{U}^{\ast }}\mathfrak{U}\mathrm{e}^{-it\mathrm{H}_{f}}P_{\mathrm{ac}%
}\left( \mathrm{H}_{f}\right) \\
& =s-\lim_{t\rightarrow \mp \infty }\mathrm{e}^{it\mathbb{U}H\mathbb{U}%
^{\ast }}\mathfrak{U}\mathrm{e}^{-it\mathrm{H}_{f}} \\
& =s-\lim_{t\rightarrow \mp \infty }\mathrm{e}^{it\mathbb{U}H\mathbb{U}%
^{\ast }}\mathrm{e}^{-itH^{(1)}}\mathfrak{U} \\
& =s-\lim_{t\rightarrow \mp \infty }\left( {\int_{\mathbb{T}^{2}}^{\oplus }}%
\mathrm{e}^{itA\left( k\right) }\,\nu \left( \mathrm{d}k\right) \right)
\left( {\int_{\mathbb{T}^{2}}^{\oplus }}\mathrm{e}^{-it\left( H^{(1)}\left(
k\right) \,\oplus \,A_{2,2}\left( k\right) \right) }\nu \left( \mathrm{d}%
k\right) \right) \mathfrak{U} \\
& =s-\lim_{t\rightarrow \mp \infty }\left( {\int_{\mathbb{T}^{2}}^{\oplus }}%
\mathrm{e}^{itA\left( k\right) }\mathrm{e}^{-it(H^{(1)}\left( k\right)
\,\oplus \,A_{2,2}\left( k\right) )}P_{\mathrm{ac}}\left( H^{(1)}\left(
k\right) \,\oplus \,A_{2,2}\left( k\right) \right) \,\nu \left( \mathrm{d}%
k\right) \right) \mathfrak{U} \\
& =\left( {\int_{\mathbb{T}^{2}}^{\oplus }}W^{\pm }\left( A\left( k\right)
,H^{(1)}\left( k\right) \oplus A_{2,2}\left( k\right) \right) \nu \left( 
\mathrm{d}k\right) \right) \mathfrak{U}\ .
\end{align*}%
Note that Lemma \ref{lemma for existence of non-bound channel copy(1)}
combined with (\ref{sddsadasd}) implies that 
\begin{equation*}
\mathrm{ran}\left( W^{\pm }\left( A\left( k\right) ,H^{(1)}\left( k\right)
\oplus A_{2,2}\left( k\right) \right) \right) =L^{2}\left( \mathbb{T}%
^{2}\right) \oplus \{0\}\ .
\end{equation*}%
In particular, 
\begin{equation*}
\mathrm{ran}\left( W^{\pm }\left( \mathbb{U}H\mathbb{U}^{\ast },\mathrm{H}%
_{f};\mathfrak{U}\right) \right) =\int_{\mathbb{T}^{2}}^{\oplus }L^{2}\left( 
\mathbb{T}^{2}\right) \oplus \{0\}\,\nu \left( \mathrm{d}k\right) \ .
\end{equation*}
\end{proof}

Observe that Lemma \ref{finite-time scattering and wave operators copy(1)}
allows one to write 
\begin{equation*}
\mathrm{e}^{itX}\mathrm{e}^{i\left( s-t\right) \left( X+Y\right) }\mathrm{e}%
^{-isX}\ ,\qquad s,t\in \mathbb{R}\ ,
\end{equation*}%
as a Dyson series for all bounded operators $X,Y$. This can be applied to $%
X=H^{(1)}$ and $Y=\mathbb{U}H\mathbb{U}^{\ast }-X$,or in each fiber $k\in 
\mathbb{T}^{2}$ to $X=H^{(1)}\left( k\right) \oplus A_{2,2}\left( k\right) $
and $Y=A\left( k\right) -X$. When $\mathrm{U}=\mathrm{V}\in \mathbb{R}%
_{0}^{+}$ and $\mathrm{v}=\mathrm{u}:\mathbb{Z}^{2}\rightarrow \mathbb{R}%
_{0}^{+}$ in $H^{(1)}$, this result is particularly advantageous because the
operator family $(Y_{t})_{t\in \mathbb{R}}$ appearing in Lemma \ref%
{finite-time scattering and wave operators copy(1)} can be represented in a
relatively simple way in this situation:

\begin{lemma}[Finite-time scattering and wave operators]
\label{finite-time scattering and wave operators copy(3)}\mbox{ }\newline
For $\mathrm{U}\in \mathbb{R}_{0}^{+}$ and all $s,t\in \mathbb{R}$,%
\begin{multline*}
\mathrm{e}^{itH^{(1)}\left( \mathrm{U},\mathrm{u}\right) }\mathrm{e}%
^{i\left( s-t\right) \mathbb{U}H\mathbb{U}^{\ast }}\mathrm{e}%
^{-isH^{(1)}\left( \mathrm{U},\mathrm{u}\right) } \\
=\int_{\mathbb{T}^{2}}^{\oplus }\left( 
\begin{array}{cc}
\cos _{\succ }\left( B_{1,2}\left( k\right) B_{2,1}\left( k\right)
;s,t\right) & -i\sin _{\succ }\left( B_{1,2}\left( k\right) B_{2,1}\left(
k\right) ;s,t\right) \\ 
-i\sin _{\succ }\left( B_{2,1}\left( k\right) B_{1,2}\left( k\right)
;s,t\right) & \cos _{\succ }\left( B_{2,1}\left( k\right) B_{1,2}\left(
k\right) ;s,t\right)%
\end{array}%
\right) \nu \left( \mathrm{d}k\right) \ ,
\end{multline*}%
where $B_{1,2}\left( k\right) $ and $B_{2,1}\left( k\right) $ are the
operator families defined by (\ref{B0bis}) for any $k\in \mathbb{T}^{2}$,
while $\cos _{\succ }$ and $\sin _{\succ }$ are respectively defined by (\ref%
{cosinus-operator}) and (\ref{sinus-operator}).
\end{lemma}

\begin{proof}
Let $H^{(1)}\equiv H^{(1)}\left( \mathrm{U},\mathrm{u}\right) $. We infer
from Lemma \ref{finite-time scattering and wave operators copy(1)} applied
to $X=H^{(1)}$ and $Y=\mathbb{U}H\mathbb{U}^{\ast }-X$ that%
\begin{equation*}
\mathrm{e}^{itH^{(1)}}\mathrm{e}^{i\left( s-t\right) \mathbb{U}H\mathbb{U}%
^{\ast }}\mathrm{e}^{-isH^{(1)}}=V_{t,s}\doteq \mathbf{1}+\sum_{n=1}^{\infty
}\left( -i\right) ^{n}\int_{s}^{t}\mathrm{d}\tau _{1}\cdots \int_{s}^{\tau
_{n-1}}\mathrm{d}\tau _{n}B^{(\tau _{1})}\cdots B^{(\tau _{n})}\ ,
\end{equation*}%
with $(B^{(t)})_{t\in \mathbb{R}}\subseteq \mathcal{B}(L^{2}(\mathbb{T}^{2},%
\mathcal{H}))$ being the norm-continuous family defined by%
\begin{eqnarray*}
B^{(t)} &=&\mathrm{e}^{itH^{(1)}}\left( \mathbb{U}H\mathbb{U}^{\ast
}-H^{(1)}\right) \mathrm{e}^{-itH^{(1)}} \\
&=&\int_{\mathbb{T}^{2}}^{\oplus }\left( 
\begin{array}{cc}
\mathrm{e}^{itA_{1,1}\left( \mathrm{U},k\right) } & 0 \\ 
0 & \mathrm{e}^{itA_{2,2}\left( k\right) }%
\end{array}%
\right) \left( 
\begin{array}{cc}
0 & A_{1,2}\left( k\right) \\ 
A_{2,1}\left( k\right) & 0%
\end{array}%
\right) \left( 
\begin{array}{cc}
\mathrm{e}^{-itA_{1,1}\left( \mathrm{U},k\right) } & 0 \\ 
0 & \mathrm{e}^{-itA_{2,2}\left( k\right) }%
\end{array}%
\right) \nu \left( \mathrm{d}k\right) \\
&=&\int_{\mathbb{T}^{2}}^{\oplus }%
\begin{pmatrix}
0 & B_{1,2}^{(t)}\left( k\right) \\[0.5em] 
B_{2,1}^{(t)}\left( k\right) & 0%
\end{pmatrix}%
\nu \left( \mathrm{d}k\right) \ ,
\end{eqnarray*}%
the operators $B_{2,1}^{(t)}\left( k\right) $ and $B_{1,2}^{(t)}\left(
k\right) $ being defined by (\ref{B0bis}) for any $t\in \mathbb{R}$ and $%
k\in \mathbb{T}^{2}$. Then, one combines explicit computations together with
Proposition \ref{pointwise convergence theorem for direct integral} and \ref%
{fubini's theorem} to arrive at the assertion. Notice that the above
integrals are Riemann ones, $(B^{(t)})_{t\in \mathbb{R}}$ being a continuous
family in the Banach space $\mathcal{B}(L^{2}(\mathbb{T}^{2},\mathcal{H}))$.
\end{proof}

\subsubsection{Bound pair scattering channel\label{d-wave pairing channel}}

We start by studying the wave operator (\ref{scattering operator1}) with
respect to the operators $X=\mathbb{U}H\mathbb{U}^{\ast }$ and $Y=M_{\mathrm{%
E}\left( \mathrm{U},\cdot \right) }$ (\ref{ME}), the identification operator 
$J$ being $\mathfrak{P}_{\mathrm{U}}$ (\ref{frac P}) for any fixed Hubbard
coupling constant $\mathrm{U}\in \mathbb{R}_{0}^{+}$.

\begin{proposition}[Wave operators in the bound pair channel]
\label{kato's wave operator for d-wave pairing}\mbox{ }\newline
Let $h_{b}\in \lbrack 0,1/2]$ and $\mathrm{U}\in \mathbb{R}_{0}^{+}$. Then, $%
\mathbb{U}H\mathbb{U}^{\ast }\mathfrak{P}_{\mathrm{U}}=\mathfrak{P}_{\mathrm{%
U}}M_{\mathrm{E}\left( \mathrm{U},\cdot \right) }$ and for every bounded
continuous function $f\in C_{b}(\mathbb{R})$,%
\begin{equation*}
\,f\left( \mathbb{U}H\mathbb{U}^{\ast }\right) \mathfrak{P}_{\mathrm{U}}=%
\mathfrak{P}_{\mathrm{U}}{\int_{\mathbb{T}^{2}}^{\oplus }}f\left( \mathrm{E}%
\left( \mathrm{U},k\right) \right) \nu \left( \mathrm{d}k\right) \text{ }.
\end{equation*}
\end{proposition}

\begin{proof}
Using Proposition \ref{direct integral decomposition of the hamiltonian} and
Theorem \ref{existence of eigenvalue for each fiber}, we note that, for any $%
\varphi \in L^{2}(\mathbb{T}^{2})$ and almost every $k\in \mathbb{T}^{2}$,%
\begin{equation*}
\left( \mathbb{U}H\mathbb{U}^{\ast }\mathfrak{P}_{\mathrm{U}}\varphi \right)
\left( k\right) =A\left( k\right) \left( \mathfrak{P}_{\mathrm{U}}\varphi
\right) \left( k\right) =\mathrm{E}\left( \mathrm{U},k\right) \varphi \left(
k\right) \left\Vert \Psi \left( \mathrm{U},k\right) \right\Vert ^{-1}\Psi
\left( \mathrm{U},k\right) =\left( \mathfrak{P}_{\mathrm{U}}M_{\mathrm{E}%
\left( \mathrm{U},\cdot \right) }\varphi \right) \left( k\right) \ ,
\end{equation*}%
i.e., $\mathbb{U}H\mathbb{U}^{\ast }\mathfrak{P}_{\mathrm{U}}=\mathfrak{P}_{%
\mathrm{U}}M_{\mathrm{E}\left( \mathrm{U},\cdot \right) }$,keeping in mind
Equations (\ref{frac P}) and (\ref{ME}). We then obtain the last assertion
by using the Stone-Weierstrass theorem and the spectral theorem.
\end{proof}

We now study the (dressed) bound pair channel of lowest energy in the
hard-core limit. This is a consequence of the following assertion:\ 

\begin{proposition}[Bound pair channel in the hard-core limit]
\label{some strong limits}\mbox{ }\newline
Fix $h_{b}\in \lbrack 0,1/2]$. Then, 
\begin{equation}
s-{\lim\limits_{\mathrm{U}\rightarrow \infty }}\mathfrak{P}(\mathrm{U})=%
\mathfrak{P}_{\infty }  \label{i.}
\end{equation}%
and for every bounded continuous function $f\in C_{b}(\mathbb{R})$,%
\begin{equation*}
s-{\lim\limits_{\mathrm{U}\rightarrow \infty }}\,f\left( \mathbb{U}H\mathbb{U%
}^{\ast }\right) \mathfrak{P}_{\mathrm{U}}=\mathfrak{P}_{\infty }{\int_{%
\mathbb{T}^{2}}^{\oplus }}f\left( \mathrm{E}\left( \infty ,k\right) \right)
\nu \left( \mathrm{d}k\right) \text{ }.
\end{equation*}
\end{proposition}

\begin{proof}
Fix $\varphi \in L^{2}(\mathbb{T}^{2})$. For any $\mathrm{U}\in \mathbb{R}%
_{0}^{+}$ and almost every $k\in \mathbb{T}^{2}$, one has that%
\begin{equation*}
\left\Vert \mathfrak{P}_{\mathrm{U}}\varphi \left( k\right) -\mathfrak{P}%
_{\infty }\varphi \left( k\right) \right\Vert \leq 2\Vert \Psi \left( \infty
,k\right) \Vert ^{-1}\Vert \Psi \left( \mathrm{U},k\right) -\Psi \left(
\infty ,k\right) \Vert \left\vert \varphi \left( k\right) \right\vert \ .
\end{equation*}%
Thus, by Lebesgue's dominated convergence theorem, we arrive at%
\begin{equation*}
{\lim\limits_{\mathrm{U}\rightarrow \infty }}\,\left\Vert \mathfrak{P}_{%
\mathrm{U}}\varphi \left( k\right) -\mathfrak{P}_{\infty }\varphi
\right\Vert _{L^{2}\left( \mathbb{T}^{2},\mathcal{H}\right) }^{2}={%
\lim\limits_{\mathrm{U}\rightarrow \infty }\int_{\mathbb{T}^{2}}}\left\Vert 
\mathfrak{P}_{\mathrm{U}}\varphi \left( k\right) -\mathfrak{P}_{\infty
}\varphi \left( k\right) \right\Vert ^{2}\,\nu \left( \mathrm{d}k\right) =0%
\text{ }.
\end{equation*}%
Take now a bounded continuous function $f\in C_{b}(\mathbb{R})$. In
particular, there exists $L\in \mathbb{R}^{+}$ such that%
\begin{equation*}
\sup_{\mathrm{U}\in \mathbb{R}_{0}^{+}}\sup_{k\in \mathbb{T}^{2}}\left\vert
f\left( \mathrm{E}\left( \mathrm{U},k\right) \right) \right\vert \leq L\ .
\end{equation*}%
By Theorem \ref{properties of effective dispersion relation with low energy}
(i) and continuity of the function $f$, one has that%
\begin{equation*}
{\lim_{\mathrm{U}\rightarrow \infty }}\,f\left( \mathrm{E}\left( \mathrm{U}%
,k\right) \right) =f\left( \mathrm{E}\left( \infty ,k\right) \right) \ ,%
\text{$\qquad $}k\in \mathbb{T}^{2}\ .
\end{equation*}%
Moreover, 
\begin{equation*}
\mathbb{T}^{2}\ni k\longmapsto f\left( \mathrm{E}\left( \mathrm{U},k\right)
\right) \in \mathcal{L}(\mathbb{C})(=\mathcal{B}(\mathbb{C}))
\end{equation*}%
is a composition of continuous functions and is in particular strongly
measurable. Using Proposition \ref{pointwise convergence theorem for direct
integral} (or Lebesgue's dominated convergence theorem) we arrive at the
limit 
\begin{equation}
s-{\lim\limits_{\mathrm{U}\rightarrow \infty }\int_{\mathbb{T}^{2}}^{\oplus }%
}f\left( \mathrm{E}\left( \mathrm{U},k\right) \right) \nu \left( \mathrm{d}%
k\right) ={\int_{\mathbb{T}^{2}}^{\oplus }}f\left( \mathrm{E}\left( \infty
,k\right) \right) \nu \left( \mathrm{d}k\right) \ .  \label{i.2}
\end{equation}%
By using Equations (\ref{i.}), (\ref{i.2}), Proposition \ref{kato's wave
operator for d-wave pairing}, Theorem \ref{borelian functional calculus of a
direct integral of operators} (iii) and the fact that $\Vert \mathfrak{P}(%
\mathrm{U})\Vert _{\mathrm{op}}=1$ for all $\mathrm{U}\in \mathbb{R}_{0}^{+}$%
, we find that%
\begin{eqnarray*}
s-{\lim\limits_{\mathrm{U}\rightarrow \infty }}\,f\left( \mathbb{U}H\mathbb{U%
}^{\ast }\right) \mathfrak{P}_{\mathrm{U}} &=&s-{\lim\limits_{\mathrm{U}%
\rightarrow \infty }}\,\mathfrak{P}_{\mathrm{U}}f\left( M_{\mathrm{E}\left( 
\mathrm{U},\cdot \right) }\right) \\
&=&s-{\lim\limits_{\mathrm{U}\rightarrow \infty }\mathfrak{P}_{\mathrm{U}%
}\int_{\mathbb{T}^{2}}^{\oplus }}f\left( \mathrm{E}\left( \mathrm{U}%
,k\right) \right) \nu \left( \mathrm{d}k\right) \\
&=&\mathfrak{P}_{\infty }{\int_{\mathbb{T}^{2}}^{\oplus }}f\left( \mathrm{E}%
\left( \infty ,k\right) \right) \nu \left( \mathrm{d}k\right) \ .
\end{eqnarray*}
\end{proof}

\section{Appendix\label{Appendix}}

\subsection{Towards a Microscopic Theory for\ Cuprate\ Superconductivity 
\label{Section physics}}

Superconductivity was discovered in 1911 through the study of the resistance
of solid mercury at very low temperatures, which was found to disappear
below the critical temperature $T_{c}=4.2$ \textrm{K}. This phenomenon was
subsequently observed in several other materials, such as lead (in this
case, $T_{c}=7$ \textrm{K}). The first microscopic explanation of this
unexpected but very interesting physical behavior was given in 1957 by J.
Bardeen, L. Cooper and J.R. Schrieffer with what is now known as the BCS
theory. They were awarded the Nobel Prize in Physics in 1972. Their theory
explained all the superconductors known at the time, named today
\textquotedblleft conventional\textquotedblright\ superconductors. For more
details, see \cite[Chapter 10]{Kittel}.

These superconductors not only have zero resistivity (below some critical
current value), but also repel magnetic fields. This is the Meissner effect
(or Meissner-Ochsenfeld effect). See, for example, the popular images
showing a superconducting piece levitating above a magnet. However, when the
magnetic field exceeds a critical value, superconductivity can be broken and
the Meissner effect disappears abruptly. This is referred to as type I
superconductivity, while type II superconductors manisfests the appearance
of vortices beyond a first critical magnetic field and the disappearance of
any Meissner effect beyond a second critical field. The BCS theory refers to
conventional superconductors but applies for both type I and II
superconductors.

Superconductors are characterized not only by the critical temperature but
also their superconducting coherence length, which quantifies the
characteristic exponent that describes variations in the density of the
superconducting component. It is often several hundred nanometers for
conventional superconductors. More precisely, from \cite[Chapter 10, Table 5]%
{Kittel} we have the following coherence lengths $\xi $ and critical
temperatures $T_{c}$ for the following conventional superconductors:%
\begin{equation}
\begin{tabular}{|l|l|l|l|}
\hline
& $\xi $ in $\mathrm{nm}$ & $T_{c}$ in $\mathrm{K}$ & Type \\ \hline
Tin (\textrm{Sn}) & \multicolumn{1}{|c|}{$230$} & $3.72$ & 
\multicolumn{1}{|c|}{I} \\ \hline
Aluminium (\textrm{Al}) & \multicolumn{1}{|c|}{$1600$} & $1.2$ & 
\multicolumn{1}{|c|}{I} \\ \hline
Lead (\textrm{Pb}) & \multicolumn{1}{|c|}{$83$} & $7.19$ & 
\multicolumn{1}{|c|}{I} \\ \hline
Cadmium (\textrm{Cd}) & \multicolumn{1}{|c|}{$760$} & $0.52$ & 
\multicolumn{1}{|c|}{I} \\ \hline
Niobium (\textrm{Nb}) & \multicolumn{1}{|c|}{$38$} & $9.26$ & 
\multicolumn{1}{|c|}{II} \\ \hline
\end{tabular}
\label{table}
\end{equation}%
Note that the superconducting coherence length of a Niobium superconductor
is much smaller than others, which is consistent with the type-II property,
which requires shorter coherence lengths compared to type I superconductors.

In 1986, there was a major breakthrough in physics with the discovery of a
new class of superconductors, by G. Bednorz and K.A. M\"{u}ller \cite{BM86}.
They were awarded the Nobel Prize in Physics in 1987. Physically, these
materials are antiferromagnetic and insulating at low temperatures, but as
with semiconductors, one dopes them with impurities that provide extra
charge carriers and break the perfect Mott insulator phase, which is
characterized by an integer number of charge carriers on each lattice site.
As doping increases, the antiferromagnetic phase turns into a
superconducting phase. This was the discovery of high-$T_{c}$
superconductors in ceramics materials, i.e., cuprates, for critical
temperatures $T_{c}$ that today range in the interval $[39,164]$ \textrm{K}
(approximately).

These superconductors are non-conventional, and the BCS theory fails to
explain their properties. Indeed, while the conventional superconductivity
results from an effective attraction between fermions (electrons or holes,
depending on the charge carriers in each material) via phonons (i.e.,
lattice excitations), it soon became apparent that this kind of explanation
could not work for cuprates, and the question of the mediator that could
produce such an attraction has remained an open problem ever since. In fact,
even if a large amount of numerical and experimental data is available,
there is no pairing mechanism firmly established (through, for instance,
antiferromagnetic spin fluctuations, phonons, etc.). See, e.g., \cite[%
Section 7.6]{tinpou1}. The debate is strongly polarized \cite{Rodgers}
between researchers using a purely electronic/magnetic microscopic mechanism
and those using electron-phonon mechanisms.

This is undoubtedly one of the most important questions in condensed matter
physics, even if current research seems to have shown less interest in this
fundamental issue in recent years. Quoting the Nobel Prize winner M\"{u}ller
in 2007 \cite{M07}: \textquotedblleft ... \textit{It is a remarkable fact
that in these 20 years since the discovery of high temperature
superconductivity no other class of materials has been found which exhibits
this property above the boiling point of liquid nitrogen. With a view to
finding another class, it would be rewarding to understand why these
exceptional properties occur, which per se are regarded as among the
important unsolved problems in present day physics.} \textquotedblright

Our theoretical approach differs from all others and stems from a
microscopic model -- first proposed in 1985 by Ranninger-Robaszkiewicz \cite%
{antibrulage3} (see also \cite{antibrulage2,ranninger} or \cite[Section 7.4.3%
]{tinpou1}) and independently by Ionov \cite{antibrulage4} -- which, before
our works, was \emph{never} investigated in the presence of strong Coulomb
repulsions.

The cuprates are a class of compounds containing copper ($\mathrm{Cu}$)
atoms in an anion and cuprate superconductors are oxide-based cuprates with
two-dimensional $\mathrm{CuO}_{2}$ layers made of $\mathrm{Cu}^{++}$ (cf.
\textquotedblleft cuprate\textquotedblright ) and $\mathrm{O}^{--}$ (cf.
\textquotedblleft oxide\textquotedblright ) ions, which generally possess
the symmetries of the square, at least for the important family of
tetragonal cuprates such as $\mathrm{La}_{2-x}\mathrm{Sr}_{x}\mathrm{CuO}%
_{4} $ (LaSr 214) and $\mathrm{La}_{2-x}\mathrm{Ba}_{x}\mathrm{CuO}_{4}$.
See, e.g., \cite[Section 9.1.2]{Tsuei2003}, \cite[Section 2.3]{Saxena} and 
\cite[Section 6.3.1]{unitcoherence}.

As stressed in \cite[Part VII]{Jan-teller}, the very strong Jahn-Teller (JT)
effect associated with copper ions ($\mathrm{Cu}^{++}$) and its consequences
for polaron formation are largely neglected in much of the physics
literature, even though it was the JT effect that led to the discovery of
superconductivity in cuprates \cite{BM86}. See also \cite{M07,65,AZencore}.
For non-experts, let us explain that the JT effect (or JT distortion) is a
spontaneous symmetry breaking of molecules and ions that occurs via a
geometrical distortion that suppresses the spatial degeneracy of the
electronic ground state and lowers the overall energy of the system. See 
\cite{Jan-teller}. In this context, it can produce JT $n$-polarons.
Polarons, bipolarons or more generally, $n$-polarons, $n\in \mathbb{N}$, are
charge carriers that are self-trapped inside a strong and local lattice
deformation that surrounds them. They are quasi-particle formed from
fermions \textquotedblleft dressed with phonons\textquotedblright . E.g., a 
\emph{bi}polaron involves \emph{two} fermions dressed with phonons. A JT
polaron is a polaron for which the local lattice deformation is associated
with the (geometrical) JT distortion. The existence of JT (bi)polarons in
cuprates is attested in numerous experiments on cuprate superconductors \cite%
{M17,M07,65,AZencore} and we have the following experimental facts:

\begin{itemize}
\item Superconducting transport in cuprates occurs in two-dimensional $%
\mathrm{CuO}_{2}$ layers and only on oxygen atoms -- a fact well established
experimentally from 1987 by Bianconi and others \cite%
{oxygene-hole1,oxygene-hole2,oxygene-hole3,oxygene-hole4,oxygene-hole3-jap,oxygene-hole3-rome,oxygene-hole5}
-- while bipolarons are related to the strong JT effect of copper ions.

\item Because of the presence of strong \textbf{antiferromagnetic}
correlations of copper-oxides, experimentally proven even outside the
antiferromagnetic phase (see, e.g., \cite{AF1,AF2} and \cite[Chap. 3]%
{tinpou1}), it can be concluded that JT bipolarons have \textbf{zero total
spin} and that other types of polaronic configuration are disadvantaged.
This is for instance stressed in \cite[Sect. 5.2]{M07}.

\item There is also an experimental evidence of the short lifetime of
polarons in cuprates \cite{lifetime}, decaying into fermions (pairs of holes
or electrons). Expressed in terms of a length $\ell $, one sees a lifetime
comparable to the lattice spacing \cite{lifetime}. Remarkably, near the
critical temperature, $\ell $ is actually close to the coherence length in
cuprate superconductors.
\end{itemize}

\begin{figure}[tbp]
\centering
\includegraphics[scale=0.2]{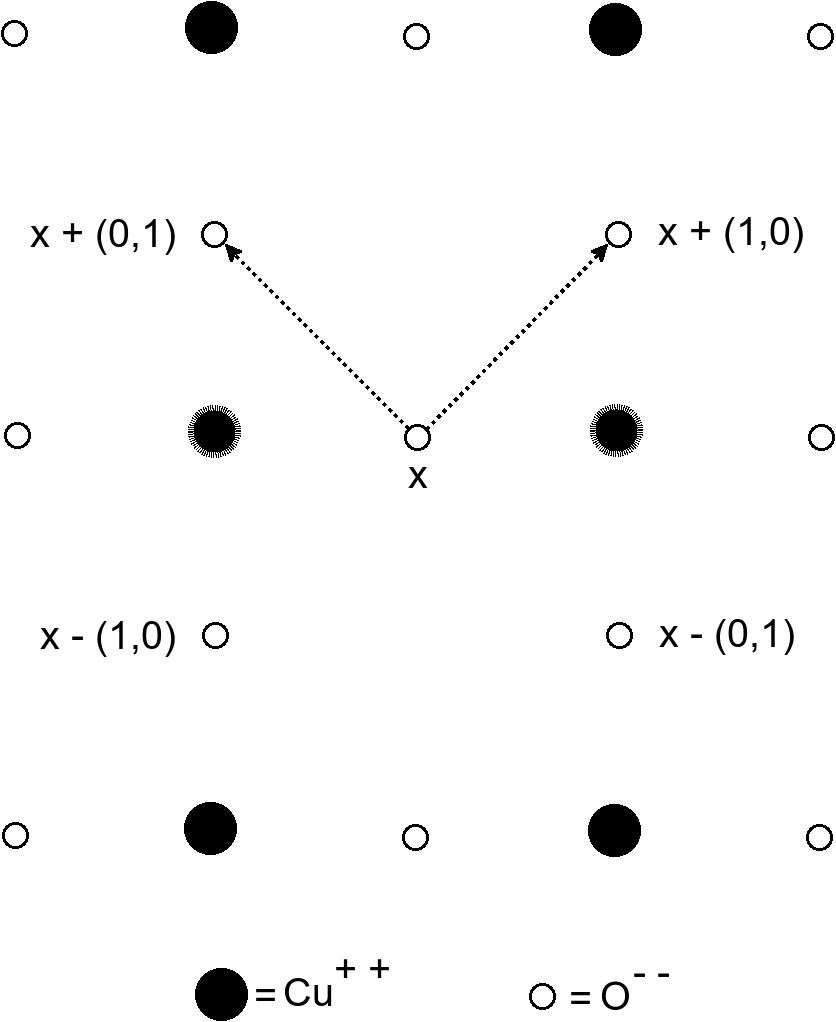}
\caption{$\mathrm{CuO}_{2}$ layer.}
\label{figureCuO.jpeg}
\end{figure}%

The most straightforward approach would therefore be to consider the JT
bipolarons as the charge carriers of cuprate superconductors. That is
exactly what Alexandrov and coauthors have done in their bipolaron theory,
based on light bipolarons \cite{superlight} as charge carriers. Quoting \cite%
[p. 4]{Alexencoreencore}: \textquotedblleft \textit{cuprate bipolarons are
relatively light because they are intersite rather than on-site pairs due to
the strong on-site repulsion, and because mainly c-axis polarized optical
phonons are responsible for the in-plane mass renormalization.}%
\textquotedblright\ See for instance \cite%
{Alexencoreencore,Alex2011,Alexbis,Alex2013} and references therein.
However, this approach does not seem consistent with superconducting
transport in cuprates occurring on oxygen ions in $\mathrm{CuO}_{2}$ layers:

In fact, a priori, (strong and local) lattice deformations (or JT
distortions) attached to $n$-polarons should barely move and this is not in
accordance with the known mobility of superconducting charge carriers. This
is confirmed in experiments:

\begin{itemize}
\item Experimental evidence (still controversial \cite{polaronsize}) of a
large mass (approx. 700 electronic masses) of polarons in cuprates \cite%
{unitmass-bipolaron}. \medskip

\item Experimental evidence (apparently not controversial) of the small mass
(approx. 3-4 electronic masses) of superconducting carriers in cuprates \cite%
[Fig. 2.]{unit-mass}.\medskip
\end{itemize}

\noindent For more recent discussions\ on the (im)mobility of (bi)polarons
in cuprates, we recommend for instance \cite%
{reviewer2-1,reviewer2-2,reviewer2-3}.

We bypass this problem by using the exchange interaction (\ref%
{Hamiltonian-fb}), while seeing the fermions as the true charge carriers. In
other words, we use exchange interactions like (\ref{Hamiltonian-fb}) to
define a simplified model for cuprates, taking into account a large mass of
bipolarons but \textbf{non-}polaronic superconducting carriers. Since the
lifetime of bipolarons in terms of a length $\ell $ is comparable to the
coherence length in cuprate superconductors near the critical temperature,
this suggests a strong exchange interaction between the (Bose-like,
zero-spin) bipolaronic state and fermion pairs (electrons or holes).

As shown in Figure \ref{figureCuO.jpeg}, bipolarons are formed around an
oxygen ion ($x$) binding an adjacent pair of copper ions, because of the JT
effect associated with $\mathrm{Cu}^{++}$. It leads to JT \textquotedblleft
intersite bipolarons\textquotedblright . That is why we consider an
annihilation (creation) operator $c_{x}$ ($c_{x}^{\ast }$) of a fermion pair
of zero total spin at $x\in \mathbb{Z}^{2}$ as defined by Equation (\ref%
{Hamiltonian-fbbis}). One simple example of such an operator is 
\begin{equation}
c_{x}\doteq \sum_{z\in \mathbb{Z}^{2},|z|\leq 1}\left( a_{x+z,\uparrow
}a_{x,\downarrow }+a_{x+z,\uparrow }a_{x-z,\downarrow }\right)
\label{bipolaron exchange term2}
\end{equation}%
with $a_{z,s}$ ($a_{z,s}^{\ast }$) being the annihilation (creation)
operator of a single fermion of spin $s\in \{\uparrow ,\downarrow \}$ at
lattice site $z\in \mathbb{Z}^{2}$. In this example we set $\mathfrak{p}%
_{2}\left( 2z\right) =\mathfrak{p}_{1}\left( z\right) =1$ when $|z|\leq 1$
and $\mathfrak{p}_{1}\left( z\right) =\mathfrak{p}_{2}\left( z\right) =0$
otherwise. Of course, one can also assign other weights to each space
configuration of fermion pairs in\ Equation (\ref{Hamiltonian-fbbis}), as
soon as at least one intersite configuration has a non-zero weight.

There is also an undeniable experimental evidence of \textbf{strong} on-site
Coulomb repulsions (cf. the Mott insulator phase at zero doping), which
forces us to consider terms like%
\begin{equation*}
\mathrm{U}\sum_{x\in \mathbb{Z}^{2}}n_{x,\uparrow }n_{x,\downarrow }\
,\qquad \mathrm{U}\gg 1\ ,
\end{equation*}%
in Equation (\ref{Hamiltonian-f}), where we recall that $n_{x,s}\doteq
a_{x,s}^{\ast }a_{x,s}$ is the number operator of fermions at $x\in \mathbb{Z%
}^{2}$ and spin $s\in \{\uparrow ,\downarrow \}$. It justifies our strong
interest in studying in this paper the hard-core limit $\mathrm{U}%
\rightarrow \infty $. See for example Theorem \ref{Maintheorem2} and, more
generally, related results that hold for $\mathrm{U}\in \left[ 0,\infty %
\right] $.

The exchange interaction as formally given by (\ref{Hamiltonian-fb}), i.e., 
\begin{equation*}
{2^{-1/2}\sum\limits_{x,y\in \mathbb{Z}^{2}}}\upsilon \left( x-y\right)
c_{y}^{\ast }\,b_{x}
\end{equation*}%
with $b_{x}$ ($b_{x}^{\ast }$) being the annihilation (creation) operator of
a JT (intersite) bipolaron, is inspired by an interband interaction proposed
by Kondo in 1963 for superconducting transition metals. In \cite%
{antibrulage3,antibrulage4}\ only an on-site version (i.e., $%
c_{y}=a_{y,\uparrow }a_{y,\downarrow }$) was proposed in 1985. Our version (%
\ref{bipolaron exchange term2}) of $c_{y}$ captures the \textquotedblleft
intersite\textquotedblright\ character of the bipolarons present in cuprates
and in \cite{articulo2} we specify the form of the coupling function $%
\upsilon $ in (\ref{Hamiltonian-fbbis}) based on the presence of large
electron-phonon anomalies\footnote{%
The so-called softening of phonon dispersion and the broadening of phonon
lines.} in cuprates at optimum doping for the following points in the
normalized Brillouin zone $\mathbb{T}^{2}\doteq \lbrack -\pi ,\pi )^{2}$: 
\begin{equation*}
(0,-\pi ),(-\pi ,0)\quad \text{\cite{phononanomaly0,phononanomaly2}\qquad
and\qquad }(0,\pm \pi /2),(\pm \pi /2,0)\quad \text{\cite%
{phononanomaly2,37,Rez} }.
\end{equation*}

The anomalies at $(0,\pm \pi /2)$ and $(\pm \pi /2,0)$ are correctly
predicted by the Density Functional Theory (DFT) involving electrons and
phonons \cite[Fig. 1 (a)]{phononanomaly3}, in contrast to those of $(0,-\pi
),(-\pi ,0)$. Moreover, when no superconducting phase appears, DFT works
very well at all quasi-momenta, including $(0,-\pi ),(-\pi ,0)$ \cite[Fig.
18 (b)]{Rez}.

The above anomalies at quasi-momenta $(0,-\pi ),(-\pi ,0)$ in the
superconducting phase, which cannot be reproduced by the DFT, is expected to
be a consequence of the existence of polaronic quasiparticles. Indeed, the
DFT used in \cite[Fig. 1 (a)]{phononanomaly3} does not take into account the
formation of compound particles out of phonons and fermions like polaronic
modes. In our theory, they are interpreted as being JT (intersite)
bipolarons which should then interact strongly with charge carriers only at
quasi-momenta $(-\pi ,0)$ and $(0,-\pi )$ (at moderate doping). The
congruence between the DFT and experimental data for phonon dispersions at $%
(\pm \pi /2,0)$ and $(0,\pm \pi /2)$ indeed makes the formation of such
quasiparticles unlikely in this region of the (normalized) Brillouin zone $%
\mathbb{T}^{2}$, and more generally in any other region relatively far from $%
(-\pi ,0)$ and $(0,-\pi )$. Consequently, the Fourier transform $\hat{%
\upsilon}$ of $\upsilon $ is chosen to take its maximum absolute value at
the points $(-\pi ,0)$ and $(0,-\pi )$. This property is fundamental to
explaining the superconductivity of cuprates in our microscopic theory.

There is indeed one very important property of superconducting carriers
(pairs) in cuprates that differs from conventional superconductors, their $d$%
-wave symmetry. The (fiber) space of a fermion pair at constant
quasimomentum $K$ is the Hilbert space $L^{2}\left( \mathbb{T}^{2},\mathbb{C}%
,\nu \right) $, see Section \ref{Model section}. Define by%
\begin{equation*}
\left[ R_{\perp }|\varphi \rangle \right] (k_{x},k_{y})\doteq \varphi
(k_{y},-k_{x})\ ,\qquad (k_{x},k_{y})\in \mathbb{T}^{2}\ ,
\end{equation*}%
the unitary operator $R_{\perp }$ implementing the $\pi /2$-rotation on $%
L^{2}\left( \mathbb{T}^{2},\mathbb{C},\nu \right) $. Then define the
mutually orthogonal projectors%
\begin{eqnarray*}
P_{s} &\doteq &\frac{R_{\perp }^{4}+R_{\perp }^{3}+R_{\perp }^{2}+R_{\perp }%
}{4}\ , \\
P_{d} &\doteq &\frac{R_{\perp }^{4}-R_{\perp }^{3}+R_{\perp }^{2}-R_{\perp }%
}{4}\ , \\
P_{p} &\doteq &\frac{R_{\perp }^{4}-R_{\perp }^{2}}{2}\ .
\end{eqnarray*}%
Since $P_{s}+P_{d}+P_{p}=\mathfrak{1}$, any wave function $\Psi _{f}\in
L^{2}\left( \mathbb{T}^{2},\mathbb{C},\nu \right) $ of a fermion pair can be
uniquely decomposed into orthogonal $s$-, $d$-\ and $p$-wave components\ as%
\begin{equation*}
\Psi _{f}=\Psi _{f}^{(s)}+\Psi _{f}^{(d)}+\Psi _{f}^{(p)}\ ,\qquad \Psi
_{f}^{(\#)}\doteq P_{\#}\Psi _{f}\ .
\end{equation*}%
In other words, an arbitrary (fermionic pair) function $\Psi _{f}$ can be
uniquely decomposed into \textquotedblleft $s$-, $d$- and $p$%
-wave\textquotedblright\ components, denoted respectively by $\Psi
_{f}^{(s)},\Psi _{f}^{(d)},\Psi _{f}^{(p)}$. Observe that%
\begin{equation}
R_{\perp }\Psi _{f}^{(s)}=\Psi _{f}^{(s)}\ ,\qquad R_{\perp }\Psi
_{f}^{(d)}=-\Psi _{f}^{(d)}\ ,\qquad R_{\perp }^{2}\Psi _{f}^{(p)}=-\Psi
_{f}^{(p)}\ .  \label{sdfgdfgdfgdfg}
\end{equation}%
So, each component has a well-defined parity with respect to the group $%
\{0,\pi /2,\pi ,3\pi /2\}$ of rotations: The $s$-wave component $\Psi
_{f}^{(s)}$ is invariant under these 4 rotations, the $d$-wave one $\Psi
_{f}^{(d)}$ is antisymmetric with respect to the $\pi /2$-rotation and the $%
p $-wave one $\Psi _{f}^{(p)}$ is antisymmetric with respect to the $\pi $%
-rotation (reflection over the origin), just like \textquotedblleft $s$%
\textquotedblright , \textquotedblleft $d$\textquotedblright\ and
\textquotedblleft $p$\textquotedblright\ atomic orbitals.

In conventional superconductivity, one has $s$-wave symmetry. For
superconducting cuprates it is more complex. It is firmly established that
fermion pairs in cuprate superconductors have zero total spin \cite{Tsuei},
which is believed to lead to $s$- or $d$-wave superconductivity, only. The $%
s $-wave symmetry is expected to correspond to fermion pairs on same lattice
sites, which should be problematic in the presence of the strong on-site
Coulomb repulsion. Therefore, $d$-wave superconductivity is anticipated in
cuprate superconductors. This prediction is experimentally confirmed. See 
\cite{Tsuei,Nature2015,tinpou1}. In cuprates, $d$-wave pairing is therefore
predominant, but experiments (involving bulk properties) still indicate the
presence of a non-negligible $s$-wave superconducting part, see \cite%
{M07,M17}. This is what our theory demonstrates, using the fact that the
Fourier transform $\hat{\upsilon}$ of $\upsilon $ is maximal at the points $%
(-\pi ,0)$ and $(0,-\pi )$, but first we need to say a few words about the
quantitative choice of its parameters. For example, if we take the hard-core
limit $\mathrm{U}\rightarrow \infty $, we get pure $d$-wave
superconductivity, as shown in the first article \cite{articulo}.

In the second paper \cite{articulo2}, we study the ground state $\Psi (%
\mathrm{U},k)\doteq (\hat{\psi}_{k}\left( \mathrm{U}\right) ,-1)$ of Theorem %
\ref{Maintheorem1} to give estimates on key features of cuprate
superconductors by using \textbf{real} parameters taken from experiments on
the prototypical cuprates based on hole-doped cuprates $\mathrm{La}_{2}%
\mathrm{CuO}_{4}$ (e.g., LaSr 214) and $\mathrm{YBa}_{2}\mathrm{Cu}_{3}%
\mathrm{O}_{7}$ (YBCO),\ near optimal doping:

\begin{itemize}
\item The hopping amplitude $\epsilon $ of charge carriers (here holes)\ in (%
\ref{Hamiltonian-f}) is accessible using the lattice spacing and the
effective mass of charge carriers. Both quantities are known for cuprates:
The lattice spacing is $\mathbf{a}=0.2672\ \mathrm{nm}$ \cite[Sect. 6.3.1]%
{unitcoherence} of the oxygen ions and the effective mass of mobile holes $%
m^{\ast }\simeq 4m_{e}$ \cite[Fig. 2.]{unit-mass}, where $m_{e}$ is the
electron mass. This corresponds to $\epsilon =\hbar ^{2}/\left( m^{\ast }%
\mathbf{a}^{2}\right) \simeq 0.266\ $\textrm{eV}.\medskip

\item In the same way, the hopping amplitude $\epsilon h_{b}$ of JT
bipolarons in (\ref{Hamiltonian-b}) is accessible using the lattice spacing
and the effective mass of bipolarons. The former is known and the latter is
estimated \cite{unitmass-bipolaron}. It leads to $h_{b}\simeq 0.00575\ll 1$.
I.e., JT bipolarons can barely move, compared to fermions.\medskip

\item The coefficient $\mathrm{U}$ in (\ref{Hamiltonian-f}) can be fixed by
using the first electronic affinity of oxygen, i.e., the energy difference
between the $\mathrm{O}^{-}$ anion state (one hole added to the $\mathrm{O}%
^{--}$ anion) and the neutral state (two holes added to $\mathrm{O}^{--}$).
These values are known with great precision: By \cite{unit-affinity}, $%
\mathrm{U}\simeq 1.461\ $\textrm{eV}. Note that $\mathrm{U}\epsilon
^{-1}\simeq 5.5$, which refers to a strong coupling regime, but it is not
the hard-core limit $\mathrm{U}\rightarrow \infty $ yet, from the
perspective of real physical estimates.

\item The intersite repulsion represented by the function $\mathrm{u}:%
\mathbb{Z}^{2}\rightarrow \mathbb{R}_{0}^{+}$ in (\ref{Hamiltonian-f})
results from the screening of the Coulomb repulsion, usually estimated via
the Thomas-Fermi screening length $\mathbf{\lambda }_{\mathrm{TF}}$.
However, in two dimensions, the decay of the screened Coulomb repulsion is
not exponential but rather polynomial \cite[Eq. (5.41)]{screening0}. In
particular, even if $\mathbf{\lambda }_{\mathrm{TF}}\leq \mathbf{a}$, we
consider the Coulomb repulsion for a few neighboring sites, with, of course,
decaying strengths (for $z\neq 0$). E.g., $\mathrm{u}\left( z\right) =0$
only when $|z|\geq r$ for some $r\leq 2$, with $\mathrm{u}\left( 0\right) =0$
and $\mathrm{u}\left( z\right) <\mathrm{U}$. \medskip
\end{itemize}

It remains to fix the exchange strength function $\upsilon :\mathbb{Z}%
^{2}\rightarrow \mathbb{R}$ in (\ref{Hamiltonian-fb}), taking into account
the special choice (\ref{bipolaron exchange term2}) for the annihilation and
creation operators of a fermion pair of zero total spin. We already know
that the absolute value of the Fourier transform $\hat{\upsilon}$ of $%
\upsilon $ takes its maximum at the points $(-\pi ,0)$ and $(0,-\pi )$, but
its precise amplitude has to be still determined. This is performed
indirectly\ through a phenomenological relationship with the density of the
superconducting charge carriers (also named superfluid): From recent
experimental data \cite{Nature2016}, for optimum doping, around $90\%$ of
the charge carriers inserted via the doping do not form superfluid. If $\hat{%
\psi}_{K}\left( \mathrm{U}\right) $ and $-1$ are respectively the fermionic
and bosonic parts of the eigenvector $\Psi (\mathrm{U},k)\doteq (\hat{\psi}%
_{K}\left( \mathrm{U}\right) ,-1)$ associated with the eigenvalue $\mathrm{E}%
\left( \mathrm{U},K\right) $ and $K=(\pi ,0),(0,\pi )$, then we can
interpret 
\begin{equation*}
\varrho =\frac{100\%}{\Vert \hat{\psi}_{K}(\mathrm{U})\Vert _{2}^{2}+1}
\end{equation*}%
as the proportion of charge carriers forming JT bipolarons. Computing this
quantity, we can identify the unique value $\hat{\upsilon}(K)\simeq 0.11\ $%
\textrm{eV} making $\varrho =90\%$. Similar to \cite{31Muller,31Mullerbis},
note that we choose $\hat{v}$ of the form 
\begin{equation*}
\left[ \alpha \left( (k_{x}-\pi )^{2}+k_{y}^{2}\right) +1\right] ^{-1}\qquad 
\text{(resp.}\quad \left[ \alpha \left( k_{x}^{2}+(k_{y}-\pi )^{2}\right) +1%
\right] ^{-1}\text{)}
\end{equation*}%
for quasimomenta $(k_{x},k_{y})\in \mathbb{T}^{2}$ near $(\pi ,0)$ or $%
(0,\pi )$, where $\alpha >0$ determines the effective mass $m^{\ast \ast
\ast }$ of (dressed) bound fermion pairs. Conversely, $\alpha $ can be
recovered from $m^{\ast \ast \ast }$.

\begin{figure}[tbp]
\centering
\includegraphics[scale=0.5]{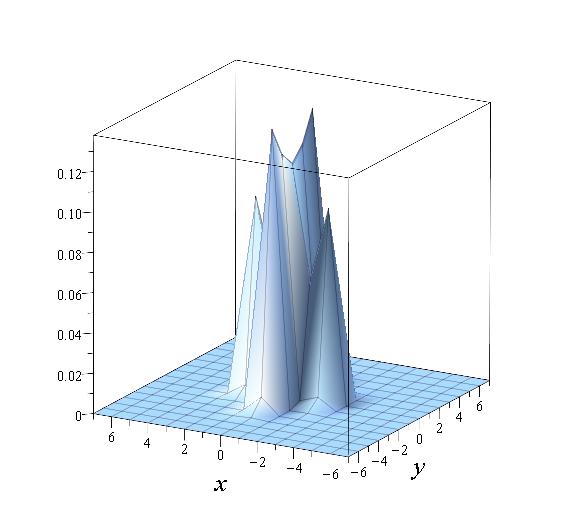}
\caption{Normalized density $|\mathcal{F}^{-1}[\hat{\protect\psi}_{1.461,(0,-%
\protect\pi )}]|^{2}$ of the dressed bound fermion pair as a function of the
(relative) position space at total quasimomentum $(0,-\protect\pi )$ for the
prototypical parameters. It is a reproduction of \protect\cite[Fig. 5]%
{articulo2}.}
\label{fig-new-JBW3}
\end{figure}

Using our mathematical results and rigorous numerical computations, we show
in \cite{articulo2} that the model gives the following quantitative
estimates in relation with properties of hole-doped cuprates $\mathrm{La}_{2}%
\mathrm{CuO}_{4}$ (LaSr 214 or LSCO) and $\mathrm{YBa}_{2}\mathrm{Cu}_{3}%
\mathrm{O}_{7}$ (YBCO):

\begin{itemize}
\item \textbf{Pairing symmetry.}

\begin{itemize}
\item Prediction: $16.5\%$ $s$-wave, $83.5\%$ $d$-wave, $0\%$ $p$-wave. See 
\cite[p. 10 and Corollary 1.1]{articulo2}

\item Experimental data: $\sim 20-25\%$ $s$-wave, $\sim 75-80\%$ $d$-wave, $%
\sim 0\%$ $p$-wave. Indirect measurement with rough estimates for the $s$%
-wave/$d$-wave ratio, see \cite{M07,M17}.
\end{itemize}

\item \textbf{Pseudogap temperature }$T_{\ast }$ (i.e., pair dissociation
energy).

\begin{itemize}
\item Prediction of the binding energy of ($d$-wave) pairs: $\mathrm{E}%
=1250\ \mathrm{K}$ found for the quasi-momenta $(-\pi ,0)$ and $(0,-\pi )$.
See Theorem \ref{Maintheorem1} (ii)--(iii) and \cite[Fig. 6]{articulo2}.

\item Experimental data:

\begin{itemize}
\item Experiments on cuprates demonstrates a pseudogap appears at $(-\pi ,0)$
and $(0,-\pi )$. See \cite[Fig. 4]{Nature2015} and references therein.

\item $T_{\ast }\simeq 100-750\ \mathrm{K}$, depending on the doping \cite[%
Fig. 26]{pseudogap}. E.g., $T_{\ast }\simeq 400\ \mathrm{K}$ around optimal
doping for $\mathrm{La}_{1.85}\mathrm{Sr}_{0.16}\mathrm{CuO}_{4}$ and $%
T_{\ast }\simeq 200\ \mathrm{K}$ for $\mathrm{La}_{1.8}\mathrm{Sr}_{0.2}%
\mathrm{CuO}_{4}$. The ratio between the theoretical bond energy (in $%
\mathrm{K}$) and the dissociation temperature\footnote{%
How to theoretically determine the dissociation temperature of a dressed
fermionic pair is not entirely clear to us. Clearly, this temperature must
be higher than the critical temperature $T_{c}$ and lower than the pseudogap
temperature $T_{\ast }$.} of dressed fermion pairs should be between $%
\mathrm{E}/T_{\ast }$ and $\mathrm{E}/T_{c}$. For $\mathrm{La}_{1.8}\mathrm{%
Sr}_{0.2}\mathrm{CuO}_{4}$, the coherence length of which perfectly matches
our prediction below, $\mathrm{E}/T_{\ast }\simeq 6.2$ and $\mathrm{E}%
/T_{c}=34.247$, with an average ratio of around $20$.

\item Binding energy of bipolarons \cite[Fig. 2]{bipolaron}: $1500\ \mathrm{K%
}$ at zero doping and $500\ \mathrm{K}$ at optimal doping for LaSr 214.

\item To compare with, for standard diatomic molecules the ratio between the
bond energy (in $\mathrm{K}$) and their dissociation temperatures ranges%
\footnote{%
For example, a quick internet search reveals that the dissociation
temperatures $T_{d}$ (in $\mathrm{K}$) and bond energies $E$ (in $\mathrm{K}$%
) of ten common diatomic molecules are as follows: $\mathrm{H}%
_{2}:T_{d}=4000\ \mathrm{K}$, $E=52438\ \mathrm{K}$, $E/T_{d}\simeq 13$; $%
\mathrm{N}_{2}:T_{d}=9500\ \mathrm{K}$, $E=1.133\,0\times 10^{5}\ \mathrm{K}$%
, $E/T_{d}\simeq 12$; $\mathrm{O}_{2}:T_{d}=6000\ \mathrm{K}$, $E=59895\ 
\mathrm{K}$, $E/T_{d}\simeq 10$; $\mathrm{F}_{2}:T_{d}=1300\ \mathrm{K}$, $%
E=19003\ \mathrm{K}$, $E/T_{d}\simeq 14.6$; $\mathrm{Cl}_{2}:T_{d}=1200\ 
\mathrm{K}$, $E=29226\ \mathrm{K}$, $E/T_{d}\simeq 24.3$; $\mathrm{Br}%
_{2}:T_{d}=800\ \mathrm{K}$, $E=23212\ \mathrm{K}$, $E/T_{d}\simeq 29$; $%
\mathrm{I}_{2}:T_{d}=700\ \mathrm{K}$, $E=18161\ \mathrm{K}$, $E/T_{d}\simeq
25.9$; $\mathrm{CO}:T_{d}=5000\ \mathrm{K}$, $E=51356\ \mathrm{K}$, $%
E/T_{d}\simeq 10.3$; $\mathrm{NO}:T_{d}=4100\ \mathrm{K}$, $E=75891\ \mathrm{%
K}$, $E/T_{d}\simeq 18.5$; $\mathrm{HCl}:T_{d}=3000\ \mathrm{K}$, $%
E=1.289\,3\times 10^{5}\ \mathrm{K}$, $E/T_{d}\simeq 43$.} from $10$ to $40$%
, with an average of around $20$.
\end{itemize}
\end{itemize}

\item \textbf{Superconducting coherence length }$\xi $ (i.e., pair radius).

\begin{itemize}
\item Prediction: $\xi _{a}=1.6\ \mathrm{nm}$ in one direction, $\xi
_{b}=2.1\ \mathrm{nm}$ in the orthogonal one at quasi-momenta $(-\pi ,0)$
and $(0,-\pi )$. See Fig. \ref{fig-new-JBW3} (in lattice units), reproducing 
\cite[Fig. 5]{articulo2}. It refers approximately to $6$ lattice sites in
one direction and $8$ lattice sites in the other one. Compare this result
with the exponential localization of the fermionic component $\mathcal{F}%
^{-1}[\hat{\psi}_{\mathrm{U},k}]$ of the eigenstate given by Theorem \ref%
{Maintheorem1} (iii).
\end{itemize}
\end{itemize}

\begin{itemize}
\item 
\begin{itemize}
\item Experimental data :

\begin{itemize}
\item $\xi _{ab}=1.6\ \mathrm{nm}$ is obtained for an optimally doped $%
\mathrm{YBa}_{2}\mathrm{Cu}_{3}\mathrm{O}_{6.9}$ for which $T_{c}=95\ 
\mathrm{K}$. See \cite[above the \textquotedblleft Summary and
conclusion\textquotedblright ]{Hwang}.

\item $\xi _{ab}=2.1\ \mathrm{nm}$ is obtained for $\mathrm{La}_{1.8}\mathrm{%
Sr}_{0.2}\mathrm{CuO}_{4}$ for which $T_{c}=36.5\ \mathrm{K}$. See \cite[%
Table II]{pseudogap2}.

\item $\xi _{ab}=3.8\ \mathrm{nm}$ is obtained for an optimally doped $%
\mathrm{La}_{1.85}\mathrm{Sr}_{0.16}\mathrm{CuO}_{4}$ for which $T_{c}=38\ 
\mathrm{K}$. See \cite[above the \textquotedblleft Summary and
conclusion\textquotedblright ]{Hwang}.

\item More generally $\xi _{ab}$ is in the nanometer range, between $1\ 
\mathrm{nm}$ and $3.8\ \mathrm{nm}$ for various other examples of $\mathrm{La%
}_{2}\mathrm{CuO}_{4}$ and $\mathrm{YBa}_{2}\mathrm{Cu}_{3}\mathrm{O}_{7}$.
See, e.g., \cite[Table 9.1]{unitcoherence} and \cite[Table 3.2 on page 60]%
{Moura}. The coherence length \textbf{is very small} compared to
conventional superconductors, for which it is generally several tens or
hundreds of nanometers. See, e.g., the table in (\ref{table}).
\end{itemize}
\end{itemize}
\end{itemize}

\noindent The pseudo-gap temperature is the temperature below which the
Fermi surface of a material exhibits a partial energy gap, in fact a gap in
a particular direction, as in the quasi-momenta $(-\pi ,0)$ and $(0,-\pi )$.
Compare with Theorem \ref{Maintheorem1} (iii). The 2-fermion 1-boson problem
studied here and in \cite{articulo2} cannot a priori explain the
superconducting phase, which is a \textbf{collective} phenomenon, but only
the pseudogap regime which is expected to be related to the formation of
fermion pairs (mainly for quasi-momenta $(-\pi ,0)$ and $(0,-\pi )$).

To conclude, this paper together with \cite{articulo,articulo2} contributes
a \textbf{mathematically rigorous} microscopic model for cuprate
superconductors that includes Jahn-Teller-type bipolarons with zero spin and
local repulsions. This model captures the following phenomenological aspects
of these materials:

\begin{itemize}
\item $d$-wave pairing not based on anisotropy.

\item Low density superconducting superfluid.

\item Pseudogap temperature.

\item Very accurate coherence lengths.

\item Solution to the \textquotedblleft large bipolaron mass vs. small mass
of superconducting carrier pairs\textquotedblright .
\end{itemize}

\noindent In addition, as proven in the Ph.D. thesis \cite{Antoniothesis},
in a mean-field-like approximation, the many body version of the model
considered here also explains another very special feature of cuprate
superconductors, namely the density waves \cite{denswave}. We therefore
think that the model we present here deserves to be studied in much more
detail, in view of a microscopic theory for cuprate superconductivity.

\subsection{The Fock-Space Formalism\label{Section Fock}}

In quantum mechanics, one generally starts with a (one-particle) Hilbert
space $\mathfrak{h}$, often realized as a space $L^{2}(\mathcal{M})$ of
square-integrable, complex-valued functions on a measure space $(\mathcal{M},%
\mathfrak{a})$. The states ofa quantum system of $n\in \mathbb{N}$ quantum
particles are then represented within the $n$-fold tensor product $\mathfrak{%
h}^{\otimes n}$ of $\mathfrak{h}$. However, identical quantum particles are
indistinguishable, meaning that they cannot be differentiated from one
another, not even in principle. In this situation, the states of these
indistinguishable quantum particles are only taken from a subspace of $%
\mathfrak{h}^{\otimes n}$.

Recall meanwhile that all quantum particles possess an intrinsic form of
angular momentum known as spin, characterized by a quantum number $\mathfrak{%
s}\in \mathbb{N}/2$. If $\mathfrak{s}$ is half-integer, then the
corresponding particles are named fermions, otherwise we have bosons. By the
celebrated spin-statistics theorem, fermionic wave functions are
antisymmetric with respect to permutations of particles, whereas the bosonic
ones are symmetric. The states of a system of $n\in \mathbb{N}$ fermions
correspond then to vectors in the subspace $\wedge ^{n}\mathfrak{h}$ of
totally antisymmetric $n$-particle wave functions in $\mathfrak{h}^{\otimes
n}$, while the states of a system of $n\in \mathbb{N}$ bosons are vectors in
the subspace $\vee ^{n}\mathfrak{h}^{n}$of totally symmetric $n$-particle
wave functions in $\mathfrak{h}^{\otimes n}$.

In most many-body quantum systems, the exact number of particles is not
known. In quantum statistical mechanics, physical properties are typically
studied in the limit $n\rightarrow \infty $ of infinite number of particles.
Quantum field theory deals with situations where the particle number and
species vary with time. The so-called Fock spaces are used to encode both
situations. For fermionic systems, the Fock space is, by definition, the
Hilbert space 
\begin{equation}
\mathfrak{F}_{-}\equiv \mathfrak{F}\left( \mathfrak{h}\right) \doteq
\bigoplus_{n=0}^{\infty }\wedge ^{n}\mathfrak{h}\ ,\qquad \wedge ^{0}%
\mathfrak{h}\doteq \mathbb{C},  \label{Fock f}
\end{equation}%
while, for bosonic systems, 
\begin{equation}
\mathfrak{F}_{+}\equiv \mathfrak{F}\left( \mathfrak{h}\right) \doteq
\bigoplus_{n=0}^{\infty }\vee ^{n}\mathfrak{h}\ ,\qquad \vee ^{0}\mathfrak{h}%
\doteq \mathbb{C}\ .  \label{Fock b}
\end{equation}%
The respective scalar products are denoted by $\left\langle \cdot ,\cdot
\right\rangle _{\mathfrak{F}_{\pm }}$. The two scalar products are the sum
over $n\in \mathbb{N}$ of each canonical scalar product on the sector $%
\wedge ^{n}\mathfrak{h}$ and $\vee ^{n}\mathfrak{h}$, respectively. In both
cases, we denote the vacuum state by $\Omega \doteq (1,0,\ldots )$.

The Fock space proved very useful, not least because it allows so-called
creation and annihilation operators: \medskip

\noindent \textbf{Fermionic case.} The annihilation operator $a\left(
\varphi \right) \in \mathcal{B}(\mathfrak{F}_{-})$ of a fermion with wave
function $\varphi \in \mathfrak{h}$ is the (linear) bounded operator
uniquely defined by the conditions $a\left( \varphi \right) \Omega =0$ and%
\begin{equation}
a\left( \varphi \right) \left( \psi _{1}\wedge \cdots \wedge \psi
_{n}\right) \doteq \frac{\sqrt{n}}{n!}\sum_{\pi \in \Pi _{n}}\mathrm{sgn}%
(\pi )\left\langle \varphi ,\psi _{\pi \left( 1\right) }\right\rangle _{%
\mathfrak{h}}\psi _{\pi \left( 2\right) }\wedge \cdots \wedge \psi _{\pi
\left( n\right) }  \label{fermion operator}
\end{equation}%
for any $n\in \mathbb{N}$ and $\psi _{1},\ldots ,\psi _{n}\in \mathfrak{h}$,
where $\Pi _{n}$ is the set of all permutations $\pi $ of $n$\ elements and $%
\mathrm{sgn}:\Pi _{n}\rightarrow \{-1,1\}$ denotes the sign of these
permutations, while $\Omega =(1,0,0,\ldots )$ is the vacuum state and $\psi
_{1}\wedge \cdots \wedge \psi _{n}$ is the orthogonal projection of $\psi
_{1}\otimes \cdots \otimes \psi _{n}\in \mathfrak{h}^{\otimes n}$ onto the
subspace of antisymmetric $n$--particle wave functions: 
\begin{equation*}
\psi _{1}\wedge \cdots \wedge \psi _{n}\doteq \frac{1}{n!}\sum_{\pi \in \Pi
_{n}}\mathrm{sgn}(\pi )\psi _{\pi \left( 1\right) }\otimes \cdots \otimes
\psi _{\pi \left( n\right) }\in \wedge ^{n}\mathfrak{h}\ .
\end{equation*}%
The creation operator of a fermion with wave function $\varphi \in \mathfrak{%
h}$ is the adjoint $a^{\ast }(\varphi )\doteq a\left( \varphi \right) ^{\ast
}$ of $a\left( \varphi \right) $, namely $a^{\ast }(\varphi )\Omega =\varphi 
$ and 
\begin{equation}
a^{\ast }(\varphi )(\psi _{1}\wedge \cdots \wedge \psi _{n})=\sqrt{n+1}%
\,\varphi \wedge \psi _{1}\wedge \cdots \wedge \psi _{n}\ .
\label{fermion operator creat}
\end{equation}%
Such operators are known to satisfy the so-called Canonical Anticommutation
Relations (CAR): For all $\varphi _{1},\varphi _{2}\in \mathfrak{h}$,%
\begin{equation*}
a(\varphi _{1})a(\varphi _{2})+a(\varphi _{2})a(\varphi _{1})=0,\quad
a(\varphi _{1})a(\varphi _{2})^{\ast }+a(\varphi _{2})^{\ast }a(\varphi
_{1})=\langle \varphi _{1},\varphi _{2}\rangle _{\mathfrak{h}}\mathfrak{1}\ .
\end{equation*}%
See \cite[p. 10]{BratteliRobinson}. Here, $\mathfrak{1}$ stands for the
identity operator on the Fock space $\mathfrak{F}_{-}$. \medskip

\noindent \textbf{Bosonic case.} The annihilation operator $b\left( \varphi
\right) $ of a boson with wave function $\varphi \in \mathfrak{h}$ is the
(linear) unbounded operator acting on $\mathfrak{F}_{+}$ and uniquely
defined by the conditions $b\left( \varphi \right) \Omega =0$ and%
\begin{equation}
b\left( \varphi \right) \left( \psi _{1}\vee \cdots \vee \psi _{n}\right)
\doteq \frac{\sqrt{n}}{n!}\sum_{\pi \in \Pi _{n}}\left\langle \varphi ,\psi
_{\pi \left( 1\right) }\right\rangle _{\mathfrak{h}}\psi _{\pi \left(
2\right) }\vee \cdots \vee \psi _{\pi \left( n\right) }
\label{boson operator}
\end{equation}%
for any $n\in \mathbb{N}$ and $\psi _{1},\ldots ,\psi _{n}\in \mathfrak{h}$,
where $\psi _{1}\vee \cdots \vee \psi _{n}$ is the orthogonal projection of $%
\psi _{1}\otimes \cdots \otimes \psi _{n}\in \mathfrak{h}^{\otimes n}$ onto
the subspace of symmetric $n$--particle wave functions: 
\begin{equation*}
\psi _{1}\vee \cdots \vee \psi _{n}\doteq \frac{1}{n!}\sum_{\pi \in \Pi
_{n}}\psi _{\pi \left( 1\right) }\otimes \cdots \otimes \psi _{\pi \left(
n\right) }\in \vee ^{n}\mathfrak{h}\ .
\end{equation*}%
As in the fermionic case, the creation operator of a boson with wave
function $\varphi \in \mathfrak{h}$ is the adjoint $b^{\ast }(\varphi
)\doteq b\left( \varphi \right) ^{\ast }$ of $b\left( \varphi \right) $,
where $b^{\ast }(\varphi )\Omega =\varphi $ and 
\begin{equation}
b^{\ast }(\varphi )(\psi _{1}\vee \cdots \vee \psi _{n})=\sqrt{n+1}\,\varphi
\vee \psi _{1}\vee \cdots \vee \psi _{n}\ .  \label{boson operatorcreat}
\end{equation}%
Such operators are known to satisfy the so-called Canonical Commutation
Relations (CCR): For all $\varphi _{1},\varphi _{2}\in \mathfrak{h}$,%
\begin{equation*}
b(\varphi _{1})b(\varphi _{2})-b(\varphi _{2})b(\varphi _{1})=0,\quad
b(\varphi _{1})b(\varphi _{2})^{\ast }-b(\varphi _{2})^{\ast }b(\varphi
_{1})=\langle \varphi _{1},\varphi _{2}\rangle _{\mathfrak{h}}\mathfrak{1}\ .
\end{equation*}%
See \cite[p. 10]{BratteliRobinson}. Here, $\mathfrak{1}$ stands again for
the identity operator on the Fock space $\mathfrak{F}_{+}$.\medskip

The interest of Fock spaces lies in the use of creation and annihilation
operators, which not only give a mathematically rigorous definition for
precesses of creation or annihilation of physical particles, but also
possess essential algebraic properties: the CAR and CCR relations given
above. Although Fock spaces and the creation and annihilation operators are
not strictly necessary for our proofs, we use them in this paper because
they allow us to define the model in a very intuitive way, which makes its
physical meaning easy to understand once the Fock-space formulation is
familiar.

\subsection{Non-autonomous Evolution Equations and Scattering Theory}

This section collects simple results on wave and scattering operators (\ref%
{scattering operator1})--(\ref{scaterring operator}) for bounded
Hamiltonians, related to their approximation by Dyson series. We start with
the following elementary lemma, resulting from the theory of non-autonomous
evolution equations:

\begin{lemma}[Finite-time scattering and wave operators]
\label{finite-time scattering and wave operators copy(1)}\mbox{ }\newline
For any self-adjoint $X,Y\in \mathcal{B}\left( \mathcal{X}\right) $ acting
on a Hilbert space $\mathcal{X}$ and all $s,t\in \mathbb{R}$,%
\begin{equation}
\mathrm{e}^{itX}\mathrm{e}^{i\left( s-t\right) \left( X+Y\right) }\mathrm{e}%
^{-isX}=\mathbf{1}+\sum_{n=1}^{\infty }\left( -i\right) ^{n}\int_{s}^{t}%
\mathrm{d}\tau _{1}\cdots \int_{s}^{\tau _{n-1}}\mathrm{d}\tau _{n}Y_{\tau
_{1}}\cdots Y_{\tau _{n}}  \label{dyson series}
\end{equation}%
with $(Y_{t})_{t\in \mathbb{R}}\subseteq \mathcal{B}(\mathcal{X})$ being the
norm-continuous family 
\begin{equation}
Y_{t}\doteq \mathrm{e}^{itX}Y\mathrm{e}^{-itX}\ ,\qquad t\in \mathbb{R}\ .
\label{sdsds}
\end{equation}
\end{lemma}

\begin{proof}
We compute that, for any $s,t\in \mathbb{R}$,%
\begin{equation*}
\partial _{t}\left\{ \mathrm{e}^{itX}\mathrm{e}^{i\left( s-t\right) \left(
X+Y\right) }\mathrm{e}^{-isX}\right\} =-i\left( \mathrm{e}^{itX}Y\mathrm{e}%
^{-itX}\right) \left( \mathrm{e}^{itX}\mathrm{e}^{i\left( s-t\right) \left(
X+Y\right) }\mathrm{e}^{-isX}\right)
\end{equation*}%
as well as%
\begin{equation*}
\partial _{s}\left\{ \mathrm{e}^{itX}\mathrm{e}^{i\left( s-t\right) \left(
X+Y\right) }\mathrm{e}^{-isX}\right\} =\left( \mathrm{e}^{itX}\mathrm{e}%
^{i\left( s-t\right) \left( X+Y\right) }\mathrm{e}^{-isX}\right) \left( i%
\mathrm{e}^{isX}Y\mathrm{e}^{-isX}\right)
\end{equation*}%
both in $\mathcal{B}(\mathcal{X})$. In other words, the family%
\begin{equation}
V_{t,s}\doteq \mathrm{e}^{itX}\mathrm{e}^{i\left( s-t\right) \left(
X+Y\right) }\mathrm{e}^{-isX}\ ,\qquad s,t\in \mathbb{R}\ ,  \label{ssssssss}
\end{equation}%
of (uniformly) bounded operators is a norm-continuous two-parameter family
of unitary operators solving the non-autonomous evolution equations%
\begin{equation}
\forall s,t\in \mathbb{R}:\qquad \partial _{t}Z_{t,s}=-iY_{t}Z_{t,s}\
,\qquad \partial _{s}Z_{t,s}=iZ_{t,s}Y_{s}\ ,\qquad Z_{s,s}=\mathbf{1}\ ,
\label{sssss1}
\end{equation}%
in $\mathcal{B}(\mathcal{X})$, where $(Y_{t})_{t\in \mathbb{R}}\subseteq 
\mathcal{B}(\mathcal{X})$ is the norm-continuous family defined by (\ref%
{sdsds}). As is well-known, there is a unique solution to this
non-autonomous evolution equations (\ref{sssss1}), which is given by the
Dyson series (\ref{dyson series}). This series is absolutely summable in $%
\mathcal{B}(\mathcal{X})$. Notice that the integrals appearing in it are
Riemann integrals, for their arguments are continuous functions taking
values in a Banach space.
\end{proof}

\begin{corollary}[Approximation of scattering and wave operators]
\label{finite-time scattering and wave operators copy(2)}\mbox{ }\newline
Let $X,Y\in \mathcal{B}\left( \mathcal{X}\right) $ be two self-adjoint
operators acting on a Hilbert space $\mathcal{X}$. Assume that the waves
operators 
\begin{equation*}
W^{\pm }\left( X+Y,X\right) \doteq s-{\lim\limits_{t\rightarrow \pm \infty }}%
\mathrm{e}^{it\left( X+Y\right) }\mathrm{e}^{-itX}P_{\mathrm{ac}}\left(
X\right)
\end{equation*}%
exist. Let $\varepsilon \in \mathbb{R}^{+}$. Then:

\begin{enumerate}
\item[i.)] For any $\varphi \in \mathrm{ran}\left( P_{\mathrm{ac}}\left(
X\right) \right) $, there is $T>0$ such that%
\begin{equation*}
T<t\implies \left\Vert \left( W^{+}\left( X+Y,X\right) -V_{0,t}\right)
\varphi \right\Vert _{\mathcal{X}}\leq \varepsilon \ ,
\end{equation*}%
whereas 
\begin{equation*}
t<-T\implies \left\Vert \left( W^{-}\left( X+Y,X\right) -V_{0,t}\right)
\varphi \right\Vert _{\mathcal{X}}\leq \varepsilon \ .
\end{equation*}

\item[ii.)] For any $\varphi ,\psi \in \mathrm{ran}\left( P_{\mathrm{ac}%
}\left( X\right) \right) $, there is $T>0$ such that%
\begin{equation*}
\left\langle \psi ,S\left( X+Y,X\right) \varphi \right\rangle _{\mathcal{X}%
}=\left\langle \psi ,W^{+}\left( X+Y,X\right) ^{\ast }W^{-}\left(
X+Y,X\right) \varphi \right\rangle _{\mathcal{X}}=\left\langle \psi
,V_{t,s}\varphi \right\rangle _{\mathcal{X}}+\mathcal{O}\left( \varepsilon
\right) \ 
\end{equation*}%
uniformly for $s<-T<T<t$.
\end{enumerate}

Here, 
\begin{equation*}
V_{t,s}\doteq \mathbf{1}+\sum_{n=1}^{\infty }\left( -i\right)
^{n}\int_{s}^{t}\mathrm{d}\tau _{1}\cdots \int_{s}^{\tau _{n-1}}\mathrm{d}%
\tau _{n}Y_{\tau _{1}}\cdots Y_{\tau _{n}}\ ,
\end{equation*}%
the norm-continuous family $(Y_{t})_{t\in \mathbb{R}}\subseteq \mathcal{B}(%
\mathcal{X})$ being defined by (\ref{sdsds}).
\end{corollary}

\begin{proof}
Assertion (i) is a direct consequence of Lemma \ref{finite-time scattering
and wave operators copy(1)}. Concerning the scattering operator, remark in
particular that, for any $r,s,t\in \mathbb{R}$, $V_{t,s}V_{s,r}=V_{t,r}$ and 
$V_{t,s}^{\ast }=V_{s,t}$. Given $\varphi ,\psi \in \mathrm{ran}\left( P_{%
\mathrm{ac}}\left( Y\right) \right) $, we have that 
\begin{multline*}
\left\vert \left\langle \psi ,\left( S\left( X+Y,X\right) -V_{t,s}\right)
\varphi \right\rangle _{\mathcal{X}}\right\vert \\
=\left\vert \left\langle \psi ,\left( W^{+}\left( X+Y,X\right) ^{\ast
}W^{-}\left( X+Y,X\right) -V_{0,t}^{\ast }V_{0,s}\right) \varphi
\right\rangle _{\mathcal{X}}\right\vert \\
=\left\vert \left\langle \psi ,\left( W^{+}\left( X+Y,X\right)
-V_{0,t}\right) ^{\ast }W^{-}\left( X+Y,X\right) \varphi \right\rangle _{%
\mathcal{X}}\right. \\
\left. +\left\langle \psi ,V_{0,t}^{\ast }\left( W^{-}\left( X+Y,X\right)
-V_{0,s}\right) \varphi \right\rangle _{\mathcal{X}}\right\vert \\
\leq \left\Vert \left( W^{+}\left( X+Y,X\right) -V_{0,t}\right) \psi
\right\Vert _{\mathcal{X}}\left\Vert W^{-}\left( X+Y,X\right) \varphi
\right\Vert _{\mathcal{X}} \\
+\left\Vert \left( W^{-}\left( X+Y,X\right) -V_{0,s}\right) \varphi
\right\Vert _{\mathcal{X}}\left\Vert \psi \right\Vert _{\mathcal{X}}\ .
\end{multline*}%
Assertion (ii) therefore follows from assertion (i).
\end{proof}

\subsection{Constant Fiber Direct Integrals\label{Constant fiber direct
integral}}

For more details, we refer to \cite[Section XIII.16]{ReedSimonIV} as well as 
\cite{Niesen-direct-integrals} for the general theory.

Let $(\mathcal{X},\mu )$ be any semifinite measure space and $\mathcal{Y}$
any separable Hilbert space. The constant fiber direct integral\ of $%
\mathcal{Y}$ over $\mathcal{X}$ is, by definition, the Hilbert space%
\begin{equation*}
\int_{\mathcal{X}}^{\oplus }\mathcal{Y}\,\mu \left( \mathrm{d}x\right)
\equiv L^{2}\left( \mathcal{X},\mathcal{Y},\mu \right) \doteq \left\{ F\in 
\mathcal{Y}^{\mathcal{X}}\,:\,\left\Vert F\left( \cdot \right) \right\Vert _{%
\mathcal{Y}}^{2}\in L^{1}\left( \mathcal{X},\mu \right) \right\}
\end{equation*}%
of equivalence classes of square-integrable $\mathcal{Y}$--valued functions
with scalar product\footnote{%
The scalar product is well-defined, by the polarization identity and the
Cauchy-Schwarz inequality. See for instance \cite[Section 7.3.2]%
{Bru-Pedra-livre}.}%
\begin{equation*}
\left\langle \varphi ,\psi \right\rangle \equiv \left\langle \varphi ,\psi
\right\rangle _{L^{2}\left( \mathcal{X},\mathcal{Y},\mu \right) }\doteq
\int_{\mathcal{X}}\left\langle \varphi \left( x\right) ,\psi \left( x\right)
\right\rangle _{\mathcal{Y}}\,\mu \left( \mathrm{d}x\right) \ ,\qquad
\varphi ,\psi \in L^{2}\left( \mathcal{X},\mathcal{Y},\mu \right) \ ,
\end{equation*}%
and the pointwise vector space operations 
\begin{equation*}
\left( \varphi +\psi \right) \left( x\right) =\varphi \left( x\right) +\psi
\left( x\right) \ ,\text{$\qquad $}\left( \text{$\alpha $}\varphi \right)
\left( x\right) =\text{$\alpha $}\varphi \left( x\right) \ ,\qquad \text{$%
\alpha \in \mathbb{C}$},\ \varphi ,\psi \in L^{2}\left( \mathcal{X},\mathcal{%
Y},\mu \right) \ .
\end{equation*}%
If $\mathcal{Y}=\mathbb{C}$ then we use the shorter notation $L^{2}\left( 
\mathcal{X},\mu \right) \equiv L^{2}\left( \mathcal{X},\mathbb{C},\mu
\right) $.

A mapping $A:\mathcal{X}\rightarrow \mathcal{B}(\mathcal{Y})$ is strongly
measurable whenever the mapping $x\mapsto \left\langle \varphi ,A(x)\psi
\right\rangle _{\mathcal{Y}}$ from $\mathcal{X}$ to $\mathbb{C}$ is
measurable for all $\varphi ,\psi \in \mathcal{Y}$. Let $L^{\infty }(%
\mathcal{X},\mathcal{Y},\mu )$ be the $C^{\ast }$-algebra of equivalence
classes of strongly measurable functions $A:\mathcal{X}\rightarrow \mathcal{B%
}(\mathcal{Y})$ with%
\begin{equation}
\left\Vert A\right\Vert _{\infty }\doteq \mathrm{ess-}\sup \left\{
\left\Vert A\left( x\right) \right\Vert _{\mathrm{op}}:x\in \mathcal{X}%
\right\} <\infty \ .  \label{norm}
\end{equation}%
Here, $\mathrm{ess-}\sup $ denotes the essential supremum and $\left\Vert
\cdot \right\Vert _{\mathrm{op}}$ stands for the operator norm. If $\mathcal{%
Y}=\mathbb{C}$ then we use the shorter notation $L^{\infty }\left( \mathcal{X%
},\mu \right) \equiv L^{\infty }\left( \mathcal{X},\mathbb{C},\mu \right) $.

A bounded operator $D$ on $L^{2}(\mathcal{X},\mathcal{Y},\mu )$ is
decomposable if there is $A\in L^{\infty }(\mathcal{X},\mathcal{Y},\mu )$
such that, for all $\psi \in L^{2}(\mathcal{X},\mathcal{Y},\mu )$,%
\begin{equation*}
\left( D\psi \right) \left( x\right) =A\left( x\right) \psi \left( x\right)
\ ,\qquad x\in \mathcal{X}\text{\quad (}\mu \text{-a.e.)}\ .
\end{equation*}%
If such an $A$ exists, then it is unique. Moreover, the mapping $A\mapsto D$
defined by the above equality is a $\ast $-homomorphism which is isometric.
See \cite[Theorem XIII.83]{ReedSimonIV}. The operators $A(x)\in \mathcal{B}(%
\mathcal{Y})$, $x\in \mathcal{X}$, are called the fibers of $D$ and we write%
\begin{equation*}
D=\int_{\mathcal{X}}^{\oplus }A\left( x\right) \,\mu \left( \mathrm{d}%
x\right) \text{ }.
\end{equation*}%
For the reader's convenience, we now give three essential properties of
decomposable operators used in the paper, referring to \cite[Theorem XIII.85
(a), (c) and (d)]{ReedSimonIV}. Note that $\sigma (X)$ denotes below the
spectrum of any operator $X$ acting on some Hilbert space, as is usual.

\begin{theorem}[Properties of decomposable operators]
\label{borelian functional calculus of a direct integral of operators}%
\mbox{
}\newline
Let $D$ be a decomposable operator on $L^{2}(\mathcal{X},\mathcal{Y},\mu )$,
the fibers $A(x)\in \mathcal{B}(\mathcal{Y})$, $x\in \mathcal{X}$, of which
are all self-adjoint. Then:

\begin{enumerate}
\item[i.)] $D$ is self-adjoint.

\item[ii.)] $\lambda \in \sigma (D)$ iff, for all $\varepsilon \in \mathbb{R}%
^{+}$,%
\begin{equation*}
\mu \left( \left\{ x\in \mathcal{X}:\sigma \left( A\left( x\right) \right)
\cap \left( \lambda -\varepsilon ,\lambda +\varepsilon \right) \neq
\emptyset \right\} \right) >0\text{ }.
\end{equation*}

\item[iii.)] For any bounded Borel function $f$ on $\mathbb{R}$, $f(D)$ is
decomposable and has fibers $f(A(x))$, $x\in \mathcal{X}$, that is,%
\begin{equation*}
f(D)=\int_{\mathcal{X}}^{\oplus }f\left( A\left( x\right) \right) \,\mu
\left( \mathrm{d}x\right) {\ .}
\end{equation*}
\end{enumerate}
\end{theorem}

The above theorem can be used to elegantly prove the following well-known
results about multiplication operators $M_{\varphi }$ by any bounded
measurable function $\varphi \in L^{\infty }(\mathcal{X},\mu )$, defined for
any $\psi \in L^{2}(\mathcal{X},\mu )$, by 
\begin{equation*}
\left( M_{\varphi }\psi \right) \left( x\right) =\varphi \left( x\right)
\psi \left( x\right) {\ ,}\qquad x\in \mathcal{X}\text{\quad (}\mu \text{%
-a.e.)}\ .
\end{equation*}

\begin{corollary}[Properties of multiplication operators]
\label{spectrum of a multiplication operator}\mbox{ }\newline
The multiplication operators $M_{\varphi }$ by $\varphi \in L^{\infty }(%
\mathcal{X},\mu )$ has the following properties:

\begin{enumerate}
\item[i.)] For any bounded Borel function $f$ on $\mathbb{R}$, one has $%
f(M_{\varphi })=M_{f\circ \varphi }$.

\item[ii.)] $\sigma (M_{\varphi })$ is the essential range $\mathrm{ess-im}%
(\varphi )$ of $\varphi $.

\item[iii.)] Its operator norm $\Vert M_{\varphi }\Vert _{\mathrm{op}}$ is
equal to $\Vert \varphi \Vert _{\infty }$.
\end{enumerate}
\end{corollary}

\begin{proof}
Noting that $M_{\varphi }$ is a decomposable operator on 
\begin{equation*}
L^{2}(\mathcal{X},\mu )=\int_{\mathcal{X}}^{\oplus }\mathbb{C}\,\mu \left( 
\mathrm{d}x\right)
\end{equation*}%
with $\varphi (x)$, seen as a linear operator on $\mathbb{C}$, being its
fibers, we can use Theorem \ref{borelian functional calculus of a direct
integral of operators} (iii) to get the equality 
\begin{equation*}
f(M_{\varphi })=\int_{\mathcal{X}}^{\oplus }f\left( \varphi \left( x\right)
\right) \,\mu \left( \mathrm{d}x\right) =M_{f\circ \varphi }\text{ }.
\end{equation*}%
This proves Assertion (i). For the second one, we use that $\sigma (\varphi
(x))=\{\varphi (x)\}$ for all $x\in \mathcal{X}$ and thus infer from Theorem %
\ref{borelian functional calculus of a direct integral of operators} (ii)
that $\lambda \in \sigma (M_{f})$ iff, for all $\varepsilon \in \mathbb{R}%
^{+}$, 
\begin{equation*}
\mu \left( \left\{ x\in \mathcal{X}:\left\vert \lambda -\varphi \left(
x\right) \right\vert <\varepsilon \right\} \right) >0\text{ }.
\end{equation*}%
In other words, one gets Assertion (ii). Assertion (iii) is an elementary
application of \cite[Theorem XIII.83]{ReedSimonIV}.
\end{proof}

Below we study the special case of multiplication operators on $L^{2}(%
\mathbb{T}^{2},\nu )$, where $\mathbb{T}^{2}\doteq \lbrack -\pi ,\pi )^{2}$
is the torus and $\nu $ is the normalized Haar measure (\ref{Haar mesure})
on $\mathbb{T}^{2}$. It is again an elementary result, used in the paper. To
this end, we recall that, for any self-adjoint operator $Y$ acting on a
Hilbert space $\mathcal{Y}$, $P_{\mathrm{ac}}(Y)$ is the orthogonal
projection onto the absolutely continuous space of $Y$, which is defined by (%
\ref{abs cont space}).

\begin{corollary}[Absolutely continuous space of multiplication operators on 
$L^{2}(\mathbb{T}^{2},\protect\nu )$]
\label{example of absolutely continuous space}\mbox{ }\newline
Let $\varphi :\mathbb{T}^{2}\rightarrow \mathbb{R}$ be a bounded Borel
function with the property that, for every Borel set $\Omega \subseteq 
\mathbb{R}$ with zero Lebesgue measure, one has $\nu (\varphi ^{-1}(\Omega
))=0$\footnotetext{%
It occurs, for instance, if $\varphi $ is non-constant and real analytic on $%
\mathbb{S}^{2}\subseteq \mathbb{R}^{2}$.}. Then, $P_{\mathrm{ac}}\left(
M_{\varphi }\right) =\mathfrak{1}$, i.e., 
\begin{equation*}
\mathrm{ran}\left( P_{\mathrm{ac}}\left( M_{\varphi }\right) \right) =L^{2}(%
\mathbb{T}^{2},\nu )\ .
\end{equation*}
\end{corollary}

\begin{proof}
Given any Borel set $\Omega \subseteq \mathbb{R}$, we deduce from Corollary %
\ref{spectrum of a multiplication operator} (i) that%
\begin{equation*}
\chi _{\Omega }\left( M_{\varphi }\right) =M_{\chi _{\Omega }\circ \varphi
}=M_{\chi _{\varphi ^{-1}\left( \Omega \right) }}\ ,
\end{equation*}%
which in turn implies that, for any $\psi \in L^{2}(\mathbb{T}^{2},\nu )$,%
\begin{equation*}
\left\langle \psi ,\chi _{\Omega }\left( M_{\varphi }\right) \psi
\right\rangle =\int_{\varphi ^{-1}\left( \Omega \right) }\left\vert \psi
\left( k\right) \right\vert ^{2}\nu \left( \mathrm{d}k\right) \ .
\end{equation*}%
Hence, if $\Omega \subseteq \mathbb{R}$ has zero Lebesgue measure, then
under the conditions of the corollary, 
\begin{equation*}
\left\langle \psi ,\chi _{\Omega }\left( M_{\varphi }\right) \psi
\right\rangle =0\ .
\end{equation*}%
In other words, for any $\psi \in L^{2}(\mathbb{T}^{2},\nu )$, $\langle \psi
,\chi _{(\cdot )}(M_{\varphi })\psi \rangle $ is absolutely continuous with
respect to the Lebesgue measure.
\end{proof}

We next provide a result on the strong operator convergence and a version of
Fubini's theorem for (constant fiber) direct integrals, which are also used
in our proofs.

\begin{proposition}[Strong operator convergence]
\label{pointwise convergence theorem for direct integral}\mbox{ }\newline
Let $(A_{n})_{n\in \mathbb{N}}$ be any bounded sequence in $L^{\infty }(%
\mathcal{X},\mathcal{Y},\mu )$. If 
\begin{equation*}
s-\lim_{n\rightarrow \infty }A_{n}\left( x\right) =A\left( x\right) ,\qquad
x\in \mathcal{X}\text{ },
\end{equation*}%
then $A\in L^{\infty }(\mathcal{X},\mathcal{Y},\mu )$ and%
\begin{equation*}
s-\lim_{n\rightarrow \infty }\int_{\mathcal{X}}^{\oplus }A_{n}\left(
x\right) \,\mu \left( \mathrm{d}x\right) =\int_{\mathcal{X}}^{\oplus
}A\left( x\right) \,\mu \left( \mathrm{d}x\right) \ .
\end{equation*}
\end{proposition}

\begin{proof}
The assertion is well-known, but for the reader's convenience we give here
its complete proof. For any $\varphi ,\psi \in \mathcal{Y}$, it follows from
the fact that $A_{n}(x)\psi \rightarrow A(x)\psi $ everywhere and the
continuity of $\left\langle \varphi ,\cdot \right\rangle \in \mathcal{Y}%
^{\ast }$ that%
\begin{equation*}
{\lim_{n\rightarrow \infty }}\,\left\langle \varphi ,A_{n}\left( x\right)
\psi \right\rangle _{\mathcal{Y}}=\left\langle \varphi ,A\left( x\right)
\psi \right\rangle _{\mathcal{Y}}\text{ },\qquad x\in \mathcal{X}\text{ }.
\end{equation*}%
This shows that $A$ is strongly measurable, because the pointwise limit of a
sequence of real-valued measurable functions is measurable as well. Now, let 
\begin{equation*}
M\doteq \sup_{n\in \mathbb{N}}\left\Vert A_{n}\right\Vert _{\infty }<\infty
\ .
\end{equation*}%
For $\mu $-a.e. $x\in \mathcal{X}$ and any $\varphi \in \mathcal{Y}$,%
\begin{equation}
\left\Vert A_{n}\left( x\right) \varphi \right\Vert _{\mathcal{Y}}\leq
\left\Vert A_{n}\left( x\right) \right\Vert _{\mathrm{op}}\left\Vert \varphi
\right\Vert _{\mathcal{Y}}\leq M\left\Vert \varphi \right\Vert _{\mathcal{Y}}%
\text{ },\qquad n\in \mathbb{N}\text{ }.  \label{ererere1}
\end{equation}%
Taking the limit $n\rightarrow \infty $, one thus gets that, for $\mu $-a.e. 
$x\in \mathcal{X}$ and any $\varphi \in \mathcal{Y}$,%
\begin{equation}
\left\Vert A\left( x\right) \varphi \right\Vert _{\mathcal{Y}%
}=\lim_{n\rightarrow \infty }\left\Vert A_{n}\left( x\right) \varphi
\right\Vert _{\mathcal{Y}}\leq M\left\Vert \varphi \right\Vert _{\mathcal{Y}}%
\text{ }.  \label{ererere2}
\end{equation}%
Hence, $M$ is an essential upper bound for $\{\Vert A(x)\Vert _{\mathrm{op}%
}\}_{x\in \mathcal{X}}$ and, therefore, $A\in L^{\infty }(\mathcal{X},%
\mathcal{Y},\mu )$. Finally, given any element $\varphi \in L^{2}(\mathcal{X}%
,\mathcal{Y},\mu )$, by (\ref{ererere1})--(\ref{ererere2}) and the triangle
inequality, we have the estimate%
\begin{equation*}
\left\Vert A_{n}\left( x\right) \varphi \left( x\right) -A\left( x\right)
\varphi \left( x\right) \right\Vert _{\mathcal{Y}}\leq 2M\left\Vert \varphi
\left( x\right) \right\Vert _{\mathcal{Y}}{\ ,}\qquad x\in \mathcal{X}\text{%
\quad (}\mu \text{-a.e.)}\ .
\end{equation*}%
Since $A_{n}\left( x\right) \varphi \left( x\right) \rightarrow A\left(
x\right) \varphi \left( x\right) $ for all $x\in \mathcal{X}$, we can
therefore apply Lebesgue's dominated convergence theorem to conclude that,
for any $\varphi \in L^{2}(\mathcal{X},\mathcal{Y},\mu )$, 
\begin{equation*}
\lim_{n\rightarrow \infty }\left\Vert \left( \int_{\mathcal{X}}^{\oplus
}A_{n}\left( x\right) \,\mu \left( \mathrm{d}x\right) \right) \varphi
-\left( \int_{\mathcal{X}}^{\oplus }A\left( x\right) \,\mu \left( \mathrm{d}%
x\right) \right) \varphi \right\Vert _{L^{2}\left( \mathcal{X},\mathcal{Y}%
,\mu \right) }=0\ .
\end{equation*}
\end{proof}

Before proving a version of Fubini's theorem for constant fiber direct
integrals, we fix some terminology concerning the Riemann integral: A
partition of the interval $[a,b]$ is a finite set $P=\{t_{0}<t_{1}<\cdots
<t_{k}\}$ where $t_{0}=a$ and $t_{k}=b$. The norm of the partition $P$ is
the number $|P|=\max_{1\leq i\leq k}(t_{i}-t_{i-1})$. A tagged partition is
a pair $P^{\ast }=(P,\xi )$ where $P$ is a partition and $\xi =(\xi
_{1},\dotsc ,\xi _{k})$ is such that $t_{i-1}\leq \xi _{i}<t_{i}$ for every $%
i=1,\dotsc ,k$. If $P^{\ast }$ is a tagged partition of $[a,b]$, the
corresponding Riemann sum for $f:[a,b]\rightarrow \mathcal{Z}$, with $%
\mathcal{Z}$ being a vector space, is 
\begin{equation*}
\Sigma (f;P^{\ast })=\displaystyle{\sum_{i=1}^{k}}\,(t_{i}-t_{i-1})f(\xi
_{i})\in \mathcal{Z}\ .
\end{equation*}

\begin{proposition}[Fubini's Theorem for direct integrals]
\label{fubini's theorem}\mbox{ }\newline
Let $A_{(\cdot )}:[a,b]\rightarrow L^{\infty }(\mathcal{X},\mathcal{Y},\mu )$
be a continuous function. Then:

\begin{enumerate}
\item[i.)] The mapping 
\begin{equation*}
\mathcal{X}\ni x\mapsto {\int_{a}^{b}}A_{t}\left( x\right) \,\mathrm{d}t\in 
\mathcal{B}(\mathcal{Y})
\end{equation*}
is an element of $L^{\infty }(\mathcal{X},\mathcal{Y},\mu )$;

\item[ii.)] Then the mapping 
\begin{equation*}
\lbrack a,b]\ni t\mapsto {\int_{\mathcal{X}}^{\oplus }}A_{t}\left( x\right)
\,\mu \left( \mathrm{d}x\right) \in \mathcal{B}\left( \int_{\mathcal{X}%
}^{\oplus }\mathcal{Y}\,\mu \left( \mathrm{d}x\right) \right)
\end{equation*}%
is continuous and 
\begin{equation*}
{\int_{a}^{b}\int_{\mathcal{X}}^{\oplus }}A_{t}\left( x\right) \,\mu \left( 
\mathrm{d}x\right) \mathrm{d}t={\int_{\mathcal{X}}^{\oplus }\int_{a}^{b}}%
A_{t}(x)\,\mathrm{d}t\,\mu \left( \mathrm{d}x\right) \text{ }.
\end{equation*}
\end{enumerate}
\end{proposition}

\begin{proof}
If $A_{(\cdot )}:[a,b]\rightarrow L^{\infty }(\mathcal{X},\mathcal{Y},\mu )$
is continuous, then so is $A_{(\cdot )}(x):[a,b]\rightarrow \mathcal{B}(%
\mathcal{Y})$ for $x\in \mathcal{X}$ $\mu $-a.e. For simplicity we may
assume that $A_{(\cdot )}(x)$ is even continuous for \emph{all} $x\in 
\mathcal{X}$. If fact, as this is true for $x\in \mathcal{X}$ $\mu $-a.e.,
for some Borel set $\mathcal{X}_{0}\subseteq \mathcal{X}$ with $\mu (%
\mathcal{X}_{0})=0$, $\mathbf{1}\left[ x\notin \mathcal{X}_{0}\right]
A_{(\cdot )}(x)$ is continuous for all $x\in \mathcal{X}$. Note that, for
all $t\in \lbrack a,b]$ the functions $A_{t}$ and $\mathbf{1}\left[ (\cdot
)\notin \mathcal{X}_{0}\right] A_{t}$ are strongly mensurable and represent
the same element (i.e., equivalence class of strongly mensurable functions $%
\mathcal{X}\rightarrow \mathcal{B}(\mathcal{Y})$) of $L^{\infty }(\mathcal{X}%
,\mathcal{Y},\mu )$. Moreover, as $[a,b]$ is compact, $\{A_{t}\}_{t\in
\lbrack a,b]}$ is bounded as a subset of the metric space $\left( L^{\infty
}(\mathcal{X},\mathcal{Y},\mu ),\Vert \cdot \Vert _{\infty }\right) $. Thus, 
\begin{eqnarray*}
\left\Vert \int_{a}^{b}A_{t}(x)\,\mathrm{d}t\right\Vert _{\mathrm{op}} &\leq
&\int_{a}^{b}\Vert A_{t}(x)\Vert _{\mathrm{op}}\,\mathrm{d}t\leq
\int_{a}^{b}\Vert A_{t}\Vert _{\infty }\,\mathrm{d}t \\
&\leq &(b-a)\sup_{t\in \lbrack a,b]}\Vert A_{t}\Vert _{\infty }<\infty \text{%
\quad (}\mu \text{-a.e.)}\ .
\end{eqnarray*}%
Let $P_{n}^{\ast }$ be a tagged partition whose norm of the corresponding
partition $P_{n}$ goes to zero as $n\rightarrow \infty $. Then for every $%
\varphi ,\psi \in \mathcal{Y}$ and $x\in \mathcal{X}$ $\mu $-a.e, 
\begin{equation*}
\left\langle \varphi ,\left( {\int_{a}^{b}}A_{t}(x)\,\mathrm{d}t\right) \psi
\right\rangle _{\mathcal{Y}}={\int_{a}^{b}}\langle \varphi ,A_{t}(x)\psi
\rangle _{\mathcal{Y}}\,\mathrm{d}t={\lim_{n\rightarrow \infty }}\,\Sigma
\left( \langle \psi ,A_{(\cdot )}(x)\varphi \rangle _{\mathcal{Y}%
};P_{n}^{\ast }\right) \text{ }.
\end{equation*}%
For the first equality we used the fact that the Riemann integral commutes
with bounded linear transformations. Observing that for $x\in \mathcal{X}$ $%
\mu $-a.e. the right-hand side is a pointwise limit of a linear combination
of continuous (hence Riemann integrable) functions, this proves assertion
(i).

Note that the mapping defined in Assertion (ii) is a composition of two
continuous functions, namely: $A_{(\cdot )}:[a,b]\rightarrow L^{\infty }(%
\mathcal{X},\mathcal{Y},\mu )$ and 
\begin{equation*}
L^{\infty }(\mathcal{X},\mathcal{Y},\mu )\ni B\mapsto \int_{\mathcal{X}%
}^{\oplus }B(x)\,\mu \left( \mathrm{d}x\right) \in \mathcal{B}\left( \int_{%
\mathcal{X}}^{\oplus }\mathcal{Y}\,\mu \left( \mathrm{d}x\right) \right) .
\end{equation*}%
Given any $\varphi ,\psi \in L^{2}(\mathcal{X},\mathcal{Y},\mu )$, observe
that the function 
\begin{equation*}
\mathcal{X}\times \lbrack a,b]\ni (x,t)\mapsto f\left( x,t\right) =\langle
\varphi (x),A_{t}(x)\psi (x)\rangle _{\mathcal{Y}}\in \mathbb{C}
\end{equation*}%
is measurable when $A_{(\cdot )}(x)\in C([a,b],\mathcal{B}(\mathcal{Y}))$
for any $x\in \mathcal{X}$.

To prove this, for each $n\in \mathbb{N}$, define the function $f_{n}:%
\mathcal{X}\times \lbrack a,b]\rightarrow \mathbb{C}$ by $%
f_{n}(x,t)=f(x,s_{t})$, with $s_{t}=\min \{m_{t,n}/n,b\}$ and $m_{t,n}\in 
\mathbb{Z}$ being such that $(m_{t,n}-1)/n\leq t<m_{t,n}/n$. In particular, 
\begin{equation*}
f_{n}(x,t)={\sum_{m\in \mathbb{Z}:m\geq na}}\mathbf{1}\left[ t\in
n^{-1}[m-1,m)\right] f(x,\min \{m/n,b\})\ .
\end{equation*}%
Characteristic functions are measurable on $[a,b]$ and $f(\cdot ,t)$ is also
measurable on $\mathcal{X}$ for every $t\in \lbrack a,b]$. So, $f_{n}$ is
measurable for all $n\in \mathbb{N}$. It is easy to check that $%
m_{t,n}/n\rightarrow t$ for any $t\in \lbrack a,b]$, which in turn implies
that $f_{n}$ pointwise converges to $f$, as $n\rightarrow \infty $, because
of the continuity of $f(x,\cdot )$ for any fixed $x\in \mathcal{X}$. The
last continuity property is a direct consequence of the assumption $%
A_{(\cdot )}(x)\in C([a,b],\mathcal{B}(\mathcal{Y}))$ together with
elementary estimates using the Cauchy-Schwarz inequality.

As a consequence, $f$ is measurable on $\mathcal{X}\times \lbrack a,b]$.
Note also that%
\begin{equation*}
{\int_{\mathcal{X}}\int_{a}^{b}}|f(x,t)|\,\mathrm{d}t\,\mu (\mathrm{d}x)\leq
(b-a)||\varphi ||_{L^{2}(\mathcal{X},\mathcal{Y},\mu )}^{2}||\psi ||_{L^{2}(%
\mathcal{X},\mathcal{Y},\mu )}^{2}\sup_{t\in \lbrack a,b]}\Vert A_{t}\Vert
_{\infty }<\infty \text{ },
\end{equation*}%
thanks to the Cauchy-Schwarz inequality for both spaces $\mathcal{Y}$ and $%
L^{2}(\mathcal{X},\mathcal{Y},\mu )$. We can then apply (usual) Fubini's
theorem to obtain%
\begin{multline*}
\left\langle \varphi ,\left( {\int_{\mathcal{X}}^{\oplus }\int_{a}^{b}}%
A_{t}(x)\,\mathrm{d}t\,\mu (\mathrm{d}x)\right) \psi \right\rangle _{L^{2}(%
\mathcal{X},\mathcal{Y},\mu )}={\int_{\mathcal{X}}}\left\langle \varphi
(x),\left( {\int_{a}^{b}}A_{t}(y)\,\mathrm{d}t\right) \psi (x)\right\rangle
_{\mathcal{Y}}\,\mu (\mathrm{d}x)= \\[0.5em]
={\int_{\mathcal{X}}\int_{a}^{b}}\langle \varphi (x),A_{t}(x)\psi (x)\rangle
_{\mathcal{Y}}\,\mathrm{d}t\,\mu (\mathrm{d}x)={\int_{a}^{b}\int_{\mathcal{X}%
}}\langle \varphi (x),A_{t}(x)\psi (x)\rangle _{\mathcal{Y}}\,\mu (\mathrm{d}%
x)\,\mathrm{d}t= \\[0.5em]
={\int_{a}^{b}}\left\langle \varphi ,\left( {\int_{\mathcal{X}}^{\oplus }}%
A_{t}(x)\,\mu (\mathrm{d}x)\right) \psi \right\rangle _{\mathcal{Y}}\,%
\mathrm{d}t=\left\langle \varphi ,\left( {\int_{a}^{b}\int_{\mathcal{X}%
}^{\oplus }}A_{t}(x)\,\mu (\mathrm{d}x)\,\mathrm{d}t\right) \psi
\right\rangle _{L^{2}(\mathcal{X},\mathcal{Y},\mu )}\ .
\end{multline*}%
As $\varphi ,\psi $ are arbitrary, we arrive at Assertion (ii).
\end{proof}

We conclude this short account on constant fiber direct integrals by
providing a representation of them as tensor products:

\begin{proposition}[Direct integrals and tensor products]
\label{direct integral as a tensor product}\mbox{ }\newline
There is a unique unitary transformation $\mathbf{V}:L^{2}(\mathcal{X},\mu
)\otimes \mathcal{Y}\rightarrow L^{2}(\mathcal{X},\mathcal{Y},\mu )$ such
that 
\begin{equation*}
\mathbf{V}\left( f\otimes \varphi \right) \left( x\right) =f\left( x\right)
\varphi \ ,\qquad f\in L^{2}\left( \mathcal{X},\mu \right) ,\ \varphi \in 
\mathcal{Y},\ x\in \mathcal{X}\text{\quad (}\mu \text{-a.e.)}\ .
\end{equation*}
\end{proposition}

\begin{proof}
See \cite[Proposition 5.2]{Niesen-direct-integrals}. 
\end{proof}

\subsection{The Birman-Schwinger Principle\label{birman-schwinger principle}}

There are various versions of the Birman-Schwinger principle in the
literature and we give below the precise version that is used in our proofs.
To this end, we first define \emph{Birman-Schwinger\ operators}: For any
operator $T$ acting on some complex vector space, recall that $\rho
(T)\subseteq \mathbb{C}$ denotes its resolvent set. Given two (bounded)
operators $T,V$ acting on some complex vector space and any $\lambda \in
\rho (T)$, we define the associated Birman-Schwinger\ operator to be%
\begin{equation}
\mathrm{B}\left( \lambda \right) \equiv \mathrm{B}\left( \lambda ,T,V\right)
\doteq V\left( T-\lambda \mathfrak{1}\right) ^{-1}V\text{ }.
\label{Birman-Schwinger operator}
\end{equation}%
It turns out that, for all $\lambda \in \rho (T)$, $1$ is an eigenvalue of $%
\mathrm{B}\left( \lambda \right) $ iff $\lambda $ is an eigenvalue of $%
T-V^{2}$:

\begin{lemma}[The eigenvalues of Birman-Schwinger operators]
\label{Lemma dfddfdfdf}\mbox{ }\newline
Let $T,V$ be two bounded operators acting on a vector space $\mathcal{X}$
over $\mathbb{C}$. Assume that $\lambda \in \rho (T)$ is an eigenvalue of $%
T-V^{2}$ and let $\{\varphi _{i}\}_{i\in I}$ denote any basis of the
corresponding eigenspace. Define $\gamma _{i}\doteq V\varphi _{i}$, $i\in I$%
. Then, 
\begin{equation}
\varphi _{i}=\left( T-\lambda \mathfrak{1}\right) ^{-1}V^{2}\varphi
_{i}=\left( T-\lambda \mathfrak{1}\right) ^{-1}V\gamma _{i}\text{ },\qquad
i\in I\text{ },  \label{dfddfdfdf}
\end{equation}%
and $\{\gamma _{i}\}_{i\in I}$ is a linearly independent set satisfying%
\begin{equation}
\mathrm{B}\left( \lambda \right) \gamma _{i}=V\varphi _{i}=\gamma _{i}\text{ 
},\qquad i\in I\text{ }.  \label{dfddfdfdf2}
\end{equation}
\end{lemma}

\begin{proof}
Suppose that $\lambda $ is an eigenvalue of $T-V^{2}$ and let $\{\varphi
_{i}\}_{i\in I}$ be a basis of the corresponding eigenspace. Set $\gamma
_{i}=V\varphi _{i}$, $i\in I$. Then%
\begin{equation*}
\left( T-\lambda \mathfrak{1}\right) \varphi _{i}=\left( T-V^{2}\right)
\varphi _{i}+\left( V^{2}-\lambda \mathfrak{1}\right) \varphi _{i}=\lambda
\varphi _{i}+\left( V^{2}-\lambda \mathfrak{1}\right) \varphi
_{i}=V^{2}\varphi _{i}
\end{equation*}%
so that (\ref{dfddfdfdf}) holds true. By (\ref{dfddfdfdf}), $\gamma _{i}$ is
a non-zero vector for any $i\in I$ since $\varphi _{i}\neq 0$ for all $i\in
I $. As linear transformations map a linearly dependent set onto a linearly
dependent set, we conclude that $\{\gamma _{i}\}_{i\in I}$ is a linearly
independent set. Equation (\ref{dfddfdfdf2}) is a direct consequence of (\ref%
{Birman-Schwinger operator}) and (\ref{dfddfdfdf}).
\end{proof}

This last lemma is explicitly used in the proof of Corollary \ref{eigenspace
of a fiber} and allows meanwhile to prove the Birman-Schwinger principle for
eigenvalues. Below, for any operator $T$, we use the notation $\mathcal{E}%
_{T}(\lambda )$ for the eigenspace associated with the eigenvalue $\lambda $
of $T$.

\begin{theorem}[Birman-Schwinger]
\label{birman-schwinger's theorem}\mbox{ }\newline
Let $T,V$ be two linear operators acting on a vector space $\mathcal{X}$
over $\mathbb{C}$ and $\lambda \in \rho (T)$. Then $\lambda $ is an
eigenvalue of $T-V^{2}$ iff $1$ is an eigenvalue of $\mathrm{B}(\lambda
)\equiv \mathrm{B}(\lambda ,T,V)$. In this case,%
\begin{equation*}
\dim \mathcal{E}_{T-V^{2}}\left( \lambda \right) =\dim \mathcal{E}_{\mathrm{B%
}\left( \lambda \right) }\left( 1\right) \text{ },
\end{equation*}%
i.e., the corresponding (geometric) multiplicities of eigenvalues are equal
to each other.
\end{theorem}

\begin{proof}
If $\lambda $ is an eigenvalue of $T-V^{2}$ then Lemma \ref{Lemma dfddfdfdf}
implies that $1$ is an eigenvalue of $\mathrm{B}(\lambda )$ and the
eigenspace of $\mathrm{B}(\lambda )$ corresponding to the eigenvalue $1$ has
at least dimension $|I|=\dim \mathcal{E}_{T-V^{2}}(\lambda )$. Conversely,
if $\{\phi _{j}\}_{j\in J}$ is a basis of the eigenspace of $\mathrm{B}%
(\lambda )$ corresponding to the eigenvalue $1$ then we set 
\begin{equation*}
\psi _{j}\doteq \left( T-\lambda \mathfrak{1}\right) ^{-1}V\phi _{j}\text{ }%
,\qquad j\in J\text{ }.
\end{equation*}%
Then, by (\ref{Birman-Schwinger operator}),%
\begin{equation*}
\phi _{j}=\mathrm{B}\left( \lambda \right) \phi _{j}=V\psi _{j}\text{ }%
,\qquad j\in J\text{ },
\end{equation*}%
which implies that $\{\psi _{j}\}_{j\in J}$ is a linearly independent set.
Thus,%
\begin{equation*}
\left( T-V^{2}\right) \psi _{j}=\left( T-\lambda \mathfrak{1}\right) \psi
_{j}+\left( \lambda \mathfrak{1}-V^{2}\right) \psi _{j}=V\phi _{j}+\left(
\lambda \mathfrak{1}-V^{2}\right) \psi _{j}=\lambda \psi _{j}\text{ },\qquad
j\in J\text{ },
\end{equation*}%
and, hence, the eigenspace of $T-V^{2}$ corresponding to the eigenvalue $%
\lambda $ has at least dimension $|J|=\dim \mathcal{E}_{\mathrm{B}(\lambda
)}(1)$.
\end{proof}

\subsection{Combes-Thomas Estimates\label{Combes-Thomas estimates}}

We give here a version of the celebrated Combes-Thomas estimates, first
proven in 1973 \cite{CT73}, which is well-adapted to our framework. For the
non-expert reader, we provide also its proof, which is relatively short and
easy to understand in the particular situation we are interested in.

Fix a countable set $\Lambda $ and a pseudometric $d:\Lambda \times \Lambda
\rightarrow \mathbb{R}_{0}^{+}$ on $\Lambda $. Let $\ell ^{2}(\Lambda )$ be
the (separable) Hilbert space of square summable functions $\Lambda
\rightarrow \mathbb{C}$. Similar to (\ref{e frac}), its canonical
orthonormal basis is defined by%
\begin{equation*}
\mathfrak{e}_{x}\left( y\right) \doteq \delta _{x,y}\ ,\qquad x,y\in \Lambda
\ ,
\end{equation*}%
where $\delta _{\mathfrak{i},\mathfrak{j}}$ is the Kronecker delta. For
simplicity, as before, we use the shorter notation $\langle \cdot ,\cdot
\rangle \equiv \langle \cdot ,\cdot \rangle _{\ell ^{2}(\Lambda )}$ for its
scalar product.

For each bounded operator $T\in \mathcal{B}(\ell ^{2}(\Lambda ))$ and
positive parameter $\mu \in \mathbb{R}_{0}^{+}$, we define the quantity%
\begin{equation}
\mathbf{S}\left( T,\mu \right) \doteq \sup_{x\in \Lambda }\sum_{y\in \Lambda
}\left( \mathrm{e}^{\mu d\left( x,y\right) }-1\right) \left\vert
\left\langle \mathfrak{e}_{x},T\mathfrak{e}_{y}\right\rangle \right\vert \in %
\left[ 0,\infty \right] \ .  \label{def combes1}
\end{equation}%
Compare with (\ref{def combes100}). By definition of a pseudometric, the
function $d$ is symmetric with respect to the variables $x$ and $y$. The
same occurs with the factor $|\langle \mathfrak{e}_{x},T\mathfrak{e}%
_{y}\rangle |$, provided that $T$ is self-adjoint. Thus, in this particular
case, $\mathbf{S}(T,\mu )$ is equal to%
\begin{equation}
\mathbf{S}\left( T,\mu \right) =\sup_{y\in \Lambda }\sum_{x\in \Lambda
}\left( \mathrm{e}^{\mu d\left( x,y\right) }-1\right) \left\vert
\left\langle \mathfrak{e}_{x},T\mathfrak{e}_{y}\right\rangle \right\vert \in %
\left[ 0,\infty \right] \ .  \label{def combes2}
\end{equation}

The lemma below provides an estimate of the operator norm of $T$ in terms of
quantities that are similar to (\ref{def combes1}) and (\ref{def combes2}):

\begin{lemma}
\label{Riesz-Thorin theorem}\mbox{ }\newline
For any bounded operator $T\in \mathcal{B}(\ell ^{2}(\Lambda ))$, 
\begin{equation*}
\left\Vert T\right\Vert _{{\mathrm{o}\mathrm{p}}}^{2}\leq \left( \sup_{y\in
\Lambda }\sum_{x\in \Lambda }\left\vert \left\langle \mathfrak{e}_{x},T%
\mathfrak{e}_{y}\right\rangle \right\vert \right) \left( \sup_{x\in \Lambda
}\sum_{y\in \Lambda }\left\vert \left\langle \mathfrak{e}_{x},T\mathfrak{e}%
_{y}\right\rangle \right\vert \right) \ .
\end{equation*}
\end{lemma}

\begin{proof}
Assume without loss of generality that the above bound is finite for $T\in 
\mathcal{B}(\ell ^{2}(\Lambda ))$. Otherwise the assertion would be trivial.
Let $V:\ell ^{1}(\Lambda )+\ell ^{\infty }(\Lambda )\rightarrow \ell
^{1}(\Lambda )+\ell ^{\infty }(\Lambda )$ be the mapping defined by 
\begin{equation*}
Vf\left( x\right) \doteq \sum_{y\in \Lambda }f\left( y\right) \left\langle 
\mathfrak{e}_{x},T\mathfrak{e}_{y}\right\rangle \ ,\qquad x\in \Lambda ,\
f\in \ell ^{1}\left( \Lambda \right) +\ell ^{\infty }\left( \Lambda \right)
\ .
\end{equation*}%
If $f\in \ell ^{\infty }(\Lambda )$ then $Vf\in \ell ^{\infty }(\Lambda )$
because%
\begin{equation*}
\sup_{x\in \Lambda }\sum_{y\in \Lambda }\left\vert f\left( y\right)
\right\vert \left\vert \left\langle \mathfrak{e}_{x},T\mathfrak{e}%
_{y}\right\rangle \right\vert \leq \left\Vert f\right\Vert _{\infty
}\sup_{x\in \Lambda }\sum_{y\in \Lambda }\left\vert \left\langle \mathfrak{e}%
_{x},T\mathfrak{e}_{y}\right\rangle \right\vert <\infty \ ,
\end{equation*}%
while, for any $f\in \ell ^{1}(\Lambda )$, we also have $Vf\in \ell
^{1}(\Lambda )$ because 
\begin{equation*}
\left\Vert Vf\right\Vert _{\ell ^{1}(\Lambda )}\sum_{x\in \Lambda
}\left\vert Vf\left( x\right) \right\vert \leq \sum_{x,y\in \Lambda
}\left\vert f\left( y\right) \right\vert \left\vert \left\langle \mathfrak{e}%
_{x},T\mathfrak{e}_{y}\right\rangle \right\vert \leq \left\Vert f\right\Vert
_{\ell ^{1}(\Lambda )}\sup_{y\in \Lambda }\sum_{x\in \Lambda }\left\vert
\left\langle \mathfrak{e}_{x},T\mathfrak{e}_{y}\right\rangle \right\vert
<\infty \ ,
\end{equation*}%
using Tonelli's theorem. It then follows from the Riesz-Thorin theorem \cite[%
Theorem 6.27]{Folland2} that, for any function $f\in \ell ^{2}(\Lambda
)\subseteq \ell ^{1}(\Lambda )+\ell ^{\infty }(\Lambda )$, 
\begin{equation*}
\left\Vert Vf\right\Vert _{\ell ^{2}(\Lambda )}^{2}\leq \left\Vert
f\right\Vert _{\ell ^{2}(\Lambda )}^{2}\left( \sup_{y\in \Lambda }\sum_{x\in
\Lambda }\left\vert \left\langle \mathfrak{e}_{x},T\mathfrak{e}%
_{y}\right\rangle \right\vert \right) \left( \sup_{x\in \Lambda }\sum_{y\in
\Lambda }\left\vert \left\langle \mathfrak{e}_{x},T\mathfrak{e}%
_{y}\right\rangle \right\vert \right) \ .
\end{equation*}%
Finally, we observe that 
\begin{equation*}
\left( Tf\right) \left( x\right) =\left\langle \mathfrak{e}%
_{x},Tf\right\rangle =\sum_{y\in \Lambda }f\left( y\right) \left\langle 
\mathfrak{e}_{x},T\mathfrak{e}_{y}\right\rangle =\left( Vf\right) \left(
x\right)
\end{equation*}%
whenever $x\in \Lambda $ and $f\in \ell ^{2}(\Lambda )$.
\end{proof}

We now state another, well-known, technical lemma, which is given here for
completeness. Recall that, here, $\rho (T)\subseteq \mathbb{C}$ and $\sigma
(T)\doteq \mathbb{C}\backslash \rho (T)$ respectively denote the resolvent
set and the spectrum of any element $T$ in some unital $C^{\ast }$-algebra
(like the space of bounded operators on some Hilbert space). Similar to (\ref%
{def combes10}), we use the notation 
\begin{equation}
\Delta \left( \lambda ;T\right) \doteq \min \left\{ \left\vert \lambda
-a\right\vert :a\in \sigma (T)\right\}  \label{def combes10bis}
\end{equation}%
for the distance between a complex number $\lambda \in \mathbb{C}$ and the
spectrum $\sigma (T)$ of any element $T$ in some unital $C^{\ast }$-algebra.

\begin{lemma}[Norm estimates of resolvents]
\label{lemma for proving thomas-combes theorem}\mbox{ }\newline
Let $\mathcal{X}$ be an unital $C^{\ast }$-algebra with norm $\left\Vert
\cdot \right\Vert $. Take $T,B\in \mathcal{X}$ with $T$ being self-adjoint
and let $\lambda \in \rho (T)$. If $\Vert B\Vert <\Delta (\lambda ;T)$ then $%
\lambda \in \rho (T+B)$ and%
\begin{equation*}
\left\Vert \left( T+B-\lambda \mathfrak{1}\right) ^{-1}\right\Vert \leq {%
\frac{1}{\Delta \left( \lambda ;T\right) -\left\Vert B\right\Vert }}\ .
\end{equation*}
\end{lemma}

\begin{proof}
Assume all conditions of the lemma, in particular that $\Vert B\Vert <\Delta
(\lambda ;T)$. Then, 
\begin{equation*}
\left\Vert \left( T-\lambda \mathfrak{1}\right) ^{-1}B\right\Vert \leq
\Delta \left( \lambda ;T\right) ^{-1}\left\Vert B\right\Vert <1
\end{equation*}%
and using the Neumann series \cite[Lemma 4.24]{Bru-Pedra-livre} for $%
-(T-\lambda \mathfrak{1})^{-1}B$, the element $\mathfrak{1}+(T-\lambda 
\mathfrak{1})^{-1}B$ is invertible with norm bounded by%
\begin{eqnarray*}
\left\Vert \left( \mathfrak{1}+\left( T-\lambda \mathfrak{1}\right)
^{-1}B\right) ^{-1}\right\Vert &\leq &\sum_{n=0}^{\infty }\left\Vert \left(
T-\lambda \mathfrak{1}\right) ^{-1}B\right\Vert ^{n}=\frac{1}{1-\left\Vert
\left( T-\lambda \mathfrak{1}\right) ^{-1}B\right\Vert } \\
&\leq &\frac{1}{1-\Delta \left( \lambda ;T\right) ^{-1}\left\Vert
B\right\Vert }=\frac{\Delta \left( \lambda ;T\right) }{\Delta \left( \lambda
;T\right) -\left\Vert B\right\Vert }\ .
\end{eqnarray*}%
Finally, one uses the equality%
\begin{equation*}
T+B-\lambda \mathfrak{1}=\left( T-\lambda \mathfrak{1}\right) \left( 
\mathfrak{1}+\left( T-\lambda \mathfrak{1}\right) ^{-1}B\right)
\end{equation*}%
for any $\lambda \in \rho (T)$ to deduce that $\lambda \in \rho (T+B)$ and 
\begin{equation*}
\left\Vert \left( T+B-\lambda \mathfrak{1}\right) ^{-1}\right\Vert
=\left\Vert \left( \mathfrak{1}+\left( T-\lambda \mathfrak{1}\right)
^{-1}B\right) ^{-1}\left( T-\lambda \mathfrak{1}\right) ^{-1}\right\Vert
\leq \frac{1}{\Delta \left( \lambda ;T\right) -\left\Vert B\right\Vert }\ .
\end{equation*}
\end{proof}

We can now prove the following version of Combes-Thomas estimates:

\begin{theorem}[Combes-Thomas estimates]
\label{combes-thomas theorem}\mbox{ }\newline
Let $T\in \mathcal{B}(\ell ^{2}(\Lambda ))$ be a self-adjoint operator.
Given $\mu \in \mathbb{R}_{0}^{+}$ and $\lambda \in \mathbb{C}$ with $\Delta
(\lambda ;T)>\mathbf{S}(T,\mu )$, the following inequality holds true:%
\begin{equation*}
\left\vert \left\langle \mathfrak{e}_{x},\left( T-\lambda \mathfrak{1}%
\right) ^{-1}\mathfrak{e}_{y}\right\rangle \right\vert \leq \frac{\mathrm{e}%
^{-\mu d\left( x,y\right) }}{\Delta \left( \lambda ;T\right) -\mathbf{S}%
\left( T,\mu \right) }\ ,\qquad x,y\in \Lambda \ .
\end{equation*}
\end{theorem}

\begin{proof}
Fix $y\in \Lambda $ and $R\in \mathbb{R}^{+}$. Define the function $\varphi
:\Lambda \rightarrow \lbrack 1,\mathrm{e}^{\mu R}]$ by%
\begin{equation*}
\varphi \left( x\right) \doteq \exp \left( \mu \min \left\{ d\left(
x,y\right) ,R\right\} \right) \ ,\qquad x\in \Lambda \ .
\end{equation*}%
Clearly, $\varphi $ and $1/\varphi $ are bounded and the inverse of the
multiplication operator $M_{\varphi }\in \mathcal{B}(\ell ^{2}(\Lambda ))$
by $\varphi $ is nothing else than $M_{1/\varphi }$. Because $\varphi $ is a
real-valued function, $M_{\varphi }^{\ast }=M_{\varphi }$ and, for any $x\in
\Lambda $, $\mathfrak{e}_{x}$ is of course an eigenvector of $M_{\varphi }$
with associated eigenvalue $\varphi (x)$. In particular, 
\begin{equation*}
\left\langle \mathfrak{e}_{x},M_{\varphi }TM_{\varphi }^{-1}\mathfrak{e}%
_{z}\right\rangle =\frac{\varphi \left( x\right) }{\varphi \left( z\right) }%
\left\langle \mathfrak{e}_{x},T\mathfrak{e}_{z}\right\rangle \ ,\qquad
x,z\in \Lambda \ .
\end{equation*}%
Since $(x,z)\mapsto \min \left\{ d\left( x,z\right) ,R\right\} $ is another
pseudometric on $\Lambda $, for all $x,z\in \Lambda $, we have that%
\begin{equation*}
\min \left\{ d\left( x,y\right) ,R\right\} -\min \left\{ d\left( z,y\right)
,R\right\} \leq \min \left\{ d\left( x,z\right) ,R\right\} \leq d\left(
x,z\right) .
\end{equation*}%
In particular, the operator $B\doteq M_{\varphi }TM_{\varphi }^{-1}-T$
satisfies the bound%
\begin{equation*}
\left\vert \left\langle \mathfrak{e}_{x},B\mathfrak{e}_{z}\right\rangle
\right\vert \leq \left( \mathrm{e}^{\mu d\left( x,z\right) }-1\right)
\left\vert \left\langle \mathfrak{e}_{x},T\mathfrak{e}_{z}\right\rangle
\right\vert \ ,\qquad x,z\in \Lambda \ .
\end{equation*}%
By Lemma \ref{Riesz-Thorin theorem} together with Equations (\ref{def
combes1}) and (\ref{def combes2}), it follows that 
\begin{equation*}
\left\Vert B\right\Vert _{{\mathrm{o}\mathrm{p}}}\leq \mathbf{S}\left( T,\mu
\right) <\Delta \left( \lambda ;T\right) \ .
\end{equation*}%
Applying now Lemma \ref{lemma for proving thomas-combes theorem}, we then
arrive at the bound%
\begin{equation*}
\left\Vert \left( M_{\varphi }TM_{\varphi }^{-1}-\lambda \mathfrak{1}\right)
^{-1}\right\Vert _{{\mathrm{o}\mathrm{p}}}\leq \frac{1}{\Delta \left(
\lambda ;T\right) -\left\Vert B\right\Vert _{{\mathrm{o}\mathrm{p}}}}\leq 
\frac{1}{\Delta (\lambda ;T)-\mathbf{S}(T,\mu )}\ .
\end{equation*}%
Finally, for any $x\in \Lambda $ such that $d(x,y)\leq R$ we observe from
the last upper bound that%
\begin{multline*}
\mathrm{e}^{\mu d\left( x,y\right) }\left\vert \left\langle \mathfrak{e}%
_{x},\left( T-\lambda \mathfrak{1}\right) ^{-1}\mathfrak{e}_{z}\right\rangle
\right\vert =\frac{\varphi \left( x\right) }{\varphi \left( 1\right) }%
\left\vert \left\langle \mathfrak{e}_{x},\left( T-\lambda \mathfrak{1}%
\right) ^{-1}\mathfrak{e}_{z}\right\rangle \right\vert =\left\vert
\left\langle \mathfrak{e}_{x},\left( M_{\varphi }TM_{\varphi }^{-1}-\lambda 
\mathfrak{1}\right) ^{-1}\mathfrak{e}_{z}\right\rangle \right\vert \\
\leq \frac{1}{\Delta (\lambda ;T)-\mathbf{S}(T,\mu )}\ .
\end{multline*}%
Since $y\in \Lambda $ and $R\in \mathbb{R}^{+}$ are arbitrary parameters,
the above inequality in fact holds true for all $x,y\in \Lambda $.
\end{proof}

\subsection{Elementary Observations}

For completeness and the reader's convenience, we conclude the appendix by
given a few elementary results related to the space of bounded operators on
a Hilbert space.

\begin{proposition}[Monotonicity of the inverse on operators]
\label{inversion is monotone decreasing}\mbox{ }\newline
Let $B$ and $C$ be two positive bounded operators on a Hilbert space with
bounded (positive) inverse. If $B\leq C$ then $C^{-1}\leq B^{-1}$.
\end{proposition}

\begin{proof}
The proof is standard. Since it is very short, we give it for completeness.
If $B$ and $C$ commute, then%
\begin{equation*}
B^{-1}-C^{-1}=B^{-1}\left( C-B\right) C^{-1}\geq 0\ ,
\end{equation*}%
because $B^{-1}$, $C^{-1}$ and $C-B$ are positive commuting operators. If $B$
and $C$ do not commute then we observe that $B\leq C$ yields 
\begin{equation*}
C^{-\frac{1}{2}}BC^{-\frac{1}{2}}\leq \mathfrak{1}\text{ },
\end{equation*}%
which in turn implies that 
\begin{equation*}
C^{\frac{1}{2}}B^{-1}C^{\frac{1}{2}}\geq \mathfrak{1}\text{ },
\end{equation*}%
because $\mathfrak{1}$ and $C^{-\frac{1}{2}}BC^{-\frac{1}{2}}$ commute. From
the last inequality it follows that $B^{-1}\geq C^{-1}$.
\end{proof}

\begin{proposition}[Monotone convergence theorem for operators]
\label{prop weak lim monotone}\mbox{ }\newline
Let $\mathcal{X}$ be any complex Hilbert space. Any increasing (decreasing)
monotone net $(A_{i})_{i\in I}$ of self-adjoint elements in $\mathcal{B}(%
\mathcal{X})$ that is bounded from above (below) has a supremum (infimum) in 
$\mathcal{B}(\mathcal{X})$. The supremum (infimum) is itself also
self-adjoint and is the strong operator limit of the net.
\end{proposition}

\begin{proof}
\cite[Proposition 2.17]{Bru-Pedra-livre} already tells us that any
increasing (decreasing) monotone net $(A_{i})_{i\in I}$ of self-adjoint
elements in $\mathcal{B}(\mathcal{X})$ that is bounded from above (below)
has a supremum (infimum) $A_{\infty }$ in $\mathcal{B}(\mathcal{X})$. The
supremum (infimum) $A_{\infty }$ is itself also self-adjoint and is the weak
operator limit of the net. This is proven by using the polarization identity
and the Riesz representation theorem together with elementary estimates. To
conclude the proof, it remains to show that $A_{\infty }\in \mathcal{B}(%
\mathcal{X})$ is the limit of the increasing net\ $(A_{i})_{i\in I}$ also in
the strong operator topology. To this end, assume that $(A_{i})_{i\in I}$ is
an increasing net and define the (decreasing) net $(B_{i})_{i\in I}$ of
positive operators by $B_{i}\doteq A_{\infty }-A_{i}\geq 0$. By
construction, this net converges in the weak operator topology to $0\in $ $%
\mathcal{B}(\mathcal{X})$, which in turn implies that the net $%
(B_{i}^{1/2})_{i\in I}$ converges in the strong operator topology to $0\in $ 
$\mathcal{B}(\mathcal{X})$. As the net $(B_{i}^{1/2})_{i\in I}$ is
norm-bounded and $B_{i}=B_{i}^{1/2}B_{i}^{1/2}$, we then conclude that also $%
(B_{i})_{i\in I}$ converges in the strong operator topology to $0\in $ $%
\mathcal{B}(\mathcal{X})$, that is, $A_{\infty }$ is the strong operator
limit of the net $(A_{i})_{i\in I}$. If $(A_{i})_{i\in I}$ is a decreasing
net then we consider the increasing net $(-A_{i})_{i\in I}$ to conclude that 
$(A_{i})_{i\in I}$ has a infimum $A_{\infty }=A_{\infty }^{\ast }\in 
\mathcal{B}(\mathcal{X})$, which is again the strong operator limit of the
net $(A_{i})_{i\in I}$.
\end{proof}

\noindent \textit{Acknowledgments:} This work is supported by the Basque
Government through the grant IT1615-22 and BERC 2022-2025 program, by the
Ministry of Science and Innovation: PID2020-112948GB-I00 funded by
MCIN/AEI/10.13039/501100011033 and by \textquotedblleft ERDF A way of making
Europe\textquotedblright , as well as by CNPq (303682/2025-6). We would also
like to express our gratitude to the reviewers for their constructive
suggestions, particularly the discussion of enhanced binding in Quantum
Field Theory in the introduction. We are also grateful for their interest in
improving the paper and for their help in correcting a calculation error by
providing an explicit proof of Proposition \ref{direct integral
decomposition of the hamiltonian}.

\end{document}